\documentclass[pdftex,a5paper, 10pt, chapterprefix, twoside, openright]{scrreprt}
\usepackage[includehead, %
includefoot, %
headsep=5mm, %
footskip=10mm, %
top=17.5mm, 
bottom=17.5mm, 
left=12.5mm, 
right=24.5mm]{geometry} 
\usepackage[utf8]{inputenc}
\usepackage[english]{babel}
\usepackage[dvipsnames]{xcolor}
\usepackage{color}
\usepackage{import}
\usepackage{amsmath, amsthm, amssymb}
\usepackage{csquotes}
\usepackage{graphicx}
\usepackage{wrapfig}
\usepackage[format=plain]{caption}
\usepackage{subcaption}
\usepackage[outercaption]{sidecap}
\usepackage{amsmath, amssymb, graphics, setspace}
\usepackage{pdflscape}
\usepackage{siunitx}
\usepackage{url}
\usepackage{stackengine}
\usepackage[toc,page]{appendix}
\usepackage[colorlinks]{hyperref}
\usepackage{hyperref}
\hypersetup{colorlinks = true,    linkcolor = blue,	citecolor = magenta}
\usepackage[numbers]{natbib}
\usepackage{multirow}
\usepackage[export]{adjustbox}
\usepackage{advdate}

\newcommand*{\doi}[1]{\href{https://doi.org/#1}{\color{magenta}{#1}}}

\newcommand\numberthis{\addtocounter{equation}{1}\tag{\theequation}}

\AtBeginDocument{%
  \let\mtcontentsname\contentsname
  \renewcommand\contentsname{\MakeUppercase\mtcontentsname}
}

\begin{document}

\begin{titlepage}
\setlength{\hoffset}{1.3cm}

\includegraphics[height = 2cm, right]{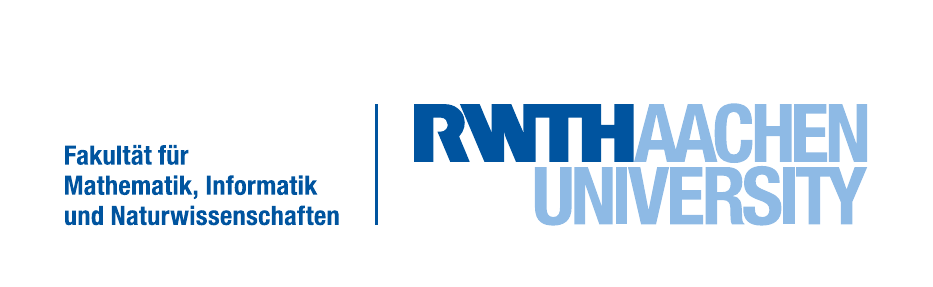}

\vspace*{0.5cm}
 			
\large
\begin{center}
DESIGN 
OF AN INDUCTIVELY SHUNTED TRANSMON QUBIT WITH TUNABLE TRANSVERSE AND LONGITUDINAL COUPLING\\
\end{center}


\vspace{.5cm} 			
 			
\normalsize 

Von der Fakultät für Mathematik, Informatik und Naturwissenschaften der RWTH Aachen
University genehmigte Dissertation zur Erlangung des akademischen Grades einer
Doktorin der Naturwissenschaften

\vspace{.5cm}

\centering

vorgelegt von\\
 			
\vspace{.5cm}
M. Sc. Susanne Richer\\
\vspace{.5cm}
aus Bocholt

\vspace{1cm}

\flushleft

Berichter:\\
\vspace{.5cm}
Prof. Dr. David DiVincenzo\\
Dr. Ioan Pop\\
Prof. Dr. Christoph Stampfer\\

\vspace{.5cm}

Tag der mündlichen Prüfung: 28. Februar 2018\\

\vspace{.5cm}

Diese Dissertation ist auf den Internetseiten der Universitätsbibliothek verfügbar.

\end{titlepage}

\normalsize
\setlength{\hoffset}{0cm}

\cleardoublepage

%

\pagenumbering{roman} 
\setcounter{page}{1} 

\chapter*{ABSTRACT}
\addcontentsline{toc}{chapter}{ABSTRACT} 


\textit{Superconducting qubits} are among the most promising
and versatile building blocks on the road to a functioning  quantum computer. 
One of the main challenges in superconducting qubit architectures is to couple qubits in a well-controlled manner, especially in circuit constructions that involve many qubits. 
In order to avoid unwanted cross-couplings, qubits are oftentimes coupled via harmonic resonators, which act as buses that mediate the interaction. 
\\
This thesis is set in the framework of superconducting transmon-type qubit architectures with special focus on two important types of coupling between qubits and harmonic resonators: \textit{transverse} and \textit{longitudinal coupling}. While transverse coupling naturally appears in transmon-like circuit constructions, longitudinal coupling is much harder to implement and hardly ever the only coupling term present.
Nevertheless, we will see that longitudinal coupling offers some remarkable advantages with respect to scalability and readout. 
\\
This thesis will focus on a design, which combines both these coupling types in a single circuit and provides the possibility to choose between pure transverse and pure longitudinal or have both at the same time.
The ability to choose between transverse and longitudinal coupling in the same circuit provides the flexibility to use one for coupling to the next qubit and one for readout, or vice versa.\\
We will start with an introduction to circuit quantization, where we will explain how to describe and analyze superconducting electrical circuits in a systematic way and discuss which characteristic circuit elements make up qubits and resonators. 
We will then introduce the two types of coupling between qubit and resonator which are provided in our design: transverse and longitudinal coupling.
In order to show that longitudinal coupling has some remarkable advantages with respect to the scalability of a circuit, we will discuss a scalable qubit architecture, which can be implemented with our design.
Translating this discussion from the Hamiltonian level to the language of circuit quantization, we will show how to design circuits with specifically tailored couplings.\\
Having introduced these basic concepts, we will focus on our circuit design that consists of an inductively shunted transmon qubit with tunable coupling to an embedded harmonic mode. Using a symmetric design, static transverse coupling terms are cancelled out, while the parity of the only remaining coupling term can be tuned via an external flux.
The distinctive feature of the tunable design is that the transverse coupling disappears when the longitudinal is maximal and vice versa. \\
Subsequently, we will turn to the implementation of our circuit design, discuss how to choose the parameters, and present an adapted alternative circuit, where coupling strength and anharmonicity scale better than in the original circuit. Furthermore, we show how the anharmonicity and the coupling can be boosted by additional flux-biasing. 
We will see that for conveniently chosen parameters longitudinal and transverse coupling have comparable values, while all other coupling terms can be suppressed.
In addition, we present a proposal for an experimental device that will serve as a prototype for a first experiment. 
\\
Coming back to the scalable architecture mentioned above, we will show how our design can be scaled up to a grid, which can be done in modular fashion with strictly local couplings. In such a grid of fixed-frequency qubits and resonators with a particular pattern of always-on interactions, coupling is strictly confined to nearest and next-nearest neighbor resonators; there is never any direct qubit-qubit coupling. 
\\
We will conclude the thesis discussing different possibilities to do readout with our circuit design, including a short discussion of the coupling between the circuit and the environment, and the influence of dissipation. 



\cleardoublepage

\chapter*{ACKNOWLEDGMENTS}
\addcontentsline{toc}{chapter}{ACKNOWLEDGMENTS}  

After four years and about 1800 cups of coffee (that's a conservative estimate), my thesis is finally done and printed. I would like to thank a few people, without whom this would not have been possible.\\
First, of course, I would like to thank my supervisor Prof. David DiVincenzo, who guided me through this time with seemingly endless patience, which I am sure was not always easy. Thank you for your advice, your inspiration and your always constructive criticism. And, of course, thank you for reading almost every word of this thesis and helping me with both physics and grammar. I feel enormously privileged for the opportunity to write my thesis in such a great institute with such an exceptional advisor.\\
I would like to thank Dr. Ioan Pop and his postdocs Natasha Maleeva and Sebastian Skacel from \textit{Karlsruhe Institute of Technology} for inspiring discussions and the opportunity to see some of my ideas translated from paper to chip. I am very grateful for this collaboration, as it helped me a lot to get an experimentalist's view on my pen-and-paper circuit drawings. I am very much looking forward to see this work advancing in the future.\\
Thanks to all my colleagues at the IQI for all those years of great coffee and companionship. Thank you Helene Barton, for being the good heart of the institute. Thank you Niko, for being the best office mate I could imagine. Thank you Martin, Stefano, Alessandro, Manuel and Dickel (had I listened to all your comments this thesis would be the most poetic thesis ever written), for reading parts of this thesis and helping me with your comments and questions. \\
Thank you Martin (I admit I still haven't explained to you what my work is about), Marcia, Verena, Bettina, Fiona, Martin, Christine, Fabian, for always being there for me. Lastly and most of all, I would like to thank my parents for (almost) thirty years of love and encouragement. I could not have done this without you!

\cleardoublepage

\phantomsection
\addcontentsline{toc}{chapter}{CONTENTS} 
\tableofcontents

\cleardoublepage

\pagenumbering{arabic} 
\setcounter{page}{1}

\chapter{INTRODUCTION}


At the end of the 19th century, many scientists believed that human knowledge of physics was nearly complete. They were convinced that \enquote{almost everything is already discovered, and all that remains is to fill a few unimportant holes}~\cite{Stein2011, Lightman2005} as Max Planck was told by one of his professors in 1878. 
There were, however, a few experimental results that could not be explained with what we now call \textit{classical physics}, as for example the photoelectric effect. 
In other cases, the theory led to absurd results such as the \textit{ultraviolet catastrophe} for black-body radiation, which clearly did not agree with experimental results. \\
In 1900, Max Planck managed to explain black-body radiation by making one crucial assumption, which was done as he later said \enquote{in an act of despair}~\cite{Gosson2017}. 
His assumption, the so-called \textit{Planck's postulate}~\cite{Planck1900}, states that the radiation energy $E$ of a certain frequency $f$ is quantized and can only appear in units of the fundamental constant $h$, that is

\begin{align}
E = h\, f,
\end{align}

where $h = 6.626 \times 10^{-34}$ is correspondingly known as \textit{Planck's constant}. While this was at first considered to be a purely formal assumption, Planck's postulate turned out to revolutionize physics and is now considered to be one of the foundations of \textit{quantum mechanics}.\\
While most, but not all macroscopic phenomena can be accurately described with classical physics, quantum mechanics evolved to be the fundamental theory to describe physics at microscopic scales. In the so-called correspondence limit (or classical limit), that is at energy scales much larger than $h$, we recover the results from classical mechanics.\\
While for many years, quantum mechanics was used to gain a deep understanding of phenomena found in nature, the new field of \textit{quantum computation}~\cite{Nielsen2000} takes up the challenge of \textit{designing} systems which exhibit controllable quantum behavior. In 1982, Richard Feynman first conceived the idea of a computer, which would \enquote{itself be built of quantum mechanical elements which obey quantum mechanical laws}~\cite{Feynman1982}. He states that while a classical computer is incapable of simulating the behavior of a large quantum-mechanical system due to its many degrees of freedom, a \textit{quantum computer} could do such a simulation, given that it exhibits the same quantum behavior as the system it wants to simulate.\\
In classical computers, information is stored in binary units, so-called \textit{bits}, which can have only two possible values, commonly represented as $0$ and $1$.
The basic idea of a quantum computer consists in substituting the classical bit by a quantum bit, that is a \textit{qubit}, which can, in principle, be any system with two possible quantum states, which we will call $| 0 \rangle$ and $| 1 \rangle$. Being a quantum-mechanical system, this means that the qubit can represent any superposition of its two basic states, that is

\begin{align}
| \psi \rangle = \alpha | 0 \rangle + \beta | 1 \rangle.
\end{align} 

Astonishingly, when these quantum systems interact, they may end up in a so-called entangled state, which means that their states cannot be described independently of each other. Even when these entangled systems are far apart, a manipulation of one of them will immediately influence the other.
This phenomenon, which is called \textit{quantum entanglement}, is so counter-intuitive that scientists found it difficult to accept it~\cite{Einstein1935}.\\
Based on the phenomena of quantum entanglement and superposition states, Peter Shor presented a quantum algorithm for integer factorization in 1994 that scaled exponentially better than any classical factorizing algorithm~\cite{Shor1997}, which made clear that a quantum computer could outperform classical computers. Ever since, the interest in quantum computers has grown and their immense potential is becoming more and more evident. \\
For the realization of a quantum computer a high number of qubits is necessary, which need to be individually accessible and controllable~\cite{Divincenzo2000}. To perform quantum computations, interactions need to be enabled at least between neighboring qubits~\cite{Nielsen2000}. 
In order to protect their quantum coherence, the qubits have to be shielded from environmental influences. These requisites constitute a dilemma: we need to be able to access and manipulate the qubits without destroying their coherence.\\ 
For the physical implementation of such a quantum computer, there are many conceivable realizations. 
One of the most promising fields among these are \textit{superconducting qubits}. As opposed to most other realizations, such as for example spin qubits, 
superconducting qubits are macroscopic objects visible to the naked eye, which consist of small electric circuits of superconducting material. Superconductivity itself is a quantum phenomenon, where electrons spontaneously form a condensate of so-called Cooper pairs, when the material is cooled below a certain temperature. One of the crucial advantages of superconducting qubits is the fact that their characteristic properties, such as frequency and anharmonicity, as well as the coupling between them, can be designed and customized for the experiment in question. 
Among superconducting qubits a distinction is drawn between charge~\cite{Bouchiat1998}, flux~\cite{Orlando1999}, and phase qubits. 
The \textit{transmon qubit} was developed from charge qubits~\cite{Koch2007} and soon became very popular due to its reduced sensitivity to charge noise.\\
While qubit coherence times have improved immensely in recent years \cite{Yan2016, Reagor2016, Minev2016}, it remains difficult to couple them in a well-controlled manner, especially in circuit constructions that involve many qubits. The challenge consists in  performing quantum gates on selected qubits without corrupting the others. 
In order to avoid unwanted cross-couplings, superconducting qubits are oftentimes coupled via harmonic resonators, which act as buses that mediate the interaction \cite{Ghosh2013}. This is supposed to make the interaction more controllable.\\
This thesis is set in the framework of coupling structures between superconducting transmon-type qubits and harmonic resonators, focusing on a circuit design that provides two inherently different types of coupling between a qubit and a resonator in the same circuit.
Most commonly, superconducting qubit architectures work with the so-called \textit{transverse coupling}~\cite{Wallraff2004, Blais2004}, which is known from cavity quantum electrodynamics.
While this well-studied coupling type is easy to implement and useful in terms of readout, it is increasingly challenging to control in larger qubit architectures \cite{Riste2015, Paik2016, Versluis2017}, as unwanted cross-couplings degrade the circuit's performance.
In contrast to this stands \textit{longitudinal coupling} \cite{Billangeon2015, Didier2015, Royer2017, Geller2015, Weber2017}, which is harder to implement but exhibits some remarkable advantages in terms of scalability and readout.

\section{OUTLINE}

This thesis will focus on these two inherently different coupling types between qubits and resonators: transverse and longitudinal coupling. The core of the thesis will be the design of an inductively shunted transmon qubit with tunable transverse and longitudinal coupling to an embedded resonator. This design was first published in Ref.~\cite{Richer2016} and further enhanced and adapted in Ref.~\cite{Richer2017}.\\
We will start with an introduction to circuit quantization in Chapt.~\ref{chapt:quantization}, which mostly follows Ref.~\cite{Devoret1997}. We will explain how to describe and analyze superconducting electrical circuits in a systematic way and discuss which characteristic circuit elements make up qubits and resonators. Going from a circuit description to a Hamiltonian, we will discuss how to incorporate external fluxes and how to treat multi-mode circuits with redundant degrees of freedom. Note that this discussion is customized for transmon-like qubits, as we work in a harmonic oscillator basis.\\
In Chapt.~\ref{chapt:coupling}, we will introduce the two types of coupling between qubit and resonator which are provided in our design, transverse and longitudinal coupling, and explain how the corresponding Hamiltonians can be diagonalized. We will then discuss a scalable qubit architecture conceived by Billangeon et al. in Ref.~\cite{Billangeon2015}, which relies on longitudinal coupling. Finally, we will show how to translate this discussion to the circuit theory language from Chapt.~\ref{chapt:quantization} in order to design circuits with specifically tailored couplings.\\
Chapter~\ref{chapt:design} will focus on our circuit design that consists of an inductively shunted transmon qubit with tunable coupling to an embedded harmonic mode, which was  presented in Refs.~\cite{Richer2016, Richer2017}.
We will demonstrate that the architecture provides the possibility to flux-choose between pure longitudinal and pure transverse coupling, or have both at the same time.
While transverse coupling naturally appears in transmon-like circuit constructions, longitudinal coupling is usually much smaller and hardly ever the only coupling term present. The distinctive feature of the tunable design is that the transverse coupling disappears when the longitudinal is maximal and vice versa. As opposed to other approaches, pure longitudinal coupling can be reached with moderate changes in the qubit frequency. Being able to choose between either kind of coupling in the same circuit provides the flexibility to use one for coupling to the next qubit and one for readout, or vice versa.\\
Subsequently, Chapt.~\ref{chapt:adaptations} will be about the implementation of the circuit design presented in Chapt.~\ref{chapt:design}. We will discuss how to choose the parameters obeying experimental constraints.
We will also present an adapted alternative circuit, where coupling strength and anharmonicity scale better than in the original circuit and show how the anharmonicity and the coupling can be boosted by additional flux-biasing. 
We will see that for conveniently chosen parameters longitudinal and transverse coupling have comparable values, while all other coupling terms can be suppressed.
Finally, we present a proposal for an experimental device that will serve as a prototype for a first experiment. The sample, most of which can be fabricated using standard thin-film aluminum, could be embedded in a 3D wave\-guide with strong coupling to the resonator mode.\\
In Chapt.~\ref{chapt:scalability} will come back to the scalable architecture mentioned in Chapt.~\ref{chapt:coupling} and show how our design can be scaled up to a grid following the scheme from Ref.~\cite{Billangeon2015}. In such a grid of fixed-frequency qubits and resonators with a particular pattern of always-on interactions, coupling is strictly confined to nearest and next-nearest neighbor resonators; there is never any direct qubit-qubit coupling. We note that just a single unique qubit frequency suffices for the scalability of this scheme.  The same is true for the resonators, if the resonator-resonator coupling constants are varied instead. We will present different circuit alternatives and show that the scale-up can be done in modular fashion with strictly local couplings.\\
Chapter~\ref{chapt:readout} will be about different possibilities to do readout with our circuit design, including a short discussion of the coupling of the circuit to the environment and the influence of dissipation. Finally, Chapt.~\ref{chapt:conclusions} contains some conclusions and an outlook on future research and open questions.\\
Please note that parts of this thesis are taken from or based on Refs.~\cite{Richer2016, Richer2017}, as also stated at the begin of the corresponding chapters. The abstract, introduction and conclusions of this thesis contain sentences taken from Refs.~\cite{Richer2016, Richer2017}.

\chapter{CIRCUIT QUANTIZATION}
\label{chapt:quantization}


Superconducting qubits are a realization of quantum bits that consist of superconducting electrical circuits which exhibit quantum behavior despite of being macroscopic objects. In superconducting materials, electrons spontaneously form Cooper pairs when the material is cooled below a critical temperature. These Cooper pairs, each consisting of two electrons with opposite spin, are bosonic objects and can therefore all occupy the same ground state. This is a macroscopic quantum phenomenon, where the Cooper pairs form a condensate that can be described by a collective degree of freedom, that is a single wave function.\\
In order to explain and analyze the quantum behavior of electrical circuits, we have to find a quantized mathematical description of them. As these circuits can be very large many-mode systems, we will need a systematic way of analyzing them. The method that will mostly be used here follows the pedagogic approach by Devoret~\cite{Devoret1997}, though we will later have a look on a similar method by Burkard et al.~\cite{Burkard2004, Burkard2005}, whose systematic circuit analysis is especially useful for circuits that include dissipative elements. A good introduction to circuit quantum electrodynamics is also given in the thesis of Bishop~\cite{Bishop2010}.\\
We will assume that the superconducting circuits in question can be described and depicted as lumped element circuits, which means that the elements of the circuit are much smaller than the wavelengths of the circuit modes. Borrowing terms from graph theory, we can describe an electrical circuit as a compound of nodes and branches, where every circuit element, such as a capacitor or an inductor, is considered as a branch that connects two nodes. 
We will start by describing two important single-mode circuits, before turning to larger many-mode systems. Having established the basic methods for circuit quantization, we will examine flux-dependent circuits and circuits with dissipative elements.

\section{QUANTUM LC RESONATOR}
\label{sec:resonator}

\begin{figure}[tb]
\centering
\captionbox{A simple harmonic resonator circuit, consisting of a capacitor in parallel to an inductor.\label{fig:resonator}}%
  [.43\linewidth]{\includegraphics[height=2.5cm]{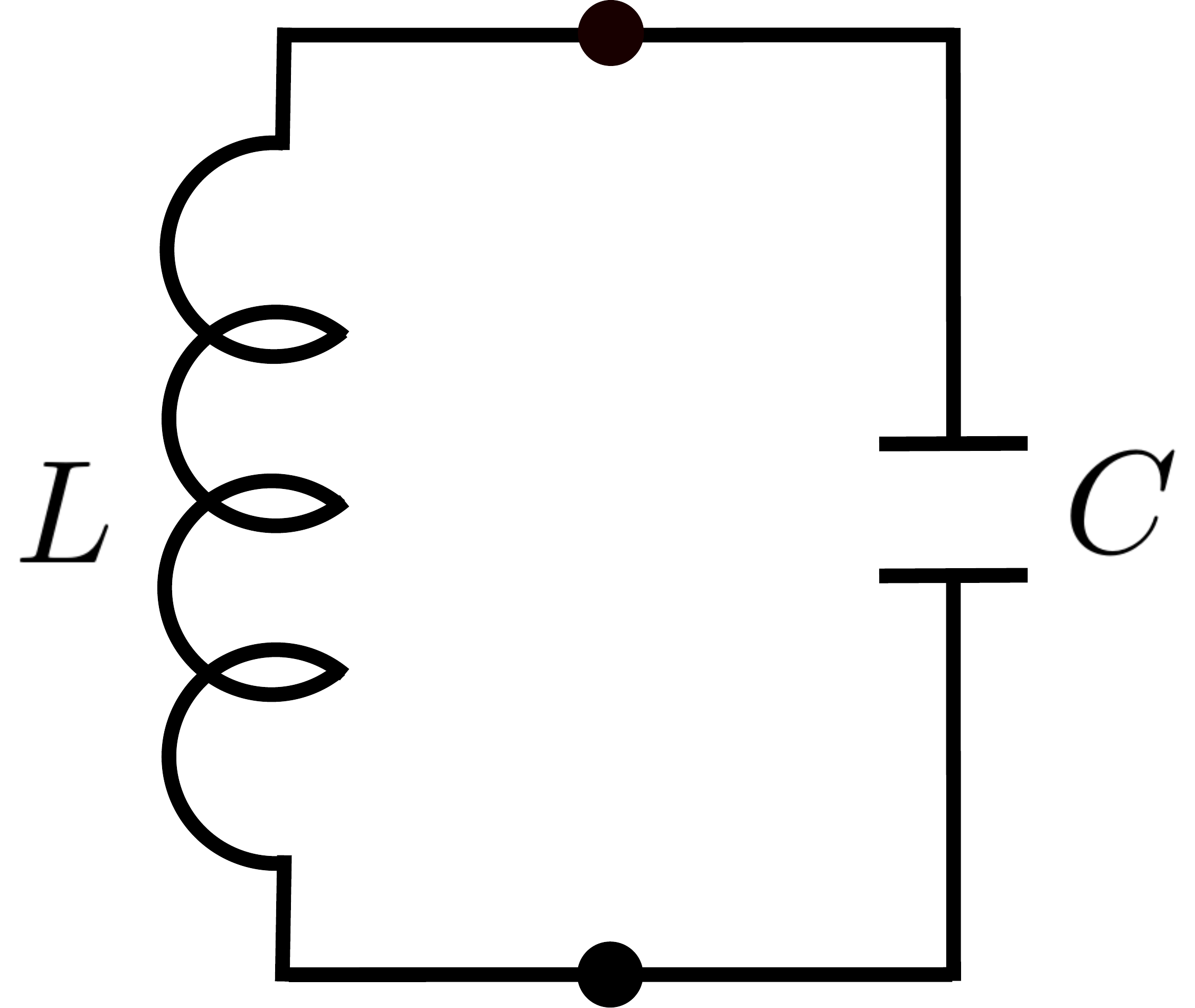}}
\hspace{5pt}
\captionbox{Harmonic oscillator potential with equidistant eigenenergies $\hbar\,\omega$.\label{fig:harmonic}}
  [.53\linewidth]{\includegraphics[height=2.5cm]{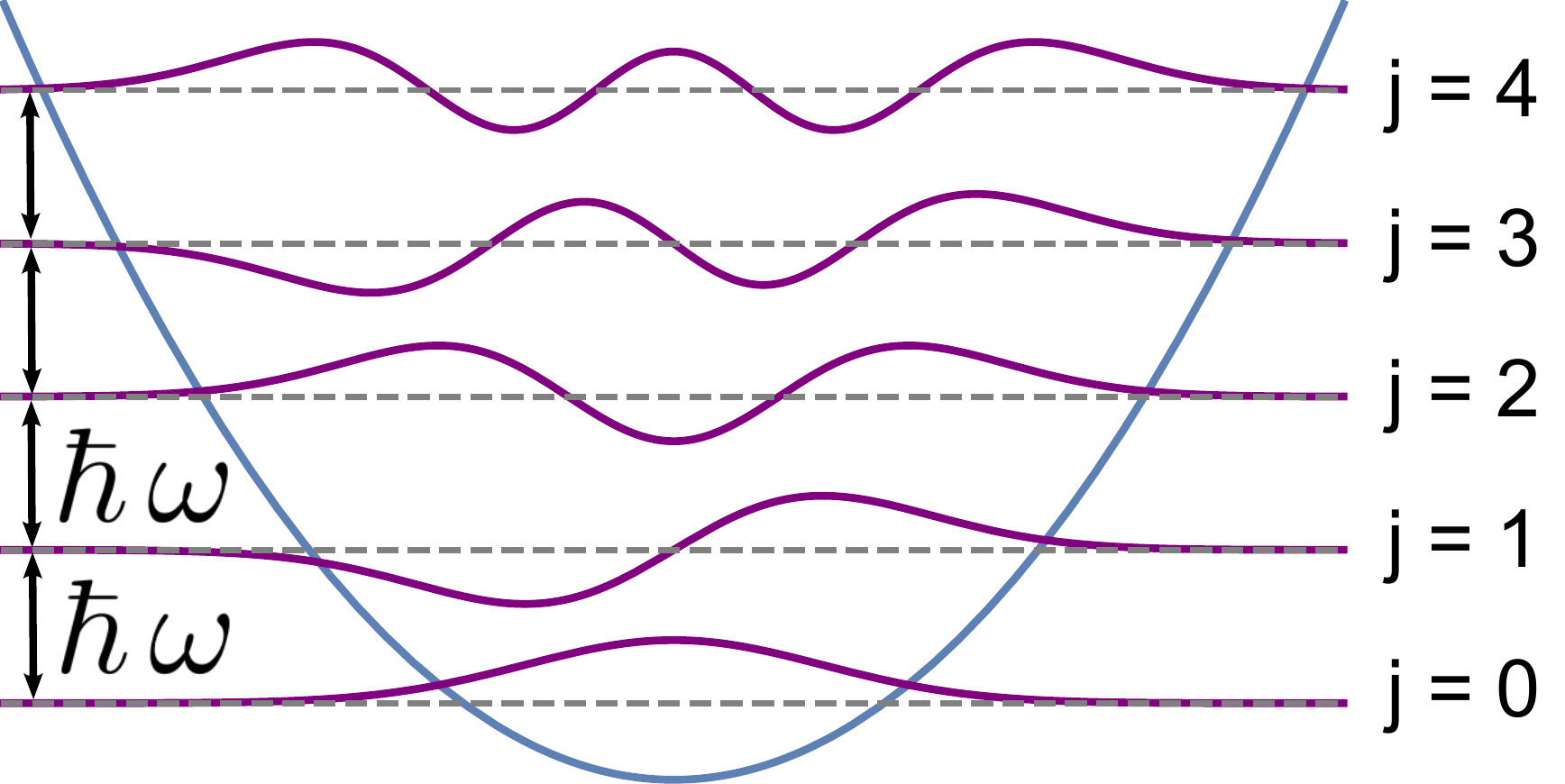}}
\end{figure}

Let us have a look at the example circuit depicted in Fig.~\ref{fig:resonator}, which is a simple $L\,C$ resonator. It consists of two nodes, which are connected by two branches, an inductor in parallel with a capacitor. A circuit with two nodes is always a single-mode system. This becomes immediately plausible when we apply Kirchhoff's voltage law, which states that the voltage differences around a closed path add up to zero. For our example this means that the voltage across one of the branches depends on the voltage across the other. In general, we can state that every closed path in a circuit results in a dependent degree of freedom. The total number of degrees of freedom in a circuit with $k$ nodes can thus never be larger than $k - 1$.
The variables we will use to describe such circuits are the magnetic flux $\Phi$ and its conjugate variable, the charge $Q$.
While the magnetic flux is the time integral over the voltage $V(t)$ across an element 

\begin{align}
\Phi (t) = \int_{-\infty}^t V(t') \, dt',
\label{eq:flux_definition}
\end{align}

its conjugate variable, the charge,  is the time integral over the current $I(t)$ flowing through the same branch

\begin{align}
Q (t) = \int_{-\infty}^t I(t') \, dt'.
\label{eq:charge_definition}
\end{align}

The current $I_L$ through an inductive element stands in a simple linear relation to the magnetic flux $\Phi$ threading the inductor. It is

\begin{align}
I_L = \frac{1}{L} \Phi,
\label{eq:il}
\end{align}

where the inductance $L$ defines the proportionality between them. In a capacitive element the capacitance $C$ defines the proportionality between the charge $Q$ on the capacitor plates and the voltage $V$ across it

\begin{align}
Q = C \, V.
\label{eq:iq}
\end{align}

In order to get a mathematical description of the circuit in terms of equations of motion, we can start by enforcing Kirchhoff's current law, saying that all currents that flow in or out a node must add up to zero. Using Eqs.~\ref{eq:flux_definition}-\ref{eq:iq}, the equation of motion for the single degree of freedom of the $L\,C$ resonator in terms of the magnetic flux is given by

\begin{align}
I_L + I_C = \frac{1}{L} \Phi + \dot Q = \frac{1}{L} \Phi + C \,\ddot \Phi = 0,
\label{eq:eqmr}
\end{align}

where the dot in $\dot Q = dQ(t)/dt$ is a shorthand that stands for the time derivative. This equation of motion corresponds to a harmonic oscillator, where the acceleration is inversely proportional to the deflection.
From the equation of motion, we can deduce the Lagrangian using the so-called \textit{Euler-Lagrange} equation

\begin{align}
\frac{d}{dt} \left(\frac{d\mathcal{L}}{d\dot\Phi_i}\right) - \frac{d\mathcal{L}}{d\Phi_i} = 0,
\label{eq:euler}
\end{align} 

where $\Phi_i$ for $i = 1, ..., \,N$ are the $N$ degrees of freedom of the system.
While the equations of motion of a system are unique, any Lagrangian that leads to these equations via Euler-Lagrange is correct. For the example of the $L\,C$ resonator, we can thus write

\begin{align*}
\mathcal{L} = \frac{C}{2} \dot \Phi^2  - \frac{1}{2L} \Phi^2. \numberthis
\label{eq:lagr_r}
\end{align*}

It is easy to see that this Lagrangian leads to the equation of motion from above (Eq.~\ref{eq:eqmr}) by using Euler-Lagrange (Eq.~\ref{eq:euler}). Note that while the inductive term depends on the flux $\Phi$, the capacitive term depends on its time derivative $\dot \Phi$. In this flux representation, it thus makes sense to interpret capacitive terms as the kinetic energy of a system and inductive terms as the potential energy. In this picture, the flux $\Phi$ is the analog to a position variable, while $\dot \Phi$ represents its velocity.
Using a Legendre transformation, we can go the Hamiltonian representation

\begin{align}
\mathcal{H} = \sum_{i=1}^N Q_i \, \dot\Phi_i - \mathcal{L} \qquad \text{with} \qquad Q_i = \frac{d\mathcal{L}}{d\dot\Phi_i},
\label{eq:legendre}
\end{align}

where we introduce the conjugate variable to the flux, the charge $Q$. The Hamiltonian thus reads

\begin{align}
\mathcal{H} = \frac{C}{2} Q^2  + \frac{1}{2L} \Phi^2,
\label{eq:ham_res}
\end{align}

where the first term corresponds to the electrostatic energy that is stored in the capacitor, while the second term stands for the magnetic energy in the inductor.
We will now employ the second quantization formalism by treating $\Phi$ and $Q$ as operators that can be expressed in terms of the creation and annihilation operators $a^\dagger$ and $a$. These are given by

\begin{align}
a^\dagger  = \sum_{j=0}^\infty \sqrt{j + 1} \, | j + 1 \rangle \langle  j | \qquad \qquad a = \sum_{j=0}^\infty \sqrt{j + 1} \, | j \rangle \langle  j + 1 |,
\label{eq:operators}
\end{align}

where the Fock states $| j \rangle$ are the eigenstates of the harmonic oscillator and $a^\dagger$ and $a$ satisfy the commutation relation

\begin{align}
[a, a^\dagger ] = 1
\end{align}

(compare App.~\ref{app:operators}).
The ansatz we take for the quantization will be

\begin{align*}
\Phi = \sqrt{\frac{\hbar  Z_{0}}{2}} (a^\dagger + a)  \qquad \qquad
Q =  \sqrt{\frac{\hbar}{2  Z_{0}}} i \, (a^\dagger - a)
\numberthis
\label{eq:quant_r}
\end{align*}

(compare~\cite{Devoret1997}), where the characteristic impedance $Z_{0}$ is given by

\begin{align}
Z_0 = \sqrt{\frac{L}{C}}.
\end{align}

The ansatz (Eq.~\ref{eq:quant_r}) is chosen such that the Hamiltonian has the form

\begin{align}
\mathcal{H} = \hbar \, \omega_r \, \left(a^\dagger a + \frac{1}{2}\right),
\label{eq:ham_res_final}
\end{align}

i.e. such that any non-diagonal terms disappear. For the eigenfrequency the ansatz yields

\begin{align}
\omega_r = \sqrt{\frac{1}{L \,C}},
\end{align}

which is equal to the classical result for an uncoupled $L\,C$ resonator. Clearly, the quantization ansatz given in Eq.~\ref{eq:quant_r} fulfills the commutation relation for the conjugate variables $\Phi$ and $Q$ as

\begin{align}
[\Phi, Q] = \frac{i \, \hbar}{2} [a^\dagger + a, a^\dagger - a] = i \,\hbar.
\end{align}

A simple $L\,C$ resonator can thus be described as a quantum harmonic oscillator. Figure~\ref{fig:harmonic} shows its potential and lowest eigenstates. A harmonic oscillator has a quadratic potential function, here given by the inductive energy of the circuit. The eigenstates of a harmonic oscillator are given by Hermite's polynomials (see for example Ref.~\cite{Brandt2003}). Importantly, its eigenenergies are all equally-spaced, being simply the multiples of the eigenenergy $\hbar \, \omega$ (see Eq.~\ref{eq:ham_res_final}). This is, in fact, the reason why such a system can not be easily used as a qubit. As stated by DiVincenzo in Ref.~\cite{Divincenzo2000}, a qubit needs to be a well-defined two-level system. A harmonic oscillator, however, is a multi-level system, where no transition between two levels is unique. It is thus impossible to pick out two levels and define them as a qubit, as we could never be sure the excitations would stay in this code space.

\section{A SIMPLE QUBIT}
\label{sec:qubit}

\begin{SCfigure}[50][tb]
\centering
    \includegraphics[width=.5\linewidth]{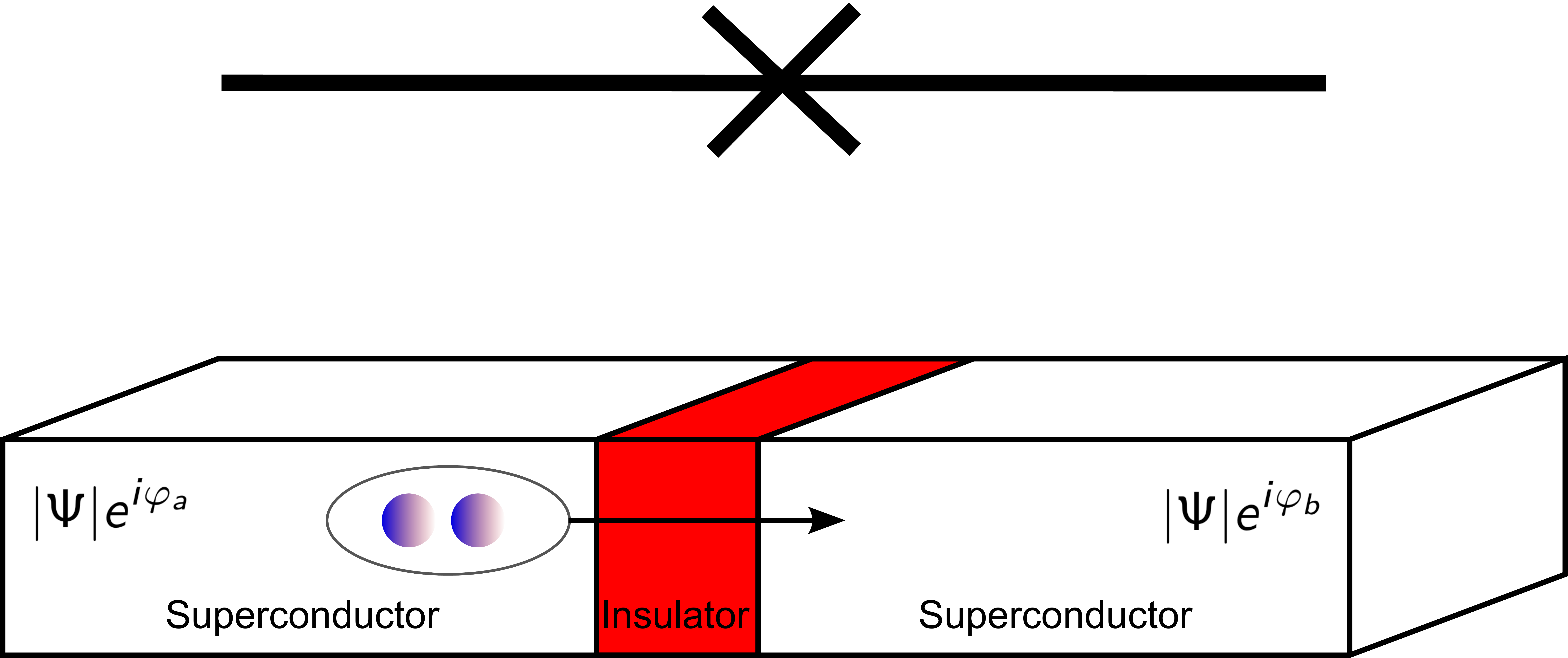}
  \caption{A Josephson junction consists of two superconductors connected by a weak link of non-superconducting material. 
  As a circuit element, a Josephson junction is depicted by the symbol above. \label{fig:josephson_junction}}   
\end{SCfigure}

Nevertheless, it is possible to use a multi-level superconducting circuit as a qubit. The only thing we need is a sufficiently anharmonic potential, in order to lift the degeneracy of the eigenenergies. This can be realized by using so-called \textit{Josephson junctions}~\cite{Josephson1962, Zagoskin2011}. While normal inductors and capacitors are known from usual high-temperature electric circuits, a Josephson junction is a circuit element that only exists in superconducting circuits. It can be made from a thin insulating barrier in between two superconductors, as depicted in Fig.~\ref{fig:josephson_junction}. Tunneling of Cooper pairs through the barrier leads to a current across the junction. 
As the Cooper pairs on both sides of the junction form a condensate that can be described by a single wave function, we can define a phase difference $\varphi$ across the barrier. The current across the junction is related to this superconducting phase difference by the first Josephson relation, that is

\begin{align}
I_J = I_\text{crit} \sin(\varphi), 
\label{eq:ij}
\end{align}

where $I_\text{crit}$ is the critical current of the junction, that is the maximal tunneling current through the junction. The second Josephson relation relates the phase difference $\varphi$ to the voltage across the junction

\begin{align}
\varphi = \frac{2\pi}{\Phi_0} \int_{-\infty}^t V(t') \, dt',
\label{eq:phase_definition}
\end{align}

where the constant $\Phi_0 = h/(2e)$ is called the \textit{magnetic flux quantum}. Comparing Eq.~\ref{eq:phase_definition} to Eq.~\ref{eq:flux_definition}, we can conclude that the superconducting phase across the junction formally corresponds to a rescaled magnetic flux $\Phi = \Phi_0/(2\pi) \, \varphi$ associated with the junction. As a circuit element, the Josephson junction can be seen as a non-linear inductance due to the non-linear relation between current and phase or flux given in Eq.~\ref{eq:ij}. 

\sidecaptionvpos{figure}{c}

\begin{figure}[tb]
\centering
\captionbox{A simple qubit circuit consisting of a Josephson junction with a capacitor in parallel.\label{fig:qubit}}%
  [.43\linewidth]{\includegraphics[height=2.5cm]{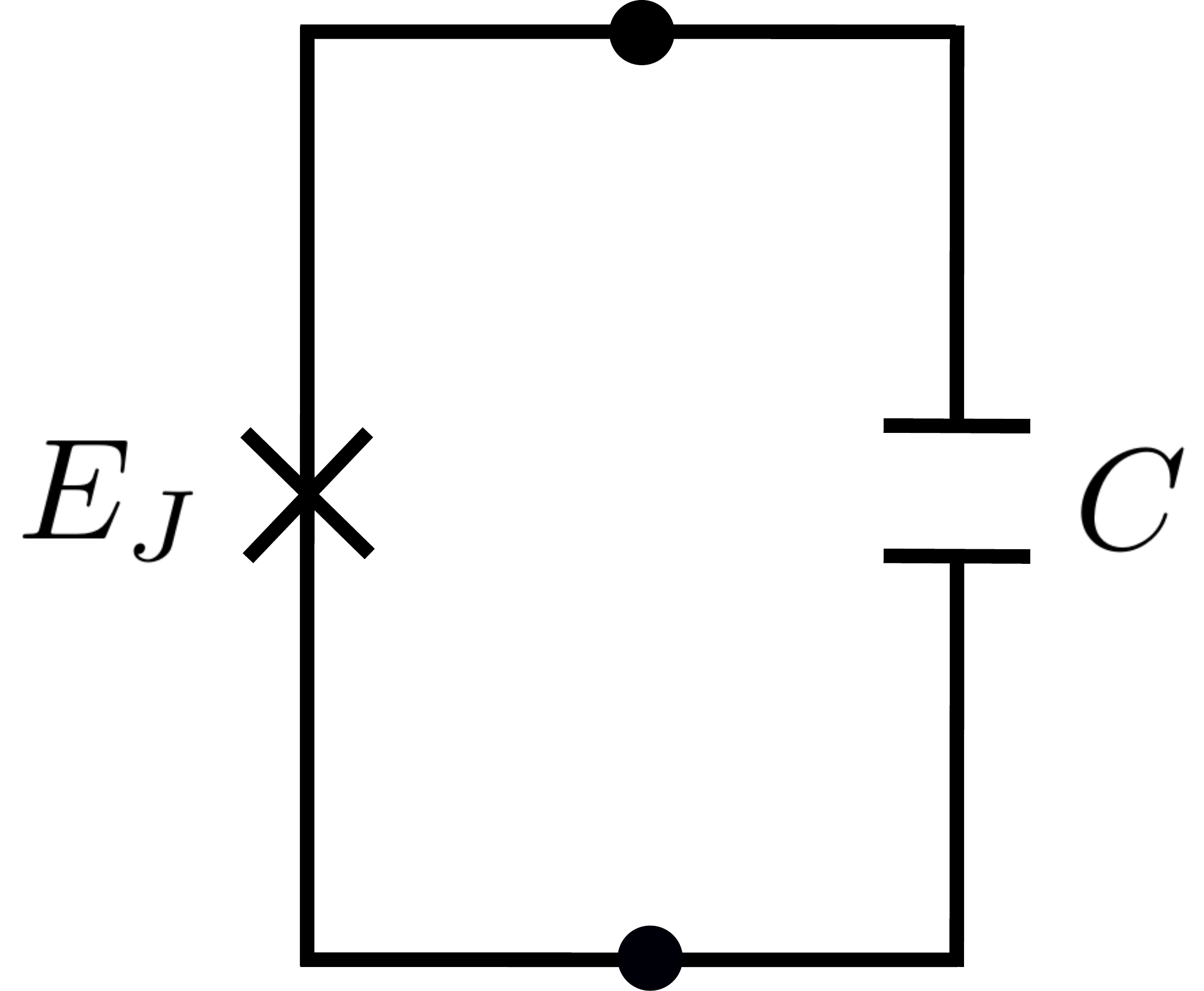}}
\hspace{5pt}
\captionbox{Potential energy and eigenenergies for a weakly anharmonic transmon-type qubit with unique transition energies between the eigenstates.
\label{fig:anharmonic}}
  [.53\linewidth]{\includegraphics[height=2.5cm]{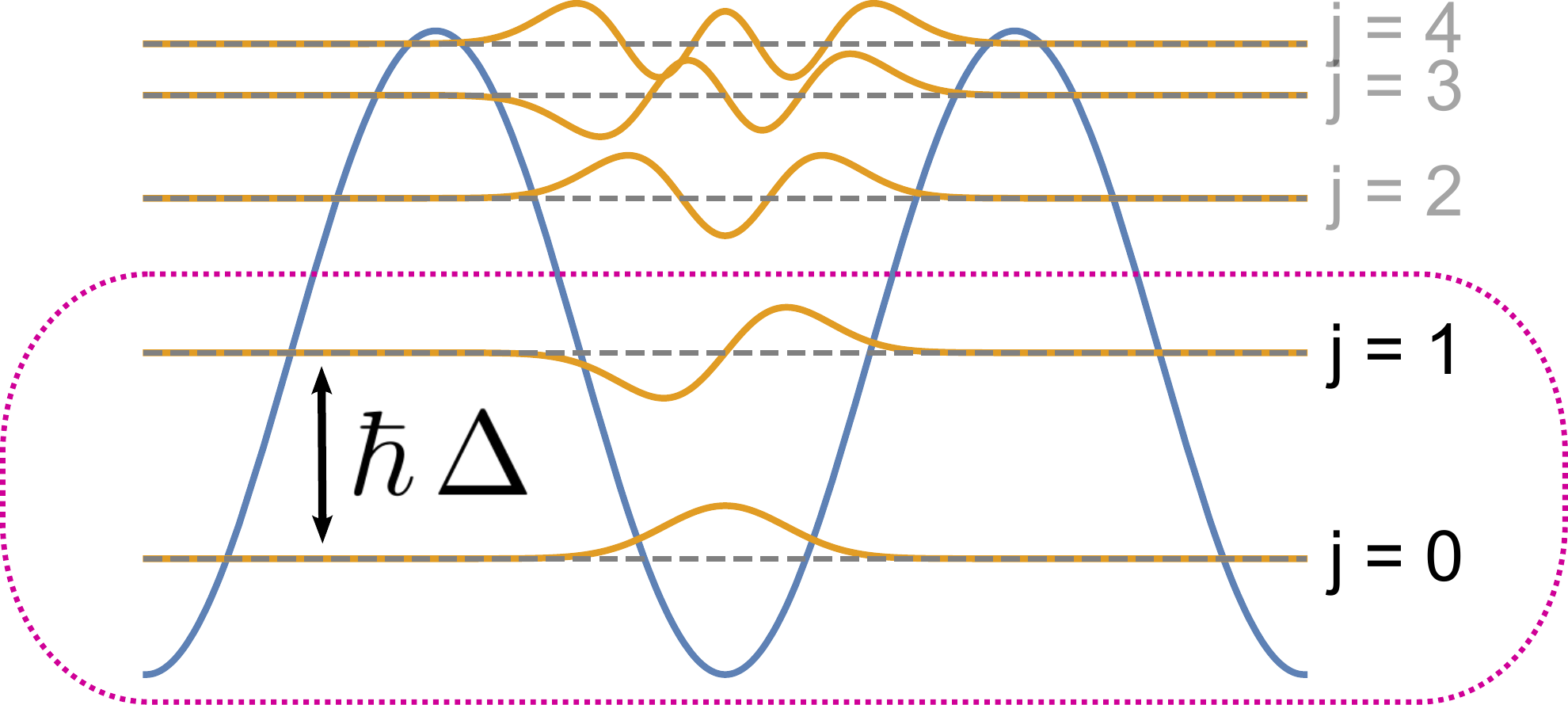}}
\end{figure}

Now the simplest superconducting qubit we can imagine consists of a single Josephson junction with a capacitor in parallel to it, as depicted in Fig.~\ref{fig:qubit}. Just as the harmonic resonator circuit described above, this circuit is a system with a single degree of freedom, which can be taken to be the phase difference between the two nodes. To get a mathematical description, we start again by writing down the equation of motion according to Kirchhoff's current law, yielding

\begin{align}
I_J + I_C = I_\text{crit} \sin(\varphi) + C \frac{\Phi_0}{2\pi}\ddot \varphi &= 0
\label{eq:eqmq}
\end{align}

in terms of the phase difference between the two nodes. 
The corresponding Lagrangian due to Euler-Lagrange (Eq.~\ref{eq:euler}) is given by

\begin{align*}
\mathcal{L} = \left(\frac{\Phi_0}{2\pi}\right)^2\frac{C}{2} \dot \varphi^2 + E_J \cos(\varphi), \numberthis
\label{eq:lagr_qu}
\end{align*}

where $E_J = I_\text{crit} \Phi_0/(2\pi)$ is the so-called Josephson energy, which is proportional to the critical current of the junction. When going to the Hamiltonian representation, we can either use the conjugate variables flux and charge as done above (Eq.~\ref{eq:ham_res}) or use unitless variables, which are very common in circuits with Josephson junctions. With the charging energy $E_C = e^2/(2C)$ of the junction, we find

\begin{align}
\mathcal{H} = \frac{Q^2}{2\,C} - E_J \cos\left(\frac{2\pi}{\Phi_0} \Phi\right) = 4 E_C \, n^2 - E_J \cos(\varphi),
\label{eq:ham_qu}
\end{align}

where we introduced the conjugate variable to the flux, the charge $Q$ (again via Eq.~\ref{eq:legendre}), and its unitless version $n = Q/(2e)$, which corresponds to the number of Cooper pairs on a node. 
This Hamiltonian in terms of the superconducting phase $\varphi$ and the number of Cooper pairs $n$ is known from transmon qubits~\cite{Koch2007}, as well as charge qubits such as the Cooper pair box~\cite{Bouchiat1998}. Now the only difference between this Hamiltonian (Eq.~\ref{eq:ham_qu}) and the one given for the harmonic resonator (Eq.~\ref{eq:ham_res}) is the anharmonic cosine potential instead of the harmonic potential. Figure~\ref{fig:anharmonic} shows the qubit potential and its lowest eigenstates for a weakly anharmonic case. The crucial difference to the resonator potential is that the eigenstates are not equally-spaced any more, due to the anharmonicity of the cosine. This is why we can use the system as a qubit. Picking out the lowest two eigenstates, we effectively have a two-level system, as the energy transition between them is unique. As long as the anharmonicity is large enough, transitions between the two qubit states can be driven without leaving this code space.\\
While this Hamiltonian (Eq.~\ref{eq:ham_qu}) is exactly solvable in terms of Mathieu functions~\cite{Bishop2010}, we will use an approximative approach here that is valid for weakly anharmonic systems such as the transmon qubit~\cite{Koch2007}. As opposed to the Cooper pair box, the transmon qubit is characterized by its large capacitance, such that $E_J \gg E_C$ in Eq.~\ref{eq:ham_qu}. While this reduces the anharmonicity, the capacitance shunts the qubit very effectively against charge noise.
In the limit of weak anharmonicity, quantization can be done by first treating the system as harmonic and then defining its anharmonicity, that is its quartic deviation from a harmonic system. 
We thus assume that the anharmonicity of the system is small enough to be captured by a Duffing oscillator approach, which is an anharmonic oscillator with a quartic potential energy term. This is valid for weakly anharmonic transmon-like qubits.
We will use 

\begin{align*}
\varphi = \sqrt[4]{\frac{2 E_C}{E_{J}}} (c^\dagger + c) \qquad \qquad n = \frac{1}{2}\sqrt[4]{\frac{E_J}{2 E_C}} i \, (c^\dagger - c)
 \numberthis
\label{eq:quant_qu}
\end{align*}

with the creation and annihilation operators as defined in Eq.~\ref{eq:operators}, which we call $c^\dagger$ and $c$ to distinguish them from the operators for the resonator. This ansatz corresponds to the one given in Eq.~\ref{eq:quant_r} with a characteristic impedance of $Z_0 = \hbar/e^2 \sqrt{E_C/(2E_J)}$. The harmonic approximation of Eq.~\ref{eq:ham_qu} is given by

\begin{align}
\mathcal{H}^{(2)} = 4 E_C \, n^2 - E_J +\frac{E_J}2 \varphi^2.
\label{eq:ham_qu2}
\end{align}

Leaving out the constant term, the quantization ansatz (Eq.~\ref{eq:quant_qu}) leads to

\begin{align}
\mathcal{H} = \hbar \, \omega_q \, \left(c^\dagger c + \frac{1}{2}\right),
\label{eq:ham_qubit}
\end{align}

where 

\begin{align}
\omega_q = \frac{\sqrt{8 E_J  E_C}}{\hbar}
\end{align}

is the harmonic approximation to the frequency of the qubit. The ansatz given in Eq.~\ref{eq:quant_qu} clearly fulfills the commutation relation for the conjugate variables $\Phi$ and $Q$, as

\begin{align}
[\Phi, Q] = \left[\frac{\Phi_0}{2\pi}\varphi, 2e \,n\right] = \frac{i \, \hbar}{2} [c^\dagger + c, c^\dagger - c] = i \,\hbar,
\end{align}

while the unitless variables fulfill $[\varphi, n] = i$. In order to define the quartic anharmonicity of the qubit, we simply have to expand the Hamiltonian (Eq.~\ref{eq:ham_qu}) up to fourth order, that is

\begin{align}
\mathcal{H}^{(4)} = 4 E_C \, n^2 - E_J +\frac{E_J}2 \varphi^2 - \frac{E_J}{24} \varphi^4
\label{eq:ham_fourthorder}
\end{align}

and insert the quantization ansatz (Eq.~\ref{eq:quant_qu}). Using 

\begin{align}
\langle j | (c^\dagger + c)^4 | j \rangle = 6j^2 + 6j +3
\label{eq:quartic_term}
\end{align}

(compare~\cite{Koch2007}) and of course $\langle j | a^\dagger a | j \rangle = j$, the energy of state $j$ up to fourth order is

\begin{align}
E_j &= - E_J + \sqrt{8E_J E_C} \left(j+\frac{1}{2}\right) - \frac{E_C}{12} (6j^2 + 6j +3).
\end{align}

The quartic anharmonicity of the qubit yields

\begin{align}
\alpha = \frac{E_{12} - E_{01}}\hbar = - \frac{E_C}\hbar
\label{eq:qubit_anh}
\end{align}

with $E_{ij} = E_j - E_i$, while the so-called relative anharmonicity is

\begin{align}
\alpha_r = \frac{E_{12} - E_{01}}{E_{01}} = - \frac{E_C}{\sqrt{8E_J E_C} - E_C}.
\end{align}

When we plug the quantization (Eq.~\ref{eq:quant_qu}) into the fourth-order Hamiltonian (Eq.~\ref{eq:ham_fourthorder}) we can use a rotating wave approximation to rewrite

\begin{align}
(c^\dagger + c)^4 \to 6 (c^\dagger c)^2 + 6 \, c^\dagger c + 3
\end{align}

(compare Eq.~\ref{eq:quartic_term}), which means we leave out the fast-rotating non-diagonal terms. The Hamiltonian thus yields

\begin{align*}
\mathcal{H}^{(4)} &= \hbar \left(\omega_q +\frac{\alpha}{2}\right) c^\dagger c + \hbar \frac{\alpha}{2} (c^\dagger c)^2 \\
&= \hbar \sum_{j=0}^\infty \left(\left(\omega_q + \frac{\alpha}{2}\right) j + \frac\alpha 2 j^2 \right) |j \rangle \langle j | \equiv \hbar \sum_{j=0}^\infty \omega_j |j \rangle \langle j |, \numberthis
\label{eq:ham4quant}
\end{align*}

where we used the definition for the creation and annihilation operators (Eq.~\ref{eq:operators}) and left out constant terms. This expression makes clear why this system, as opposed to a harmonic oscillator, can be used as a qubit. The frequency difference $\omega_{j+1} - \omega_j$ of adjacent energy states depends on the level $j$. It is thus possible to pick out the two lowest levels and treat them as a qubit since their transition frequency is unique.
Now, assuming the system is anharmonic enough, the so-called two-level approximation is justified. We can thus make the transition

\begin{align*}
&\sum_{j = 0}^\infty \omega_j | j \rangle \langle j | \to \sum_{j = 0}^1 \omega_j | j \rangle \langle j |
=(\omega_q + \alpha) | 1 \rangle \langle 1| \\
= &(\omega_q + \alpha) \left(\frac{|1 \rangle \langle 1| - |0 \rangle \langle 0|}{2} + \frac{|1 \rangle \langle 1| + |0 \rangle \langle 0|}{2}\right) \\
= &(\omega_q + \alpha) \left(\frac{\sigma_z}2 + \frac{\sigma_0}{2}\right)  \numberthis
\label{eq:trans}
\end{align*}

(compare App.~\ref{app:operators}), that is truncating to the first two levels. In this two-level approximation the Pauli matrices $\sigma_x \mp i\,\sigma_y$ replace the creation and annihilation operators $a^\dagger$ and $a$.
Ignoring the constant term in Eq.~\ref{eq:ham_qubit}, as well as the identity operator $\sigma_0$ in Eq.~\ref{eq:trans}, we can write

\begin{align}
\mathcal{H} = \hbar \,\frac{\Delta}{2} \sigma_z
\end{align}

for the Hamiltonian of the qubit with the effective qubit transition frequency

\begin{align}
\Delta = \omega_q + \alpha = \frac{E_{01}}{\hbar} = \frac{\sqrt{8 E_J E_C} - E_C}{\hbar}.
\end{align}

Note that the transition frequency between the two qubit levels is not the pure harmonic frequency, but experiences a shift due to the anharmonicity (compare Ref.~\cite{Poletto2012}).

\section{MULTIDIMENSIONAL CIRCUITS}
\label{sec:multidimensional_circuits}
We have thus introduced the two most important building blocks in superconducting qubits - a simple qubit and a harmonic resonator. Both example circuits were systems with only one degree of freedom. This makes the circuit analysis very easy. However, as soon as a system has more than one degree of freedom, we have the choice how to define the variables. When treating systems with multiple degrees of freedom, it is convenient to work with so-called node phases or, equivalently, node fluxes. This means that we associate each node in the circuit with a superconducting phase $\varphi_i$ or an associated magnetic flux $\Phi_i = \Phi_0/(2\pi) \varphi_i$. Remembering that the phase or flux are given by the integral over the voltage across an element (Eqs.~\ref{eq:flux_definition} and \ref{eq:phase_definition}), it becomes clear that the real variables are the differences between these node variables. Equivalently, we could say that the overall phase is undefined and a node phase or flux only makes sense in relation to another. The concept of node fluxes is thus a bit artificial, but it gives us a lot of flexibility in choosing the variables in which we want to describe the circuit. It is also very useful from a pedagogic point of view, as it allows us to write down any circuit Lagrangian in terms of the node fluxes simply by sight, however complicated the circuit may be. 

\sidecaptionvpos{figure}{c}
\begin{SCfigure}[50][t]
\centering
    \includegraphics[height=4cm]{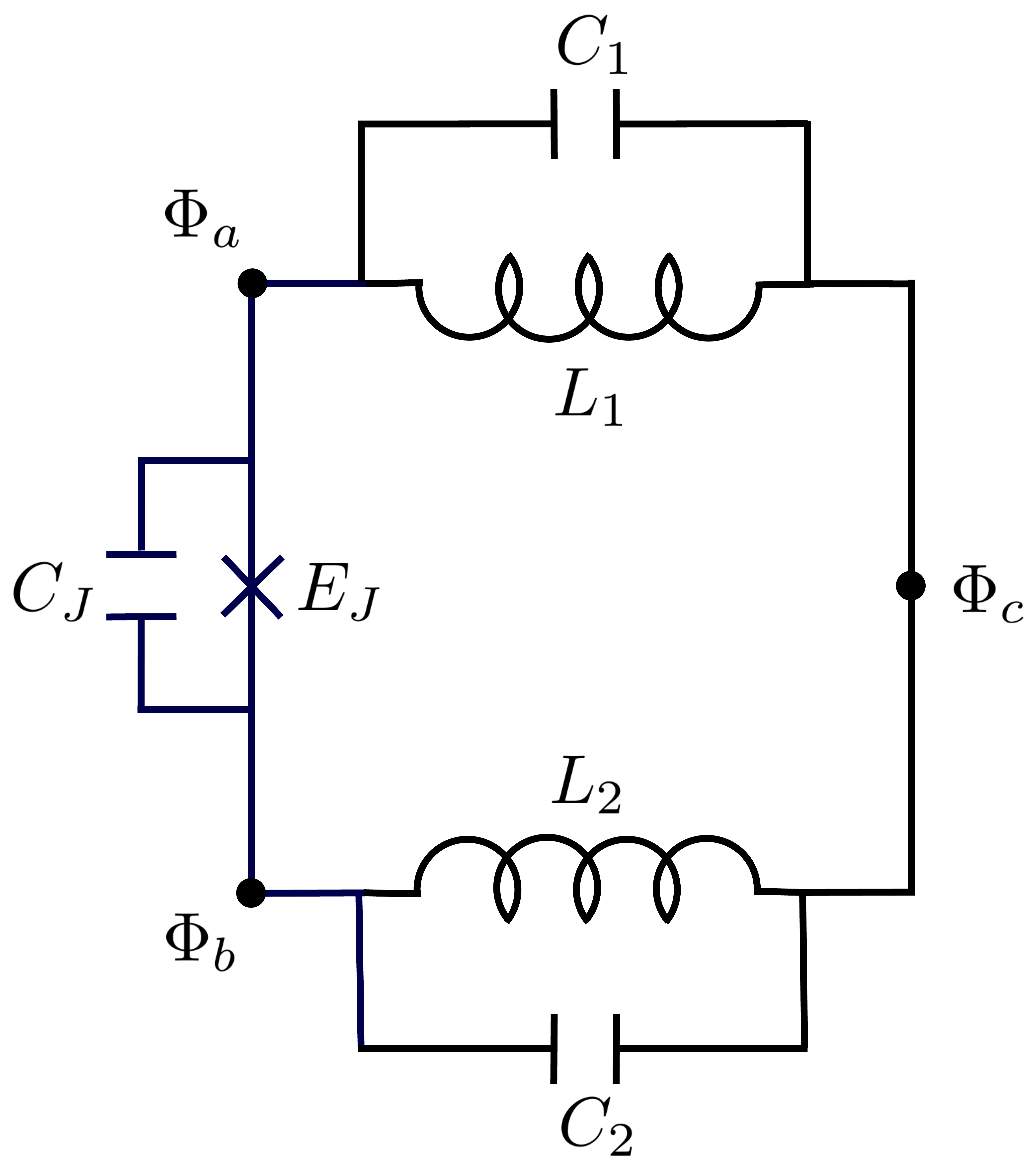}
  \caption{Example circuit with two degrees of freedom. The circuit consists of two inductors and a Josephson junction with capacitors in parallel to them. It can be interpreted as a qubit coupled to a resonator or as two coupled qubits, depending on the choice of variables.\label{fig:example0}}
\end{SCfigure}

Let us consider the example circuit depicted in Fig.~\ref{fig:example0}. The circuit has three nodes, which corresponds to a maximum of two independent variables. We could also say that it has six branches forming four closed paths, which leaves two independent variables. Remember that this is due to the fact that the voltage differences around a closed path must add up to zero, as stated in Kirchhoff's voltage law. Looking at the Lagrangians of the example circuits above (Eqs. \ref{eq:lagr_r}, \ref{eq:lagr_qu}), we can directly write down the Lagrangian for this circuit in terms of the node fluxes, that is

\begin{align*}
\mathcal{L} &=  \frac{C_J}{2} (\dot \Phi_a - \dot\Phi_b)^2 + \frac{C_1}{2} (\dot \Phi_a - \dot\Phi_c)^2 + \frac{C_2}{2} (\dot \Phi_b - \dot\Phi_c)^2 \\
&- \frac{1}{2L_1} (\Phi_a - \Phi_c)^2 - \frac{1}{2L_2} (\Phi_b - \Phi_c)^2 + E_J \cos\left(\frac{2\pi}{\Phi_0}(\Phi_a - \Phi_b)\right)\numberthis.
\label{eq:lagr_nodeflux}
\end{align*}

Each circuit element corresponds to one term in this Lagrangian in terms of the node fluxes on either side of the branch. Note that we did not need to assign directions to the branches, given that the sign of the fluxes does not play a role, as long as no external fluxes are applied. It is obvious that one node flux variable must depend on the others, as the circuit can only have two independent variables.\\
As mentioned above, we have the freedom to choose the variables in which we want to represent the circuit. We can thus choose any two independent flux differences in the circuit, as well as any linearly independent superpositions of these. This condition of linear independence implies that the chosen flux differences do not build a closed path in the circuit, as the flux differences around a closed path can never be independent. \\
This freedom of choice in the variables means of course that there will be several equivalent descriptions of the circuit. Consider the following choice of variables

\begin{align}
\Phi_1 = \Phi_a - \Phi_c \qquad \Phi_2 = \Phi_b - \Phi_c,
\label{eq:var}
\end{align}

that is the flux differences across the two inductors. When inserting these in the Lagrangian (Eq.~\ref{eq:lagr_nodeflux}), it becomes clear that the system is inherently two-dimensional, as the third variable simply disappears. We find

\begin{align*}
\mathcal{L} &= \frac{C_J}{2} (\dot \Phi_1 - \dot\Phi_2)^2 + \frac{C_1}{2} \dot \Phi_1^2 + \frac{C_2}{2}\dot\Phi_2^2- \frac{1}{2L_1} \Phi_1^2 - \frac{1}{2L_2} \Phi_2^2 \\
&+ E_J \cos\left(\frac{2\pi}{\Phi_0}(\Phi_1 - \Phi_2)\right)\numberthis.
\label{eq:lagr_example0}
\end{align*}

We can see that this choice of variables (Eq.~\ref{eq:variables}) leads to a symmetric Lagrangian in the variables $\Phi_1$ and $\Phi_2$. Note that equivalently, we could define one node as ground without loss of generality and consider the remaining nodes as so-called \textit{active nodes} as done by Devoret in Ref.~\cite{Devoret1997}. This would lead to the same Lagrangian as the one given here (Eq.~\ref{eq:lagr_example0}).\\
Now, what can we say about this system by looking at the Lagrangian? Remembering that the distinction between the two basic building blocks, that is qubits and resonators, was made via the anharmonicity in the potential energy, we note that in this description both variables $\Phi_1$ and $\Phi_2$ have an anharmonic potential energy term via the Josephson junction in addition to the harmonic terms via the inductances.  They can thus be classified as anharmonic systems, though their anharmonicity is reduced due to the presence of the harmonic inductance terms. Whether they can be used as qubits, depends of course on the concrete parameters. We can also see that the two modes interact both via the kinetic (first term in Eq.~\ref{eq:lagr_example0}) and the potential energy (last term in Eq.~\ref{eq:lagr_example0}). The circuit can thus be seen as a system of two qubits that interact via different coupling terms. \\
However, as mentioned above, this description of the system is not unique. To see this, consider a different choice of variables, that is

\begin{align}
\Phi_q = \Phi_a - \Phi_b \qquad \Phi_r = \Phi_a + \Phi_b - 2\,\Phi_c.
\label{eq:varqr}
\end{align}

These variables are clearly independent of each other, as none can be represented as a superposition of the other. Note that these variables are simply the sum and the difference of the variables used above (Eq.~\ref{eq:var}).
When inserting the new variables in the Lagrangian (Eq.~\ref{eq:lagr_nodeflux}), we find

\begin{align*}
\mathcal{L} &= \frac{C_1 + C_2 + 4\, C_J}{8} \dot\Phi_q^2 + \frac{C_1 - C_2}{4} \dot\Phi_q \dot\Phi_r + \frac{C_1 + C_2}{8} \dot\Phi_r^2\\
&- \frac{L_1 + L_2}{8L_1 L_2} \left(\Phi_q^2 + \Phi_r^2\right) - \frac{L_2 - L_1}{4L_1 L_2} \Phi_q \Phi_r + E_J \cos\left(\frac{2\pi}{\Phi_0}\Phi_q\right)\numberthis.
\label{eq:lagr_qu_res}
\end{align*}

We can see that with this choice of variables, only one of the two degrees of freedom has an anharmonic potential. While $\Phi_q$ could thus be treated as a qubit mode, the potential for $\Phi_r$ is purely harmonic. The system can thus be seen as a qubit coupled to a resonator. What is interesting about this choice of variables, is that the coupling between the two variables depends only on the asymmetry between the two inductances $L_1$ and $L_2$ and the two capacitances $C_1$ and $C_2$. For a symmetric choice of parameters, that is $L_1 = L_2$ and $C_1 = C_2$, the coupling terms disappear. 
\\
Now, how do we get an accurate description of  such a multi-dimensional system in second quantization? The strategy used here will be to do the quantization separately for each variable, assuming that all others are in their equilibrium positions, that is at $Q_i = \Phi_i = 0$. This will be done for the harmonic approximation, while treating all anharmonic terms as a perturbation, as done in Sec.~\ref{sec:qubit} for the qubit. This procedure is similar to the black-box quantization approach presented in Ref.~\cite{Nigg2012}.\\
By way of example, we will do the quantization for the resonator within the coupled system.
Using a Legendre transformation (Eq.~\ref{eq:legendre}), we go to the Hamiltonian representation and find

\begin{align*}
\mathcal{H} &= \frac{(C_1 + C_2) (Q_q^2 + Q_r^2) + 2 (C_2 - C_1) Q_q Q_r + 4\,C_J Q_r^2}{2( C_1 C_2 + C_1 C_J + C_2 C_J)} \\
&+ \frac{L_1 + L_2}{8L_1 L_2} \left(\Phi_q^2 + \Phi_r^2\right) + \frac{L_2 - L_1}{4L_1 L_2} \Phi_q \Phi_r - E_J \cos\left(\frac{2\pi}{\Phi_0}\Phi_q\right)\numberthis.
\label{eq:ham_example0}
\end{align*}

We will now consider only the quadratic terms in $Q_r$ and $\Phi_r$, while setting $Q_q = \Phi_q = 0$. We find

\begin{align}
\mathcal{H}_r = \frac{C_1 + C_2 + 4 \,C_J}{2( C_1 C_2 + C_1 C_J + C_2 C_J)} Q_r^2 + \frac{L_1 + L_2}{8L_1 L_2} \Phi_r^2 ,
\end{align}

The ansatz from equation \ref{eq:quant_r} gives again

\begin{align}
\mathcal{H}_r = \hbar \,\omega_r \left(a^\dagger a + \frac{1}{2}\right)
\label{eq:ham_coupled_res}
\end{align}

with

\begin{align}
Z_{0,r} = 2\sqrt{\frac{L_1 \,L_2 (C_1 + C_2 + 4\,C_J)}{(L_1+L_2)(C_1 C_2 + C_1 C_J + C_2 C_J)}}
\end{align}

for the characteristic impedance and 

\begin{align}
\omega_r = \frac{1}{2}\sqrt{\frac{(L_1+L_2)(C_1 + C_2+4\,C_J)}{L_1 \,L_2 (C_1 C_2 + C_1 C_J + C_2 C_J)}}
\end{align}

for the frequency of the coupled resonator. 
For the entirely symmetric case with inductances $L_1 = L_2$ and $C_1 = C_2$, the coupling terms in Eq.~\ref{eq:ham_example0} disappear and $\omega_r$ converges to

\begin{align}
\omega_r = \sqrt{\frac{1}{L\,C}},
\end{align}

the frequency of the uncoupled harmonic resonator. The coupling to the qubit thus yields in a rescaling of the frequency of the resonator. When the coupling disappears, we regain the expression for the uncoupled system.\\
The quantization for the qubit can be done in the same fashion, again assuming that the resonator is in its equilibrium position at $Q_r = \Phi_r = 0$. Having done the quantization for both degrees of freedom, we take a look at the coupling terms. There are two terms in the Hamiltonian (Eq.~\ref{eq:ham_example0}) that involve both degrees of freedom, a capacitive term and an inductive term. In order to get an expression for the coupling terms, we need to plug in the quantization as given in Eqs.~\ref{eq:quant_r} and \ref{eq:quant_qu} with the prefactors adapted to the coupled system and apply the two-level approximation for the qubit. For the first term, we find

\begin{align}
 \frac{2 (C_2 - C_1) Q_q Q_r}{2( C_1 C_2 + C_1 C_J + C_2 C_J)} \to \hbar \,g_y \, \sigma_{y} (a^\dagger - a),
\end{align}

where we took

\begin{align*}
i(c^\dagger - c) &= i \sum_{j=0}^\infty \sqrt{j + 1} (| j + 1 \rangle \langle j | - |j \rangle \langle j + 1 | ) \\
&\to i (| 1 \rangle \langle 0 | - | 0 \rangle \langle 1 | ) = \sigma_y
\label{eq:trans_sigmay} \numberthis
\end{align*}

for the qubit operators in the two-level approximation (compare again App.~\ref{app:operators}). The coupling strength $g_y$ includes the coefficient of $Q_q Q_r$ from the Hamiltonian (Eq.~\ref{eq:ham_example0}) as well as the prefactors from the quantization ansatz (Eqs.~\ref{eq:quant_r} and \ref{eq:quant_qu}).
In the same way, we can write

\begin{align}
\frac{L_2 - L_1}{4L_1 L_2} \Phi_q \Phi_r \to \hbar \,g_x \, \sigma_{x} (a^\dagger + a),
\end{align}

with

\begin{align*}
c^\dagger + c &= \sum_{j=0}^\infty \sqrt{j + 1} (| j +1 \rangle \langle j | + |j \rangle \langle j + 1 | ) \\
&\to | 1 \rangle \langle 0 | + | 0 \rangle \langle 1 |  = \sigma_x
\label{eq:trans_sigmax} \numberthis
\end{align*}

for the inductive coupling term. The full Hamiltonian for the system depicted in Fig.~\ref{fig:example0} for the variables as chosen in Eq.~\ref{eq:varqr} is thus given by

\begin{align}
\mathcal{H} = \hbar \,\omega_r a^\dagger a +  \hbar\, \frac{\omega_{q}}{2} \sigma_{z} + \hbar\, g_{x} \,\sigma_{x} (a^\dagger + a) + \hbar\,g_{y} \,\sigma_{y} (a^\dagger - a)
\end{align}

Note that we neglected the constant term from Eq.~\ref{eq:ham_coupled_res}.

\section{FLUX-DEPENDENT CIRCUITS}

Let us now examine what happens, when external magnetic fields come into play. We have stated above that the Cooper pairs in a superconductor form a condensate that can be described by a collective many-particle wave function $\Psi \sim e^{i\varphi}$ with a coherent superconducting phase $\varphi$. As the Cooper pairs are charged particles, they will acquire a phase shift $\Delta \varphi$ when traveling in a region with non-zero magnetic potential $\mathbf{A}$. This effect is called the \textit{Aharonov-Bohm effect}~\cite{Aharonov1959}, which states that the acquired phase shift is given by

\begin{align}
\Delta \varphi = \frac{q}{\hbar} \int \mathbf{dr} \cdot \mathbf{A},
\label{eq:aharonov}
\end{align} 

where $q$ denotes the charge of the particle.
Remarkably, this is true even if the magnetic field $\mathbf{B} = \nabla \times \mathbf{A} = 0$, as the phase shift depends on the vector potential $\mathbf{A}$, not on the magnetic field itself. Note that this phase shift depends on the path of the integral, not only on the start and end point. \\
Let us consider a closed loop of superconducting material in a region with non-zero magnetic potential.
%
For the wavefunction to be single-valued, we have to demand that the phase acquired in a contour integral around the closed loop must be a multiple of $2\pi$, that is

\begin{align}
\Psi(\varphi(\mathbf{r})) = \Psi(\varphi(\mathbf{r}) + 2\pi\,m)
\end{align}

for integer $m$.
This means that the enclosed magnetic flux $\Phi$ corresponding to the phase shift $\Delta \varphi$ must be a multiple of the magnetic flux quantum $\Phi_0 = h/(2e)$. 
This flux quantization condition is given by

\begin{align}
\Delta \varphi = \frac{2e}{\hbar} \oint \mathbf{dr} \cdot \mathbf{A} = \frac{2\pi}{\Phi_0} \Phi = 2\pi \, m.
\end{align}


%
%

\begin{figure}[t]
\centering
\captionbox{The loop formed by the inductor and the Josephson junction is threaded by an external flux. Note the directions of the branch phases $\varphi_1$ and $\varphi_2$.\label{fig:example1}}%
  [.43\linewidth]{\includegraphics[height=2.5cm]{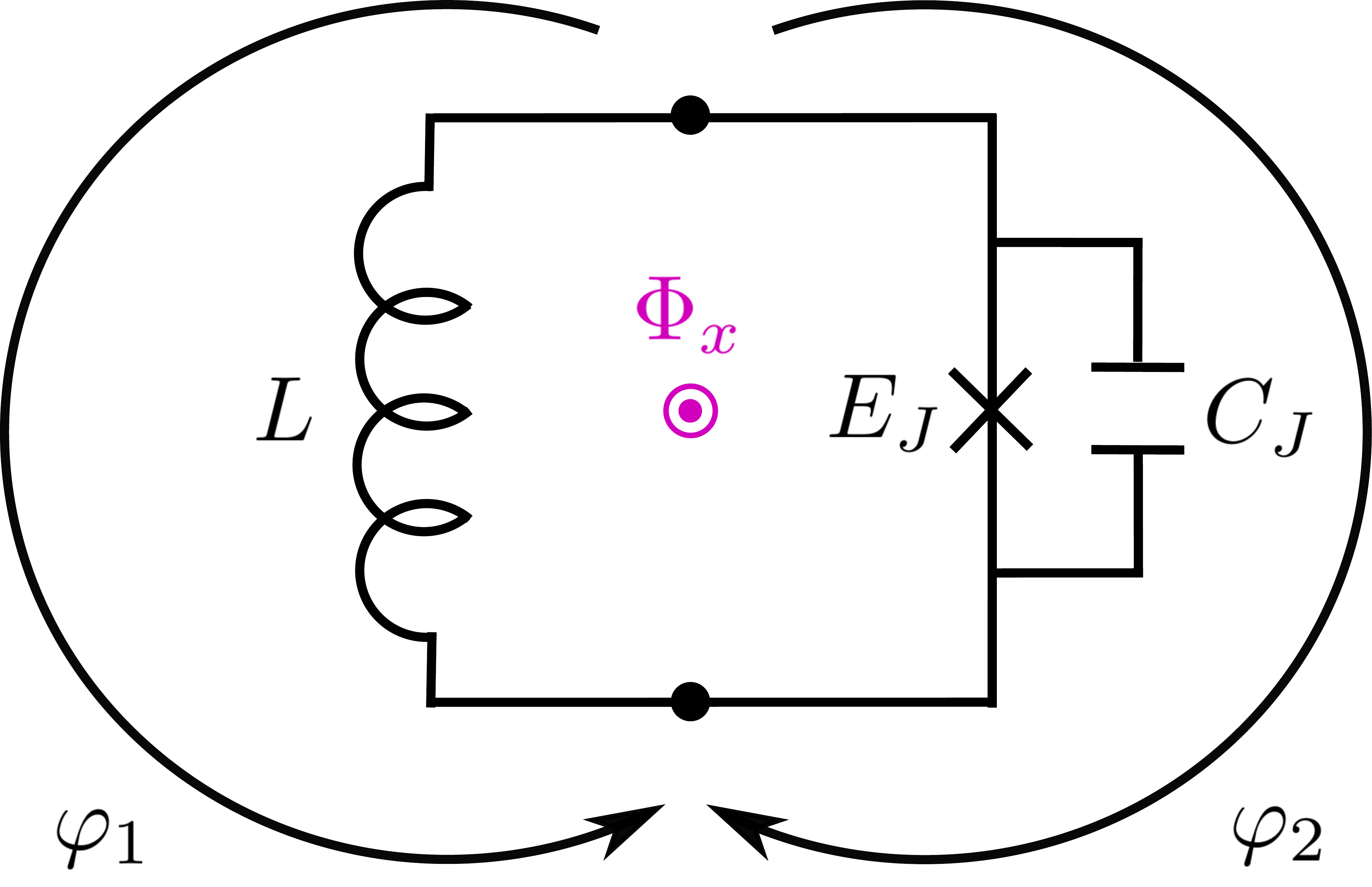}}
\hspace{5pt}
\captionbox{The two symbols describe the orientation of the magnetic field in relation to the superconducting loop where it is applied. The arrows always show in direction of the magnetic north pole.\label{fig:magnetic_field}}
  [.53\linewidth]{\includegraphics[height=2.5cm]{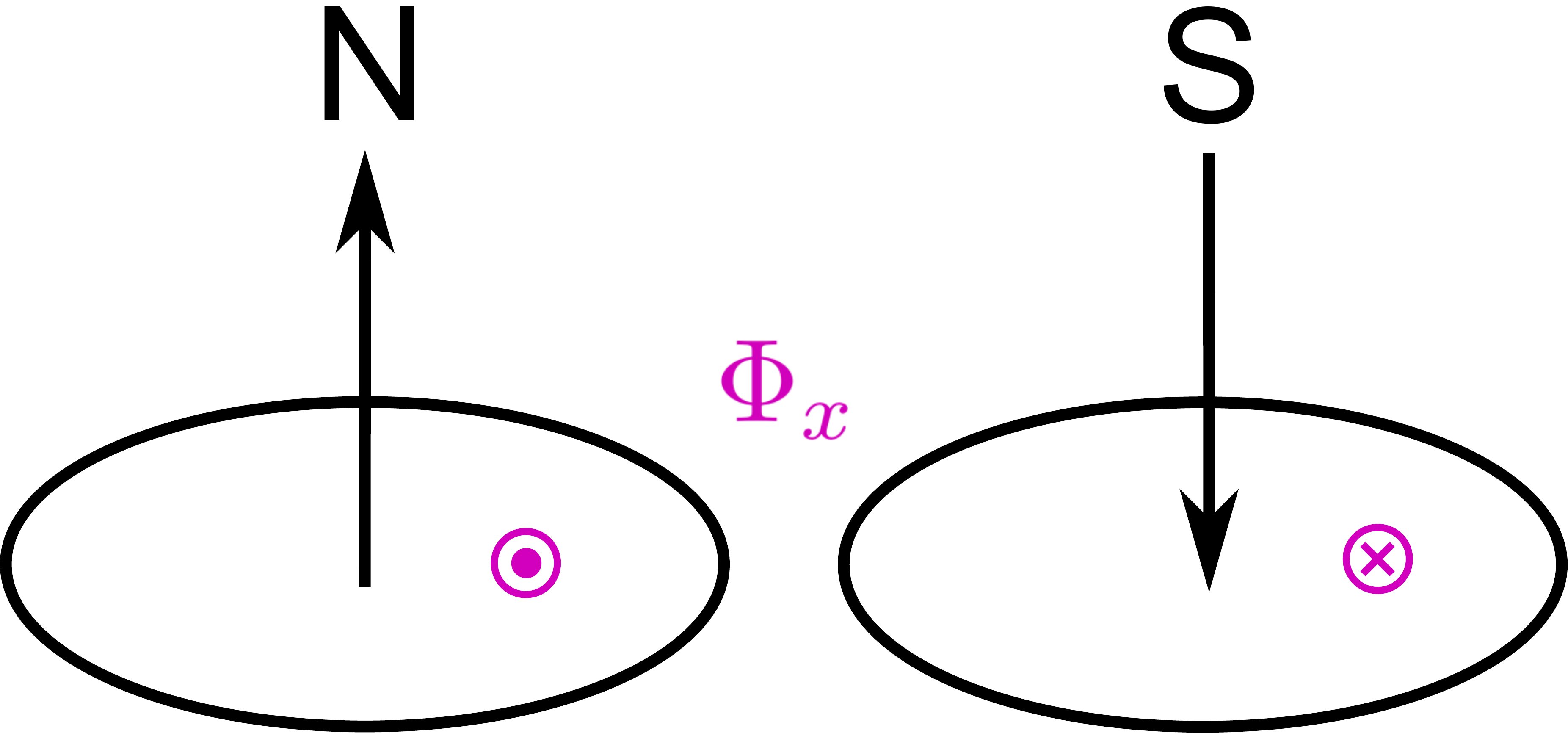}}
\end{figure}

The circuit shown in Fig.~\ref{fig:example1} contains such a loop of superconducting material, formed by the Josephson junction and the inductor. When we apply an external flux through that loop, the concept of node phases used in the previous sections breaks down due to the Aharonov-Bohm effect. It makes sense to use branch phases instead, as denoted in the figure. Intuitively, the branch phase $\varphi_1$ across the inductor should be equal to the branch phase $\varphi_2$ across the Josephson junction, as both circuit elements connect the same nodes. However, though their start and end points are the same, these branch phases correspond to different path integrals according to Eq.~\ref{eq:aharonov}.
The flux quantization rule now demands that 

\begin{align}
\varphi_2 - \varphi_1 - 2\pi \, \Phi_x/\Phi_0 = 2\pi \, m 
\label{eq:flux_quantization}
\end{align}

for integer $m$, where $\Phi_x$ denotes the external magnetic flux threading the loop between the inductor and the Josephson junction. This means that the deviation $\varphi_1 - \varphi_2$ between the two branch phases on either side of the loop is equal to the rescaled external flux $2\pi \, \Phi_x/\Phi_0$ up to an integer multiple of $2\pi$. \\
In terms of signs note that the magnetic field is taken to point out of the drawing, as explained in Fig.~\ref{fig:magnetic_field}. The branch phase $\varphi_2$ across the Josephson junction goes around the loop in clockwise direction, while the branch phase $\varphi_1$ across the inductor goes in counterclockwise direction. Hence, $\varphi_2$ appears in Eq.~\ref{eq:flux_quantization} with a positive sign, while $\varphi_1$ appears with a negative. Even though $\varphi_1 \neq \varphi_2$ due to the external flux, we can use Eq.~\ref{eq:flux_quantization} to eliminate $\varphi_2$. The potential energy for this circuit can thus be written as

\begin{align*}
\mathcal{U} &= \left(\frac{\Phi_0}{2\pi}\right)^2 \frac{\varphi_1^2}{2 L} - E_J \cos(\varphi_2) \\
&= \left(\frac{\Phi_0}{2\pi}\right)^2 \frac{\varphi_1^2}{2 L} - E_J \cos\left(\varphi_1 + \frac{2\pi}{\Phi_0} \Phi_x\right). \numberthis
\label{eq:pot_flux}
\end{align*}

In order to save the concept of node fluxes, we can follow again Devoret~\cite{Devoret1997} and define a \textit{spanning tree} in the circuit, that is a collection of branches that connects all the nodes in the circuit, but has no closed paths. For a circuit with $k$ nodes, the spanning tree has $k - 1$ elements. The phase differences across the circuit elements in the tree are taken to be the independent variables, in which the circuit is described. They are simply the differences between the corresponding node phases. Each remaining branch now defines a unique closed path in the circuit when added to the spanning tree. The phase differences across these elements include the external flux, in addition to the difference between the node phases. With this strategy, we ensure that the external fluxes are not included twice, while the number of independent variables still corresponds to the number of nodes minus one. Importantly, the sign of the external flux depends on its orientation in relation to the phase variables, see Figs.~\ref{fig:example1} and \ref{fig:magnetic_field}. \\
To understand the role of the external flux on a system such as the one depicted in Fig.~\ref{fig:example1}, we have to remember that the derivations of frequencies, anharmonicities and coupling terms as done above, all rely on series expansions in $\Phi_i = \Phi_0/(2\pi)\,\varphi_i$ around its equilibrium position, while fixing the other variables $\Phi_{j\neq i}$ at their equilibrium positions. Without external fluxes these equilibrium positions, which correspond to the minimum of the potential energy, are always at $\Phi_i = 0$. Looking at the potential energy of the system described above (Eq.~\ref{eq:pot_flux}), it is clear that the position of the potential minimum depends on the external flux, as well as on the parameters $L$ and $E_J$.

\begin{figure}[t]
\centering
\captionbox{Potential energy of the circuit depicted in Fig.~\ref{fig:example1} for a weakly anharmonic case with $E_J < E_L/2$ for external fluxes of $\Phi_x = 0, \Phi_0/4, \Phi_0/2$. The form of the potential stays roughly the same.\label{fig:shifting_minimum}}%
  [.48\linewidth]{\includegraphics[width=.48\linewidth]{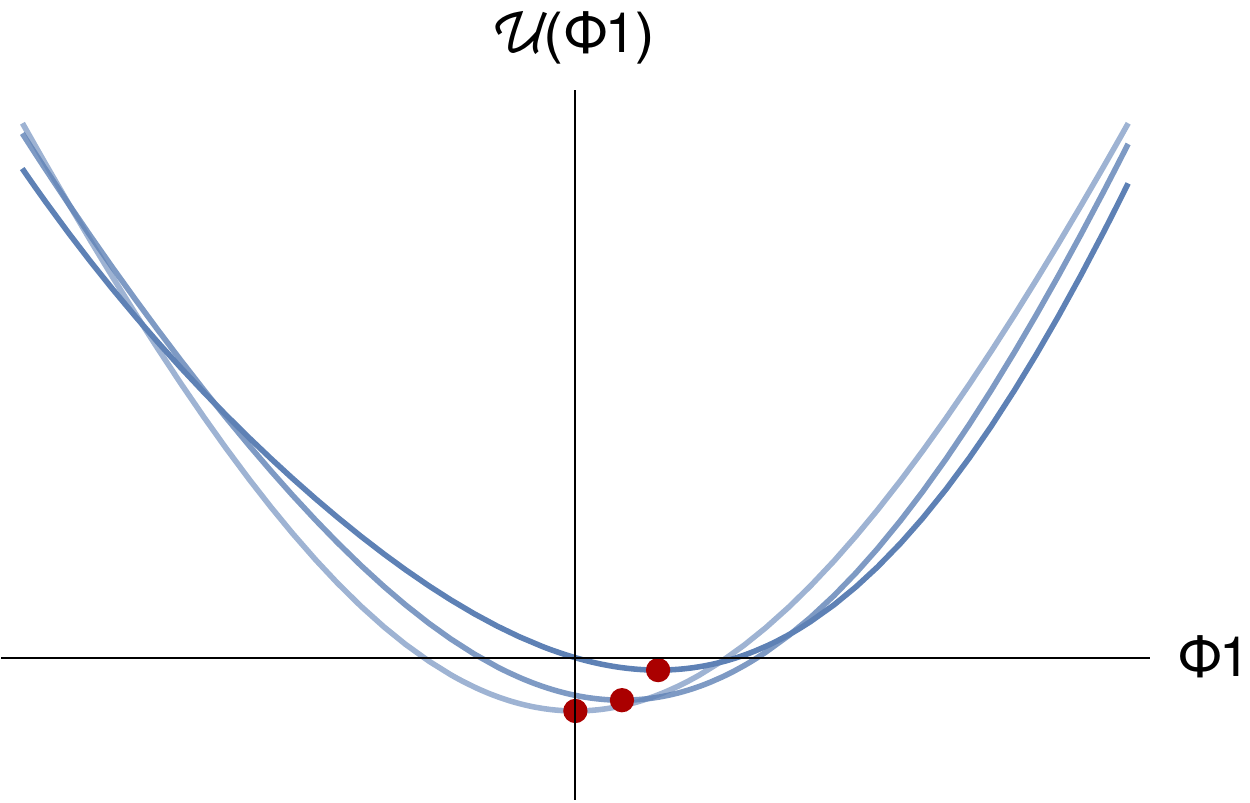}}
\hspace{5pt}
\captionbox{Potential energy of the same circuit for a more anharmonic case with $E_J > E_L/2$ for an external flux of $\Phi_x = 0$ (single well potential) and $\Phi_x = \Phi_0/2$ (double-well potential).\label{fig:double_well}}
  [.48\linewidth]{\includegraphics[width=.48\linewidth]{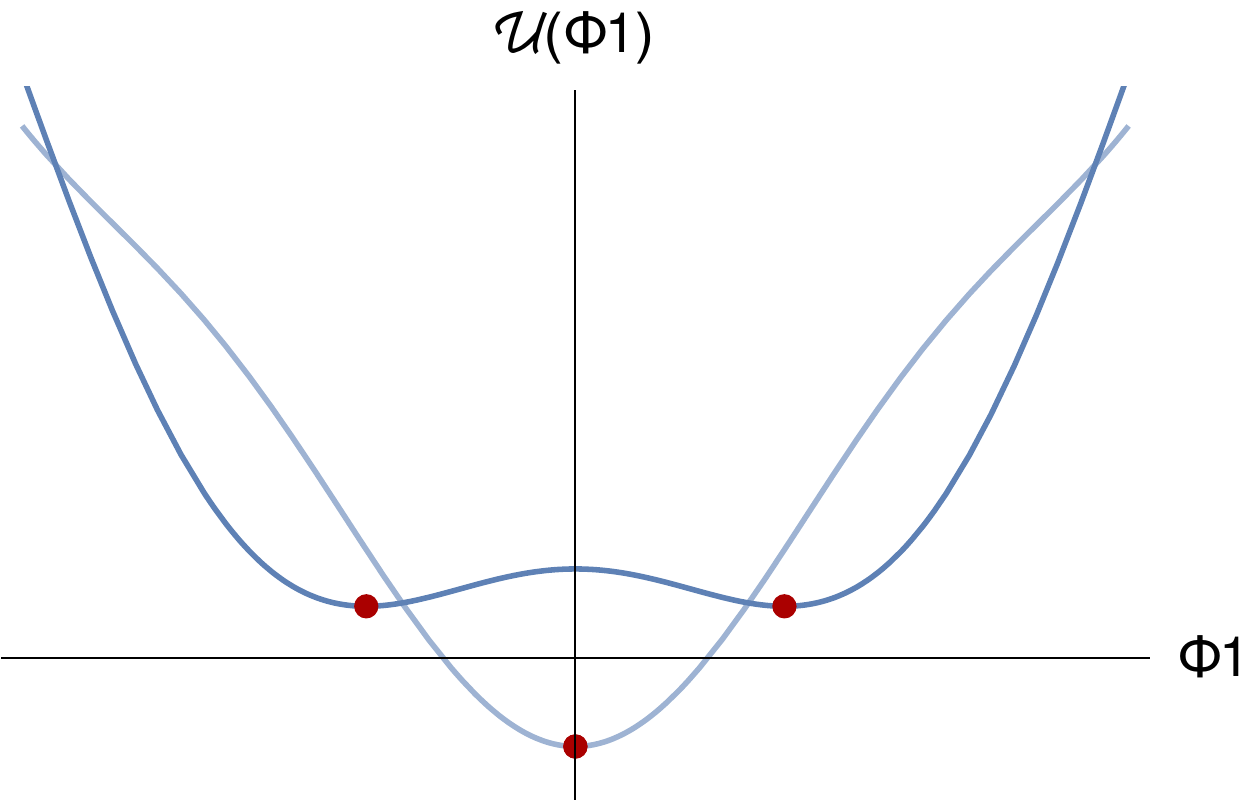}}
\end{figure}

Figure~\ref{fig:shifting_minimum} shows the potential energy of this system for different values of the external flux $\Phi_x$ for a weakly anharmonic case, where the energy of the linear inductance $E_L = (\Phi_0/(2\pi))^2/L$ is dominant compared to the energy of the Josephson junction $E_J$. In such weakly anharmonic cases the potential energy is governed by the harmonic potential due to the linear inductance and has a form similar to a parabola with a single minimum. In this case, the external flux shifts the potential minimum, but leaves the general form of the potential roughly the same. In more anharmonic cases, where $E_J$ is dominant, the influence of the anharmonic cosine term in the potential energy becomes evident. If the influence of the cosine is strong enough, the potential will have more than one minimum. However, as long as there are no external fluxes, the deepest one will always be at $\varphi_1 = 0$ due to the linear inductance. This strong influence of the cosine potential has the effect that external fluxes can change the potential drastically - for example from a single well potential at zero flux to a double-well potential at a flux of $\Phi_x = \Phi_0/2$, as shown in Fig.~\ref{fig:double_well}. Explicitly, the potential given in Eq.~\ref{eq:pot_flux}, has a double-well form with two symmetric minima left and right of $\varphi_1 = 0$ as soon as $E_J \geq E_L/2$.\\
While such a double-well potential can also be used to implement a qubit, it would be a whole different kind of qubit than the one presented in Sec.~\ref{sec:qubit}. So-called flux qubits~\cite{Orlando1999} employ double-well potential with each well representing a qubit state. As opposed to that, weakly anharmonic transmon-type qubits are defined by a single well potential or a selected well within a cosine potential, as done in Sec.~\ref{sec:qubit}. They can thus be analyzed using harmonic approximations, while treating the anharmonicity as a perturbation. 
Depending on the parameter regime, the circuit in Fig.~\ref{fig:example1} can thus either implement a single-well transmon-type qubit or a double-well flux-qubit as shown in Figs.~\ref{fig:shifting_minimum} and \ref{fig:double_well}, respectively.

\section{ELIMINATION OF DEGREES OF FREEDOM}
\label{sec:cholesky}

We will now come to another example circuit in order to introduce a very useful elimination technique for superfluous degrees of freedom. While we stated that the number of independent degrees of freedom in a circuit corresponds to the number of nodes minus one, we can oftentimes reduce this number by applying simple addition rules for parallel or series connections of inductances or capacitances. This leads to a simplified equivalent circuit with fewer nodes that behaves exactly as the original circuit.

\sidecaptionvpos{figure}{c}
\begin{SCfigure}[50][tb]
\centering
    \includegraphics[height=4.5cm]{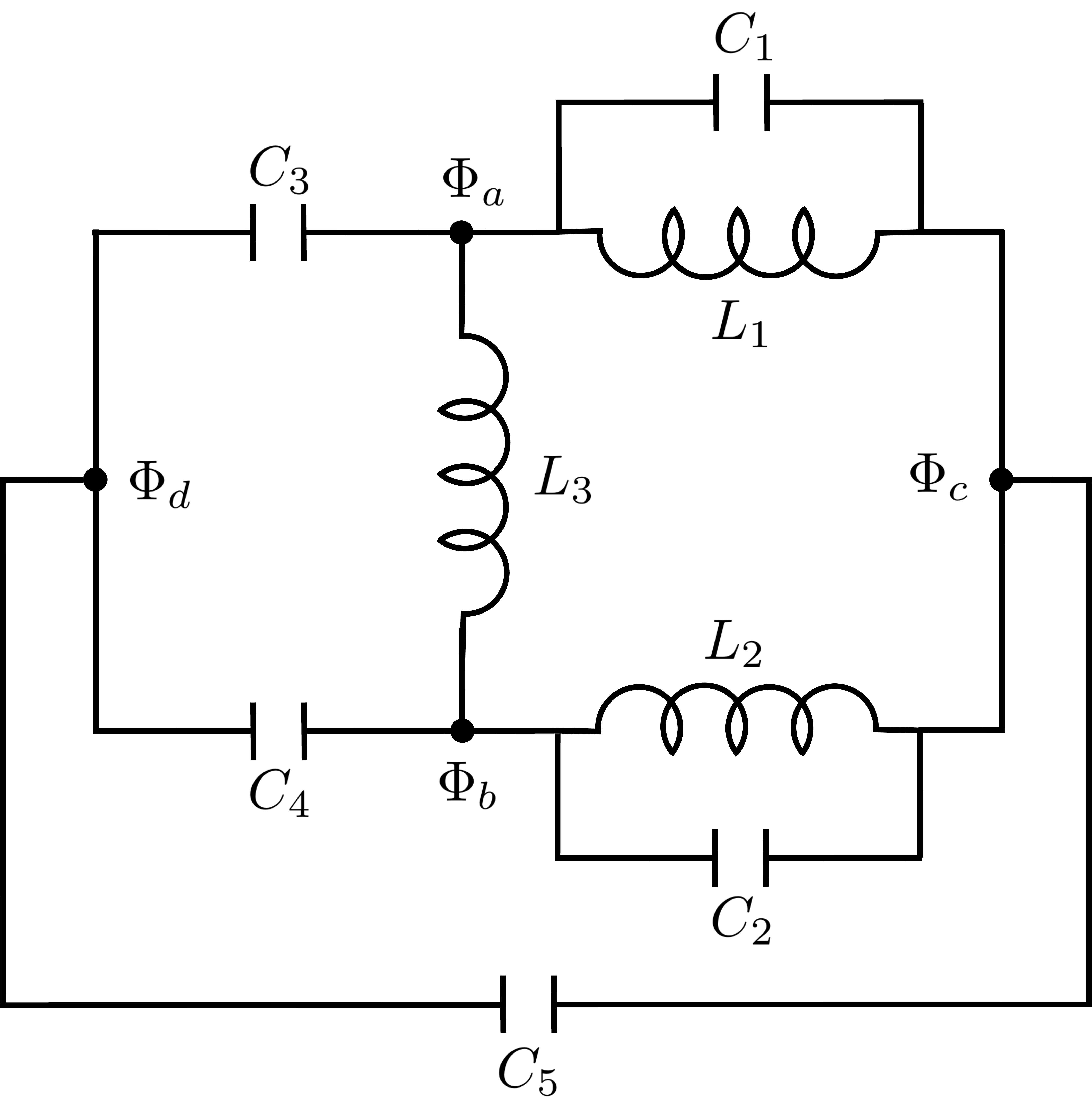}
  \caption{Example circuit with three degrees of freedom. One degree of freedom does not have a potential energy term and can therefore be decoupled from the others with the simple elimination technique presented here. As it decouples, it will not influence the evolution of the other degrees of freedom and can thus be discarded.\label{fig:example2}}
\end{SCfigure}

Even when these simple rules do not apply, it is at times possible to further reduce the number of degrees of freedom. In some cases, one can use a Born Oppenheimer approximation to eliminate degrees of freedom~\cite{Divincenzo2006}, arguing that some degrees of freedom move much faster than others. 
However, the method presented here does not need any approximations.
It is usually applicable whenever a degree of freedom appears in the kinetic but not in the potential energy, or vice versa.
The example circuit shown in Fig.~\ref{fig:example2} has four nodes, that is at maximum three independent degrees of freedom. We will show here that one degree of freedom decouples from the others and can thus be discarded. \\
As soon as circuits get larger, it makes sense to write the Lagrangian in a matrix representation, such as

\begin{align}
\mathcal{L} = \frac{1}{2}\dot{\mathbf{\Phi}}^T \mathbf{C} \, \dot{\mathbf{\Phi}} - \frac{1}{2}\mathbf{\Phi}^T \mathbf{M}_0 \mathbf{\Phi} + \sum_i E_{Ji} \cos\left(\frac{2\pi}{\Phi_0} \Phi_i\right),
\label{eq:lag_matrix}
\end{align}

where $\mathbf{\Phi}$ is a vector that represents the variables in which the circuit is described, $\mathbf{C}$ is the capacitance matrix and $\mathbf{M}_0$ a matrix containing the linear inductances. For simplicity, the chosen example here does not contain non-linear inductances. We will describe the circuit in terms of the following variables

\begin{align}
\Phi_1 = \Phi_a - \Phi_c \qquad \Phi_2 = \Phi_b - \Phi_c \qquad \Phi_3 = \Phi_a - \Phi_d,
\end{align}

which are three flux differences in the circuit that do not form a closed path and therefore are linearly independent. Using 

\begin{align}
\mathbf{\Phi}^T = (\Phi_3, \Phi_2, \Phi_1),
\label{eq:order}
\end{align}

the capacitance matrix for the circuit depicted in Fig.~\ref{fig:example2} yields

\begin{align}
\mathbf{C}=
\begin{pmatrix}
C_3 + C_4 + C_5 & C_4 & -C_4 - C_5 \\
C_4 & C_2 + C_4 & -C_4 \\
-C_4 - C_5 & -C_4 & C_1 + C_4 + C_5
\end{pmatrix},
\label{eq:capacitance_matrix}
\end{align}

while the linear inductance matrix is given by

\begin{align}
\mathbf{M}_0=
\begin{pmatrix}
0 & 0 & 0 \\
0 & \frac{1}{L_2} + \frac{1}{L_3} & -\frac{1}{L_3} \\
0 & -\frac{1}{L_3} & \frac{1}{L_1} + \frac{1}{L_3}
\end{pmatrix}.
\end{align}

Looking at the inductance matrix, we see that it does not have full rank, as there is no term including $\Phi_3$ in the potential energy. This special situation, where one degree of freedom does not appear in either the kinetic or the potential energy, gives us the possibility to decouple it with an easy transformation. There are several linear transformations that can be used to diagonalize a matrix. One of them is the so-called \textit{Cholesky decomposition} (see App.~\ref{app:cholesky}), where a Hermitian positive-definite matrix is represented as the product of an upper triangular matrix and its transpose. We will exploit the triangular form of the Cholesky decomposition by defining the following transformation matrix

\begin{align}
\mathbf{R}=
\begin{pmatrix}
1 & \frac{C_4}{C_3 + C_4 + C_5} &  \frac{- C_4 - C_5}{C_3 + C_4 + C_5} \\
0 & 1 & 0 \\
0 & 0 & 1
\end{pmatrix}.
\end{align}

This matrix consists of the first line of the Cholesky decomposition of $\mathbf{C}$ (Eq.~\ref{eq:capacitance_matrix}), rescaled to be dimensionless, while the rest of the matrix is a simple identity transformation.
This defines a variable transformation $\mathbf{\tilde\Phi} = \mathbf{R \, \Phi}$ with $\mathbf{\tilde\Phi}^T = (\Phi_*, \Phi_2, \Phi_1)$ that leaves $\Phi_1$ and $\Phi_2$ unchanged, while decoupling the first variable. Note that we have chosen the order of the variables in Eq.~\ref{eq:order}, such that the variable we want to discard is the first one.
The capacitance matrix transforms to

\begin{align*}
\mathbf{\tilde C} &= \left(\mathbf{R}^{-1}\right)^T \mathbf{C} \, \mathbf{R}^{-1} \\
&= \begin{pmatrix}
C_3 + C_4 + C_5 & 0 &0 \\
0 & C_2 + \frac{C_4 (C_3 + C_5)}{C_3 + C_4 + C_5} & - \frac{C_3 C_4}{C_3 + C_4 + C_5} \\
0 & - \frac{C_3 C_4}{C_3 + C_4 + C_5} & C_1 + \frac{C_3 (C_4 + C_5)}{C_3 + C_4 + C_5}
\end{pmatrix}, \numberthis
\end{align*}

whereby the first variable is decoupled from the other two. As the transformation $\mathbf{R}$ acts as an identity transformation on $\Phi_1$ and $\Phi_2$, the inductance matrix remains the same, that is

\begin{align}
\mathbf{\tilde M}_0 = \left(\mathbf{R}^{-1}\right)^T \mathbf{M}_0 \, \mathbf{R}^{-1} = \mathbf{M}_0.
\end{align}

The transformation corresponds to the mapping 

\begin{align}
\Phi_3 = \Phi_* + \frac{(C_4 + C_5) \Phi_1 - C_4 \Phi_2}{C_3 + C_4 + C_5} ,
\end{align}

which is a linear, invertible transformation. In terms of the original node variables, the new variable $\Phi_*$ is given by

\begin{align}
\Phi_* = \frac{C_3 \Phi_a + C_4 \Phi_b + C_5 \Phi_c}{C_3 + C_4 + C_5} - \Phi_d.
\end{align}

In these new variables, the Lagrangian has the following form

\begin{align*}
\mathcal{L} &= \left(\frac{C_1}{2} + \frac{C_3 C_5}{2(C_3 + C_4 + C_5)}\right) \dot \Phi_1^2 + \left(\frac{C_2}{2} + \frac{C_4 C_5}{2(C_3 + C_4 + C_5)}\right) \dot \Phi_2^2 \\
&+\frac{C_3 C_4}{2(C_3 + C_4 + C_5)} \left(\dot \Phi_1 - \dot \Phi_2\right)^2 + \frac{C_3 + C_4 + C_5}{2} \dot \Phi_*^2\\
&- \frac{1}{2L_1} \Phi_1^2 - \frac{1}{2L_2} \Phi_2^2 - \frac{1}{2L_3} \left(\Phi_1 - \Phi_2\right)^2\numberthis.
\label{eq:lagr3}
\end{align*}

In this description, it becomes clear that the variable $\Phi_*$ decouples from the other two and therefore does not influence their evolution. It can be discarded, thus treating the system as a system with two degrees of freedom only. This technique can be very valuable when treating systems with many degrees of freedom, as it leads to a description where only the relevant degrees of freedom play a role. Importantly, the degrees of freedom we want to keep are unaffected by the transformation. Note that no approximation had to be made for this elimination procedure.

\section{DISSIPATIVE CIRCUITS}

The example circuits treated until here all consisted of dissipation-free elements such as capacitors, inductors, and Josephson junctions. When we want to take lossy elements such as resistances into account, we have to adapt our treatment of the circuits to the resulting dissipation. A very useful method for the treatment of dissipative circuits is presented in Refs.~\cite{Burkard2004, Burkard2005} by Burkard and others. \\
In order to describe dissipation, they rely on the so-called \textit{Caldeira-Leggett} formalism~\cite{Caldeira1983}. 
In this formalism, a dissipative circuit is divided into two parts: a dissipation-free system and the environment it interacts with. This environment is depicted as a bath with infinitely many degrees of freedom. While in principle, the Hamiltonian formalism cannot capture dissipation, as the energy is always conserved, the irreversible loss of energy due to dissipation can be formally described as a transition of energy from the system to the bath. \\
Burkard describes electrical circuits as directed graphs with arbitrarily assigned directions. The relevant degrees of freedom are chosen in terms of a spanning-tree (as also done by Devoret in Ref.~\cite{Devoret1997}), a subgraph that connects all the nodes in the circuit without any closed paths. Ref.~\cite{Burkard2005} defines clear rules on how this spanning-tree should be chosen.
Lossy elements, such as external impedances are represented as a bath that interacts with the dissipation-free part of the circuit.
While Burkard's method is very systematic, it misses the pedagogy from Devoret's approach, as many intermediate results are needed before the Hamiltonian can be written down. However, with these intermediate results, it is possible to write down simple formulas for relaxation and decoherence times. Appendix~\ref{app:burkard} contains a discussion of Burkard's method by means of an explicit example, showing how the system-bath coupling leads to finite relaxation and decoherence times for the modes of the dissipation-free system.

\chapter{COUPLING SCHEMES}
\label{chapt:coupling}




One of the main challenges in superconducting qubit architectures is to couple qubits in a well-controlled manner, especially in circuit constructions that involve many qubits. The difficulty consists in  performing quantum gates on selected qubits without corrupting the others. In order to avoid unwanted cross-couplings, qubits are oftentimes coupled via harmonic resonators, which act as buses that mediate the interaction \cite{Ghosh2013}. This is supposed to make the interaction more controllable.\\
In this chapter, we will examine different coupling types between qubits and harmonic resonators, with a focus on
two inherently different types of coupling, which will be designated as \textit{transverse} and \textit{longitudinal coupling}. Note that parts of this chapter were already published in Ref.~\cite{Richer2016}. \\
Most commonly, superconducting qubit architectures work with transverse coupling \cite{Wallraff2004, Blais2004}, which is known from cavity quantum electrodynamics and involves coupling of the displacement degree of freedom of a resonator to the $\sigma_x$ degree of freedom of the qubit. 
In contrast to this stands longitudinal coupling \cite{Billangeon2015, Didier2015, Royer2017, Geller2015, Weber2017}, which means coupling of the displacement degree of freedom of a resonator to the $\sigma_z$ degree of freedom of the qubit. 
While transverse coupling naturally appears in transmon-like circuit constructions, longitudinal coupling is much harder to implement and hardly ever the only coupling term present.
Nevertheless, we will see that longitudinal coupling offers some remarkable advantages, for example with respect to scalability \cite{Billangeon2015} and readout \cite{Didier2015}. 


\section{TRANSVERSE COUPLING}
\label{sec:transverse}
Let us consider the case of transverse coupling, which involves coupling of the displacement degree of freedom of a resonator to the $\sigma_x$ degree of freedom of the qubit. The corresponding Hamiltonian is the \textit{Rabi Hamiltonian}

\begin{align}
\mathcal{H} = \hbar \,\omega_r a^\dagger a + \hbar \,\frac{\Delta}{2} \sigma_z + \hbar\,g \,  \sigma_x (a^\dagger + a),
\label{eq:Rabi}
\end{align}

where $\omega_r$ is the frequency of the resonator and $\Delta$ is the gap of the qubit, which is taken to be a two-level system. This Hamiltonian is known from cavity quantum electrodynamics, where it describes the interaction of a two-level atom with a quantized mode of an electro-magnetic field in a cavity. The atom can change between its ground and excited state by spontaneous emission or absorption of a photon from the field. In the case treated here, the qubit is an artificial two-level system, playing the role of the atom. It is therefore sometimes referred to as an artificial atom~\cite{Bishop2010}, while the physics of such artificial atoms interacting with resonators is designated circuit quantum electrodynamics. \\
In order to work with the Rabi Hamiltonian, it is oftentimes simpler to go to a diagonal frame. If the coupling $g$ between the qubit and the resonator is small compared to their detuning $\Delta - \omega_r$ (this is called the \textit{dispersive regime}), the Rabi Hamiltonian can be approximately diagonalized using the \textit{Schrieffer-Wolff} unitary transformation 

\begin{align}
\mathcal{U}_{SW} = \exp{\left(\gamma (a^\dagger \sigma_- - a \sigma_+) - \bar{\gamma} (a^\dagger \sigma_+ - a \sigma_-)\right)},
\label{eq:SW}
\end{align}

where the qubit raising and lowering operators are given by

\begin{align}
\sigma_+ = \frac{\sigma_x - i \,\sigma_y}{2} \qquad \sigma_-  = \frac{\sigma_x + i \,\sigma_y}{2}
\end{align}

(see also App.~\ref{app:operators}), while $\gamma = g/(\Delta - \omega_r)$, and $\bar\gamma = g/(\Delta + \omega_r)$ (see \cite{Winkler2003}, \cite{Zueco2009} and \cite{Richer2013}). 
To second order in $g$ we find 

\begin{align}
\mathcal{H}' = \hbar \,\omega_r a^\dagger a + \hbar \,\frac{\Delta}{2} \sigma_z + \hbar \,\frac{\bar\chi}2 \, \sigma_z (a^\dagger + a)^2 
\label{eq:dispersive}
\end{align}

with $\bar\chi = g\,(\gamma + \bar\gamma)$. Note that this is a perturbative treatment. The last term in Eq. \ref{eq:dispersive} is the so-called \textit{dispersive shift} that makes the transition frequency of the resonator dependent on the qubit's state and vice versa. While this shift is useful for read-out, it also means that the always-on interaction between the resonator and the qubit will inevitably entangle them. 
Another undesirable effect of the dispersive shift is that it makes the qubit relaxation rate dependent on the photon lifetime (the so-called \textit{Purcell effect} \cite{Sete2014}).\\
If we couple two qubits to the same resonator and use it as a quantum bus, a perturbative transformation similar to Eq. \ref{eq:SW} will indicate direct always-on coupling between the qubits. In this manner, the resonator can be used to mediate the coupling between the qubits.
However, the always-on coupling can be problematic when we want to scale such a system up to a larger grid, as the interactions can have a very long range and it will be difficult to address the individual qubits independently.  A thorough analysis of such a two-qubit system with transverse coupling to the same resonator is provided in Ref.~\cite{Billangeon2015}.\\
A very common simplification of the Rabi Hamiltonian is the Jaynes-Cummings Hamiltonian, which we obtain by neglecting the fast-rotating terms in Eq. \ref{eq:Rabi}. To understand that, we move to a rotating frame using the unitary transformation $\mathcal{U} = \exp{(i \,\mathcal{H}_0/\hbar \,t)}$ with

\begin{align}
\mathcal{H}_0 = \hbar \,\omega_r a^\dagger a + \hbar \,\frac{\Delta}{2} \sigma_z,
\end{align}

where $\mathcal{H}_0$ includes the non-interacting terms in Eq. \ref{eq:Rabi}. In this rotating frame we find the coupling terms rotating at either the sum or the difference of the frequencies of qubit and resonator, namely

\begin{align*}
\mathcal{H}_\text{rot} &= \mathcal{U}\, \mathcal{H}\, \mathcal{U}^\dagger + i \, \dot{\mathcal{U}} \,\mathcal{U}^\dagger \\
&= \hbar \,g \left(a^\dagger \sigma_- e^{i (\Delta - \omega_r) t} + a \sigma_+ e^{-i (\Delta - \omega_r) t} \right. \\
&\left. + a^\dagger \sigma_+ e^{i (\Delta + \omega_r) t} + a \sigma_- e^{-i (\Delta + \omega_r) t}\right), \numberthis
\end{align*}

where the $\sigma_x$ operator was rewritten in terms of raising and lowering operators. Assuming that $|\Delta - \omega_r| \ll \Delta + \omega_r$, we can neglect the fast rotating terms in a so-called \textit{rotating wave approximation}. Transforming back to the original non-rotating frame, the Hamiltonian is given by

\begin{align}
\mathcal{H}_{JC} = \hbar \,\omega_r a^\dagger a + \hbar \,\frac{\Delta}{2} \sigma_z + \hbar \,g \,  (\sigma_- a^\dagger + \sigma_+ a),
\label{eq:JC}
\end{align}
 
the Jaynes-Cummings Hamiltonian. As opposed to the Rabi Hamiltonian, the Jaynes-Cummings Hamiltonian has an exact analytic solution~\cite{Bishop2010} due to its block-diagonal  form. In the dispersive limit, it can be diagonalized by a transformation similar to the one done above (Eq.~\ref{eq:SW}). It yields

\begin{align}
\mathcal{H}' = \hbar \,\left(\omega_r + \chi\, \sigma_z\right) a^\dagger a + \hbar\,\frac{\Delta + \chi}{2} \sigma_z 
\label{eq:dispersive2}
\end{align}

with $\chi = g \, \gamma = g^2/(\Delta - \omega_r)$. The dispersive shift is still present in the Jaynes-Cummings model. The term proportional to $\bar\gamma$ in Eq.~\ref{eq:dispersive}, which is neglected due to the rotating wave approximation, is a so-called \textit{Bloch-Siegert shift}~\cite{Bloch1940}.

\section{LONGITUDINAL COUPLING}
\label{sec:longitudinal}
We will now turn our attention to an inherently different way of coupling a qubit to a resonator, which is designated as \textit{longitudinal coupling}. Consider the Hamiltonian

\begin{align}
\mathcal{H} = \hbar \,\omega_r \, a^\dagger a + \hbar \,\frac{\Delta}{2} \,\sigma_z + \hbar \,g \, \sigma_z (a^\dagger + a),
\label{eq:long}
\end{align}

where the displacement degree of freedom of the resonator is coupled to the $\sigma_z$ degree of freedom of the qubit, instead of the usual (\textit{transverse}) $\sigma_x$-type coupling of the Rabi Hamiltonian (Eq. \ref{eq:Rabi}). A detailed analysis of this coupling type can be found in Ref.~\cite{Billangeon2015}. 
Note that the term \textit{longitudinal coupling} is not uniquely defined and sometimes refers to a coupling type that is diagonal in both degrees of freedom.
As opposed to the case of transverse coupling, there is an exact unitary transformation that diagonalizes this Hamiltonian.
As shown in Ref.~\cite{Billangeon2015}, the so-called \textit{Lang-Firsov transformation} \cite{Firsov1963}

\begin{align}
\mathcal U = e^{\,\theta \,\sigma_z (a^\dagger - a)}
\label{eq:lang-firsov}
\end{align}

with $\theta= g/\omega_r$ directly leads to

\begin{align}
\mathcal{H}' = \mathcal U \mathcal{H}\, \mathcal U^\dagger = \hbar \,\omega_r \, a^\dagger a + \hbar \,\frac{\Delta}{2} \sigma_z - \hbar \, \frac{g^2}{\omega_r} \mathbf{1}.
\label{eq:long_result}
\end{align}

In contrast to Eq.~\ref{eq:dispersive}, there is no dispersive shift, that is no residual interaction between qubit and resonator.
Remarkably, the energies associated with the qubit and the resonator remain unaffected by the transformation.
Note that while the treatment in the transverse case (Eq. \ref{eq:SW}) is only perturbative, the Lang-Firsov transformation is \textit{exact}, with no restrictions on the coupling strength or detuning of the system. This has the advantage that even for large coupling strengths, the qubit relaxation rate is not degraded by a dependence on the finite photon lifetime. We have thus gone to a frame where the Hamiltonian is diagonal without any residual coupling.\\
Such a frame where qubit and resonator are uncoupled, may seem unhelpful, remembering that the resonator is supposed to mediate the coupling between neighboring qubits. However, the coupling between qubit and resonator can be turned on in the diagonal frame by driving the qubit. A transverse drive on the qubit with amplitude $\Omega$ and frequency $\omega_d$

\begin{align}
\mathcal{H}_d (t) = \Omega \cos(\omega_d\, t + \phi) \sigma_x
\end{align}

has to be transformed to the diagonal frame, again using Eq. \ref{eq:lang-firsov}, where it reintroduces a coupling between qubit and resonator (see again Ref.~\cite{Billangeon2015}). The coupling can thus be turned on and off simply by driving. This is clearly advantageous for scalability, as it is difficult to control always-on coupling in larger architectures. In a diagonal frame such as the one given by Eq.~\ref{eq:long_result}, qubit and resonator do not interact, unless we want them to.
As shown explicitly in Ref.~\cite{Billangeon2015}, a drive at the qubit's frequency $\omega_d = \Delta$, enables single-qubit operations (within the rotating wave approximation), while a drive at $\omega_d = |\Delta \pm  \omega_r|$ leads to sideband transitions between the qubit and the resonator that can be used to implement a controlled-phase gate between neighboring qubits (see Sec.~\ref{sec:gate}).

\section{SCALABLE ARCHITECTURE}
\label{sec:scalable_architecture}

\begin{SCfigure}[50][tb]
\centering
    \includegraphics[height=4cm]{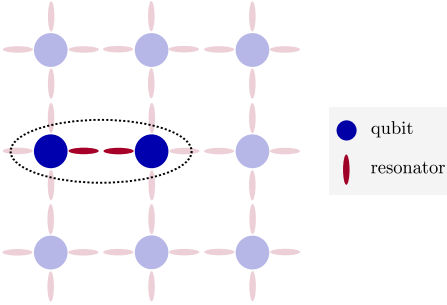}
  \caption{In Ref. \cite{Billangeon2015}, a 2D lattice of qubits is proposed in which each qubit couples to four resonators via its longitudinal degree of freedom and every resonator couples to a resonator of the next unit cell via a conjugate degree of freedom. The encircled unit corresponds to Eq.~\ref{eq:long2}.
    \label{fig:grid}}
\end{SCfigure}

As proposed in Ref.~\cite{Billangeon2015}, this idea can be extended to a grid (see Fig.~\ref{fig:grid}), where a unit cell consists of a qubit coupled longitudinally to four resonators and every resonator is coupled to a resonator of the neighboring unit cell.  The proposed Hamiltonian for two qubits coupled via two resonators is

\begin{align*}
\mathcal{H} &= \sum_{i=1}^2  \hbar \,\omega_{r,i}\, a_i^\dagger a_i +  \hbar \,\frac{\Delta_i}{2} \, \sigma_{i}^z + \hbar \,g_i \, \sigma_i^z \, (a_i^\dagger + a_i)  \\
&- \hbar \,g_c \,(a_1^\dagger - a_1)(a_2^\dagger - a_2). \numberthis
\label{eq:long2}
\end{align*}

\sidecaptionvpos{figure}{c}

Note that the two adjacent resonators are coupled through a conjugate degree of freedom, that is $i (a^\dagger - a)$ instead of $(a^\dagger + a)$. As shown explicitly in Ref.~\cite{Billangeon2015}, the Hamiltonian can be exactly diagonalized by a series of unitary transformations leading to

\begin{align}
\mathcal{H}' = \hbar \, \omega_+ a_1^\dagger a_1 + \hbar \, \omega_- a_2^\dagger a_2 + \sum_{i=1}^2 \hbar \, \frac{\Delta_i}{2} \sigma_i^z - \hbar \, \left(\frac{g_1^2}{\omega_{r,1}} + \frac{g_2^2}{\omega_{r,2}}\right) \mathbf{1}.
\label{eq:long_result2}
\end{align}

While the qubit frequencies are unaffected by the transformations, the resonator frequencies get rescaled to

\begin{align}
\omega_\pm^2 = \frac{\omega_{r,1}^2+\omega_{r,2}^2}{2} \pm \frac{1}{2} \sqrt{(\omega_{r,1}^2-\omega_{r,2}^2)^2+16 \,g_c^2 \, \omega_{r,1} \,\omega_{r,2}}.
\label{eq:rescaledfreq}
\end{align}

As the qubit Hamiltonian commutes with the longitudinal coupling term and the diagonalizing transformation (Eq.~\ref{eq:lang-firsov}), it is clear that adding more resonators coupled to the same qubit does not interfere with this diagonalization procedure. The fact that the Hamiltonian with two qubits coupled via two resonators (Eq.~\ref{eq:long2}) is diagonalizable thus means that the Hamiltonian for a grid as depicted in Fig.~\ref{fig:grid} must also be diagonalizable. As described above, a transverse drive on one of the qubits

\begin{align}
\mathcal{H}_d(t) = \Omega \cos(\omega_d \, t + \phi) \sigma_1^x
\label{eq:drive}
\end{align}

has to be transformed to the diagonal frame as well. The interaction induced by this drive will be rigorously confined to a small neighborhood of the qubit being driven, without generating any direct qubit-qubit coupling (see Ref. \cite{Billangeon2015}). 
These sideband transitions are possible due to the absence of the dispersive shift in Eqs. \ref{eq:long_result} and \ref{eq:long_result2}, since a certain transition will stay resonant irrespective of the number of photons in the resonators. \\
Within the rotating wave approximation, we can frequency-select the gate we want to drive, by choosing the appropriate driving frequency $\omega_d$ in Eq.~\ref{eq:drive}.
While single-qubit operations on qubit $i$ are implemented by a drive on the qubit at $\omega_d = \Delta_i$, sideband transitions between the qubit and either one or both neighboring resonators can be driven using respectively $\omega_d = |\Delta_i \pm \omega_\pm|$ or $\omega_d = |\Delta_i \pm \omega_+ \pm \omega_-|$, where $\omega_\pm$ are the effective frequencies of the resonators in the diagonal frame (see Eq.~\ref{eq:rescaledfreq}). There is rigorously no qubit-qubit interaction.

\sidecaptionvpos{figure}{c}
\begin{SCfigure}[50][tb]
\centering
    \includegraphics[height=4cm]{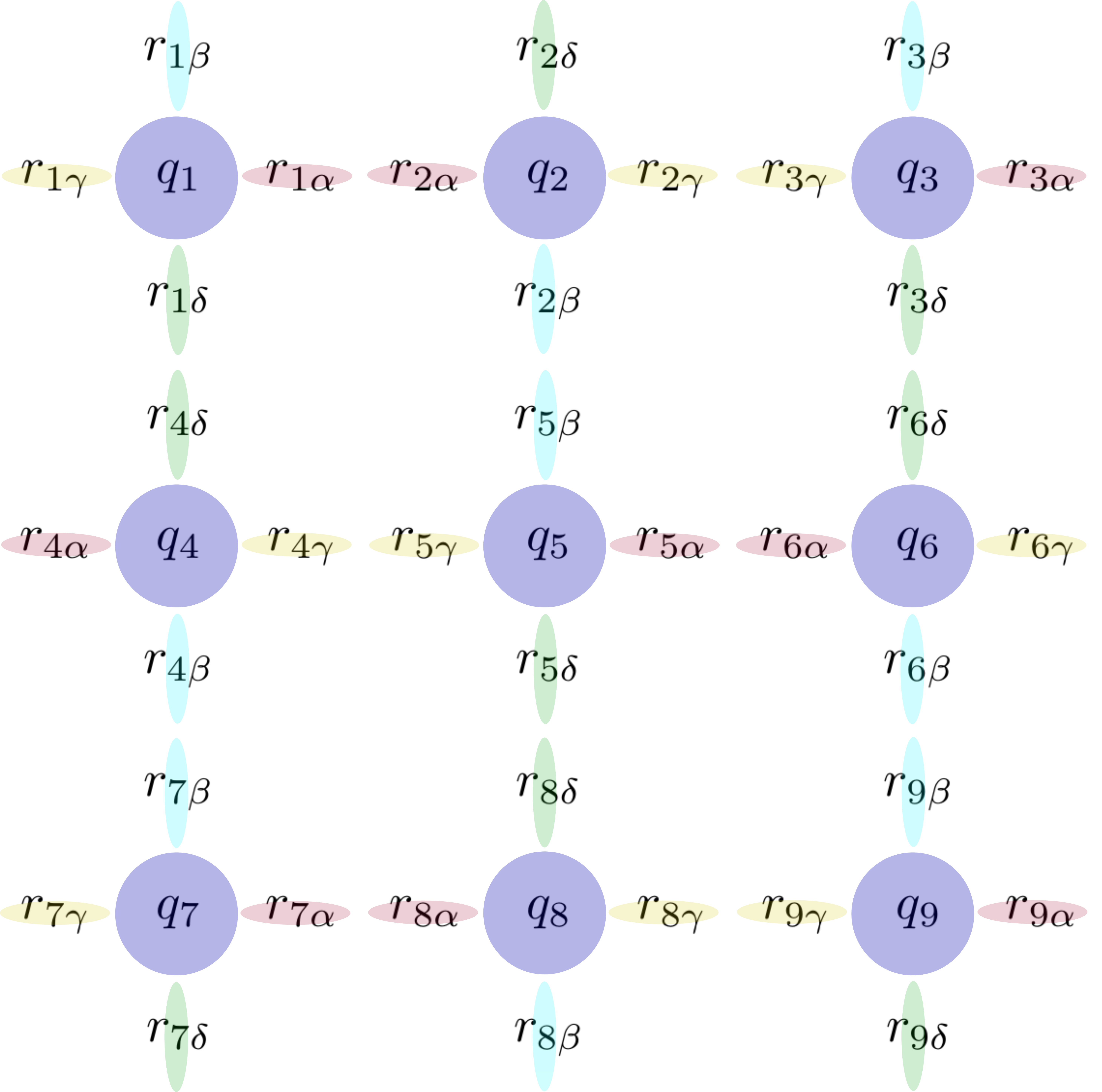}
  \caption{In order to enable unambiguous frequency-selection of sideband transitions between a qubit and its four nearest and four next-nearest resonators, we need at least eight different effective resonator frequencies $\omega_{\pm,l}$, $l=\alpha,...,\delta$. Even if all eight original frequencies are equal, this can be realized just by having four different values of $g_{c,l}$ (these correspond to the different colors in the figure). All qubits could, in principle, have the same frequency. 
    \label{fig:frequency-selection}}
\end{SCfigure}

On a grid, where every qubit has four nearest neighbor resonators and four next-nearest neighbor resonators (see Fig.~\ref{fig:grid}), we need eight different effective frequencies $\omega_{\pm,l}$, $l=\alpha,...,\delta$, for unequivocal frequency selection. 
Remarkably, as can be seen from Eq.~\ref{eq:rescaledfreq}, this can be realized even if all eight original frequencies were equal, just by having four different values of $g_{c,l}$, $l=\alpha,...,\delta$.
As coupling will be strictly restricted to the nearest and next-nearest neighbor resonators of each qubit, all qubits could, in principle, have the same frequency. 
This allows us to choose unambiguously which sideband transition we want to drive (see Fig. \ref{fig:grid}).

\subsection{TWO-QUBIT PHASE GATE}
\label{sec:gate}
As shown explicitly in Ref.~\cite{Billangeon2015}, read-out (see Sec.~\ref{sec:readout_billangeon}) and a controlled-phase gate between two neighboring qubits are possible via a series of sideband transitions between either qubit and one or both resonators. There is never any direct qubit-qubit coupling needed. The fact that all other qubits are unaffected by these actions is a significant advantage concerning the scalability of this scheme.
Their strategy consists of three steps. We first map the state of qubit two on resonator two (that is, its adjacent resonator) and then apply a sideband transition between qubit one and both resonators, which leads to a selective phase accumulation. At last, we map the state of resonator two back on qubit two.\\
Following closely Ref.~\cite{Billangeon2015}, we will assume an arbitrary state for two neighboring qubits, while both resonators between them are in their ground state, that is

\begin{align}
|\psi \rangle &= A_{00} | 0_{q,1} 0_{r,1} 0_{r,2} 0_{q,2}\rangle + A_{01} | 0_{q,1} 0_{r,1} 0_{r,2} 1_{q,2}\rangle \\&
+ A_{10} | 1_{q,1} 0_{r,1} 0_{r,2} 0_{q,2}\rangle + A_{11} | 1_{q,1} 0_{r,1} 0_{r,2} 1_{q,2}\rangle.
\end{align}

Applying the sideband transition $(a_2^\dagger \sigma_2^- + a_2 \sigma_2^+ )$ for half a period maps the state of qubit two on resonator two, that is

\begin{align}
|\psi \rangle &= A_{00} | 0_{q,1} 0_{r,1} 0_{r,2} 0_{q,2}\rangle + i\,A_{01} | 0_{q,1} 0_{r,1} 1_{r,2} 0_{q,2}\rangle \\&
+ A_{10} | 1_{q,1} 0_{r,1} 0_{r,2} 0_{q,2}\rangle + i\,A_{11} | 1_{q,1} 0_{r,1} 1_{r,2} 0_{q,2}\rangle.
\end{align}

Now, we apply the sideband transition $(\sigma_1^+ a_1^\dagger a_2 + \sigma_1^- a_1  a_2^\dagger)$, which couples only to the state $| 0_{q,1} 0_{r,1} 1_{r,2} 0_{q,2}\rangle$, but to none of the others. This leads to a selective phase accumulation, namely

\begin{align}
|\psi \rangle &= A_{00} | 0_{q,1} 0_{r,1} 0_{r,2} 0_{q,2}\rangle - i\,A_{01} | 0_{q,1} 0_{r,1} 1_{r,2} 0_{q,2}\rangle \\&
+ A_{10} | 1_{q,1} 0_{r,1} 0_{r,2} 0_{q,2}\rangle + i\,A_{11} | 1_{q,1} 0_{r,1} 1_{r,2} 0_{q,2}\rangle.
\end{align}

Finally, we map the state of resonator two back to qubit two by applying $(a_2^\dagger \sigma_2^- + a_2 \sigma_2^+ )$ again for half a period. We find

\begin{align}
|\psi \rangle &= A_{00} | 0_{q,1} 0_{r,1} 0_{r,2} 0_{q,2}\rangle + A_{01} | 0_{q,1} 0_{r,1} 0_{r,2} 1_{q,2}\rangle \\&
+ A_{10} | 1_{q,1} 0_{r,1} 0_{r,2} 0_{q,2}\rangle - A_{11} | 1_{q,1} 0_{r,1} 0_{r,2} 1_{q,2}\rangle.
\end{align}

As we can see, this sequence of sideband transitions leads effectively to a two-qubit phase gate for two neighboring qubits. Importantly, this is possible without any direct coupling between the two qubits.

\section{IMPLEMENTING LONGITUDINAL COUPLING}
\label{sec:long}

Having introduced different possibilities to couple a qubit and a resonator in Secs.~\ref{sec:transverse} and \ref{sec:longitudinal}, we would like to translate these to the circuit language from Chapt.~\ref{chapt:quantization}. How do these coupling terms look like in a circuit description? And what kind of circuit could reproduce the coupling terms described above?\\
The resonators and qubits introduced in Sec.~\ref{sec:resonator} are described as harmonic oscillators and weakly anharmonic Duffing oscillators, respectively. In the second quantization formalism, the conjugate variables flux and charge are replaced by the creation and annihilation operators (Eqs.~\ref{eq:quant_r} and \ref{eq:quant_qu}). In a flux representation, we can associate the flux $\Phi \sim a^\dagger + a$ with a displacement degree of freedom and the charge $Q \sim i(a^\dagger - a)$ with a momentum degree of freedom. For weakly anharmonic transmon-type qubits as introduced in Sec.~\ref{sec:qubit}, these correspond to the qubit operators $\sigma_x$ and $\sigma_y$ in the two-level approximation, as shown in Eqs.~\ref{eq:trans_sigmay} and \ref{eq:trans_sigmax}. The coupling terms that naturally appear in multidimensional circuits with inductances or capacitances (see Sec.~\ref{sec:multidimensional_circuits}) are always bilinear in flux or charge, such as 

\begin{align}
\Phi_q \Phi_r \sim \sigma_x (a^\dagger + a).
\end{align}

This implies that the transverse coupling term in Eq.~\ref{eq:Rabi} is the natural coupling term that appears in coupled circuits between resonators and transmon-type qubits. 
The $\sigma_z$ operator needed for longitudinal coupling, however, corresponds to the number operator $c^\dagger c$ (compare Eqs.~\ref{eq:ham4quant} and \ref{eq:trans}) in terms of creation and annihilation operators and is therefore a second order term in charge or flux. A quadratic term in charge or flux corresponds thus to a diagonal term in the operator sense, as $\sigma_z$ or $c^\dagger c$ do not change the state of the qubit or resonator. We can state that operators with odd parity correspond to linear, that is odd terms in the circuit variables, while operators with even parity correspond to quadratic, that is even terms.
The only possibility to create such a second order coupling term in transmon-type qubits is to employ a non-linear circuit element: the Josephson junction. Such a coupling term between a qubit and a resonator variable is given by

\begin{align*}
E_J \cos(\varphi_q - \varphi_r) = E_J \left(\sin(\varphi_q) \sin(\varphi_r) + \cos(\varphi_q) \cos(\varphi_r)\right) \numberthis
\label{eq:josephson_coupling}
\end{align*} 

(compare Eq.~\ref{eq:lagr_example0}) in terms of the unitless phase variables $\varphi_i = 2\pi\, \Phi_i/\Phi_0$. 
Up to second order in $\varphi_i$, this leads to two different coupling terms.
Following again Eqs.~\ref{eq:quant_r} and \ref{eq:quant_qu}, we can associate

\begin{align}
\varphi_q \,\varphi_r \sim\sigma_x (a^\dagger + a),
\end{align}

which is again a transverse coupling term, and 

\begin{align}
\varphi_q^2 \,\varphi_r^2 \sim\sigma_z (a^\dagger + a)^2.
\end{align}

While this is a $\sigma_z$ coupling term, it is clearly not the longitudinal coupling term introduced in Eq.~\ref{eq:long}. The trigonometric expansion in Eq.~\ref{eq:josephson_coupling} reveals that a Josephson coupling term involves terms which are even in both variables and terms which are odd in both variables. Longitudinal coupling as defined in Eq.~\ref{eq:long}, however, is a coupling term that has even parity in the qubit variable and odd parity in the resonator variable. This can be realized by applying an external magnetic flux $\Phi_x$ through a loop containing a Josephson junction, which gives

\begin{align}
E_J \cos\left(\varphi_q - \varphi_r + \frac{2\pi}{\Phi_0} \Phi_x \right).
\end{align} 

At an external flux of $\Phi_x = \Phi_0/4$ the parity of the coupling term changes to

\begin{align*}
E_J \left(\cos(\varphi_q) \sin(\varphi_r) - \sin(\varphi_q) \cos(\varphi_r)\right) \numberthis.
\label{eq:josephson_coupling2}
\end{align*} 

The first term in Eq.~\ref{eq:josephson_coupling2} is even in $\varphi_q$ and odd in $\varphi_r$, which corresponds to the desired longitudinal coupling. Using the rules from above, we can associate

\begin{align}
\varphi_q^2 \,\varphi_r \sim\sigma_z (a^\dagger + a)
\label{eq:coup_long}
\end{align}

as the longitudinal coupling term as defined above, and 

\begin{align}
\varphi_q \,\varphi_r^2 \sim\sigma_x (a^\dagger + a)^2
\label{eq:coup_counter_long}
\end{align}

as its counterpart with reversed parity. Higher order series expansions would of course lead to more coupling terms. Depending on the parameters, we can however assume that these will be smaller than the ones presented here. The circuit presented in Chapt.~\ref{chapt:design} uses Josephson junctions and flux-tuning to create longitudinal coupling and a specific symmetric design, which ensures that all other coupling terms can be canceled out.\\
The parity observations made here stay of course true when coupling two qubits or two resonators. Coupling terms created by inductances or capacitances will always be linear in both variables, while coupling terms with even parity can only be created by Josephson junctions. For coupling terms with asymmetric parity (odd in one variable, even in the other, such as in Eqs.~\ref{eq:coup_long} and \ref{eq:coup_counter_long}), external fluxes are necessary.

\chapter{CIRCUIT DESIGN: INDUCTIVELY SHUNTED TRANSMON QUBIT}
\label{chapt:design}

This chapter will be about the design of an inductively shunted transmon qubit with flux-tunable coupling to an embedded harmonic mode, as first presented in Ref. \cite{Richer2016} and further refined and adapted in Ref. \cite{Richer2017}. Note that most of this chapter was already published in Ref.~\cite{Richer2017}. This design is the core of the thesis and while this chapter provides a detailed analysis of its characteristics, the next chapters will focus on adaptations and their implementation (Chapt.~\ref{chapt:adaptations}), as well as possible applications (Chapts.~\ref{chapt:scalability} and \ref{chapt:readout}) of the circuit presented here. \\
The circuit construction offers the possibility to flux-choose between the two inherently different coupling types introduced in Chapt.~\ref{chapt:coupling}: transverse and longitudinal coupling, that is coupling of the displacement degree of freedom of the resonator to the $\sigma_x$ or $\sigma_z$ degree of freedom of the qubit, respectively. We will see that by applying an external magnetic flux we can change the parity of the coupling between qubit, and resonator mode in order to flux-choose between pure longitudinal and pure transverse coupling, or have both at the same time. Being able to choose between either kind of coupling in the same circuit provides the flexibility to use one for coupling to the next qubit and one for readout, or vice versa. As opposed to other approaches, pure longitudinal coupling can be reached with moderate changes in the qubit frequency.\\
While transverse coupling naturally appears in transmon-like circuit constructions, longitudinal coupling is usually much smaller and hardly ever the only coupling term present. The distinctive feature of the tunable design presented here is that the transverse coupling disappears when the longitudinal is maximal and vice versa. For conveniently chosen parameters, we show that longitudinal and transverse coupling have comparable values, while all other coupling terms can be suppressed.
Using the methods introduced in Chapt.~\ref{chapt:quantization}, we will start by having a closer look at the circuit and explicitly derive the relevant quantities (frequencies, couplings and anharmonicities) as a function of the external flux. \\
Throughout this and the following chapters, we will describe all circuits in terms of the superconducting phase, which is a rescaled flux or, equivalently, the time integral over a voltage

\begin{align}
\varphi_i = \left(\frac{2\pi}{\Phi_0}\right) \Phi_i = \left(\frac{2\pi}{\Phi_0}\right) \int_{-\infty}^t V_i(t') dt'
\end{align}

(compare Eq. \ref{eq:phase_definition}). This dimensionless variable is advantageous as the circuits involve numerous Josephson junction whose behavior is governed by the superconducting phase. For simplicity, we will work with so-called node phases as introduced in Chapt.~\ref{chapt:quantization}. Note that, as the overall phase is undefined, the real variables are the phase differences between the nodes, which reduces the number of independent variables by one. 
The circuit shown in Fig.~\ref{fig:qubit_resonator} has three nodes, which corresponds to two independent variables: a qubit coupled to a resonator. We will start by explaining the characteristic design of the circuit on the basis of its Lagrangian in Sec.~\ref{sec:design}, go to second quantization  in Sec.~\ref{sec:quantization} in order to find the Hamiltonian from which we draw the frequencies and anharmonicities and turn to the different resulting coupling terms in Sec.~\ref{sec:coupling}.

\section{QUBIT-RESONATOR SYSTEM WITH FLUX-TUN\-ABLE COUPLING}
\label{sec:design}

\begin{SCfigure}[50][tb]
\centering
    \includegraphics[width = .5\textwidth]{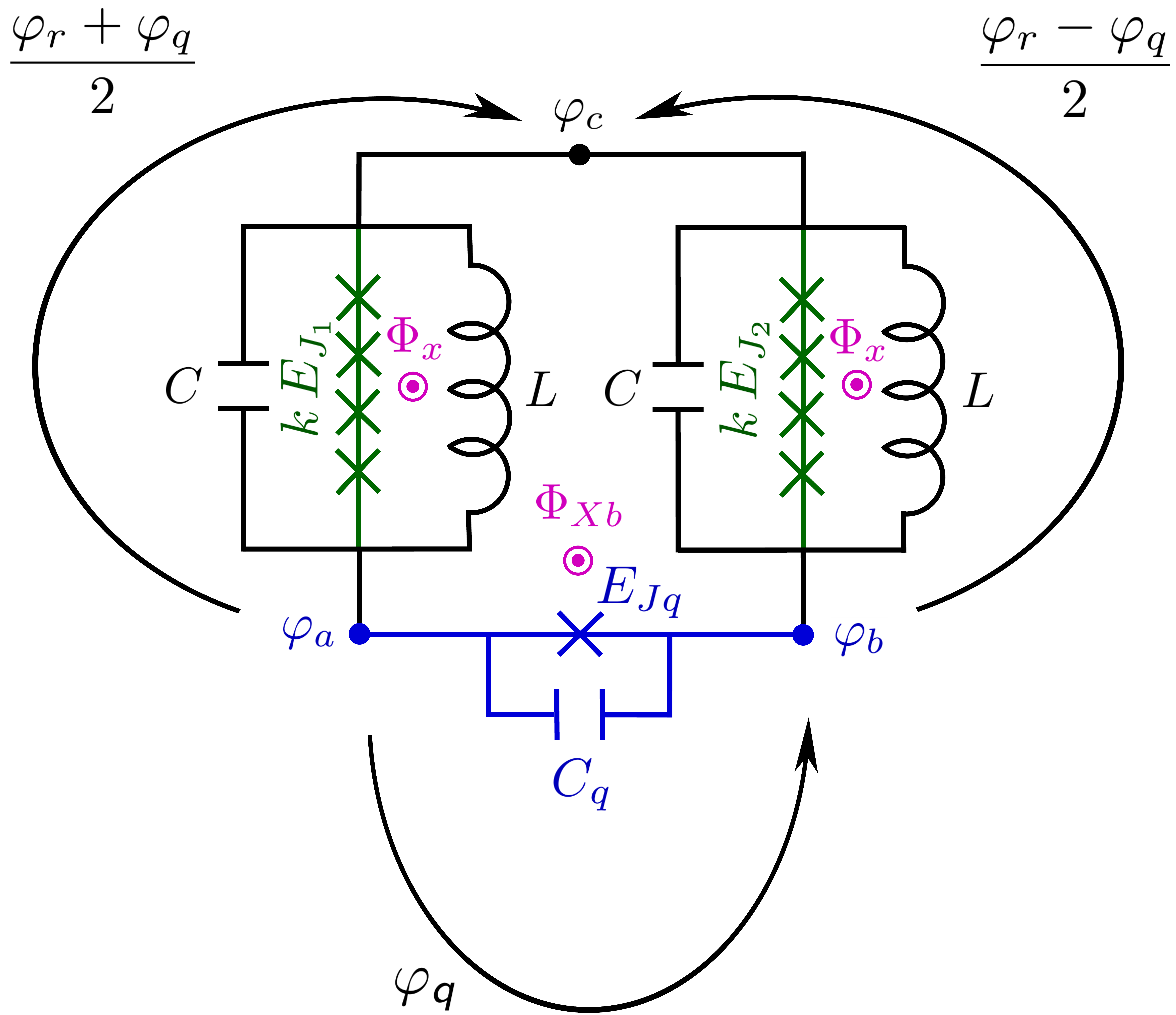}
  \caption{Inductively shunted transmon qubit with the possibility to flux-choose between longitudinal and transverse coupling to an embedded resonator. The qubit mainly consists of a single Josephson junction (depicted in blue).    \label{fig:qubit_resonator}}
\end{SCfigure}

Figure~\ref{fig:qubit_resonator} shows the circuit that implements an inductively shunted transmon qubit with flux-tunable coupling to an embedded resonator. The qubit essentially consists of a single Josephson junction with energy $E_{Jq}$, with a capacitance $C_q$ in parallel. We include the parallel plate capacitance of the qubit junction in $C_q$. The rest of the circuit is made up of two symmetric branches, each consisting of one or several Josephson junctions in parallel with a capacitor and an inductor. Similarly to the fluxonium qubit \cite{Manucharyan2009, Pop2014}, the inductive shunting protects the qubit from charge noise.
The qubit and resonator variables are chosen such that the superconducting phase differences across these two coupling branches are the sum and the difference of qubit and resonator variables, that is

\begin{align}
\varphi_q = \varphi_a - \varphi_b \qquad \varphi_r = \varphi_a + \varphi_b - 2\,\varphi_c,
\label{eq:variables}
\end{align}

\sidecaptionvpos{figure}{c}

where $\varphi_{abc}$ are the phases at the nodes of the circuit as depicted in Fig.~\ref{fig:qubit_resonator}.
Because of the left-right symmetry of the design, all coupling terms via the capacitances and inductances identically cancel out, and the coupling between qubit and resonator is only created by the coupling junctions (or junction arrays) $E_{J1}$ and $E_{J2}$. We will see that the external flux $\Phi_x$ through the two coupling loops can be used to change the parity of the coupling term in order to implement longitudinal coupling. We will show here that the external flux $\Phi_x$ can also be used to tune between pure longitudinal and pure transverse coupling at conveniently chosen realistic parameters, and we analyze the system as a function of this flux. Furthermore, we will allow for an additional external flux in the big loop $\Phi_{Xb}$ and show how the coupling and anharmonicity are boosted at $\Phi_{Xb} = \Phi_0/2$, where $\Phi_0$ is the magnetic flux quantum.
The kinetic energy of the qubit-resonator system is given by

\begin{align}
\mathcal T &= \left(\frac{\Phi_0}{2\pi}\right)^2 \left(\frac{2\,C_q+C}{4} \, \dot\varphi_q^2 + \frac{C}{4} \, \dot\varphi_r^2\right)
\label{eq:kin}
\end{align}

with the dimensionless phase variables as defined in Eq.~\ref{eq:variables}. Clearly, there is no coupling between qubit and resonator via the kinetic energy. The corresponding potential energy can be written as

\begin{align*}
\mathcal U &= \left(\frac{\Phi_0}{2\pi}\right)^2 \frac{1}{4L} \, (\varphi_q^2 + \varphi_r^2) - E_{Jq} \cos(\varphi_q + \varphi_{Xb})\\
& - k\,E_{J1} \cos\left(\frac{\varphi_r + \varphi_q}{2 k} + \frac{\varphi_x}{k}\right) - k\,E_{J2} \cos\left(\frac{\varphi_r - \varphi_q}{2 k} + \frac{\varphi_x}{k}\right),\numberthis
\label{eq:pot}
\end{align*}

where $\varphi_x = 2\pi \,\Phi_x/\Phi_0$ is the external flux through the two coupling loops and $\varphi_{Xb} =2\pi \,\Phi_{Xb}/\Phi_0$ is the external flux through the big loop, both rescaled to be dimensionless. We will assume $\varphi_{Xb} = 0$ in this section and consider its effect in Sec.~\ref{sec:fluxbias}. As depicted in Fig.~\ref{fig:qubit_resonator}, we might want to use arrays of $k$ equal Josephson junctions for the coupling branches in order to suppress the nonlinearity of the resonator as well as higher-order coupling terms. In Sec.~\ref{sec:array}, we will have a closer look at these arrays.  \\
The resonator is designed symmetrically, such that the coupling between qubit and resonator is only defined by the two coupling junction arrays $k\,E_{Ji}$, that is the second line in Eq.~\ref{eq:pot}. A trigonometric expansion leads to four different coupling terms, which we will classify by their parity (see Tab.~\ref{tb:coupling}). These are given by

\begin{align*}
\mathcal{U}_\text{coup} &= k\,E_{J\Delta} \sin\left(\frac{\varphi_q}{2 k}\right)\sin\left(\frac{\varphi_r }{2 k}\right) \cos\left(\frac{\varphi_x}{k}\right) \\
& + k\,E_{J\Sigma} \cos\left(\frac{\varphi_q}{2 k}\right)\sin\left(\frac{\varphi_r }{2 k}\right) \sin\left(\frac{\varphi_x}{k}\right)\\
&+ k\,E_{J\Delta} \sin\left(\frac{\varphi_q}{2 k}\right)\cos\left(\frac{\varphi_r }{2 k}\right) \sin\left(\frac{\varphi_x}{k}\right) \\
& - k\,E_{J\Sigma} \cos\left(\frac{\varphi_q}{2 k}\right)\cos\left(\frac{\varphi_r }{2 k}\right) \cos\left(\frac{\varphi_x}{k}\right) \numberthis,
\label{eq:expansion}
\end{align*}

where $E_{J\Sigma} = E_{J1} + E_{J2}$ is the sum of the coupling junctions, while $E_{J\Delta} = E_{J1} - E_{J2}$ is their difference.
For the qubit, there are two terms with odd parity and two terms with even parity, meaning that these terms are an odd or even function in the qubit variable $\varphi_q$. The same is true for the resonator.\\
Longitudinal coupling involves the coupling of the displacement degree of freedom of the resonator to the $\sigma_z$ degree of freedom of the qubit. This means that the coupling term is an odd function in the resonator variable, and an even function in the qubit variable, which is true for the second term in Eq.~\ref{eq:expansion}.
The longitudinal coupling $g_{zx}$ is maximal at $\varphi_x = k \, \pi/2$, when it is given by

\begin{align*}
 k\,E_{J\Sigma} \cos\left(\frac{\varphi_q}{2 k}\right)\sin\left(\frac{\varphi_r }{2 k}\right) \hat{=} \,\hbar \, g_{zx} \, \sigma_z (a^\dagger + a), \numberthis
\label{eq:coupz}
\end{align*}

which is an even function in the qubit variable and an odd function in the resonator variable. The longitudinal coupling term is proportional to the sum of the coupling junctions $E_{J\Sigma}$.\\
Transverse coupling on the other hand involves the coupling of the displacement degree of freedom of the resonator to the $\sigma_x$ degree of freedom of the qubit. This means that the coupling term is an odd function in both the resonator variable and the qubit variable, which is true for the first term in Eq.~\ref{eq:expansion}. The transverse coupling $g_{xx}$ is maximal at zero flux, $\varphi_x = 0$, when it is given by

\begin{align*}
 k\,E_{J\Delta} \sin\left(\frac{\varphi_q}{2 k}\right)\sin\left(\frac{\varphi_r }{2 k}\right) \hat{=} \,\hbar \, g_{xx} \, \sigma_x (a^\dagger + a), \numberthis
\label{eq:coupx}
\end{align*}

which is an odd function both in the qubit and the resonator variable.
As opposed to the longitudinal coupling, the transverse coupling is proportional to the junction asymmetry $E_{J\Delta}$, which may be designed to be about 3~-~8~\% of $E_{J\Sigma}$.
It is important to notice that the transverse term disappears at $\varphi_x = k \, \pi/2$, where the longitudinal coupling has its maximum, while the longitudinal coupling, on the other hand, disappears at zero flux. We will see later that for favorably chosen parameters, the other two coupling terms in Eq.~\ref{eq:expansion}, $g_{xz}$ and $g_{zz}$, will be negligible, such that we can flux-choose between pure longitudinal and pure transverse coupling. 
Having both types of coupling in the same circuit gives us the flexibility to use one for coupling to the next qubit and one for readout, or vice versa.

\section{QUANTIZATION}
\label{sec:quantization}

Let us take a step back and do an explicit derivation of the kinetic and potential energy given above (Eqs.~\ref{eq:kin} and \ref{eq:pot}) using the methods introduced in Chapt.~\ref{chapt:quantization}. The kinetic energy in terms of the node phases depicted in Fig.~\ref{fig:qubit_resonator} simply yields

\begin{align}
\mathcal T &= \left(\frac{\Phi_0}{2\pi}\right)^2 \left(\frac{C_q}{2} \, \left(\dot\varphi_a - \dot\varphi_b\right)^2 + \frac{C}{2} \, \left(\dot\varphi_a - \dot\varphi_c\right)^2 + \frac{C}{2} \, \left(\dot\varphi_b - \dot\varphi_c\right)^2\right).
\end{align}

\begin{SCfigure}[50][tb]
\centering
    \includegraphics[height=4cm]{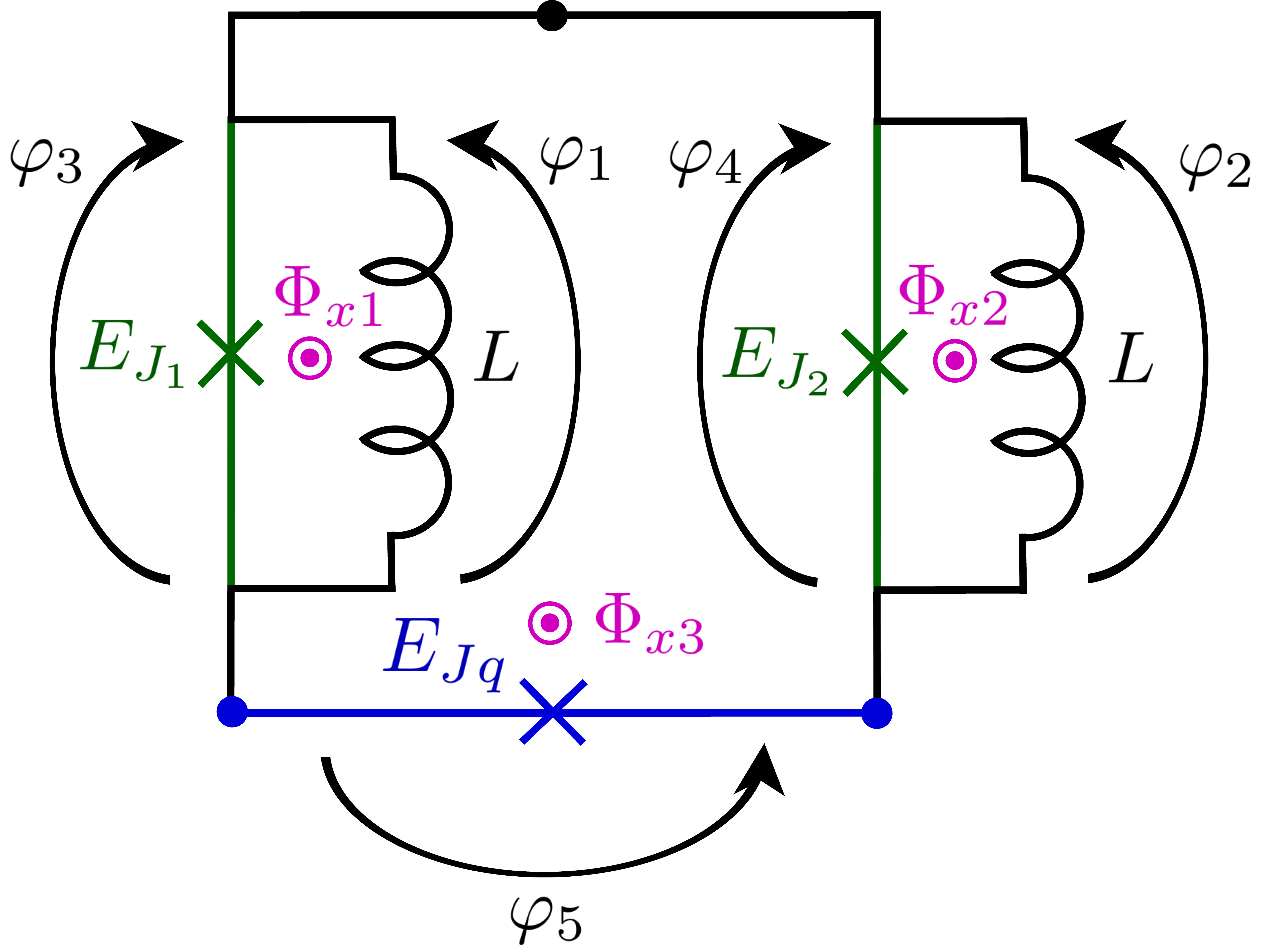}
  \caption{Simplified circuit with branch phases instead of node phases in order to understand how the external fluxes enter in the Lagrangian. The magnetic field points out of the drawing (compare Fig.~\ref{fig:magnetic_field}). Only two of these five branch phases are independent variables. 
    \label{fig:qubit_resonator_simplified}}
\end{SCfigure}

Inserting the variables for qubit and resonator mode as defined in Eq.~\ref{eq:variables} directly leads to Eq.~\ref{eq:kin}. \\
Deriving the potential energy is more complicated due to the different external fluxes and their directions. Figure~\ref{fig:qubit_resonator_simplified} shows a simplified version of the circuit from Fig.~\ref{fig:qubit_resonator} employing branch phases instead of node phases to emphasize the role of the external fluxes. We will focus on single coupling junctions instead of the array, in order to derive the potential energy of the circuit for $k = 1$. The simplified figure includes the five circuit elements that take part in the potential energy. The branch phases across these elements are called $\varphi_i$. They form three superconducting loops, each threaded by an external flux $\Phi_{xj}$. The external fluxes all point out of the drawing (compare Fig.~\ref{fig:magnetic_field}). For each of these loops, we can formulate the flux quantization condition, that is

\sidecaptionvpos{figure}{c}

\begin{align*}
&\varphi_3 - \varphi_1 - 2\pi\, \Phi_{x1}/\Phi_0 = 0\\
&\varphi_4 - \varphi_2 - 2\pi\, \Phi_{x2}/\Phi_0 = 0\\
&\varphi_1 - \varphi_4 - \varphi_5 - 2\pi\, \Phi_{x3}/\Phi_0 = 0, \numberthis
\label{eq:flux_quant3}
\end{align*}

as introduced in Eq.~\ref{eq:flux_quantization} from Chapt.~\ref{chapt:quantization}.
Note that the branch phases with clockwise orientation around the external fluxes enter with positive sign, while the branch phases with counterclockwise orientation enter with negative sign.
It is clear that Eq.~\ref{eq:flux_quant3} allows us to eliminate three of the five variables. We will keep the branch phases across the two inductors $\varphi_1$ and $\varphi_2$ and thus write the potential energy as

\begin{align*}
\mathcal{U} &= \left(\frac{\Phi_0}{2\pi}\right)^2 \left(\frac{\varphi_1^2}{2L} + \frac{\varphi_2^2}{2L}\right) - E_{J1} \cos(\varphi_3) - E_{J2} \cos(\varphi_4) - E_{Jq} \cos(\varphi_5)\\
&= \left(\frac{\Phi_0}{2\pi}\right)^2 \left(\frac{\varphi_1^2}{2L} + \frac{\varphi_2^2}{2L}\right) - E_{Jq} \cos\left(\varphi_1 - \varphi_2 - \frac{2\pi}{\Phi_0} (\Phi_{x2}+ \Phi_{x3})\right)\\
& - E_{J1} \cos\left(\varphi_1 + \frac{2\pi}{\Phi_0} \Phi_{x1}\right) - E_{J2} \cos\left(\varphi_2 + \frac{2\pi}{\Phi_0} \Phi_{x2}\right) . \numberthis
\end{align*}

For simplicity, we define

\begin{align}
\Phi_{Xb} = - \Phi_{x2} - \Phi_{x3}
\end{align}

as the effective external flux appearing in the term for the qubit junction, such that the potential energy yields

\begin{align*}
\mathcal{U} &= \left(\frac{\Phi_0}{2\pi}\right)^2 \left(\frac{\varphi_1^2}{2L} + \frac{\varphi_2^2}{2L}\right) - E_{Jq} \cos\left(\varphi_1 - \varphi_2 + \varphi_{Xb}\right)\\
& - E_{J1} \cos\left(\varphi_1 + \varphi_{x1}\right) - E_{J2} \cos\left(\varphi_2 +\varphi_{x2}\right), \numberthis
\end{align*}

now using unitless variables for the external fluxes. Rewriting this with

\begin{align}
\varphi_1 = \varphi_a - \varphi_c = \frac{\varphi_r + \varphi_q}{2} \qquad \varphi_2 = \varphi_b - \varphi_c = \frac{\varphi_r - \varphi_q}{2}
\end{align}

leads to the potential given in Eq.~\ref{eq:pot} for $k = 1$. We have seen that the external flux $\Phi_{Xb}$ that appears in the term for the qubit junction actually depends on the flux $\Phi_{x2}$ through one of the coupling loops. The case $\Phi_{Xb} = 0$ thus requires a compensating flux through the big loop.\\

In order to find expressions for the frequencies, anharmonicities, and couplings, we will employ the second quantization formalism. We go to  the Hamiltonian representation and start by having a look at the quadratic terms of one variable, while the other is fixed at zero as done in Chapt.~\ref{chapt:quantization}. (More accurately, it should be fixed at the potential minimum, which depends on the flux, as done in Chapt.~\ref{chapt:adaptations}. We will see, however, that the formulas given here are a good approximation.) \\
In order to quantize the qubit, we will at first treat it as a harmonic system and later on calculate its anharmonicity, that is its quartic deviation from a harmonic system, following the strategy introduced in Chapt.~\ref{chapt:quantization}.
A series expansion of the Hamiltonian around $\varphi_q = 0$ (at $\varphi_r = 0$) up to second order yields

\begin{align*}
\mathcal{H}_q = \frac{(2 e \, n_q)^2}{2\, C_q + C} &+ \frac{E_{Jq}}{2} \varphi_q^2 + \left(\frac{\Phi_0}{2\pi}\right)^2 \frac{1}{4L} \varphi_q^2 + \frac{E_{J\Sigma}}{8k} \cos\left(\frac{\varphi_x}{k}\right) \, \varphi_q^2 \numberthis,
\label{eq:Hq1}
\end{align*}

where 

\begin{align}
n_q = \frac{1}{\hbar} \frac{\partial \mathcal{T}}{\partial \dot \varphi_q}
\label{eq:conj}
\end{align}

is a rescaled charge and the conjugate variable to $\varphi_q$. The potential energy of the qubit is governed by the Josephson energy of the qubit junction $E_{Jq}$, which is shunted by the inductance $L$.
In order to describe the flux dependence of Eq.~\ref{eq:Hq1}, we introduce the dimensionless coefficient 

\begin{align}
\eta =\frac{E_{J\Sigma}}{2k} \left(\frac{2\pi}{\Phi_0}\right)^2 L \cos\left(\frac{\varphi_x}{k}\right),
\label{eq:eta}
\end{align}

which disappears for pure longitudinal coupling, that is for $\varphi_x = k \, \pi/2$. This flux-dependence parameter is governed by the ratio between the Josephson energy of the coupling array and the energy of the inductance in parallel to it. 
Using Eq.~\ref{eq:eta} we define

\begin{align}
E_{Jq}^* = E_{Jq} +  \left(\frac{\Phi_0}{2\pi}\right)^2 \frac{1 + \eta}{2L}
\label{eq:EJeffective}
\end{align}

as the inductively shunted effective Josephson energy and 

\begin{align}
E_C = \frac{e^2}{2\,C_q + C} = \frac{e^2}{2\,C_\text{tot}}
\label{eq:charging_energy}
\end{align}

as the charging energy of the qubit, where $C_\text{total} = C_q + C/2$  is the total capacitance in parallel to the qubit junction. With these abbreviations, Eq.~\ref{eq:Hq1} can be rewritten as

\begin{align*}
\mathcal{H}_q &= 4\,E_C \, n_q^2 + \frac{E_{Jq}^*}{2} \varphi_q^2\numberthis.
\label{eq:Hq}
\end{align*}

Due to the effective Josephson energy, this is equivalent to the expression from Chapt.~\ref{chapt:quantization} for the uncoupled qubit (Eq.~\ref{eq:ham_qu2}).
We go to second quantization using

\begin{align*}
\varphi_q = \sqrt[4]{\frac{2 E_C}{E_{Jq}^*}} (c^\dagger + c) \qquad
n_q = \frac{1}{2}\sqrt[4]{\frac{E_{Jq}^*}{2 E_C}} i \, (c^\dagger - c) \numberthis
\label{eq:phiq}
\end{align*}

(compare Eq.~\ref{eq:quant_qu}) in Eq.~\ref{eq:Hq}, which yields 

\begin{align}
\mathcal{H}_q = \hbar \, \omega_q \, \left(c^\dagger c + \frac{1}{2}\right)
\end{align}

with the harmonic qubit frequency

\begin{align*}
\omega_q &= \frac{\sqrt{8 E_C E_{Jq}^*}}\hbar. \numberthis
\label{eq:omega_q}
\end{align*}

The quantization rules for the qubit given in Eq.~\ref{eq:phiq}, fulfill the commutation relation for the conjugate variables flux $\Phi_q$ and charge $Q_q$

\begin{align}
[\Phi_q, Q_q] = \left[\frac{\Phi_0}{2\pi}\varphi_q, 2e \,n_q\right] = \frac{i \, \hbar}{2} [c^\dagger + c, c^\dagger - c] = i \,\hbar
\end{align}

with $\Phi_0 = \pi \hbar/e$.
In order to determine whether the system can be treated as a qubit, we need to know its anharmonicity, that is its quartic deviation from a harmonic system. The fourth-order term in the potential energy (Eq.~\ref{eq:pot}) for $\varphi_r = 0$ is

\begin{align*}
&-\frac{1}{24} \left(E_{Jq} + \left(\frac{\Phi_0}{2\pi}\right)^2 \frac{\eta}{8\,k^2L}\right) \varphi_q^4 \\
=&-\frac{E_C}{12 \, E_{Jq}^*} \left(E_{Jq} + \left(\frac{\Phi_0}{2\pi}\right)^2 \frac{\eta}{8\,k^2L}\right) (c^\dagger + c)^4\numberthis,
\end{align*}

using Eqs.~\ref{eq:eta} and \ref{eq:phiq}. Since

\begin{align}
\langle j | (a^\dagger + a)^4 | j \rangle = 6j^2 + 6j +3
\label{eq:quartic}
\end{align}

(see Ref.~\cite{Koch2007}), where $|j \rangle$ are the Fock state eigenvectors, the energy of state $j$ up to fourth order is

\begin{align*}
E_j^{(q)} &= \sqrt{8E_C E_{Jq}^*} \left(j+\frac{1}{2}\right)  \\
&-\frac{E_C}{12 \, E_{Jq}^*} \left(E_{Jq} + \left(\frac{\Phi_0}{2\pi}\right)^2 \frac{\eta}{8\,k^2L}\right) (6j^2 + 6j +3). \numberthis
\label{eq:Ej}
\end{align*}

The quartic anharmonicity of the qubit is given by

\begin{align}
\alpha^{(q)} = \frac{E_{12}^{(q)} - E_{01}^{(q)}}\hbar = - E_C \, \frac{E_{Jq} + \left(\frac{\Phi_0}{2\pi}\right)^2 \frac{\eta}{8\, k^2L}}{\hbar E_{Jq}^*}
\label{eq:alpha}
\end{align}

with $E_{ij}^{(q)} = E_{j}^{(q)} - E_i^{(q)},$ which leads to a correction to the qubit frequency, that is

\begin{align}
\Delta = \frac{E_{01}^{(q)}}{\hbar} = \omega_q + \alpha^{(q)}.
\label{eq:delta}
\end{align} 

We see that the qubit anharmonicity is governed by the charging energy of the qubit $E_C$ and the ratio between $E_{Jq}$ and $E_{Jq}^*$. Remembering again that $\eta = 0$ for pure longitudinal coupling at $\varphi_x = k \, \pi/2$, this is the same expression as the one given by Koch et al. in Ref.~\cite{Koch2007} for the transmon anharmonicity, apart from the rescaling of the Josephson energy due to the inductive shunting (Eq.~\ref{eq:EJeffective}). In Ref.~\cite{Koch2007} Koch et al. give an estimation for a minimal required relative anharmonicity of 

\begin{align}
\alpha_r^{(q)} = \frac{E_{12}^{(q)} - E_{01}^{(q)}}{E_{01}^{(q)}} \geq \frac{1}{200 \pi}.
\end{align}

As shown below, we can reach relative qubit anharmonicities which are one order of magnitude higher than this. As we will show in Sec.~\ref{sec:fluxbias}, the qubit anharmonicity can be significantly boosted using an additional flux bias $\varphi_{Xb}$ through the big loop (see Fig.~\ref{fig:qubit_resonator}).\\
In the two-level approximation, the qubit Hamiltonian is given by

\begin{align}
\mathcal{H}_q = \hbar \,\frac{\Delta}{2} \sigma_z.
\end{align}

For the resonator we follow the strategy used above and do a series expansion up to second order around $\varphi_r = 0$ (at $\varphi_q = 0$), that is

\begin{align*}
\mathcal{H}_r =\frac{(2 e \, n_r)^2}{C} +  \left(\frac{\Phi_0}{2\pi}\right)^2 \frac{1 + \eta}{4L}  \varphi_r^2  
=\frac{Q_r^2}{C} +   \frac{1 + \eta}{4L}  \Phi_r^2 
 \numberthis,
\label{eq:Hr}
\end{align*}

where the flux $\Phi_r$ and the charge $Q_r = 2e \,n_r$ are again conjugate variables that fulfill the commutation relation

\begin{align}
[\Phi_r, Q_r] = \left[\frac{\Phi_0}{2\pi}\varphi_r, 2e \,n_r\right] = \frac{i \, \hbar}{2} [a^\dagger + a, a^\dagger - a] = i \,\hbar.
\end{align}

The quantization step is done by inserting

\begin{align*}
\Phi_r = \frac{\Phi_0}{2\pi}\varphi_r= \sqrt{\frac{\hbar  Z_{0}}{2}} (a^\dagger + a)  \qquad
Q_r =  2e \, n_r = \sqrt{\frac{\hbar}{2  Z_{0}}} i \, (a^\dagger - a)\numberthis
\label{eq:phir}
\end{align*}

(compare Eq.~\ref{eq:quant_r}) in Eq.~\ref{eq:Hr} and choosing the characteristic impedance $Z_{0}$ such that the Hamiltonian has the form

\begin{align}
\mathcal{H}_r = \hbar \, \omega_r \, \left(a^\dagger a + \frac{1}{2}\right),
\end{align}

i.e. such that the non-diagonal terms disappear. This is satisfied for

\begin{align}
Z_{0} = 2\sqrt{\frac{L}{C(1+\eta)}},
\label{eq:Z_0}
\end{align}

which directly gives

\begin{align}
\omega_r = \sqrt{\frac{1 + \eta}{L \,C}}
\label{eq:omegar}
\end{align} 

for the resonator frequency. We see that $\omega_r$ acquires a flux-dependence due to $\eta$ (see Eq. \ref{eq:eta}). However, the effect of $\eta$ can be suppressed by increasing the number of junctions $k$ in the array (see Sec.~\ref{sec:array}).\\ 
In order to investigate whether our resonator can really be treated as a harmonic system, we will calculate its anharmonicity. Using Eq.~\ref{eq:phir}, the fourth-order term in the potential energy (Eq.~\ref{eq:pot}), again for $\varphi_q = 0$, can be written as

\begin{align*}
- \left(\frac{\Phi_0}{2\pi}\right)^2 \frac{\eta}{192\, k^2 L} \varphi_r^4
= - \left(\frac{2\pi}{\Phi_0}\right)^2 \frac{\hbar^2\eta}{192\, k^2 C(1 + \eta)} (a^\dagger + a)^4 \numberthis.
\end{align*}

Using again Eq.~\ref{eq:quartic}, the energy of state $j$ up to fourth order is

\begin{align*}
E_j^{(r)} = \hbar \, \omega_r \left(j+\frac{1}{2}\right) - \left(\frac{2\pi}{\Phi_0}\right)^2 \frac{\hbar^2\eta \, (6j^2 + 6j +3)}{192\, k^2 C(1 + \eta)}\numberthis. 
\end{align*}

The anharmonicity of the resonator is then given by

\begin{align*}
\alpha^{(r)} &= \frac{E_{12}^{(r)} - E_{01}^{(r)}}\hbar = \frac{\eta \,\pi^2 \hbar }{\eta \,\pi^2 \hbar  - 4 k^2 (1+\eta)^\frac{3}{2} \sqrt{C/L}} \numberthis,
\label{eq:alphares}
\end{align*}

where again $E_{ij}^{(r)} = E_{j}^{(r)} - E_i^{(r)}$. We see that the anharmonicity of the resonator is proportional to the parameter $\eta$. Remarkably, $\eta$ is zero at the longitudinal coupling point $\varphi_x = k \, \pi/2$, where the resonator anharmonicity goes through zero and changes its sign. This remains true when we include higher order terms, as all series terms in the potential energy in $\varphi_r$ from third order onward are proportional to $\eta$. If we want to work with a static system with pure longitudinal coupling, we can thus assume our resonator to be perfectly harmonic. At any other point in flux, though, we should choose our parameters carefully, in order to ensure that $\alpha^{(r)}$ remains small.\\
For the implementation of the scalable architecture discussed in Sec.~\ref{sec:scalable_architecture}, it is very important that the resonator is a harmonic system, as both the controlled-phase gate and the readout scheme (see Sec.~\ref{sec:readout_billangeon}) rely on sideband transitions. These are possible due to the absence of the dispersive shift for longitudinal coupling (see Sec.~\ref{sec:longitudinal}), which implies that a certain sideband transition stays resonant irrespective of the number of photons in the resonator. In Chapt.~\ref{chapt:adaptations}, we will see that while the point in flux where the transverse coupling disappears is slightly shifted away from $\varphi_x = k \, \pi/2$ due to the flux dependence of the potential minimum. However, the same is true for the zero-crossing of the resonator anharmonicity. We can thus assume that pure longitudinal coupling always goes along with an almost perfectly harmonic resonator.

\section{ANALYZING THE COUPLING TERMS}
\label{sec:coupling}

Now, we would like to have a look at the coupling terms, taking into account the four terms with different parities shown in Tab.~\ref{tb:coupling} (compare also Sec.~\ref{sec:long}). We will do a series approximation of the potential energy up to second order in both $\varphi_r$ and $\varphi_q$ around zero, which is assumed to be the potential minimum (see Chapt.~\ref{chapt:adaptations} for a more exact numerical treatment).
For two identical coupling junctions (or coupling arrays), only $\sigma_z$-type coupling terms are possible, as all uneven terms in $\varphi_q$ cancel out. This means that the $\sigma_z$ coupling terms are proportional to the sum of the coupling junctions $E_{J\Sigma}$, while the $\sigma_x$ coupling terms are proportional to their difference $E_{J\Delta}$. \\
At zero flux $\varphi_x = 0$, we find the transverse coupling, which we call $g_{xx}$ as it has odd parity both in $\varphi_q$ and $\varphi_r$. It is

\begin{align*}
\frac{E_{J\Delta}}{4\,k}   \, \varphi_q \, \varphi_r \cos\left(\frac{\varphi_x}{k}\right) \hat{=} \, \hbar \, g_{xx} \, \sigma_x (a^\dagger + a) \numberthis
\end{align*}

with

\begin{align}
g_{xx} =\frac{E_{J\Delta}}{2\,k \sqrt{\hbar}}   \sqrt[4]{\frac{2 E_C}{E_{Jq}^*}}  \frac{\pi}{\Phi_0} \sqrt[4]{\frac{L}{C}\frac{1}{1 + \eta}} \cos\left(\frac{\varphi_x}{k}\right).
\label{eq:gxx}
\end{align}

\begin{table}[t]
\begin{center}
\begin{tabular}{ c | c | c | c | c}
Symbol & Coupling type &  Resonator & Qubit & Dependency  \\ [\smallskipamount]
\hline
&&&&\\
$g_{xx}$ & transverse & \multirow{2}{*}{odd $\sim \varphi_r$} &  odd $\sim \varphi_q$ & $\sim E_{J\Delta} \cos\left(\frac{\varphi_x}{k}\right)$ \\ [\smallskipamount]
$g_{zx}$ & longitudinal &  &  even $\sim \varphi_q^2$ & $\sim E_{J\Sigma} \sin\left(\frac{\varphi_x}{k}\right)$
\\ [\smallskipamount]
$g_{xz}$ & unwanted & \multirow{2}{*}{even $\sim \varphi_r^2$} &  odd $\sim \varphi_q$ & $\sim E_{J\Delta} \sin\left(\frac{\varphi_x}{k}\right)$ \\ [\smallskipamount]
$g_{zz}$ & unwanted &  &  even $\sim \varphi_q^2$ & $\sim E_{J\Sigma} \cos\left(\frac{\varphi_x}{k}\right)$
\end{tabular}
    \captionof{table}{The coupling can be split up into four terms with different parity. The $x$ stands for terms with odd parity, the $z$ for terms with even parity in the qubit and resonator variables $\varphi_q$ and $\varphi_r$. While the terms with even qubit parity are proportional to the sum of the coupling junctions $E_{J\Sigma}$, the terms with odd qubit parity are proportional to the junction asymmetry $E_{J\Delta} \ll E_{J\Sigma}$. Note that in a series expansion the odd terms are first order terms, while the even ones are of second order. Therefore $g_{xx}$ is the lowest order term, while $g_{zz}$ is the highest. \label{tb:coupling}}   
\end{center}
\end{table}

There is a competing term with a similar flux dependence, which has even parity in both $\varphi_q$ and $\varphi_r$, namely

\begin{align*}
-\frac{E_{J\Sigma}}{64\,k^3}  \, \varphi_q^2 \, \varphi_r^2 \cos\left(\frac{\varphi_x}{k}\right) \hat{=} \, \hbar \,  g_{zz} \, \sigma_z \, (a^\dagger + a)^2 \numberthis
\end{align*}

with

\begin{align}
g_{zz} &=-\frac{E_{J\Sigma}}{16\,k^3}   \sqrt{\frac{2 E_C}{ E_{Jq}^*}}  \left(\frac{\pi}{\Phi_0}\right)^2 \sqrt{\frac{L}{C}\frac{1}{1 + \eta}} \cos\left(\frac{\varphi_x}{k}\right).
\label{eq:gzz}
\end{align}

At $\varphi_x = k \, \pi/2$ both $g_{xx}$ and $g_{zz}$ vanish, while two other coupling terms are at their joint maximum. One is the longitudinal coupling term

\begin{align*}
-\frac{E_{J\Sigma}}{16\,k^2}   \, \varphi_q^2 \, \varphi_r \sin\left(\frac{\varphi_x}{k}\right) \hat{=}\, \hbar \, g_{zx} \, \sigma_z (a^\dagger + a) \numberthis
\end{align*}

with

\begin{align}
g_{zx} =-\frac{E_{J\Sigma}}{8\,k^2 \sqrt{\hbar}}   \sqrt{\frac{2 E_C}{E_{Jq}^*}}  \frac{\pi}{\Phi_0} \sqrt[4]{\frac{L}{C}\frac{1}{1 + \eta}} \sin\left(\frac{\varphi_x}{k}\right),
\label{eq:gzx}
\end{align}

where the $z$ in $g_{zx}$ stands for even qubit parity and the $x$ for odd resonator parity.
The competing $\sigma_x$ term has opposite parity

\begin{align*}
-\frac{E_{J\Delta}}{16\,k^2}  \, \varphi_q \, \varphi_r^2 \sin\left(\frac{\varphi_x}{k}\right) \hat{=} \, \hbar \, g_{xz} \, \sigma_x \, (a^\dagger + a)^2 \numberthis
\end{align*}

with

\begin{align}
g_{xz} &=-\frac{E_{J\Delta}}{4\,k^2}   \sqrt[4]{\frac{2 E_C}{ E_{Jq}^*}}  \left(\frac{\pi}{\Phi_0}\right)^2 \sqrt{\frac{L}{C}\frac{1}{1 + \eta}} \sin\left(\frac{\varphi_x}{k}\right).
\label{eq:gxz}
\end{align}

We will see later that for conveniently chosen parameters, only the transverse coupling $g_{xx}$ (which is the lowest order term) and the longitudinal coupling $g_{zx}$ play a role. The $g_{zz}$ term is the highest order term and therefore much smaller than the others. The $g_{xz}$ term is of the same order as the longitudinal coupling, but is suppressed by the small junction asymmetry $E_{J\Delta} \ll E_{J\Sigma}$. As the flux dependences of longitudinal and transverse coupling have a quadrature relation to one another, each is at its maximum when the other disappears and vice versa. \\
In Ref.~\cite{Didier2015}, similar expressions are obtained for transverse and longitudinal coupling between a qubit and a resonator. 
However, the qubit considered there is a split transmon with a single flux loop, in which the two Josephson junctions play the role of qubit junctions and coupling junctions at the same time. 
Ref.~\cite{Hutchings2017} includes a thorough treatment of the flux-dependence of split transmon qubits such as the one used in Ref.~\cite{Didier2015}. The harmonic frequency of a split transmon is given by

\begin{align}
\omega_q = \sqrt{8 E_C E_{J\Sigma}  \sqrt{\cos\left(\frac{\varphi_x}2\right)^2 + d^2 \sin\left(\frac{\varphi_x}2\right)^2}},
\end{align}

where $d = E_{J\Delta}/E_{J\Sigma}$ is the relative junction asymmetry (compare Refs. \cite{Koch2007} and \cite{Hutchings2017}) and $\varphi_x$ is the rescaled external flux through the loop formed by the junctions. Clearly, the frequency has a minimum at $\varphi_x = \pi$, which deepens with decreasing asymmetry $d$. 
For small junction asymmetries ($d = 0.02$ in Ref.~\cite{Didier2015}), switching to pure longitudinal coupling would be accompanied by a much larger change in the qubit frequency compared to our design, where the roles of the qubit and coupling junctions are separated.
However, the goal in Ref.~\cite{Didier2015} is to perform readout using a time-dependent flux with a small amplitude of up to $0.05\,\Phi_0$, thus avoiding immoderate frequency changes. In a time-dependent frame, the transverse coupling due to spurious junction asymmetry can argued to be non-relevant with rotating wave arguments. \\ 
Our concept, however, is to apply a much higher flux of $\Phi_0/4$ in order to switch to pure longitudinal coupling. As in our approach there is a qubit junction $E_{Jq}$ in addition to the coupling junctions $E_{J1}$ and $E_{J2}$, the qubit frequency will vary much less over the full tuning range than in a split transmon (compare Eq.~\ref{eq:omega_q}). The additional qubit junction in our design thus ensures a moderate flux dependence of the frequency, independent of the coupling junction asymmetry.\\
As recently shown by Hutchings et al. in Ref.~\cite{Hutchings2017}, the qubit dephasing rate of split transmons is proportional to the sensitivity of the qubit frequency to the external flux. This work comes to the conclusion that restricting the tunability of the qubit frequency to a few hundred MHz over the full tuning range leads to dephasing times which are nearly independent of flux noise. Staying in a regime with moderate flux tunability in the range of hundreds of MHz (see Chapt.~\ref{chapt:adaptations}), our qubit should be nearly unaffected by flux noise. Note that this is true independent of the coupling junction asymmetry. \\
It is important to note that in our design the coupling can be tuned by applying an external flux through the two smaller coupling loops, while an additional flux through the big qubit loop can be used to boost the anharmonicity (see Sec.~\ref{sec:fluxbias}). For the bigger loop, we consider the two cases $\Phi_{Xb} = 0$ and $\Phi_{Xb} = \Phi_0/2$, both of which are sweet-spots with respect to flux noise. 
\chapter{ADAPTATIONS, PARAMETERS, AND IMPLEMENTATION}
\label{chapt:adaptations}



This chapter will be about the implementation of the circuit design presented in Chapt.~\ref{chapt:design}. We will discuss how to choose the parameters and present two adaptations of the original circuit from Fig.~\ref{fig:qubit_resonator}, as well as a proposal for the physical implementation of a prototype device. Note that most of this chapter was already published in Ref.~\cite{Richer2017}. The section concerning the physical implementation of the design (that is Sec.~\ref{sec:implementation} and Fig.~\ref{fig:phys_impl}) was part of Ref.~\cite{Richer2017} and provided by my co-authors Nataliya Maleeva, Sebastian~T. Skacel and Ioan~M. Pop.\\
A possible experiment to verify the model could be a measurement of the qubit-resonator dispersive shift as a function of the external flux through the coupling loops $\varphi_x$. While the transverse coupling $g_{xx}$ leads to a qubit-state dependence of the resonator frequency, the longitudinal coupling $g_{zx}$ does not. This means that the qubit-resonator dispersive shift should disappear at $\varphi_x = k \, \pi/2$, where $g_{xx}$ goes through zero and we have pure longitudinal coupling, assuming that all other coupling terms are negligible. \\
We will treat the cases of single coupling junctions and coupling junction arrays separately as they require different restrictions on the parameters. In addition, for both cases we will examine the effect of a flux-biasing of $\varphi_{Xb} = \pi$ in the big loop. As the design strongly relies on symmetry, we will have a look on the effect of asymmetries in Sec.~\ref{sec:asymmetry}. In Sec.~\ref{sec:adaptation} we will come to an adapted circuit, where the coupling and the anharmonicity scale better than for the original circuit.\\
One approximation we made in deriving the formulas for frequencies, anharmonicities, and couplings in Chapt.~\ref{chapt:design} was to assume that the potential energy minimum is always at $\varphi_q = \varphi_r = 0$. This is crucial as all important quantities were derived using series approximations around the potential minimum. However, looking closely at the potential function given in Eq.~\ref{eq:pot}, we see that this is only true at fluxes $\varphi_x = \mu \, k \, \pi$ for integer multiples $\mu$, but nowhere in between. The exact position of the minimum depends of course strongly on the chosen parameters. The solution is thus to numerically determine the potential energy minimum for a given set of parameters (including the external fluxes) and calculate the frequencies, anharmonicities, and couplings again by series approximations around this potential minimum. As we will see in the next section, the formulas given above are a good approximation. Though they cannot capture the flux dependence exactly, they always give the right values at fluxes $\varphi_x = \mu \, k \, \pi$.
This numerical treatment becomes especially important, when we allow for a flux in the big loop $\varphi_{Xb}$, which can be used to boost the anharmonicity (see Secs.~\ref{sec:fluxbias} and \ref{sec:adaptation}).

\section{CASE ONE: SINGLE COUPLING JUNCTIONS}
\label{sec:single}

\begin{table}[t]
\begin{center}
    \begin{tabular}{ c | c || c | c }  
    \multicolumn{2}{c}{Parameters} & \multicolumn{2}{c}{Results}\\ [\smallskipamount]
    \hline
    & & & \\ 
    $E_{Jq}$ & $h$ 10 GHz & $\omega_r/(2\pi)$ & 6.2 - 8 GHz \\ [\smallskipamount]
    $E_{J\Sigma}$ & $h$ 20 GHz & $\Delta/(2\pi)$ & 5.4 - 6.4 GHz\\ [\smallskipamount]
     $E_{J\Delta}/E_{J\Sigma}$ & 0.08 & $g_{zx}^\text{max}/(2\pi)$ & 53 MHz \\ [\smallskipamount]
     $C$ & 114 fF & $g_{xx}^\text{max}/(2\pi)$ & 49 MHz \\ [\smallskipamount]
     $C_q$ & 70 fF & $g_{zz}^\text{max}/(2\pi)$ & 5 MHz \\ [\smallskipamount]
   $L$ & 4.5 nH & $g_{xz}^\text{max}/(2\pi)$ &  6 MHz \\ [\smallskipamount]
   $L_\text{max}$ & 4.9 nH & $|\alpha_r^{(q)}|$ & 0.8 - 1.1\% \\ [\smallskipamount]
   $L_\text{crit}$ & 5.6 nH & $|\alpha_r^{(r)}|$ &  $\leq$ 0.5\% \\ [\smallskipamount]
    \end{tabular}
    \captionof{table}{A good choice of parameters for the single coupling junction case ($k=1$) at zero flux through the big loop $\varphi_{Xb} = 0$. $L$ needs to be less than or equal to $L_\text{max}$ to ensure that the resonator frequency stays in the 6 - 8 GHz range and less than $L_\text{crit}$ in order to avoid a double-well potential for all possible values of flux (compare Sec.~\ref{sec:fluxbias}).
    On the right we show the frequencies, anharmonicities, and couplings, which vary with the flux in the coupling loops.}
    \label{tb:k1}
\end{center}
\end{table}

In this section we will discuss how to favorably choose the parameters for the experiment described above given the constraints of the real system. We will start with the case of single coupling junctions, which means we set $k = 1$ in all the formulas from Chapt.~\ref{chapt:design}. There are several restrictions we have to obey, in order to find a convenient set of parameters.
For example we will require the resonator frequency to always stay in the range of $\omega_r/(2\pi) = 6 - 8$ GHz, which is a convenient microwave range, recently used in the setup of Ref.~\cite{Kou2017} to perform multiplexed quantum readout. We want the qubit frequency to be well separated from the resonator frequency, as any overlap could lead to unwanted cross talk. 
As mentioned in Sec.~\ref{sec:coupling}, it is advisable to stay in a regime with moderate flux tunability of the qubit frequency in order to avoid dephasing due to flux noise.
These being hard constraints, our goal is that the qubit anharmonicity should be as high as possible, while the anharmonicity of the resonator should be negligible.
As we are aiming here for a system where we can flux-choose between transverse and longitudinal coupling, we will choose our parameters such that the longitudinal coupling is as high as possible, while the transverse coupling should be comparable. Note that the transverse coupling can be easily controlled via the junction asymmetry. All other coupling terms should be negligible in order to have pure longitudinal or transverse coupling.\\
Looking at the expression for the resonator frequency given in Eq.~\ref{eq:omegar}, it becomes clear that the absolute value of $\eta$ should not become bigger than $1$, since within our series approximation $\omega_r$ would not be well-defined. Following from Eq.~\ref{eq:eta}, it is clear that $\omega_r$ will have a maximum at zero flux and a minimum at $\varphi_x = \pi$. In order to ensure that it always stays between 6 and 8 GHz, we can fix the capacitance $C$ in terms of $L$ and $|\eta|$, such that $\omega_r/(2\pi) = 8$ GHz at its maximum, and then define a maximal inductance $L_\text{max}$, such that $\omega_r/(2\pi) \geq 6$ GHz at its minimum. With these hard constraints and the goal of having high coupling and qubit anharmonicity, while keeping the qubit frequency well separated from the resonator frequency, we tried out different parameter values until we found the \textit{optimal} solution. The junction asymmetry $d = E_{J\Delta}/E_{J\Sigma}$ is chosen such that the maximal transverse coupling $g_{xx}$ is approximately as big as the maximal longitudinal coupling $g_{zx}$.

\sidecaptionvpos{figure}{c}
\begin{SCfigure}[50][tb]
\centering
    \includegraphics[width=.5\linewidth]{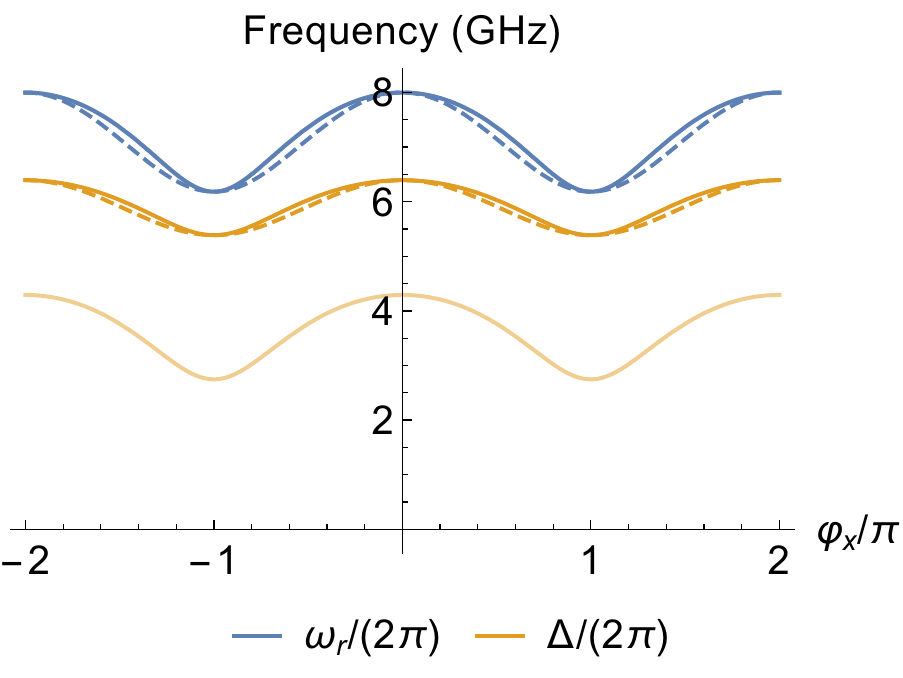}
  \caption{The qubit frequency $\Delta$ and the resonator frequency $\omega_r$ as a function of the (reduced) flux through the coupling loops $\varphi_x$. Solid lines show accurate numerical results, dashed lines show the predictions using the formulas from Sec.~\ref{sec:quantization}, the lighter color curve shows results at a flux $\varphi_{Xb} = \pi$ in the big loop. While the resonator frequency is not affected by the large-loop flux-biasing, the qubit frequency experiences a drop. \label{fig:freqk1}}   
\end{SCfigure}

Table~\ref{tb:k1} shows the chosen parameters and the frequencies, anharmonicities, and couplings they lead to. Figure~\ref{fig:freqk1} shows the frequencies of qubit and resonator as a function of the flux through the coupling loops. As we can see, they always stay well separated.
Solid lines in the figure show our accurate numerical results including the effect of the flux-dependent potential energy minimum (see above), dashed lines show the predictions using the formulas from Chapt.~\ref{chapt:design}. While the predictions are always accurate at the maxima and minima, they deviate slightly in between. The lighter color curve shows results at a flux $\varphi_{Xb} = \pi$ through the big loop (see Sec.~\ref{sec:fluxbias}). Note that the formulas from Chapt.~\ref{chapt:design} are only applicable for $\Phi_{Xb}=0$.

\begin{figure}[tb]
\centering
\subcaptionbox{Longitudinal ($g_{zx}$) and transverse coupling ($g_{xx}$).\label{fig:coupk1_1}}
  [.48\linewidth]{\includegraphics[width=.48\linewidth]{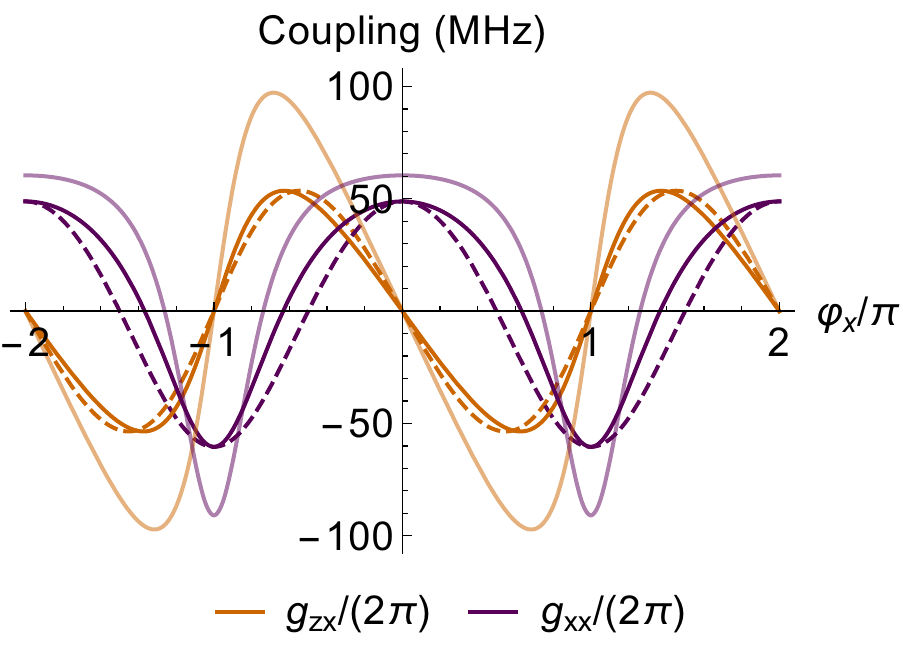}}
\hspace{5pt}
\subcaptionbox{Unwanted coupling terms, note the change of scale.\label{fig:coupk1_2}}
  [.48\linewidth]{\includegraphics[width=.48\linewidth]{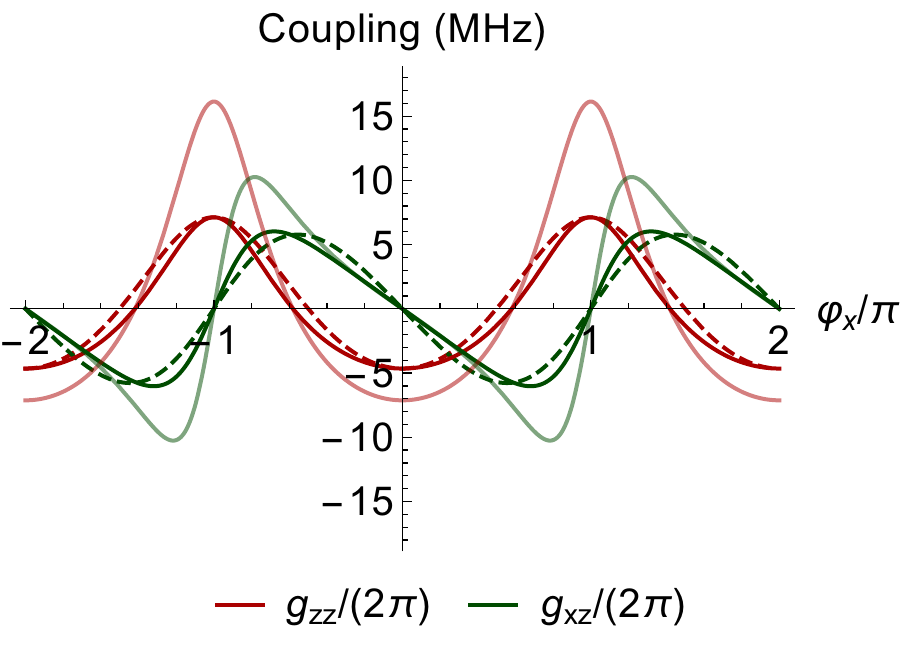}}
\caption{The four most important coupling terms as a function of the (reduced) flux through the coupling loops $\varphi_x$. Solid lines show accurate numerical results, dashed lines show the predictions using the formulas from Chapt.~\ref{chapt:design}, the lighter color curves show results at a flux $\varphi_{Xb} = \pi$ in the big loop. The longitudinal coupling is almost doubled due to the large-loop flux-biasing, while the point where the transverse coupling disappears is considerably shifted. The longitudinal coupling always disappears exactly at multiples of $\varphi_x = k\, \pi$. \label{fig:coupk1}}
\end{figure}

While both the longitudinal and the transverse coupling reach values above 50 MHz, the spurious terms are far from being negligible (see Fig.~\ref{fig:coupk1}). The unwanted $g_{xz}$ coupling reaches 11~\% of the longitudinal coupling $g_{zx}$ at their joint maximum, while the unwanted $g_{zz}$ reaches 9~\% of the transverse coupling $g_{xx}$ at their maximum. The dashed lines show again the predictions using the formulas from Chapt.~\ref{chapt:design}. We see that they are a good but not perfect approximation. In particular, the point where the transverse coupling disappears and the longitudinal coupling peaks is shifted. The lighter color curves show what happens when we consider the large-loop flux-biasing (see Sec.~\ref{sec:fluxbias}).

\sidecaptionvpos{figure}{c}
\begin{SCfigure}[50][tb]
\centering
    \includegraphics[width=.5\linewidth]{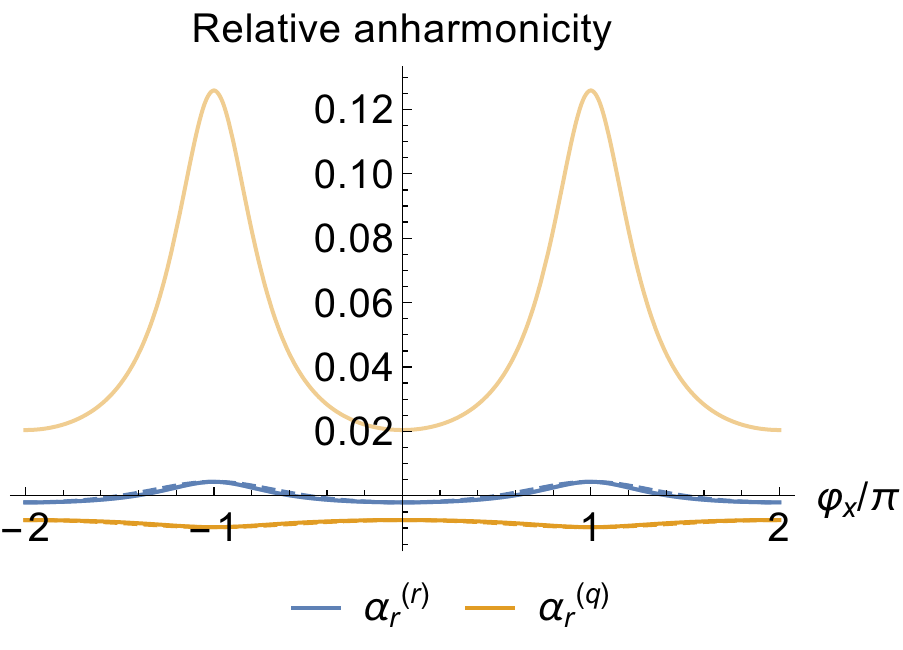}
  \caption{The relative anharmonicities of qubit and resonator
     as a function of the (reduced) flux through the coupling loops $\varphi_x$. Solid lines show accurate numerical results, dashed lines show the predictions using the formulas from Chapt.~\ref{chapt:design}. Note that in this plot the predictions are indistinguishable from the numerical results. The lighter color curve shows results at a flux $\varphi_{Xb} = \pi$ in the big loop.    \label{fig:anhk1}}   
\end{SCfigure}

The resonator anharmonicity is clearly a problem, as it is almost as large as the qubit anharmonicity (see Fig.~\ref{fig:anhk1}). While it goes through zero almost exactly when the transverse coupling disappears, it is much too high at all other values of flux. The lighter color curve shows again results at a flux $\varphi_{Xb} = \pi$ in the big loop, to be discussed now.

\section{FLUX-BIASING}
\label{sec:fluxbias}
Applying a flux of $\varphi_{Xb} = \pi$ through the big loop as suggested in Eq.~\ref{eq:pot} has a very interesting effect on the circuit's behavior. It changes the qubit spectrum, but does not affect the resonator. As shown in Fig.~\ref{fig:freqk1}, the qubit frequency drops to 2.5 - 4~GHz, while the resonator frequency remains unchanged. The qubit anharmonicity is now positive and boosted up to 17~\% (see Fig.~\ref{fig:anhk1}). The coupling also gets a boost and is approximately doubled (see Fig.~\ref{fig:coupk1}). The zero-crossing of the transverse coupling is shifted away from $\varphi_x = k \, \pi/2$, along with the maximum of the longitudinal coupling.\\
Now, what exactly happens, when we put a flux through the big loop? Looking again at the expression for the qubit frequency (Eq.~\ref{eq:omega_q}) and its derivation, it becomes clear that a flux $\varphi_{Xb} = \pi$ through the big loop corresponds to the transition $E_{Jq} \to - E_{Jq}$ in the potential function. The qubit potential thus consists primarily of a parabola with its minimum at $\varphi_q = 0$ (due to the inductive part) and a cosine with a maximum at $\varphi_q = 0$ (due to the qubit junction). Clearly, this could lead to a double-well potential, similar to flux qubits \cite{Yan2016, Chiorescu2003}. This is a case which we want to avoid - we have chosen not to explore flux (i.e. persistent-current) qubits, and thus all our analysis is designed for a single-well treatment. As long as we stay out of this double-well regime, the minimum of the total potential will still be at $\varphi_q = 0$ (this is without the effect of the flux through the coupling loops) and we can reuse the expressions from Chapt.~\ref{chapt:design} with $E_{Jq} \to - E_{Jq}$.
For the harmonic qubit frequency, we thus find

\begin{align}
\omega_q^\pi = \frac{\sqrt{8 E_C (E_L(1 + \eta)  - E_{Jq})}}{\hbar},
\label{eq:omega_pi}
\end{align}

where $E_L = (\Phi_0/(2\pi))^2/(2L)$ is the energy stored in the inductor. This explains the drop in the qubit frequency shown in Fig.~\ref{fig:freqk1}. It also shows that we need to make sure that $E_L (1+\eta)$ is always bigger than $E_{Jq}$, such that the qubit frequency remains well-defined and we avoid the double-well potential. This implies another critical (maximal) inductance, which is

\begin{align}
L_\text{crit} = \left(\frac{\Phi_0}{2\pi}\right)^2 \frac{1 + \eta}{2 E_{Jq}}.
\label{eq:lcrit}
\end{align}

For the parameters used here (see Tab.~\ref{tb:k1}), $L_\text{crit}$ is, however, bigger than the $L_\text{max}$ defined in Sec.~\ref{sec:single}, such that it does not limit the permitted parameter space any further. The qubit anharmonicity also changes considerably due to the flux in the big loop. It is

\begin{align}
\alpha^{(q) \pi} = - E_C \frac{E_L \frac{\eta}{4k^2} - E_{Jq} }{\hbar (E_L (1 + \eta) - E_{Jq})}.
\label{eq:anhpi}
\end{align}

\sidecaptionvpos{figure}{c}
\begin{SCfigure}[50][tb]
\centering
    \includegraphics[width=.55\linewidth]{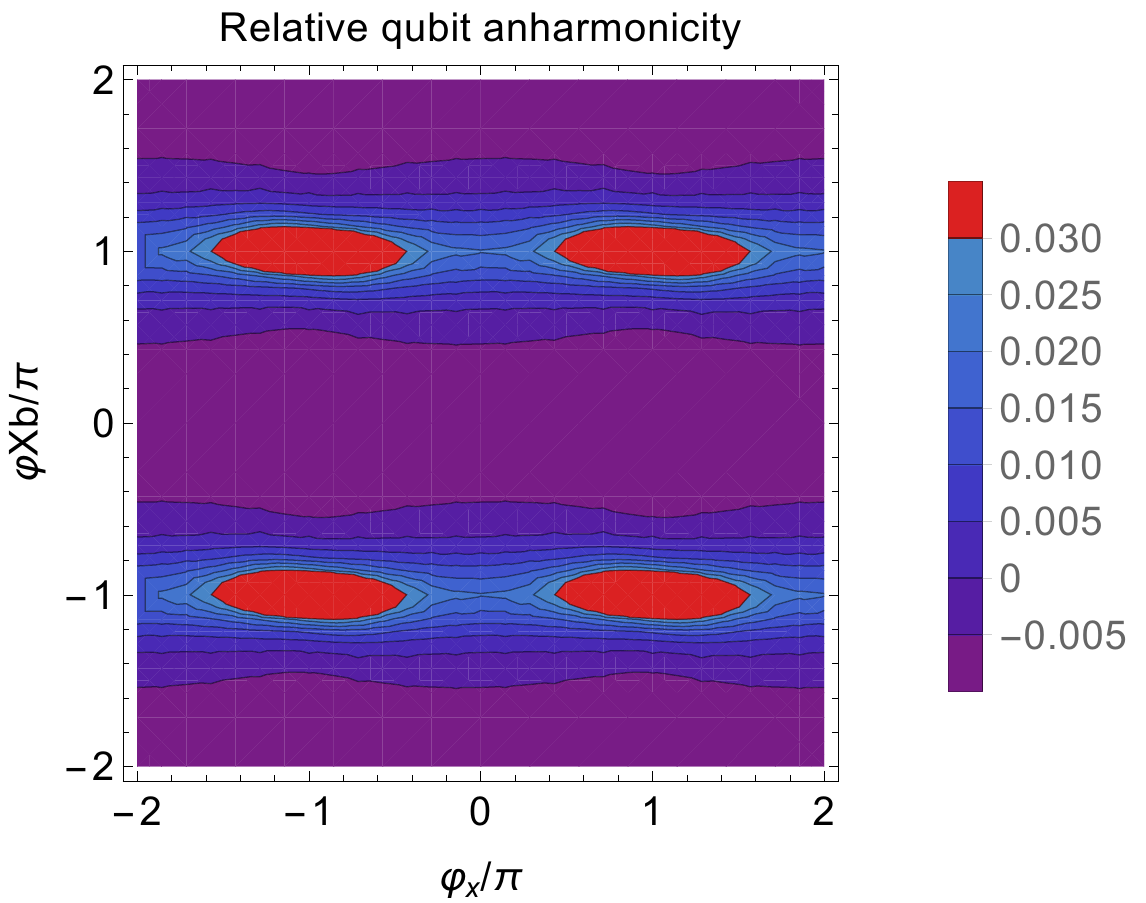}
  \caption{Relative anharmonicity of the qubit as a function of the (reduced) fluxes through the coupling loops $\varphi_x$ and through the big loop $\varphi_{Xb}$. While the anharmonicity is negative in a large range around $\varphi_{Xb} = 0$ (dark blue region), it changes sign when approaching $\varphi_{Xb} = \pi$ and has a steep maximum at $\varphi_{Xb} = \varphi_x = \pi$. \label{fig:anh3dk1}}   
\end{SCfigure}

\begin{SCfigure}[50][tb]
\centering
    \includegraphics[width=.5\linewidth]{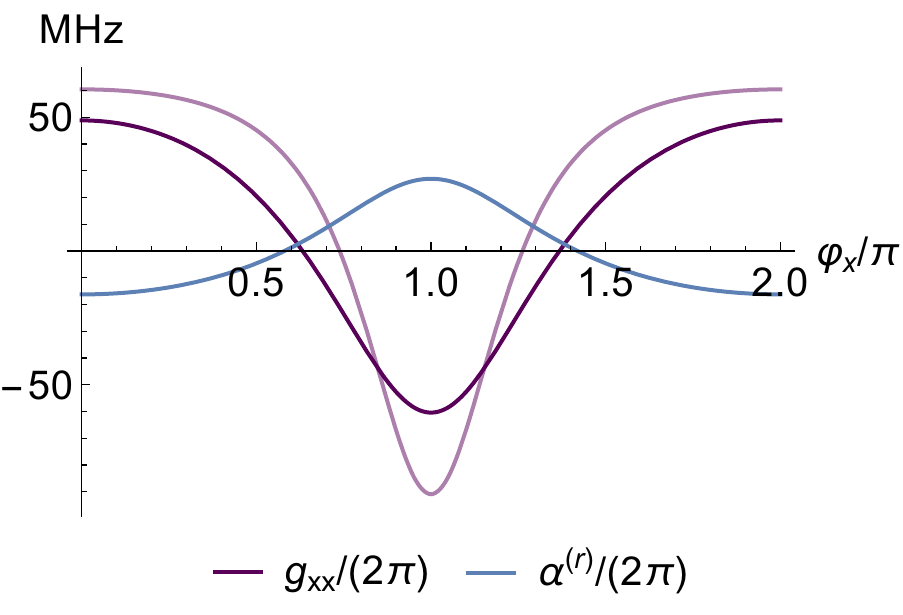}
  \caption{Due to the flux-biasing, the zero-crossing of the transverse coupling gets shifted away from the zero-crossing of the resonator anharmonicity. The lighter purple curve shows the transverse coupling with flux-biasing.  \label{fig:shift}}   
\end{SCfigure}

With the parameters from Tab.~\ref{tb:k1}, $E_{Jq}$ is approximately half of $E_L$, which means that the denominator in Eq.~\ref{eq:anhpi} is positive (this is actually required by Eq.~\ref{eq:lcrit}), while the numerator is negative, yielding a positive anharmonicity. Figure~\ref{fig:anh3dk1} shows that the qubit anharmonicity has a steep maximum at $\varphi_{Xb} = \varphi_x = \pi$. While it is negative in a large range around $\varphi_{Xb} = 0$, it changes sign when approaching $\varphi_{Xb} = \pi$. This implies that there is a point in between where the qubit anharmonicity is zero. The qubit with $\varphi_{Xb}=\pi$ is reminiscent of the capacitively shunted flux qubit (CSFQ) (see Ref.~\cite{Yan2016}). Note that due to the positive anharmonicity, the effective qubit frequency (see Eq.~\ref{eq:delta}) does not drop as much as Eq.~\ref{eq:omega_pi} suggests.\\
We have stated above that the point where the transverse coupling disappears is considerably shifted due to the flux-biasing. As the maximum of the longitudinal coupling experiences the same shift, this does not seem very important. However, this also means that the zero-crossing of the transverse coupling gets shifted away from the zero-crossing of the resonator anharmonicity, as can be seen in Fig.~\ref{fig:shift}. It shows the transverse coupling with and without the flux-biasing (as also shown in Fig.~\ref{fig:coupk1_1}), along with the resonator anharmonicity. The lighter purple curve shows the transverse coupling at a flux of $\varphi_{Xb} = \pi$ in the big loop, while the dark purple curve shows the transverse coupling at $\varphi_{Xb} = 0$. The blue curve shows the resonator anharmonicity, which has its zero-crossing almost at the same point as the dark purple curve. Note that as opposed to Fig.~\ref{fig:anhk1}, this is not a relative anharmonicity. The resonator anharmonicity is not affected by the flux-biasing, which only influences the qubit. \\
With respect to the implementation of the scalable architecture discussed in Sec.~\ref{sec:scalable_architecture}, this effect of the flux-biasing is problematic. As also mentioned in Sec.~\ref{sec:quantization}, the scalable architecture relies on sideband transitions, both for the controlled-phase gate described in Sec.~\ref{sec:gate} and the readout scheme discussed in Sec.~\ref{sec:readout_billangeon}. These are possible due to the absence of the dispersive shift for longitudinal coupling (see Sec.~\ref{sec:longitudinal}), which implies that a certain sideband transition stays resonant irrespective of the number of photons in the resonator. Without the flux-biasing, we can assume that pure longitudinal coupling always goes along with an almost perfectly harmonic resonator. Figure~\ref{fig:shift} shows, however, that this is not true anymore when we consider flux-biasing.

\sidecaptionvpos{figure}{t}

In the following, we will try out different adaptations of the original circuit, one simply being arrays of Josephson junctions instead of the single coupling junctions. We will see that this leads to a significant suppression of both the unwanted coupling terms $g_{xz}$ and $g_{zz}$ and the resonator anharmonicity. However, we will first have a look at the effect of asymmetries in the next section.


\section{EFFECT OF ASYMMETRIES}
\label{sec:asymmetry}
The capacitances and inductances in the design shown in Fig.~\ref{fig:qubit_resonator} are supposed to be symmetric, such that the coupling between qubit and resonator is only created by the coupling junctions $E_{J1}$ and $E_{J2}$. This has the advantage that the resulting transverse coupling (Eq.~\ref{eq:gxx}) is flux-dependent and goes through zero at $\varphi_x = k\, \pi/2$, which leads to pure longitudinal coupling. Transverse coupling terms caused by asymmetric inductances or capacitances would, however, be independent of the external flux. While the capacitances in our design (see Sec.~\ref{sec:implementation}) can be fabricated very accurately, the asymmetry in the inductances could be in the neighborhood of $\delta L = (L_1 - L_2)/(L_1 + L_2) \sim 0.01$. \\
To first order in $\delta L$, the frequencies and anharmonicities of qubit and resonator are unaffected by this asymmetry. The same is true for  the coupling terms with even qubit parity, such as the longitudinal coupling $g_{zx}$. The coupling terms with odd qubit parity, such as the transverse coupling $g_{xx}$, contain, however, a term proportional to $\delta L$. To first order in $\delta L$, the total transverse coupling is given by

\begin{figure}[tb]
\centering
\subcaptionbox{The dotted lines show how the transverse coupling gets shifted for different inductance asymmetries of $\delta L = 0.5 \%,  1.0\%, 1.5 \%$. The solid lines show the longitudinal and transverse coupling at $\delta L = 0$. All other parameters are as given in Tab.~\ref{tb:k1}, notably $d = 8 \%$. \label{fig:asyml}}
  [.48\textwidth]{\includegraphics[width=.48\textwidth]{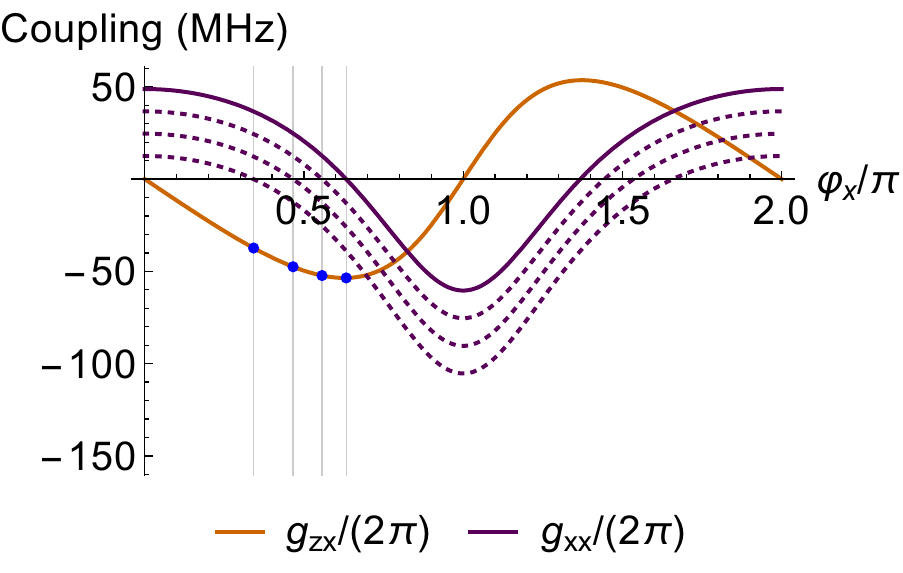}}
\hspace{.01\textwidth}
\subcaptionbox{The dotted lines show the transverse coupling for inductance asymmetries of $\delta L = 0.5 \%,  1.0\%, 1.5 \%$, while keeping the ratio between coupling junction asymmetry and inductance asymmetry fixed at $d = 10\, \delta L$. The solid lines show the longitudinal and transverse coupling at $\delta L = 0$, $d = 8\%$. \label{fig:asymd}}
  [.48\textwidth]{\includegraphics[width=.48\textwidth]{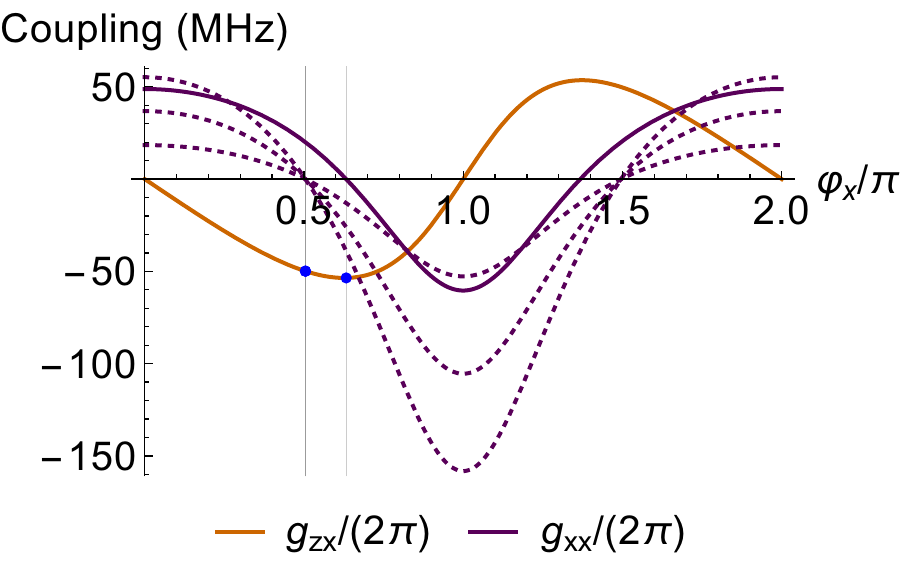}}
\caption{The blue points in both plots mark the value of the longitudinal coupling at the point where the transverse coupling disappears. Increasing the junction asymmetry $d$ helps to keep the zero-crossing of the transverse coupling near the maximum of the longitudinal coupling.\label{fig:asym}}
\end{figure}

\begin{align}
g_{xx}^\text{total} = g_{xx}^\text{asym} + |g_{xx}^\text{sym}| \cos\left(\frac{\varphi_x}k\right),
\label{eq:coup_asym}
\end{align}

where the symmetric flux-dependent part is given by Eq.~\ref{eq:gxx}.
The asymmetric part due to the unequal inductances is a constant offset independent of flux. From Eq.~\ref{eq:coup_asym} it is clear that as long as $|g_{xx}^\text{sym}| / |g_{xx}^\text{asym}| > 1$, there is still a point in flux where the total transverse coupling $g_{xx}^\text{total}$ goes through zero. For a significant $\delta L$ of a few percent, this point might, however, be considerably shifted from the ideal value of $\varphi_x = k \, \pi/2$, where the longitudinal coupling is maximal (compare Fig.~\ref{fig:asyml}). To first order in $\delta L$, the point in flux where the transverse coupling disappears obeys

\begin{align}
\cos\left(\frac{\varphi_x}k\right) \big |_{g_{xx}^\text{total} = 0} = \frac{g_{xx}^\text{asym}}{|g_{xx}^\text{sym}|} = \frac{\delta L}{d} \frac{k\,\Phi_0^2}{\pi^2 E_{J\Sigma} (L_1 + L_2)}.
\label{eq:trans_zero}
\end{align}

\begin{SCfigure}[50][tb]
\centering
    \includegraphics[width=.5\textwidth]{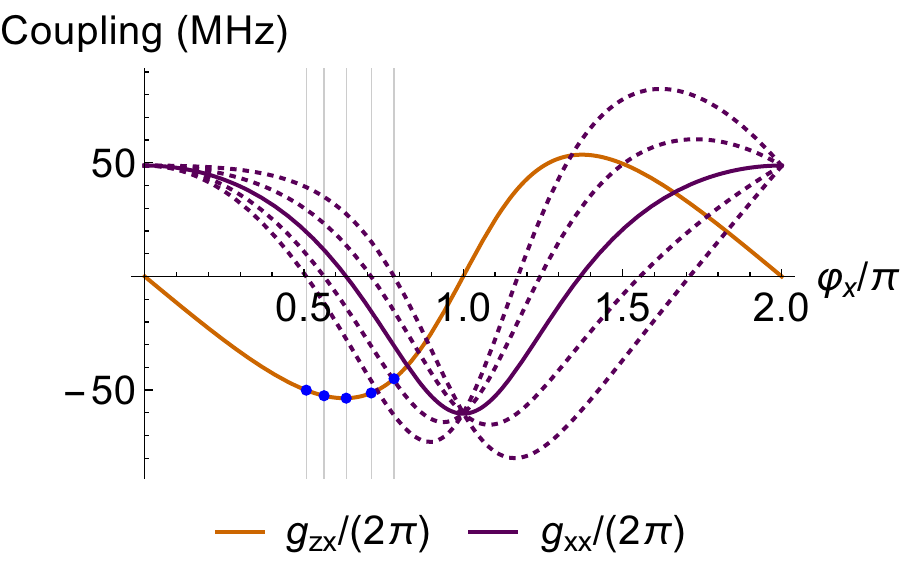}
  \caption{An asymmetry in the external fluxes through the coupling loops leads to a reshaping of the transverse coupling, which shifts the zero crossing. The dotted lines show the transverse coupling for $\delta \varphi_x = (\varphi_{x1} - \varphi_{x2})/(\varphi_{x1} + \varphi_{x2}) = \pm 0.5\%, \pm 1.0\%$. \label{fig:asymphi}}   
\end{SCfigure}

Note that while $g_{xx}^\text{asym}$ is proportional to $\delta L$, $g_{xx}^\text{sym}$ is proportional to the coupling junction asymmetry $d = E_{J\Delta}/E_{J\Sigma}$. It is clear from Eq.~\ref{eq:trans_zero} that for a fixed ratio of $\delta L/d$, the transverse coupling will always disappear at the same point in flux. If the asymmetry in the inductances should be problematically high, this could be compensated by an increased coupling junction asymmetry $d$ (compare Fig.~\ref{fig:asymd}).\\
An asymmetry in the external fluxes through the two coupling loops mainly has an impact on the coupling terms with odd qubit parity, such as the transverse coupling, while its effect on all other quantities is negligible. In the transverse coupling, it leads to a reshaping of the flux dependence and thereby also slightly shifts the zero-crossing point as shown in Fig.~\ref{fig:asymphi}. Assuming that the flux asymmetry can be kept below one percent, it would not lead to any significant degradation of the system's behavior. In the case of asymmetric inductances, an asymmetric flux through the coupling loops could even be beneficial, as it could be used to move the zero-crossing of the transverse coupling back to the maximum of the longitudinal coupling.

\sidecaptionvpos{figure}{t}

\section{CASE TWO: COUPLING JUNCTION ARRAYS}
\label{sec:array}
When we want to substitute the single coupling junctions by coupling arrays as suggested in Fig.~\ref{fig:qubit_resonator}, there are a few things we have to take into account.  Assuming that all junctions in such an array are equal, we can describe the potential energy of an array of $k$ junctions as

\begin{align}
\mathcal{U}_k = - \sum_{i=1}^k E_{Ji} \cos(\varphi_i) = - k E_J \cos\left(\frac{\varphi + m \,2\pi}{k}\right),
\label{eq:u_array}
\end{align}

where $\varphi = \sum_i \varphi_i$ is the total phase over the junction array and $m \in \mathbb{Z}$ is the integer number of flux quanta in a loop formed by the junctions, thereby numbering the metastable solutions for $\varphi$. For the coupling scheme to work, we require $m$ to be constant in time over long durations.
As described in Ref.~\cite{Masluk2012, Matveev2002}, so-called phase-slip events, that is integer changes in $m$, can be detected by jumps in the frequency of the system. However, phase slips are suppressed by choosing a large $E_{Ji}/E_{Ci}$ ratio for each individual junction, and time spans on the order of hours or days with constant $m$ can be realistically achieved \cite{Pop2010}. We will take $m$ to be zero here and expand for large $k$, finding

\begin{align}
\mathcal{U}_k \approx - k E_J + \frac{E_J}{2 k} \varphi^2  = - k E_J + \left(\frac{\Phi_0}{2\pi}\right)^2 \frac{1}{2L_J} \, \varphi^2
\end{align}
 
with an effective inductance of $L_J = k\, (\Phi_0/(2\pi))^2 /E_J$. The effective inductance of such an array is thus proportional to the number of junctions~$k$.

\begin{figure}[tb]
\centering
\subcaptionbox{Longitudinal ($g_{zx}$) and transverse coupling ($g_{xx}$). \label{fig:coupkn_1}}
  [.48\linewidth]{\includegraphics[width=.48\linewidth]{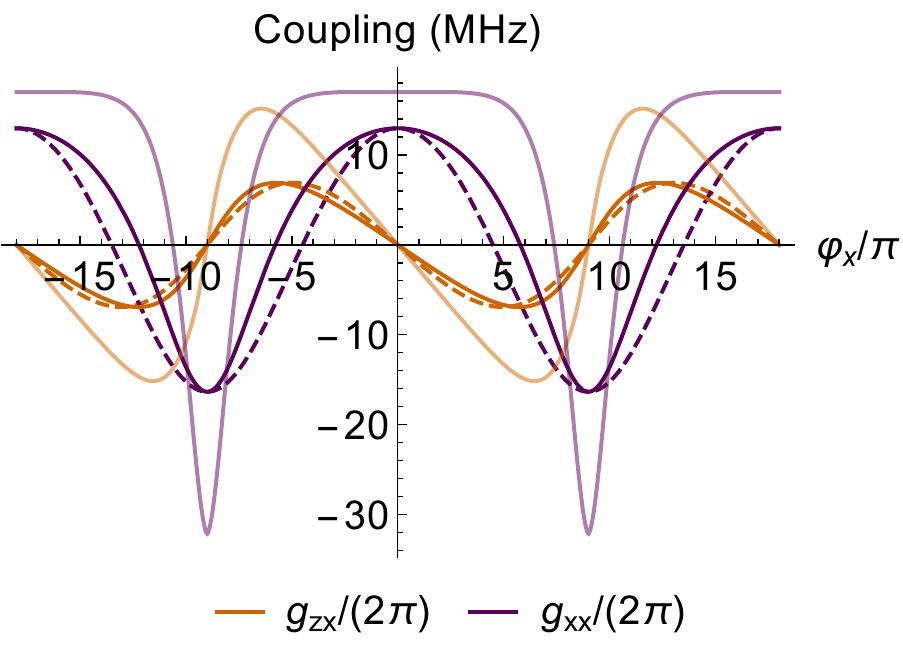}}
\hspace{5pt}
\subcaptionbox{Unwanted coupling terms, note the change of scale. \label{fig:coupkn_2}}
  [.48\linewidth]{\includegraphics[width=.48\linewidth]{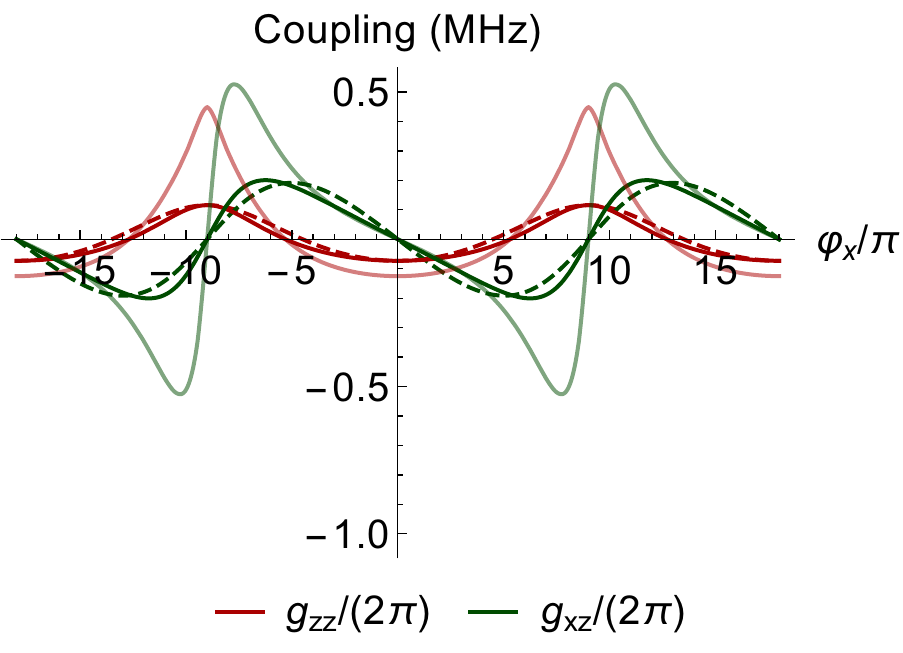}}
\caption{The four most important coupling terms as a function of the (reduced) flux through the coupling loops $\varphi_x$ for a coupling junction array of $k = 9$ junctions per array. Solid lines show accurate numerical results, dashed lines show the predictions using the formulas from Chapt.~\ref{chapt:design}, the lighter color curves show what happens at a flux $\varphi_{Xb} = \pi$ in the big loop. The longitudinal coupling is almost doubled due to the large-loop flux-biasing, while the point where the transverse coupling disappears is considerably shifted. The longitudinal coupling always disappears exactly at multiples of $\varphi_x = k\, \pi$.    \label{fig:coupkn}}
\end{figure}

With such a treatment, we are of course neglecting the dynamics of the internal degrees of freedom of the array \cite{Viola2015, Hutter2011}. This is justified as long as the energies of these degrees of freedom are far enough separated from the relevant energies of our system, that is the frequencies of qubit and resonator. Explicitly, we have to require their plasma frequencies $\sqrt{8 E_{Ci} E_{Ji}}/h$ to be above 20 GHz in order to push the self-resonant modes of the array well above the resonator mode. Apart from that, we require that $E_{Ji}/E_{Ci} \geq$ 100 to prevent phase slips \cite{Matveev2002, Pop2010}, where $E_{Ci}$ is the charging energy of each individual junction. Putting these two constraints together, we can conclude that each coupling junction in such an array needs to have a Josephson energy larger than $E_{Ji} = h $ 70 GHz, which is a lot bigger than what we assumed for the single-junction case. Apart from this restriction, we proceed just as in the previous section, trying out different parameter values, now including the number of junctions $k$ per coupling array, until we find the \textit{optimal} solution.

\begin{table}[t]
\begin{center}
    \begin{tabular}{ c | c || c | c}
    \multicolumn{2}{c}{Parameters} & \multicolumn{2}{c}{Results}\\
    \hline
    & & & \\
    $E_{Jq}$ & $h$ 10 GHz & $\omega_r/(2\pi)$ & 6 - 8 GHz \\ [\smallskipamount]
    $E_{J\Sigma}$ & $h$ 160 GHz & $\Delta/(2\pi)$ & 5.3 - 6.3 GHz\\ [\smallskipamount]
    $E_{J\Delta}/E_{J\Sigma}$ & 0.02 & $g_{zx}^\text{max}/(2\pi)$ & 6 MHz \\ [\smallskipamount]
         $C$ & 102 fF & $g_{xx}^\text{max}/(2\pi)$ & 13 MHz \\ [\smallskipamount]
     $C_q$ & 60 fF & $g_{zz}^\text{max}/(2\pi)$ & 0.07 MHz \\ [\smallskipamount]
   $L$ & 5.0 nH & $g_{xz}^\text{max}/(2\pi)$ &  0.2 MHz \\ [\smallskipamount]
   $L_\text{max}$ & 5.0 nH & $|\alpha_r^{(q)}|$ & 0.9 - 1.5\% \\ [\smallskipamount]
   $L_\text{crit}$ & 5.6 nH & $|\alpha_r^{(r)}|$ &  $\leq$ 0.007\% \\ [\smallskipamount]
    \end{tabular}
    \captionof{table}{The chosen parameters for the case of coupling junction arrays, here for $k=9$, at zero flux through the big loop $\varphi_{Xb} = 0$. $L$ needs to be less or equal to $L_\text{max}$ to ensure that the resonator frequency stays in the 6 - 8 GHz range and less than $L_\text{crit}$ in order to avoid a double-well potential for all possible values of flux.
    On the right the frequencies, anharmonicities, and couplings, which vary with the flux in the coupling loops.}
    \label{tb:kn}
\end{center}
\end{table}

Table~\ref{tb:kn} shows the chosen parameters for the multi-junction case, here for $k = 9$ junctions per array. While we ascertain that the unwanted coupling terms are considerably suppressed compared to the longitudinal and the transverse coupling, the coupling is smaller in general (see Fig.~\ref{fig:coupkn}). The longitudinal coupling is suppressed by almost one order of magnitude compared to the single-junction case. While the unwanted $g_{xz}$ coupling reaches approximately 4~\% of the longitudinal coupling $g_{zx}$ at their joint maximum, the unwanted $g_{zz}$ coupling reaches only 0.4~\% of the transverse coupling $g_{xx}$ at their maximum.

\begin{figure}[tb]
\centering
\includegraphics[width=.48\linewidth]{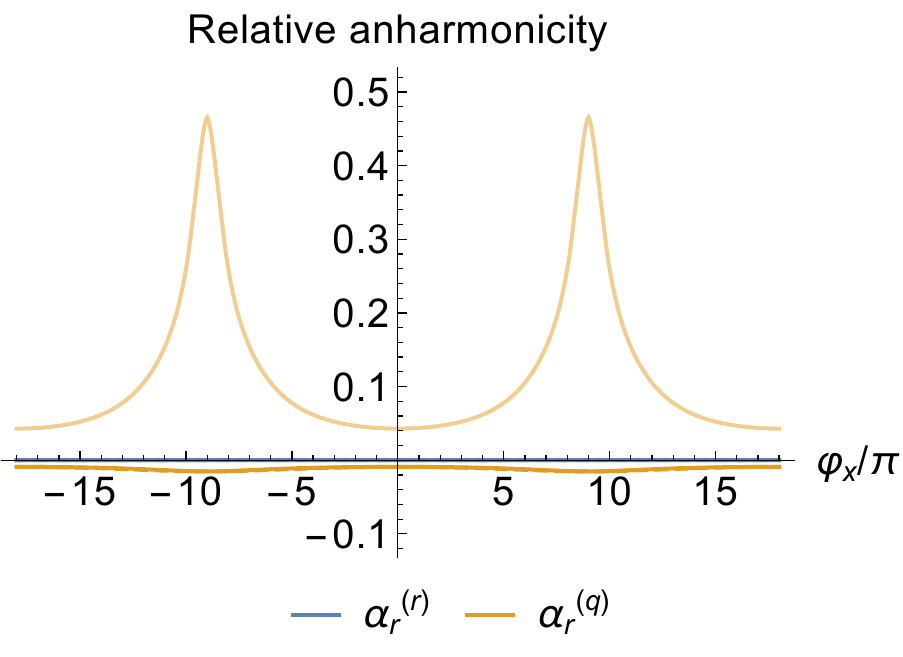}
\hspace{5pt}
\includegraphics[width=.48\linewidth]{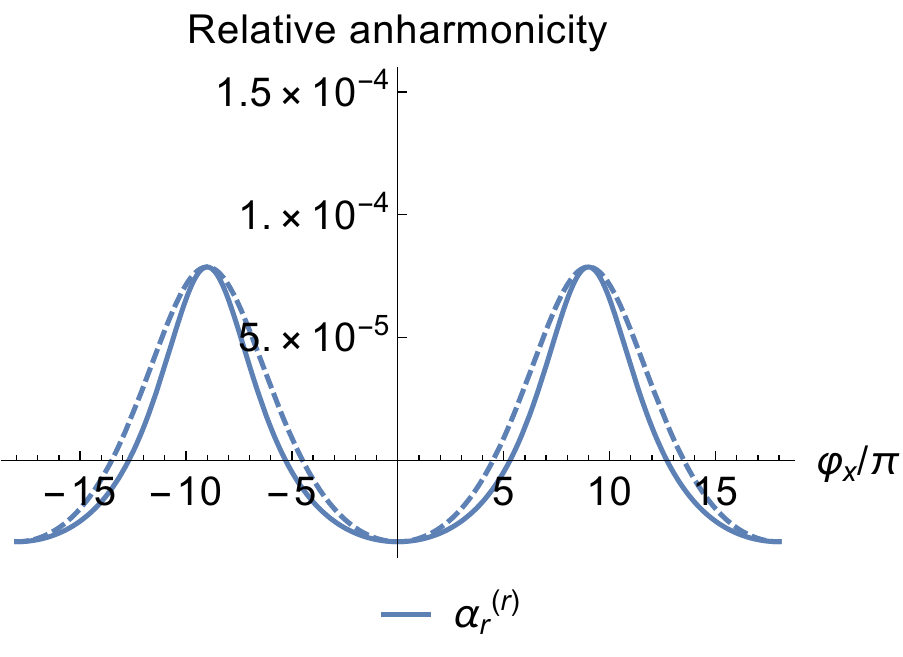}
\caption{The relative anharmonicities of qubit and resonator as a function of the (reduced) flux through the coupling loops $\varphi_x$ for a coupling junction array with $k = 9$ junctions per array. Solid lines show accurate numerical results, dashed lines show the predictions using the formulas from Chapt.~\ref{chapt:design}. Note that in this plot the predictions are indistinguishable from the numerical results. The lighter color curve shows results at a flux $\varphi_{Xb} = \pi$ in the big loop. The smaller plot on the right shows again the relative anharmonicity of the resonator, note the change of scale.    \label{fig:anhkn}}
\end{figure}

The resonator anharmonicity is considerably suppressed to less than 0.007~\%, while the qubit anharmonicity stays roughly the same (see Fig.~\ref{fig:anhkn}). Putting a flux of $\varphi_{Xb} = \pi$ through the big loop has a similar effect as before. The qubit anharmonicity changes sign and is boosted to up to more than 30~\%, while the resonator anharmonicity is not affected by the large-loop flux-biasing. The zero-crossing of the transverse coupling is shifted due to the flux-biasing, along with the maximum of the longitudinal coupling.
Even though the suppression of the unwanted coupling and the resonator anharmonicity are a considerable improvement over the single-junction case, the simultaneous suppression of the longitudinal coupling is unfortunate. We will therefore try out another adaptation of the original circuit, as described in Sec.~\ref{sec:adaptation}.

\section{CASE THREE: CIRCUIT WITH ADDITIONAL INDUCTANCE}
\label{sec:adaptation}

\begin{figure}[tb]
\centering
\captionbox{Adapted qubit-resonator system with additional inductances $L_{a}$ in the coupling branches. Just as in the previous designs, the coupling is controlled via the flux through the coupling loops $\Phi_x$, while the flux through the big loop $\Phi_{Xb}$ can be used to boost the anharmonicity.    \label{fig:add}}
  [.56\linewidth]{\includegraphics[height=4.8cm]{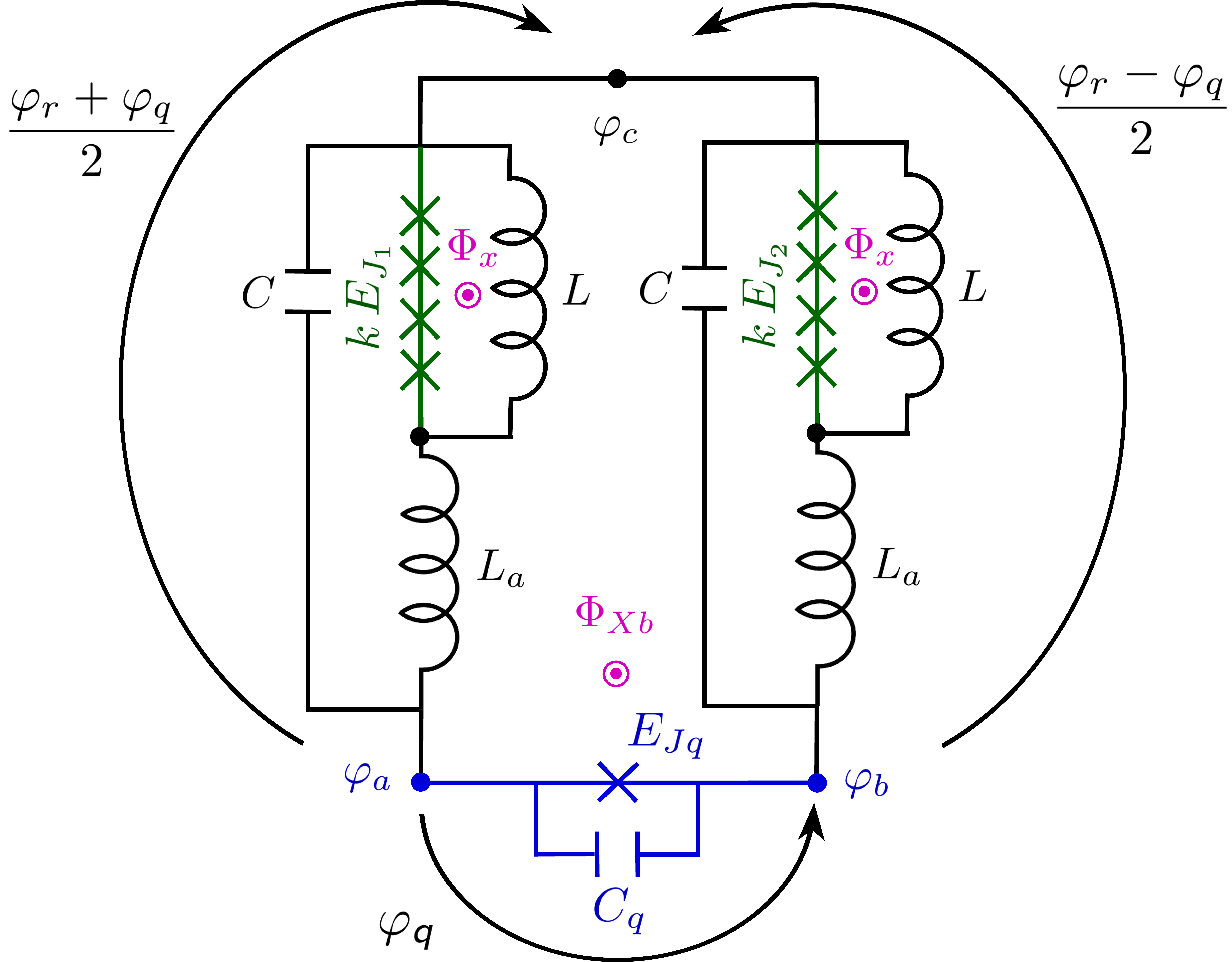}}
\hspace{5pt}
\captionbox{Detail of a single coupling branch from Fig.~\ref{fig:add}. The phase difference across the junction array $\varphi_d$ has no dynamics on its own but depends on the phase difference $\varphi$ across the whole device. \label{fig:junc_ind}}
  [.4\linewidth]{\includegraphics[height=4.8cm]{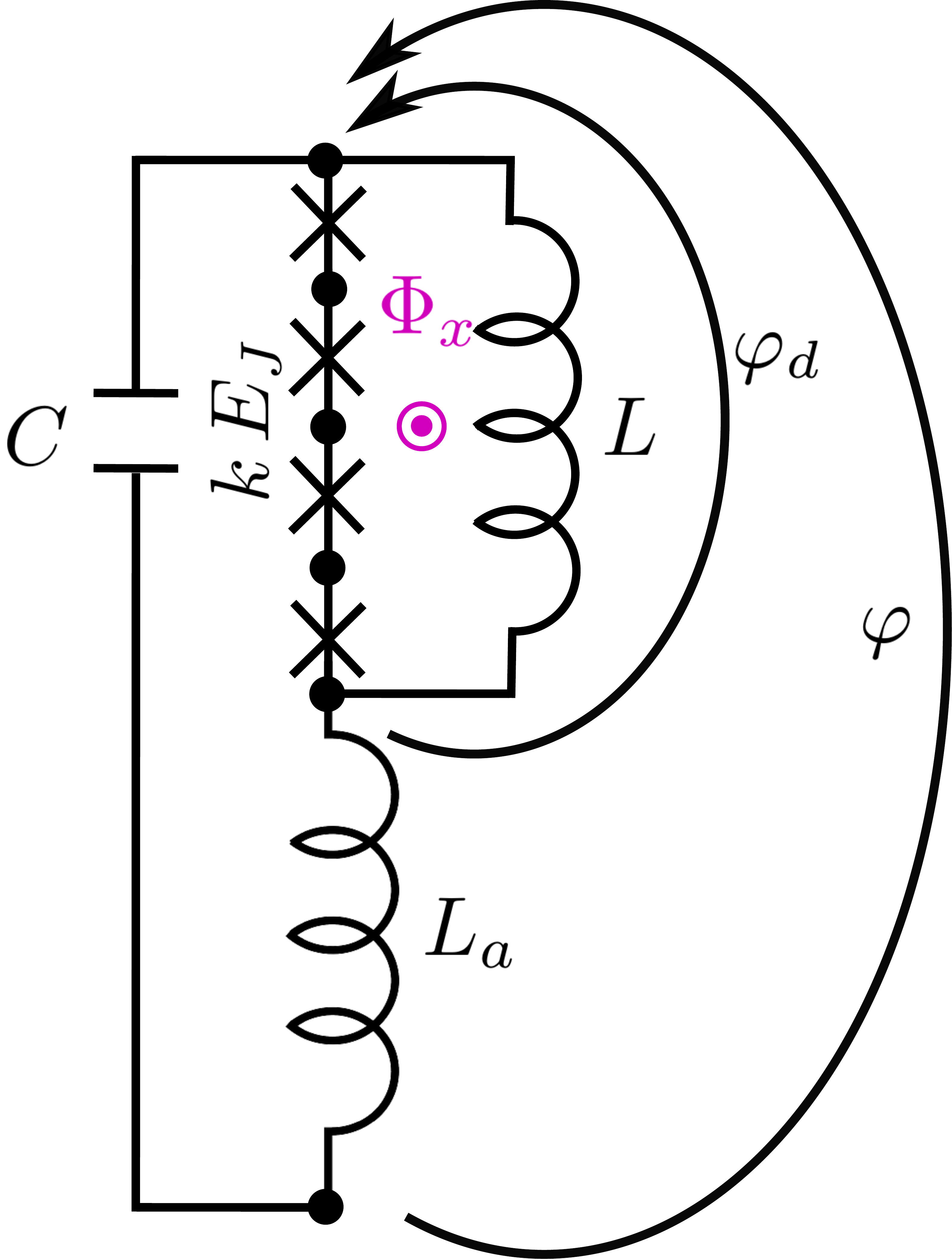}}
\end{figure}

Figure~\ref{fig:add} shows an adaptation of the original circuit in which we have added an additional inductor in each coupling branch, in series with both the coupling junction array and the already existing inductor. 
Clearly, this adds an additional degree of freedom to each coupling branch. However, this additional variable can be considered to be a dependent variable (just like the internal degrees of freedom within the coupling array) that can be eliminated, as it does not have a significant capacitive term and therefore no low-frequency dynamics on its own.

To explain this, we will start with a description of a single coupling branch as shown in Fig.~\ref{fig:junc_ind}. 
This is a system with $n=k+2$ nodes, where $k$ is the number of junctions in the array. That makes $n-1=k+1$ degrees of freedom, $k$ of them without their own dynamics. We define $\varphi$ as the phase difference across the whole device and denote the dependent phase difference over the junction array as $\varphi_d$ as depicted in Fig.~\ref{fig:junc_ind}, the phase difference over a single junction being $\varphi_d/k$, assuming the junctions are all equal and $m = 0$ in Eq.~\ref{eq:u_array}. This already eliminates all phases inside the junction array. The phase difference across the inductance $L_a$ must then be $\varphi-\varphi_d$. The Lagrangian for the system shown in Fig.~\ref{fig:junc_ind} yields

\begin{align*}
\mathcal{L} = \left(\frac{\Phi_0}{2\pi}\right)^2 &\left(\frac{C}{2} \dot \varphi^2 - \frac{1}{2L_a} (\varphi - \varphi_d)^2 - \frac{1}{2L} \varphi_d^2\right)\\
& + k \, E_J \cos\left(\frac{\varphi_d + \varphi_x}{k}\right), \numberthis
\end{align*}

where $\varphi_x = 2\pi \, \Phi_x/\Phi_0$ is again the reduced external flux through the coupling loop. From here we can deduce the equations of motion for $\varphi$ and $\varphi_d$, being

\begin{align*}
C \ddot \varphi &=  \frac{1}{L_a} (\varphi_d - \varphi)  \\
0 &= \frac{1}{L_a} (\varphi_d - \varphi) + \frac{1}{L} \varphi_d + \left(\frac{2\pi}{\Phi_0}\right)^2 E_J \sin\left(\frac{\varphi_d + \varphi_x}{k}\right).\numberthis
\end{align*}

As noted above, there is no capacitive term in the second equation. It is thus not a differential equation, but simply a non-linear algebraic equation in $\varphi$ and $\varphi_d$, which can be used to eliminate $\varphi_d$. However, it is not analytically possible to solve the second equation for $\varphi_d$. Our strategy will therefore be to solve it for $\varphi$ and invert this function numerically for a given set of parameters in order to eliminate $\varphi_d$. Solving for $\varphi$ thus yields

\begin{align}
\varphi(\varphi_d) = \gamma \, \varphi_d + \beta \sin\left(\frac{\varphi_d + \varphi_x}{k}\right)
\label{eq:invert}
\end{align}

with the abbreviations $\beta = (2\pi/\Phi_0)^2 L_a E_J$ and $\gamma = 1 + L_a/L$, where $\beta$ corresponds to the screening parameter known from SQUID terminology \cite{Clarke2004}, that is the ratio between Josephson energy and inductive energy. 

\sidecaptionvpos{figure}{c}
\begin{SCfigure}[50][tb]
\centering
    \includegraphics[width=.5\linewidth]{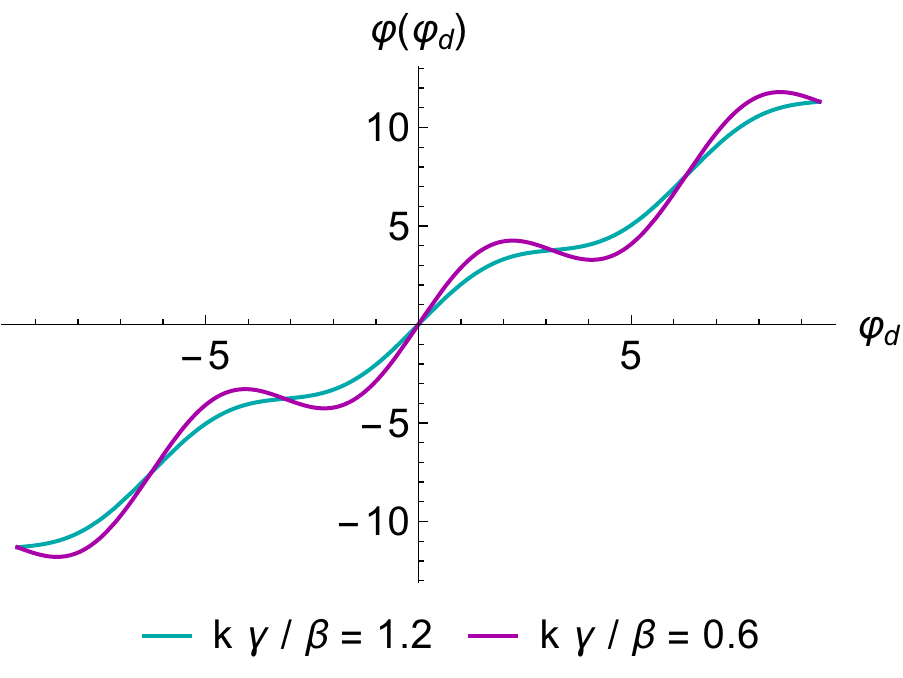}
  \caption{$\varphi$ as a function of the dependent variable $\varphi_d$ (referring to Fig.~\ref{fig:junc_ind}) for different parameters. While the blue curve with $k\, \gamma/\beta > 1$ is invertible, the magenta one with $k\, \gamma/\beta < 1$ is not.    \label{fig:invertible}}   
\end{SCfigure}

In terms of the parameters $k$, $\beta$ and $\gamma$, we can distinguish two different cases. The function is invertible as long as $k\, \gamma/\beta$ is above the critical value of one, compare Fig.~\ref{fig:invertible}. If the function is not invertible, the potential becomes multi-valued. This is a parameter regime we want to avoid. In order to see what that condition means, we can rewrite it as

\begin{align}
\gamma > \beta/k \quad \Leftrightarrow \quad E_{L} + E_{La} > E_J/k,
\label{eq:crit}
\end{align}

where $E_L = (\Phi_0/(2\pi))^2/L$ is the energy associated with the inductance and $E_{La}$ is the same for the additional inductance $L_a$. The condition given in Eq.~\ref{eq:crit} thus means that the energy of the two inductances must be $k$ times bigger than the energy of each junction in the array, in order to ensure that Eq.~\ref{eq:invert} is invertible and there is a well-defined potential.\\
The potential energy for the complete qubit-resonator system depicted in Fig.~\ref{fig:add} is given by

\begin{align*}
\mathcal{U} &= \left(\frac{\Phi_0}{2\pi}\right)^2 \biggl(\frac{1}{2 L_{a}} f\left(\frac{\varphi_r + \varphi_q}{2}, \varphi_x, \beta_1, k, \gamma\right)  \\
&+\frac{1}{2 L_{a}} f\left(\frac{\varphi_r - \varphi_q}{2}, \varphi_x, \beta_2, k, \gamma\right)\biggr) \\
&- E_{Jq} \cos(\varphi_q + \varphi_{Xb}), \numberthis
\end{align*}

where

\begin{align*}
f(\varphi, \varphi_x, \beta, k, \gamma) = \varphi^2 &- 2\, \varphi\, \varphi_d + \gamma \,\varphi_d^2 \\
&- 2 \,k \,\beta \cos\left(\frac{\varphi_d + \varphi_x}{k}\right) \numberthis
\end{align*}

is a function that describes one coupling branch as depicted in Fig.~\ref{fig:junc_ind}, in which the dependent variable $\varphi_d$ must be replaced by the numerical inversion of Eq.~\ref{eq:invert}. The kinetic energy is the same as for the original circuit (Eq.~\ref{eq:kin}). In analogy to what we described in Sec.~\ref{sec:fluxbias}, we want this circuit to be also usable at a flux-biasing of $\varphi_{Xb} = \pi$ through the big loop. We thus have to make sure that we do not go into parameter ranges, where the potential is a double well. This can be done by determining the curvature of the potential in the $\varphi_q$ direction at a flux $\varphi_{Xb} = \pi$ through the big loop and $\varphi_x = k\, \pi$ through the coupling loops. If the curvature is positive here, it will always be positive. While we can not define a critical inductance as done in Sec.~\ref{sec:fluxbias} (Eq.~\ref{eq:lcrit}), the equivalent in this case is a critical (minimum) number of junctions $k_\text{crit}$ that ensures a positive curvature of the potential.

\begin{table}[t]
\begin{center}
    \begin{tabular}{ c | c || c | c }
    \multicolumn{2}{c}{Parameters} & \multicolumn{2}{c}{Results}\\ [\smallskipamount]
    \hline
    & & & \\
    $E_{Jq}$ & $h$ 5 GHz & $\omega_r/(2\pi)$ & 6 - 8 GHz \\ [\smallskipamount]
    $ E_{J\Sigma}$ & $h$ 155 GHz & $\Delta/(2\pi)$ & 4.8 - 5.8 GHz\\ [\smallskipamount]
    $E_{J\Delta}/E_{J\Sigma}$ & 0.02 & $g_{zx}^\text{max}/(2\pi)$ & 10 MHz \\ [\smallskipamount]
    $C$ & 65 fF & $g_{xx}^\text{max}/(2\pi)$ & 9 MHz \\ [\smallskipamount]
    $C_q$ & 50 fF & $g_{zz}^\text{max}/(2\pi)$ & 0.06 MHz \\ [\smallskipamount]
    $L$ & 4.5 nH & $g_{xz}^\text{max}/(2\pi)$ & 0.5 MHz \\ [\smallskipamount]
    $L_a$ & 3 nH &  $|\alpha_r^{(q)}|$ & 1.1 - 2\%  \\ [\smallskipamount]
     $k_\text{crit}$ & 3.3 & $|\alpha_r^{(r)}|$ &  $\leq$ 0.003\%  \\ [\smallskipamount]
    \end{tabular}
    \captionof{table}{The chosen parameters for the case of the adapted circuit with the added inductance, here for $k=5$ junctions per coupling array, at zero flux through the big loop $\varphi_{Xb} = 0$. $k_\text{crit}$ defines a lower threshold for the number of junctions $k$ in order to avoid a double-well potential for all possible values of flux.
    On the right the frequencies, anharmonicities, and couplings, which vary with the flux in the coupling loops.}
    \label{tb:add}
\end{center}
\end{table}

\begin{SCfigure}[50][tb]
\centering
    \includegraphics[width=.5\linewidth]{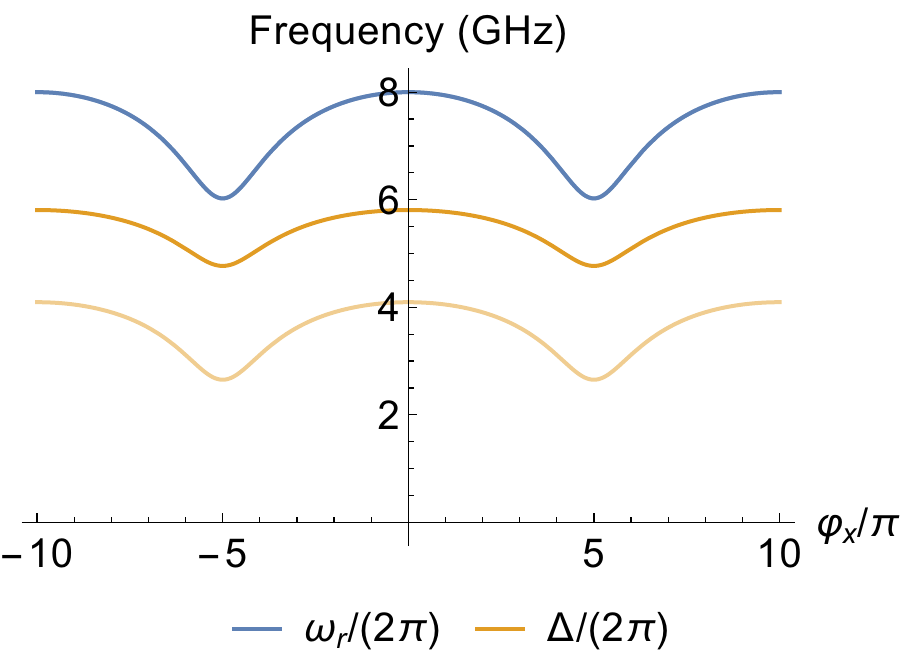}
  \caption{The qubit frequency $\Delta$ and the resonator frequency $\omega_r$ as a function of the (reduced) flux through the coupling junctions $\varphi_x$ for the adapted circuit with $k = 5$ junctions per array. The lighter color curve shows results at a flux $\varphi_{Xb} = \pi$ in the big loop.\label{fig:addfreq}}   
\end{SCfigure}

From here on, our strategy is the one described in Sec.~\ref{sec:single}. For a given set of parameters (including the external fluxes), we first determine the position of the potential energy minimum in $\varphi_q$ and $\varphi_r$ and then calculate the frequencies, anharmonicities, and couplings using series approximations around that minimum. To choose the best parameters, we again fix the capacitance $C$ in terms of the other variables, such that the resonator frequency is $\omega_r/(2\pi) = 8$ GHz at zero flux, where $\omega_r$ has its maximum. Then we try out different values for the other parameters until we find the solution that gives the highest longitudinal coupling and anharmonicity, while satisfying all the conditions mentioned above. The chosen parameters for this circuit are shown in Tab.~\ref{tb:add}. \\
Figure~\ref{fig:addfreq} shows the frequencies of qubit and resonator as a function of the flux through the coupling loops $\varphi_x$. Their flux dependence looks a lot like in the two cases described above. The lighter color curve shows results at a flux $\varphi_{Xb} = \pi$ in the big loop. The qubit frequency again experiences a drop due to the large-loop flux-biasing, while the resonator is unaffected by this.

\sidecaptionvpos{figure}{c}

\begin{figure}[tb]
\centering
\subcaptionbox{Longitudinal ($g_{zx}$) and transverse coupling ($g_{xx}$).    \label{fig:addcoup_1}}
  [.48\linewidth]{\includegraphics[width=.48\linewidth]{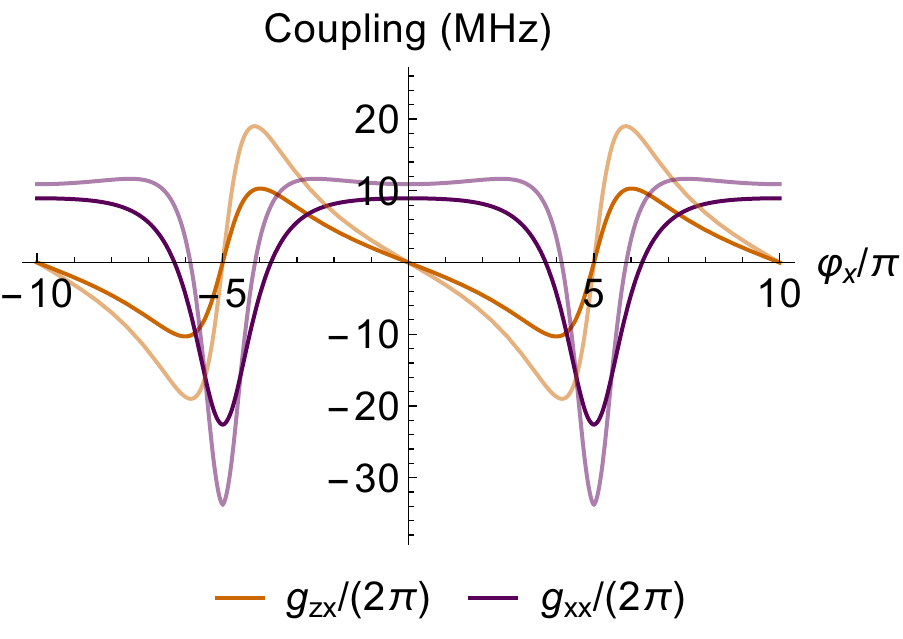}}
\hspace{5pt}
\subcaptionbox{Unwanted coupling terms, note the change of scale. \label{fig:addcoup_2}}
  [.48\linewidth]{\includegraphics[width=.48\linewidth]{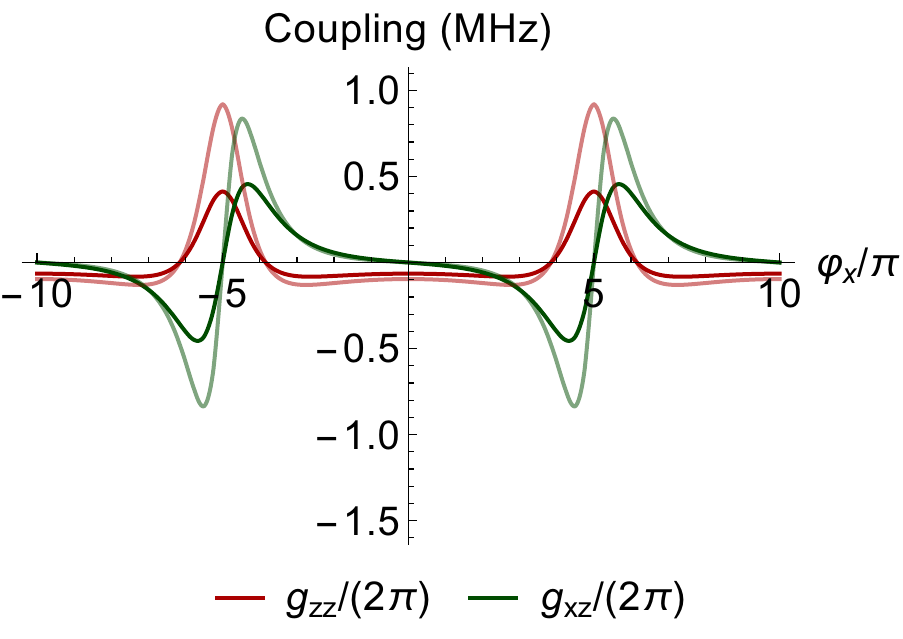}}
\caption{The four most important coupling terms as a function of the (reduced) flux through the coupling junctions $\varphi_x$ for the adapted circuit with $k = 5$ junctions per array. The lighter color curves show what happens at a flux $\varphi_{Xb} = \pi$ in the big loop. Due to the large-loop flux-biasing, all coupling terms get slightly bigger (the longitudinal coupling is almost doubled), while the point where the transverse coupling disappears is considerably shifted.\label{fig:addcoup}}
\end{figure}

Figure~\ref{fig:addcoup} shows the four most important coupling terms, again as a function of the flux through the coupling loops $\varphi_x$. Compared to the single-junction case, the coupling is smaller and loses its resemblance to the trigonometric functions in the formulas given in Chapt.~\ref{chapt:design}. The unwanted coupling terms are suppressed. The unwanted $g_{xz}$ coupling reaches 4~\% of the longitudinal coupling at their joint maximum, while the unwanted $g_{zz}$ coupling is about 0.5~\% of the transverse coupling at zero flux. Though the longitudinal coupling $g_{zx}$ is smaller than in the single-junction case, it is slightly bigger than in the case of the coupling junction array without the additional inductor. The lighter color curves show what happens at a flux $\varphi_{Xb} = \pi$ in the big loop. Due to the large-loop flux-biasing, the longitudinal coupling is almost doubled. The point where the transverse coupling disappears is considerably shifted, along with the maximum of the longitudinal coupling. 

\begin{figure}[tb]
\centering
\includegraphics[width=.48\linewidth]{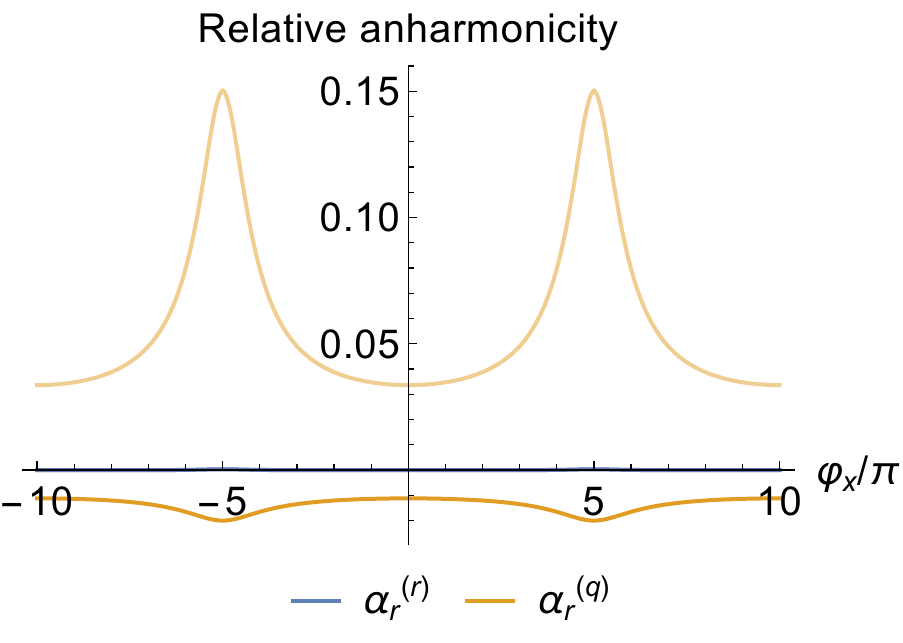}
\hspace{5pt}
\includegraphics[width=.48\linewidth]{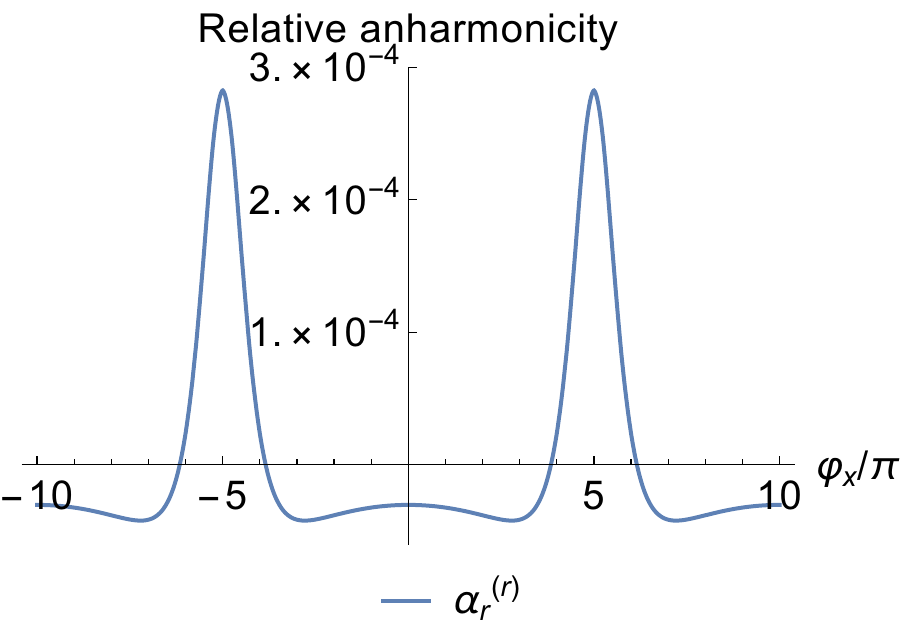}
\caption{The relative anharmonicities of qubit and resonator as a function of the (reduced) flux through the coupling loops $\varphi_x$ for the adapted circuit with $k = 5$ junctions per array. The lighter color curve shows results at a flux $\varphi_{Xb} = \pi$ in the big loop. While the resonator anharmonicity is unchanged, the qubit anharmonicity is now positive and boosted to up to 18~\%. The plot on the right shows again the relative anharmonicity of the resonator, note the change of scale. \label{fig:addanh}}
\end{figure}

The qubit anharmonicity is slightly bigger than in the single-junction or the coupling-array case, while the resonator anharmonicity is suppressed to less than 0.03~\% (see Fig.~\ref{fig:addanh}). The large-loop flux-biasing leads again to a boost in qubit anharmonicity, here to up to 18~\%. \\
We can conclude that the adapted circuit with the additional inductor works better than the circuit using only the junction array. The single-junction case seems problematic due to the high resonator anharmonicity. In all cases flux-biasing with $\varphi_{Xb} = \pi$ in the big loop leads to a boost in anharmonicity and to an increase of coupling strength of almost a factor of two.

\section{PHYSICAL IMPLEMENTATION}
\label{sec:implementation}

\begin{figure*}[ht]
	\centering
	\vspace{-2mm}
	\includegraphics[width=.95\textwidth]{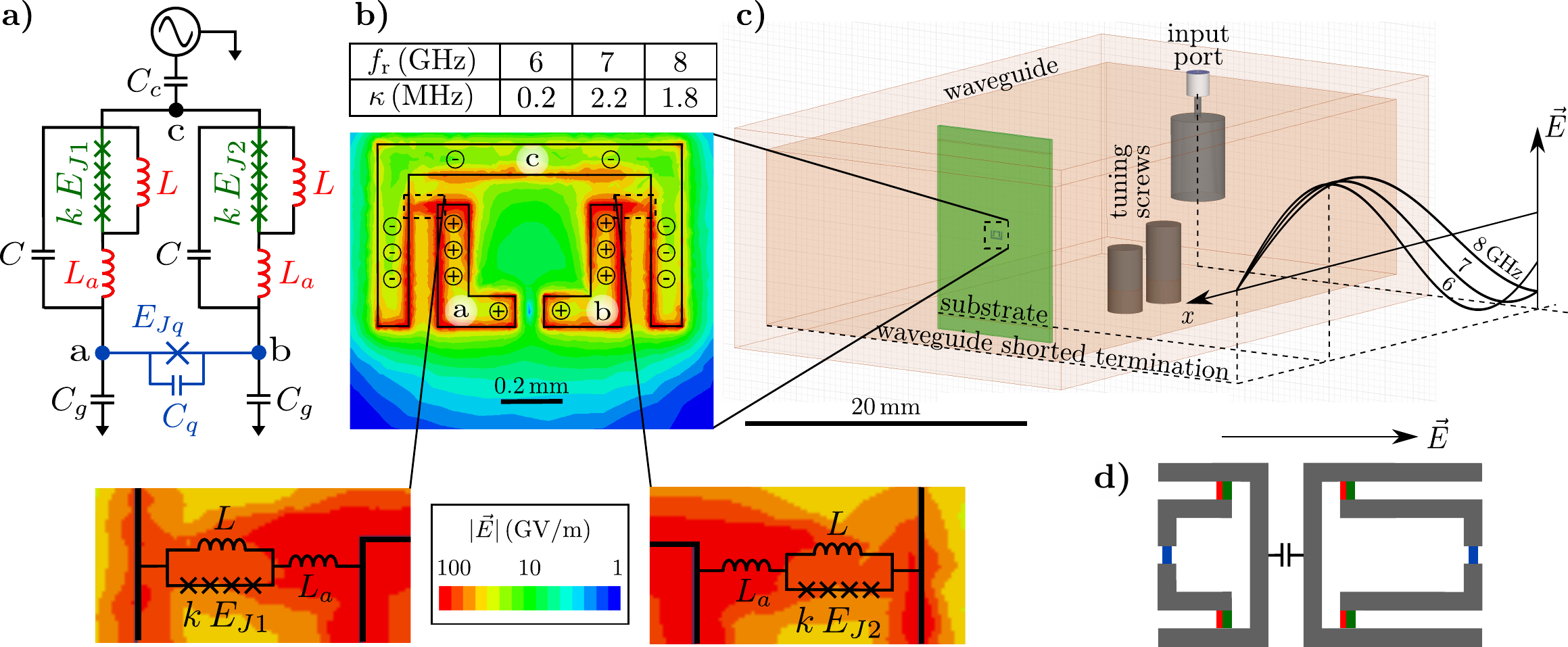}
	\vspace{-2mm}
	\caption{\small Proposal for the physical implementation of the inductively shunted transmon qubit with tunable transverse and longitudinal coupling. \textbf{a)} Electrical schematic of the qubit-resonator circuit coupled to the control and readout microwave environment. The qubit dynamics is dominated by the Josephson junction with energy $E_{Jq}$ and capacitance $C_q$ (colored in blue). The frequency of the resonator mode is given by the equivalent inductance formed by $L$, $L_a$ (colored in red), the inductances of the Josephson junction arrays (colored in green), and the shunting capacitances $C$. \textbf{b)} Finite element model used to simulate the resonator coupling to the microwave drives. The color scale indicates the magnitude of the computed electric field at the surface of the thin film superconducting electrodes, for a total energy stored in resonator mode of $1\,\mathrm{J}$. The $\mathbf{+}$ and $\mathbf{-}$ symbols represent the polarity of the electric field. The capacitors $C$ and $C_q$ are implemented using so-called finger capacitors, while $C_c$ and $C_g$ are given by the stray field coupling to the rectangular waveguide sample holder shown in Panel \textbf{c}. The inductive elements of the circuit are introduced in the model as lumped elements connecting the pads (shown in the insets below). The table shows the resulting linewidth values $\kappa$ for three different frequencies of the resonator mode, chosen in the pass-band of the waveguide. \textbf{c)} Finite element model used to simulate the 3D waveguide sample holder. Recently, a similar sample holder geometry has been used to perform multiplexed quantum readout \cite{Kou2017}. The qubit-resonator circuit is deposited on a sapphire substrate which is indicated by the green rectangle. The electric field magnitude along the waveguide is frequency dependent, and its profile is schematically shown for $6$, $7$, and $8\,\mathrm{GHz}$. The impedance and the mode profile between the waveguide and the coaxial cable connected to the input port are matched using the tuning screws. \textbf{d)} Direct extension of the proposed physical implementation for two capacitively coupled qubit-resonator systems (compare Sec.~\ref{sec:two_blocks}). The resonators are designed to have different eigenmode frequencies, and they can be individually addressed using the collective waveguide mode represented by the direction of the $\vec{E}$ field.}
\label{fig:phys_impl}
\end{figure*}

Figure~\ref{fig:phys_impl} shows a possible physical implementation of the inductively shunted transmon qubit. One of the main challenges is to realize compact, low-loss and linear inductances, in the range of several $\mathrm{nH}$, required for the shunting inductors $L$ and $L_a$ (see Fig.~\ref{fig:phys_impl}\textbf{a}). For this purpose, we propose the use of a superconducting strip consisting of a high kinetic inductance material such as granular aluminum, or niobium and titanium nitrides, which have been shown to achieve inductances in the range of $\mathrm{nH}/\square$ \cite{Rotzinger2017, Annunziata2010, Samkharadze2016, Vissers2010}. The rest of the circuit, including all Josephson junctions, can be fabricated using standard thin-film aluminum. The electrical connections between these different metallic layers can be realized using recently developed argon ion cleaning and contacting techniques which preserve the coherence of the circuit \cite{Dunsworth2017, Wu2017, Grunhaupt2017}. \\
The capacitances required for shunting the qubit, $C_q$, and the resonator, $C$, as well as the coupling capacitors, $C_c$ and $C_g$, can all be implemented by the relatively simple structure shown in Fig.~\ref{fig:phys_impl}\textbf{b}. For clarity, the three superconducting island phases are labeled using the same notation as in Fig.~\ref{fig:qubit_resonator}. The structure is designed to couple to the first propagating mode of a 3D wave\-guide, following the sample-holder geometry described in Ref.~\cite{Kou2017}.  The electric-field magnitude is indicated by the color scale. The maximum values, in the range of $100\,\mathrm{GV/m}$ for an energy of $1\,\mathrm{J}$ stored in the mode, are comparable to the electric field values reported in Ref.~\cite{Grunhaupt2017}, which enabled the measurement of microwave resonators with internal quality factors exceeding $10^6$ in the quantum regime. Notice that the proposed implementation satisfies the required left-right symmetry of the schematics in Fig.~\ref{fig:phys_impl}\textbf{a}, with comfortable margins of error, below $1\,\%$, for either optical or electron-beam lithography.\\
The 3D wave\-guide model shown in Fig.~\ref{fig:phys_impl}\textbf{c} offers the advantage of strong coupling for the resonator mode inside the designed pass band between $6$ and $8\,\mathrm{GHz}$, as indicated by the table in Fig.~\ref{fig:phys_impl}\textbf{b}, while the qubit mode can be efficiently decoupled from the microwave environment. The finite-element simulations indicate a qubit mode coupling quality factor as high as $10^8$.\\
The magnetic field required to tune the fluxes $\Phi_x$ and $\Phi_{Xb}$ (see Fig.~\ref{fig:qubit_resonator}) can be controlled using a direct current coil, which can be attached to the exterior of the sample holder, with the current flowing in a plane perpendicular to the x-axis. In the simplest implementation, the same coil can bias both fluxes, making use of a large ratio between the areas of the superconducting loops enclosing $\Phi_x$ and $\Phi_{Xb}$. Thus, small field variations can be used to tune $\Phi_{Xb}$, quasi-independently from $\Phi_x$.\\
The currently proposed physical implementation is meant as a prototype to test the tunability of the transverse and longitudinal coupling, nevertheless the design shown in Fig.~\ref{fig:phys_impl}\textbf{c} could be adapted for a higher density of qubits. In Fig.~\ref{fig:phys_impl}\textbf{d} we show a direct extension of the concept for two qubits using capacitive coupling between the resonators. With more involved RF designs, it is possible to enlarge the qubit matrix, and add strictly local qubit and resonator drives by using recent advancements in flip-chip and micromachined superconducting circuit technology \cite{Minev2016, Brecht2017, Rosenberg2017}.
\chapter{APPLICATION: SCALABLE DESIGN}
\label{chapt:scalability}

An application of longitudinal coupling is the scalable design discussed in Sec.~\ref{sec:scalable_architecture}, which was conceived by Billangeon  et al. in Ref.~\cite{Billangeon2015}. In the architecture proposed there, each unit cell consists of a qubit longitudinally coupled to four resonators, while each resonator is coupled to a resonator from the next unit cell via a conjugate degree of freedom (see Eq.~\ref{eq:long2} and Fig.~\ref{fig:grid}). The design introduced in Chapt.~\ref{chapt:design} implements pure longitudinal coupling to a resonator for a certain value of the external flux. By tuning this flux, the coupling to the resonator can be switched between pure transverse and pure longitudinal coupling. In this chapter, we will show how the design from Chapt.~\ref{chapt:design} can be scaled up to a grid that implements the architecture from Ref.~\cite{Billangeon2015}. We will focus here on the case of single coupling junctions, but the extensions discussed here are applicable to all adaptations from Chapt.~\ref{chapt:adaptations}.
Note that parts of this chapter were already published in Ref.~\cite{Richer2016}.

\section{EXTENSION TO n RESONATORS}
\label{sec:n_resonators}

\sidecaptionvpos{figure}{c}
\begin{SCfigure}[50][tb]
\centering
    \includegraphics[height=5cm]{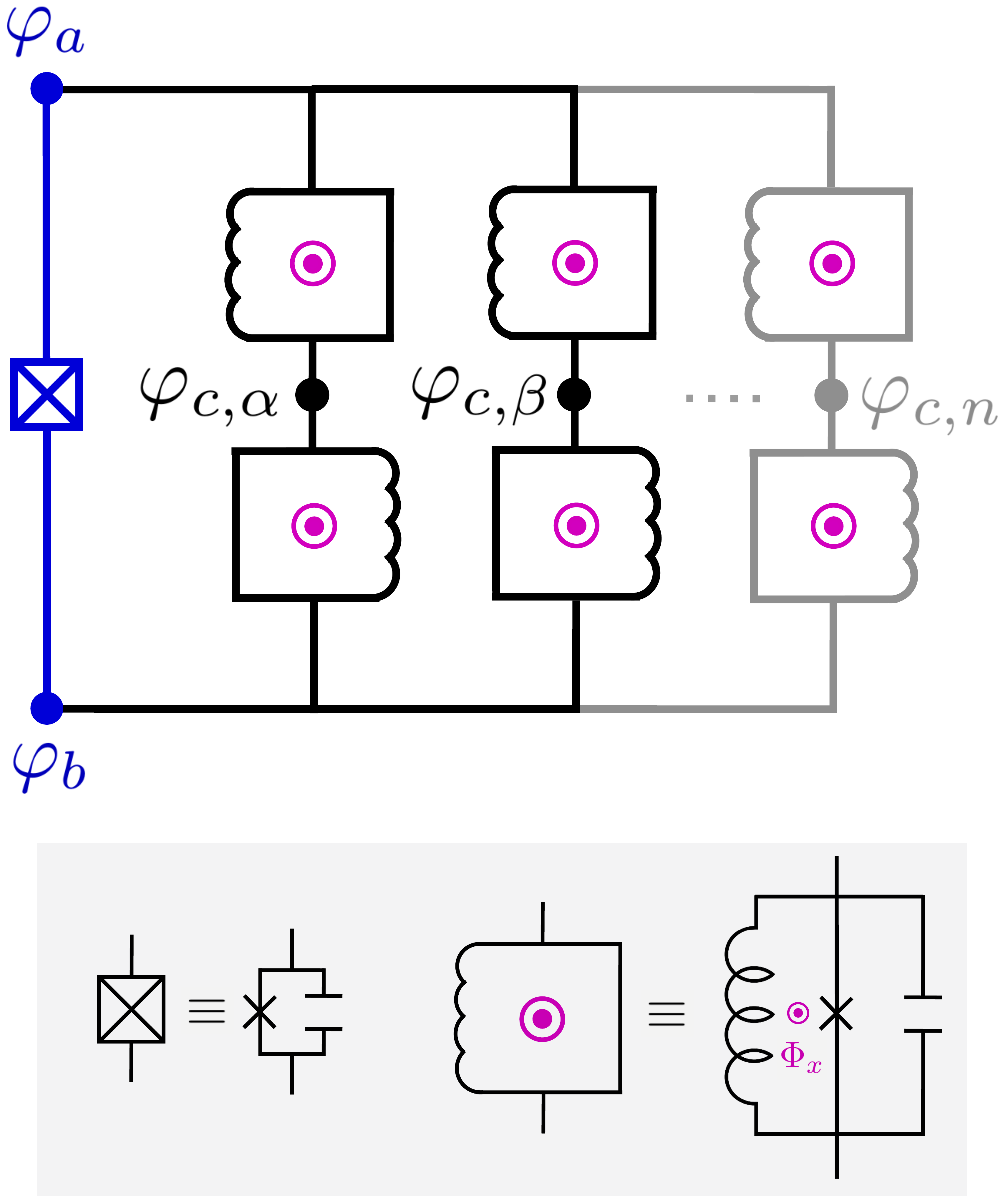}
  \caption{Qubit coupled to $n$ resonators. The box below explains the abbreviations used in the figure. The circuit is drawn from the design presented in Chapt.~\ref{chapt:design} (Fig.~\ref{fig:qubit_resonator}) by duplicating the structure in parallel to the qubit junction. The coupling between the qubit and each resonator is independently tunable via the external fluxes. There is inherently no coupling between the individual resonators. 
\label{fig:qubit_many_resonators}}
\end{SCfigure}

The design depicted in Fig.~\ref{fig:qubit_resonator} is easily extendable to a block of one qubit coupled separately to any number of resonators as depicted in Fig.~\ref{fig:qubit_many_resonators}. This is done by multiplying the resonator structure in parallel to the qubit.
Fortunately, adding another resonator arm to the system does not have any effect on the first resonator and the only coupling terms are the ones between the qubit and each resonator. As shown in Fig.~\ref{fig:qubit_many_resonators}, each resonator adds a node $\varphi_{c,l}$ to the system. The variables for qubit and resonators are taken to be 

\begin{align}
\varphi_q = \varphi_a - \varphi_b \qquad \varphi_{r,l} = \varphi_a + \varphi_b - 2 \,\varphi_{c,l},
\end{align}

so that the qubit is still defined by the phase difference across the qubit junction, as in Sec.~\ref{sec:design}. We assume that each resonator arm contains two symmetric capacitances $C_l$ and two symmetric inductances $L_l$, as well as two slightly asymmetric coupling junctions $E_{J1,l}$ and $E_{J2,l}$. As the resonators are independent of each other, there is no need for any symmetries between two resonators. For simplicity reasons, we will restrict this discussion to the case of single coupling junctions instead of coupling junction arrays (see Secs.~\ref{sec:design} and \ref{sec:array}) and ignore fluxes through the big loops.
The kinetic energy of the multi-resonator system is given by

\begin{align}
\mathcal T &= \sum_{l} \left(\frac{\Phi_0}{2\pi}\right)^2 \left(\frac{2\,C_q+C_l}{4} \, \dot\varphi_q^2 + \frac{C_l}{4} \, \dot\varphi_{r,l}^2\right),
\end{align}

which clearly includes no coupling terms. The potential energy yields


\begin{align*}
\mathcal U &=  \sum_{l} \left(\frac{\Phi_0}{2\pi}\right)^2 \frac{1}{4L_l} \, (\varphi_q^2 + \varphi_{r,l}^2)  - E_{Jq} \cos(\varphi_q)\\
& - E_{J1,l} \cos\left(\frac{\varphi_{r,l} + \varphi_q}{2} + \varphi_{x,l}\right) - E_{J2,l} \cos\left(\frac{\varphi_{r,l} - \varphi_q}{2} + \varphi_{x,l}\right),\numberthis
\label{eq:pot_n}
\end{align*}

where $\varphi_{x,l}$ refers to the (rescaled) flux through the coupling loops of resonator $l$. Note that the coupling between the qubit and a resonator can be controlled separately for each resonator by tuning this flux. Transverse and longitudinal coupling can be chosen depending on the application.\\
The frequencies of the resonators will still be given by Eq.~\ref{eq:omegar}, as the resonators are independent of each other. The qubit frequency and anharmonicity, however, will be influenced by each resonator the qubit couples to. Hence, the effective Josephson energy and the charging energy of the qubit are obtained by the substitutions $C \to \sum_l C_l$ and $1/L \to \sum_l 1/L_l$ in Eq.~\ref{eq:EJeffective} and Eq.~\ref{eq:charging_energy}. With these substitutions, all the results from Chapt.~\ref{chapt:design}, including the qubit frequency and anharmonicity, as well as the coupling terms, apply also here.
As each resonator arm adds another harmonic term to the qubit's potential, it will be important to set $E_{Jq}$ to be sufficiently large compared with $(\Phi_0/(2\pi))^2 \sum_l 1/(4L_l)$ to maintain the qubit's anharmonicity. Note that the parameters shown in Chapt.~\ref{chapt:adaptations} are adjusted for a single resonator and should be adapted when coupling more than one resonator to a qubit.

\section{TWO COUPLED BLOCKS}
\label{sec:two_blocks}

\sidecaptionvpos{figure}{c}
\begin{SCfigure}[50][tb]
\centering
    \includegraphics[width=.55\linewidth]{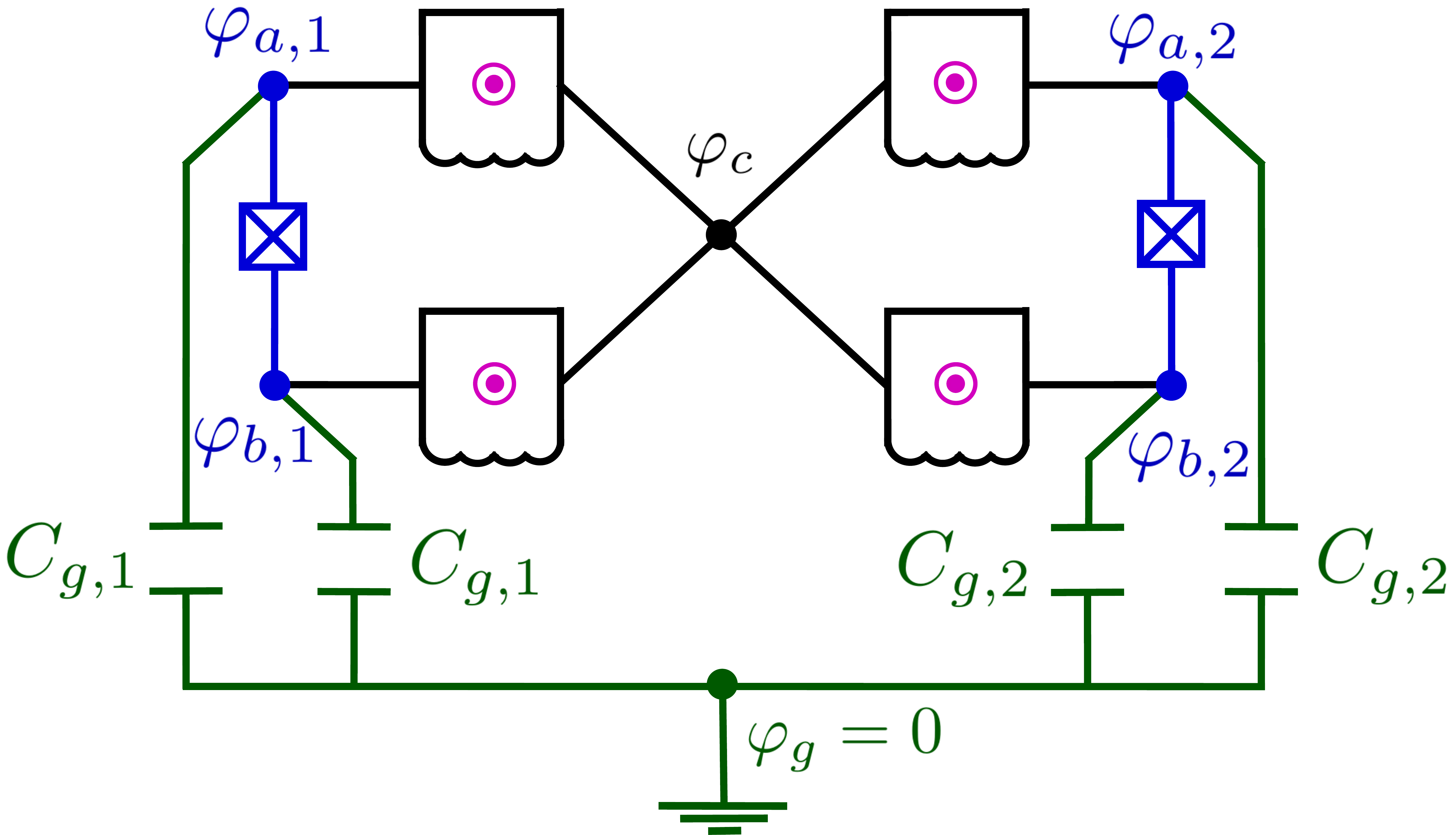}
  \caption{This circuit realizes two qubits and two resonators 	coupled  according to Eq.~\ref{eq:long2}. See Fig.~\ref{fig:qubit_many_resonators} for the abbreviations used here. Fig~\ref{fig:two_blocks_cc} shows an alternative circuit, which avoids large superconducting loops when scaled up to a grid.\label{fig:two_blocks}}
\end{SCfigure}

In order to implement Eq.~\ref{eq:long2}, we need to couple one of these blocks to the next one via the charge degree of freedom of the resonator. Fig.~\ref{fig:two_blocks} shows the simplest way we have found to obtain the desired coupling between two blocks. A slightly more complicated alternative, which might be better suited for experimental reasons, will be shown in the next section.
We connect two neighboring resonator branches by tying their $\varphi_{c}$ nodes together and connect all qubit nodes to a common ground node $\varphi_g$ via capacitances $C_{g,1}$ and $C_{g,2}$. Note that the two blocks are uncoupled for $C_{g,i} = 0$. As the node phases are only defined up to an overall phase, we have the freedom to set $\varphi_g = 0$ (\textit{ground node} \cite{Devoret1997}) without loss of generality. However, we will keep $\varphi_g$ here in order to show that possible fluctuating charge offsets between ground and the rest of the circuit can not influence our system.
The kinetic energy of this coupled system in terms of the node phases marked in Fig.~\ref{fig:two_blocks} is

\begin{align*}
\mathcal T &=  \sum_{i=1}^2 \left(\frac{\Phi_0}{2\pi}\right)^2 \biggl(\frac{C_{q,i}}{2} (\dot\varphi_{a,i} - \dot\varphi_{b,i})^2 + \frac{C_i}{2} (\dot\varphi_{a,i} - \dot\varphi_{c})^2 + \frac{C_i}{2} (\dot\varphi_{b,i} - \dot\varphi_{c})^2\\
& +\frac{C_{g,1}}{2} \left( (\dot\varphi_{a,i} - \dot\varphi_g)^2\right) + \frac{C_{g,2}}{2}\left((\dot\varphi_{b,i} - \dot\varphi_g)^2 \right) \biggr) \numberthis.
\label{eq:T_two_blocks}
\end{align*}

Inserting the variables for the qubits and resonators, that is

\begin{align*}
\varphi_{q,i} = \varphi_{a,i} - \varphi_{b,i} \qquad 
\varphi_{r,i} = \varphi_{a,i} + \varphi_{b,i} - 2 \,\varphi_{c}, \numberthis
\label{eq:variables_two_blocks}
\end{align*}

leads to

\begin{align*}
\mathcal T &=  \sum_{i=1}^2  \left(\frac{\Phi_0}{2\pi}\right)^2 \biggl(\frac{C_{q,i}}{2} \dot\varphi_{q,i}^2 + \frac{C_i + C_{g,i}}4 \, (\dot\varphi_{q,i}^2 + \dot\varphi_{r,i}^2)   \\
&+ (C_{g,1} + C_{g,2}) \, \dot{\bar\varphi}^2 + \left(C_{g,1} \,\dot\varphi_{r,1} + C_{g,2}\, \dot\varphi_{r,2}\right)  \dot{\bar\varphi}\biggr) \numberthis
\label{eq:T_two_blocks_2}
\end{align*}

with $\bar\varphi = \varphi_{c} - \varphi_g$.
We can see that, apart from the two qubit and resonator variables, a fifth variable $\bar\varphi$ appears in Eq.~\ref{eq:T_two_blocks_2} that mediates the coupling between the resonator variables $\varphi_{r,1}$ and $\varphi_{r,2}$. We will show here that this variable decouples from the others and can thus be discarded. In the potential energy there is no coupling between the two blocks

\begin{align*}
\mathcal U &= \sum_{i=1}^2  \left(\frac{\Phi_0}{2\pi}\right)^2  \frac{1}{4L_i} (\varphi_{q,i}^2 + \varphi_{r,i}^2) - E_{Jq,i} \cos(\varphi_{q,i}) \\
&  - E_{J1,i} \cos\left(\frac{\varphi_{r,i} + \varphi_{q,i}}{2} + \varphi_{x,i}\right) - E_{J2,i} \cos\left(\frac{\varphi_{r,i} - \varphi_{q,i}}{2} + \varphi_{x,i}\right)\numberthis
\label{eq:pot_two_blocks}
\end{align*}

and the fifth variable $\bar\varphi$ does not appear. This situation, where a variable appears in the kinetic but not in the potential energy, allows for a variable elimination following the recipe described in Sec.~\ref{sec:cholesky} and App.~\ref{app:cholesky}.
We use the substitution

\begin{align}
\bar \varphi = \varphi_* - \frac{C_{g,1}\,\varphi_{r,1} + C_{g,2}\, \varphi_{r,2}}{2(C_{g,1} + C_{g,2})},
\label{eq:eliminate}
\end{align}

which is a linear, invertible transformation that leaves the variables for qubits and resonators (Eq.~\ref{eq:variables_two_blocks}) unchanged. It leads to

\begin{align*}
\mathcal T = \sum_{i=1}^2 & \left(\frac{\Phi_0}{2\pi}\right)^2 \biggl(\frac{2\, C_{q,i} + C_i + C_{g,i}}4 \, \dot\varphi_{q,i}^2 + \frac{C_i}4 \, \dot\varphi_{r,i}^2   \\
&+ (C_{g,1} + C_{g,2}) \, \dot\varphi_*^{2} + \frac{C_{g\mu} }{8} \, (\dot\varphi_{r,1} - \dot\varphi_{r,2})^2 \biggr) \numberthis
\label{eq:T_two_blocks_final}
\end{align*}

with the abbreviation

\begin{align}
C_{g\mu} = 2 \frac{C_{g,1} C_{g,2}}{C_{g,1} + C_{g,2}},
\label{eq:Cgmu}
\end{align}

which makes clear that there is a direct capacitive coupling between the two resonator variables as desired, while the unwanted variable $\varphi_*$ decouples. As $\varphi_*$ does not appear at all in the potential energy of the coupled system and is decoupled in the kinetic energy, it can be safely discarded.\\
We have thus found a system that implements the Hamiltonian proposed by Billangeon (Eq.~\ref{eq:long2}) for two qubits and two resonators and is scalable to a grid (see Sec.~\ref{sec:n_resonators}).\\
Note that as $\varphi_g$ appears only in the discarded variable $\varphi_*$, a charge offset between ground and the rest of the circuit can not influence our system variables (Eq.~\ref{eq:variables_two_blocks}). The rest of the circuit is protected against charge noise due to the inductive shunting. However, a grid made of the circuit in Fig.~\ref{fig:two_blocks} would include large superconducting loops. As this is experimentally problematic, because the flux cannot be controlled precisely over large areas, we will present an alternative in the next section, where the superconducting loops are broken by an additional capacitor between the two blocks.

\subsection{ALTERNATIVE CIRCUIT}
\label{sec:alternative}

\begin{SCfigure}[50][tb]
\centering
    \includegraphics[width=.55\linewidth]{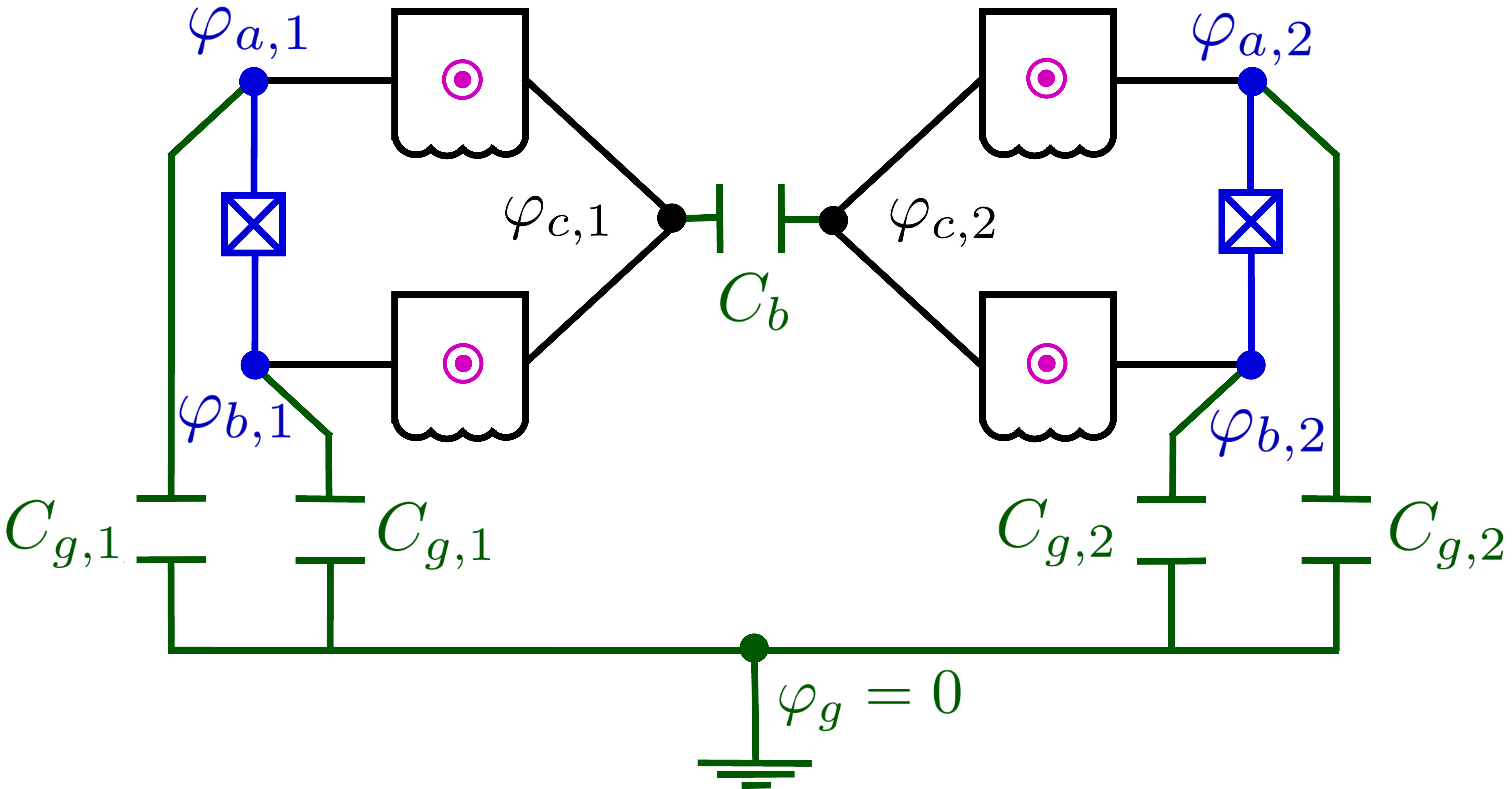}
  \caption{Alternative circuit, which implements two qubits coupled via two coupled resonators according to Eq.~\ref{eq:long2}. As opposed to Fig.~\ref{fig:two_blocks}, this circuit includes the additional capacitance $C_b$ between the blocks, in order to avoid large superconducting loops. \label{fig:two_blocks_cc}}
\end{SCfigure}

Fig.~\ref{fig:two_blocks_cc} shows an alternative circuit to the one shown in Fig.~\ref{fig:two_blocks}, which has the advantage that large superconducting loops are avoided. Instead of fusing the two blocks' $\varphi_c$ nodes into one, they are now connected via a capacitance $C_b$. Apart from that, the circuit is equal to the one presented in the last section. Due to the additional capacitor, it clearly has one node more than before. When the capacitance between the $\varphi_{c,i}$ nodes goes to infinity, which means fusing the two nodes together, version two smoothly converges to version one. We will have a look at the quantization of the two-qubit-two-resonator system for both possible versions.

\sidecaptionvpos{figure}{c}

The potential energy of the circuit is clearly the same as before (Eq.~\ref{eq:pot_two_blocks}), where the phase variables for the qubits and resonators are now defined as

\begin{align}
\varphi_{q,i} = \varphi_{a,i} - \varphi_{b,i} \qquad \varphi_{r,i} = \varphi_{a,i} + \varphi_{b,i} - 2\,\varphi_{c,i}
\label{eq:vartwo}
\end{align}

Note that in terms of branch variables, these are equal to the variables used above (Eq.~\ref{eq:variables_two_blocks}).
The potential energy only contains the four variables for the qubits and the resonators. Clearly, there is no coupling between the two blocks via the potential energy.
In the kinetic energy, however, two additional variables appear, which we would like to eliminate in a similar way as in Sec.~\ref{sec:two_blocks}. It is

\begin{align*}
\mathcal T &= \sum_{i=1}^2 \left(\frac{\Phi_0}{2\pi}\right)^2 \bigg(\frac{2\,C_{q,i}+C_i+C_{g,i}}{4} \, \dot\varphi_{q,i}^2  + \frac{C_i }{4} \, \dot\varphi_{r,i}^2 \\
 & +\frac{C_{g,i}}{4} (\dot\varphi_{r,i} + 2 \dot{\bar\varphi}_i)^2 + \frac{C_b}{2} (\dot{\bar\varphi}_1 - \dot{\bar\varphi}_2)^2\bigg)\numberthis \label{eq:Trr0}, 
\end{align*}

with $\bar\varphi_i = \varphi_{c,i} - \varphi_g$. Just as in the last section in Eq.~\ref{eq:T_two_blocks_2}, the additional variables mediate the interaction between the two resonators.\\ 
A simple transformation similar to the one done above (Eq.~\ref{eq:eliminate}) decouples the two unwanted variables, as explicitly shown in App.~\ref{app:cholesky}. 
Remarkably, this can be done without changing any of the system variables (Eq.~\ref{eq:vartwo}). Therefore, the potential energy remains again unaffected by the elimination procedure. We are left with a four-dimensional system of two qubits and two resonators, as expected. 
The kinetic energy is now given by

\begin{align*}
\mathcal T &= \sum_{i=1}^2 \left(\frac{\Phi_0}{2\pi}\right)^2 \bigg(\frac{2\,C_{q,i}+C_i+C_{g,i}}{4} \, \dot\varphi_{q,i}^2  \\
 &+ \frac{C_i }{4} \, \dot\varphi_{r,i}^2 + \frac{C_{b\mu}}{16} (\dot\varphi_{r,1} - \dot \varphi_{r,2})^2\bigg), \numberthis
\label{eq:Trr}
\end{align*}

with the abbreviation (compare Eq.~\ref{eq:Cgmu})

\begin{align}
C_{b\mu} = 2 \frac{C_b C_{g\mu}}{C_b + C_{g\mu}} \quad \text{with} \quad 
C_{g\mu} = 2 \frac{C_{g,1} C_{g,2}}{C_{g,1} + C_{g,2}}.
\label{eq:Cmu}
\end{align}

When the capacitance $C_b$ between the blocks goes to infinity, which means fusing the $\varphi_{c,i}$ nodes together, we find $C_{b\mu} = 2\,C_{g\mu}$. This means both possible versions are correctly described by Eq.~\ref{eq:Trr} (compare Eq.~\ref{eq:T_two_blocks_final} after discarding $\varphi_*$). For unconnected blocks $C_{b\mu}$ disappears, that is when either $C_b$ or $C_{g,1}$ or $C_{g,2}$ are equal to zero. If $C_{g,1} = C_{g,2} = C_g$, we simply find $C_{g\mu} = C_g$.

\subsection{QUANTIZATION}
\label{sec:twoblocksquant}

The expressions for the qubits in second quantization can be easily adap\-ted for the system of two coupled blocks by the transition $C_i \to C_i + C_{g,i}$ in the charging energy for the qubit (Eq.~\ref{eq:charging_energy}). For the resonator, however, we will redo the quantization procedure here for the circuit with the capacitor between the blocks presented in Sec.~\ref{sec:alternative}. As mentioned above, the transition to the system without the capacitor (Sec.~\ref{sec:two_blocks}) is done by setting $C_{b\mu} = 2\,C_{g\mu}$ (compare Eq.~\ref{eq:Cmu}).
In order to go to second quantization, we need to go to the Hamiltonian representation. The Legendre transformation (compare Eq.~\ref{eq:legendre}) of the kinetic energy given in Eq.~\ref{eq:Trr} can be written as

\begin{align*}
\mathcal T &= \sum_{i=1}^2 \frac{(2e\, n_{q,i})^2}{2C_{q,i} + C_i + C_{g,i}} + \frac{2 C_\mu}{C_{b\mu} + 2 C_\mu} \frac{(2e\, n_{r,i})^2}{C_i}  \\
&+ \frac{C_{b\mu}}{C_{b\mu} + 2 C_\mu} \frac{(2e\, n_{r,1} + 2e\,n_{r,2})^2}{C_1 + C_2}, \numberthis
\label{eq:Tq}
\end{align*}

where we introduced the abbreviation

\begin{align}
C_\mu = 2 \frac{C_1 C_2}{C_1 + C_2}.
\end{align}

It is clear that for unconnected blocks, that is for $C_{b\mu} = 0$, the coupling term in Eq.~\ref{eq:Tq} disappears, which directly leads back to the uncoupled case (Eqs.~\ref{eq:Hq1} and \ref{eq:Hr} in Chapt.~\ref{chapt:design}).\\
Following the strategy for the quantization from Chapt.~\ref{chapt:quantization}, we look at the quadratic terms in the Hamiltonian for one of the resonators, setting all other variables to zero. It is

\begin{align*}
\mathcal{H}_{r,i} &= \frac{C_i C_{b\mu} +  2 C_\mu (C_1 + C_2)}{C_i (C_1 + C_2)(C_{b\mu} + 2 C_\mu)} (2 e \, n_{r,i})^2 +   \frac{1 + \eta_i}{4L_i}  \Phi_{r,i}^2 \numberthis,
\end{align*}

where 

\begin{align}
\eta_i =\frac{E_{J\Sigma_i}}{2k_i} \left(\frac{2\pi}{\Phi_0}\right)^2 L_i \cos\left(\frac{\varphi_{x,i}}{k_i}\right)
\end{align}

was defined as in Eq.~\ref{eq:eta}.

\begin{align*}
\mathcal{H}_{r,i} &= \frac{2C_j + C_\mu}{C_\mu (C_1 + C_2) + 2C_1 C_2} Q_{r,i}^2  +   \frac{1 + \eta_i}{4L_i}  \Phi_{r,i}^2 \numberthis,
\label{eq:Hrr}
\end{align*}

with the conjugate variables $\Phi_{r,i} = \frac{\Phi_0}{2\pi} \varphi_{r,i}$ and $Q_{r,i} = 2e \, n_{r,i}$.
We go to second quantization using again the ansatz from Eq.~\ref{eq:quant_r}. Choosing the characteristic impedance such that all non-diagonal terms disappear as done in Sec.~\ref{sec:resonator} leads to 

\begin{align}
\omega_{r,i} = \sqrt{\frac{(4\, C_\mu (C_1+C_2)+2 C_i C_{b\mu})(1 + \eta_1)}{\left(2 \,C_i^2 \,C_{b\mu}+C_\mu (C_1+C_2) (4 C_i+C_{b\mu})\right) L_i}}.
\label{eq:omegarr}
\end{align} 

and 

\begin{align}
Z_{0r,i} = \frac{2 L_i}{1 + \eta_i} \omega_{r,i}.
\end{align}

For zero coupling, that is $C_{b\mu} = 0$, both these expressions converge to the ones from Chapt.~\ref{chapt:design} (Eqs.~\ref{eq:Z_0} and \ref{eq:omegar}).
The Hamiltonian in second quantization yields of course

\begin{align}
\mathcal H_{r,i} = \hbar \, \omega_{r,i} \, a_i^\dagger a_i.
\end{align}

The coupling between the two resonators is given by

\begin{align*}
&\frac{2 C_{b\mu} Q_{r,1} Q_{r,2}}{C_{b\mu} (C_1+C_2) + 4C_1 C_2}  = \hbar \, g_{c} \, (a_1^\dagger - a_1)(a_2^\dagger - a_2) \numberthis
\end{align*}

with

\begin{align}
g_{c} = - \frac{C_{b\mu}}{2} \frac{\sqrt[4]{\frac{(1+\eta_1)(1+\eta_2)}{(4C_1 + C_{b\mu})(4C_2+C_{b\mu})L_1 L_2}}}{\sqrt{C_{b\mu} (C_1+C_2) + 4C_1 C_2}}.
\label{eq:gc}
\end{align}

Clearly, this is directly proportional to $C_{b\mu}$ and goes smoothly to zero when the blocks are uncoupled.

\subsection{STRAY CAPACITANCES}
\label{sec:stray}

\begin{SCfigure}[50][tb]
\centering
    \includegraphics[width=.55\linewidth]{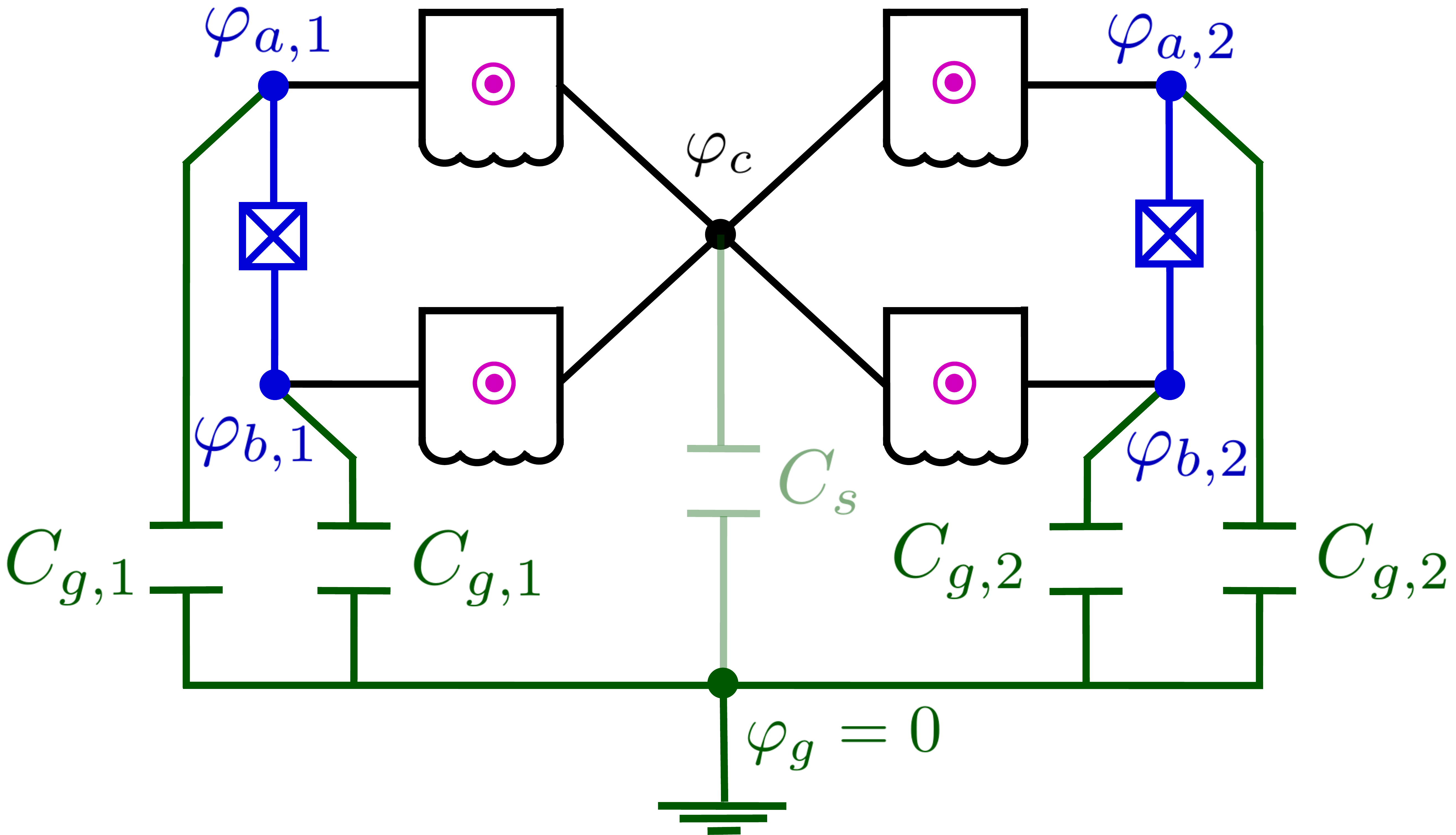}
  \caption{A stray capacitance between $\varphi_c$ and ground leads to a rescaling of the coupling. This is true for both the examples presented above, that is Figs.~\ref{fig:two_blocks} and \ref{fig:two_blocks_cc}. For the abbreviations used here, compare again Fig.~\ref{fig:qubit_many_resonators}. \label{fig:stray}}
\end{SCfigure}

Stray capacitances might appear between the resonator nodes $\varphi_{c,i}$ and ground as shown in Fig.~\ref{fig:stray}. As we will see, this only leads to a rescaling of the coupling but does not affect its form or its strict locality. While this is true for both the circuits introduced above, we will focus here on the simpler circuit introduced in Sec.~\ref{sec:two_blocks} for reasons of brevity. 
Such a stray capacitance will add a term $\sim \,C_s \dot{\bar\varphi}^2$ with $\bar\varphi = \varphi_c - \varphi_g$ to the kinetic energy given in Eq.~\ref{eq:T_two_blocks}. In order to compensate for this extra term, we have to adapt the transformation given in Eq.~\ref{eq:eliminate} to

\sidecaptionvpos{figure}{c}

\begin{align}
\bar \varphi = \varphi_* - \frac{C_{g,1}\,\varphi_{r,1} + C_{g,2}\, \varphi_{r,2}}{2(C_{g,1} + C_{g,2}) + C_s}.
\label{eq:eliminate_stray}
\end{align}

This leads to rescaling of the coupling term to

\begin{align*}
 \left(\frac{\Phi_0}{2\pi}\right)^2 \frac{C_{g,1} C_{g,2}}{2(2\, C_{g,1} + 2\, C_{g,2} + C_s)} \, (\dot\varphi_{r,1} - \dot\varphi_{r,2})^2. \numberthis
\label{eq:T_two_blocks_stray}
\end{align*}

Note again that Eq.~\ref{eq:eliminate_stray} leaves the system variables unchanged. Thus, the main effect of the stray capacitance $C_s$ is the shunting of the coupling term in Eq.~\ref{eq:T_two_blocks_stray}.

\section{EXTENSION TO GRID}
\label{sec:grid}

\begin{SCfigure}[50][tb]
\centering
    \includegraphics[width=.55\linewidth]{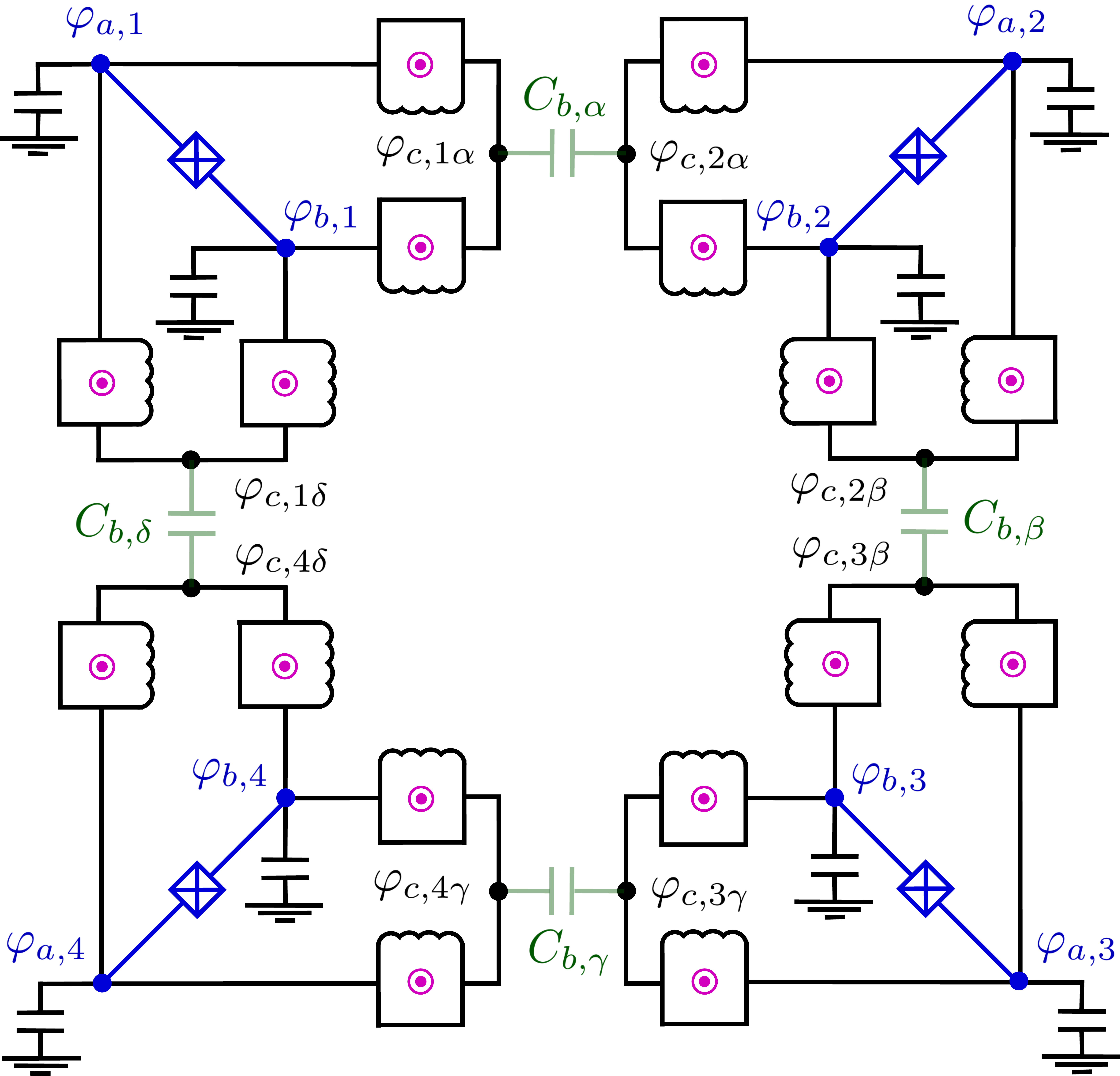}
  \caption{This circuits realizes a plaquette of four qubits coupled according to the scheme from Ref.~\cite{Billangeon2015}. Compare Fig.~\ref{fig:qubit_many_resonators} for the abbreviations used here. Two adjacent resonators are coupled by a capacitance $C_{b,l}$ between the nodes $\varphi_{c,il}$ and $\varphi_{c,jl}$, where $l = \alpha, \beta, ...$ numbers the connection between two resonators and $i,j = 1,2, ...$ the qubit they belong to.
  \label{fig:ring}}
\end{SCfigure}

It is straightforward to extend this idea to the grid proposed by Billangeon \cite{Billangeon2015} (see Fig.~\ref{fig:grid}), where every qubit couples longitudinally to its resonators and every resonator couples capacitively to a resonator of the next block.
In analogy to the two coupled blocks described in the last section, Fig.~\ref{fig:ring} shows a plaquette of four coupled blocks, which corresponds to the circuit shown in Fig.~\ref{fig:two_blocks_cc}. Every two adjacent resonators are coupled by a capacitance $C_{b,l}$ between the nodes $\varphi_{c,il}$ and $\varphi_{c,jl}$, where $l = \alpha, \beta, ...$ labels the connection between the two resonators and $i,j = 1,2, ...$ the qubit they belong to as indicated in Fig.~\ref{fig:ring}. When the capacitance between two resonators goes to infinity $C_{b,l} \to \infty$, the nodes on either side of it are fused to a single node $\varphi_{c,l}$, which corresponds to the circuit shown in Fig.~\ref{fig:two_blocks}.
All qubit nodes (blue colored nodes) are connected to the same ground node $\varphi_g = 0$ (compare Sec.~\ref{sec:two_blocks}) via capacitances $C_{g,i}$. \\
In order to see why this coupling is entirely local, we have to remember that there is no coupling between two resonators coupled to the same qubit, as shown in Sec.~\ref{sec:n_resonators}. While the qubits are coupled to their resonators via the potential energy (see Eq.~\ref{eq:pot_n}), the coupling between two adjacent resonators is via the kinetic energy. For every unit of two qubits coupled via two resonators, we will find the kinetic energy to be equal to Eq.~\ref{eq:Trr0}, or Eq.~\ref{eq:T_two_blocks_2} in the case of $C_{b,l} \to \infty$. 
In order to emphasize this locality, we can rewrite the kinetic energy for the ring of four coupled qubits depicted in Fig.~\ref{fig:ring} as

\sidecaptionvpos{figure}{c}

\begin{align}
\mathcal T = \sum_{i=1,...} \sum_{l=\alpha,...} \left(\frac{\Phi_0}{2\pi}\right)^2 \biggl(\frac{2 C_{q,i} + C_{il} + C_{g,l}}2 \, \dot\varphi_{q,i}^2 +\frac12 \, \boldsymbol{\dot\varphi}_l^T \mathbf{C}_l \, \boldsymbol{\dot\varphi}_l \biggr),
\label{eq:T_matrix}
\end{align}

with $\boldsymbol{\varphi}_l^T = (\bar\varphi_{il}, \bar\varphi_{jl},\varphi_{r,il}, \varphi_{r,jl})$, where $\varphi_{r,il}$ and $\varphi_{r,jl}$ (compare Eq.~\ref{eq:vartwo}) correspond to two adjacent resonators, which belong to the qubits $i$ and $j$ and are connected by the capacitance $C_{b,l}$. For every unit of two coupled resonators, there are two superfluous variables $\bar\varphi_{il} = \varphi_{c,il} - \varphi_g$, which will be eliminated following the strategy from Sec.~\ref{sec:two_blocks}. Note that there is no connection between two such units of coupled resonators and none of the variables in $\boldsymbol{\varphi}_\alpha$ appears in $\boldsymbol{\varphi}_\beta$.
The capacitance matrix $\mathbf{C}_l$ in Eq.~\ref{eq:T_matrix} is given by

\begin{align}
\mathbf{C}_l =\begin{pmatrix}
C_{b,l} + 2\, C_{g,i} & - C_{b,l} & C_{g,i} & 0 \\
 - C_{b,l} & C_{b,l} + 2\, C_{g,j} & 0 & C_{g,j} \\
C_{g,i} & 0 & \frac{C_{il} + C_{g,i}}2 & 0 \\
0 & C_{g,j} & 0 & \frac{C_{jl} + C_{g,j}}2 
\end{pmatrix}.
\end{align}


The elimination of the superfluous variables (compare App.~\ref{app:cholesky}) can be done separately for each such unit of two coupled resonators, as done in Sec.~\ref{sec:two_blocks}. Note again that this variable elimination leaves the variables for qubits and resonators unchanged. In the new basis $\boldsymbol{\tilde\varphi}_l^T = (\varphi_{*,il}, \varphi_{*,jl},\varphi_{r,il}, \varphi_{r,jl})$, the capacitance matrix is given by

\begin{align*}
\mathbf{\tilde C}_l =\begin{pmatrix}
C_{b,l} + 2\, C_{g,i} & 0 & 0 & 0 \\
0 & \frac{2(C_{g,i} + C_{g,j})(C_{b,l}+C_{g\mu,l})}{C_{b,l} + 2C_{g,i}} & 0 & 0 \\
0 & 0 & \frac{C_{il}}{2} + \frac{C_{b\mu,l}}4 & -\frac{C_{b\mu,l}}{4} \\
0 & 0 & -\frac{C_{b\mu,l}}{4} & \frac{C_{jl}}{2} + \frac{C_{b\mu,l}}4 
\end{pmatrix}, 
\hspace{-5pt}\numberthis
\end{align*}

where $C_{b\mu,l}$ is defined as in Eq.~\ref{eq:Cgmu} for every two resonators $i$ and $j$ coupled via a capacitance $C_{b,l}$. In this new frame, the resonators are directly coupled to each other, while the superfluous variables $\varphi_{*,il}$ and $\varphi_{*,jl}$ are decoupled and can thus be discarded.
The kinetic energy thus transforms to

\begin{align}
\mathcal T = \sum_{i=1,...} \sum_{l=\alpha,...} \left(\frac{\Phi_0}{2\pi}\right)^2 \biggl(\frac{2 C_{q,i} + C_{il} + C_{g,l}}2 \, \dot\varphi_{q,i}^2 +\frac12 \, \boldsymbol{\dot{\tilde\varphi}}_l^T \mathbf{\tilde C}_l \, \boldsymbol{\dot{\tilde\varphi}}_l \biggr),
\end{align}

with $\boldsymbol{\varphi}_l^T = (\varphi_{*,il}, \varphi_{*,jl},\varphi_{r,il}, \varphi_{r,jl})$, where $\varphi_{r,il}$ and $\varphi_{r,jl}$ are left unchanged.
For the case of $C_{b,l} \to \infty$, we find

\begin{align}
\mathbf{\tilde C}_l \to \begin{pmatrix}
\infty & 0 & 0 & 0 \\
0 & 2(C_{g,i} + C_{g,j}) & 0 & 0 \\
0 & 0 & \frac{C_{il}}{2} + \frac{C_{g\mu,l}}4 & -\frac{C_{g\mu,l}}{4} \\
0 & 0 & -\frac{C_{g\mu,l}}{4} & \frac{C_{jl}}{2} + \frac{C_{g\mu,l}}4 
\end{pmatrix}
\end{align}

with $C_{g\mu,l}$ as defined in Eq.~\ref{eq:Cgmu}. This corresponds again to the circuit from Fig.~\ref{fig:two_blocks}, that is Eq.~\ref{eq:T_two_blocks_final}, which has one superfluous variable less.\\
To sum up, we can say that while the potential energy of a grid corresponding to Fig.~\ref{fig:ring} contains the longitudinal coupling between a qubit and its resonators with no connection between two blocks $i$ and $j$, the kinetic energy contains the transverse coupling between two resonators, with no connection between two such units $\alpha$ and $\beta$. The coupling thus remains entirely local, when the circuit is scaled up to a grid.\\
As discussed in Sec.~\ref{sec:scalable_architecture}, we need eight different frequencies $\omega_{\pm,l}$ with $l=\alpha,...,\delta$ for unequivocal frequency selection, where $\omega_\pm$ are the effective frequencies in the diagonal frame, as defined in Eq.~\ref{eq:rescaledfreq}.
In principle, this can be realized even if all eight original frequencies were equal, just by having four different values of $g_{c,l}$. As there is rigorously no coupling between neighboring qubits, there is no restriction on the qubit frequencies and they could all chosen to be the same.\\
For our implementation, this could mean that we fabricate all blocks of a qubit and its resonators identically and vary only the coupling capacitance $C_{b,l}$ between them. Looking at Figs.~\ref{fig:frequency-selection} and Fig.~\ref{fig:ring} (and again Eq.~\ref{eq:rescaledfreq}), it is clear that only four different coupling capacitances are needed on a grid to achieve eight different effective frequencies $\omega_{\pm,l}$ with $l=\alpha,...,\delta$ for the nearest and next-nearest neighbor resonator of a qubit. While this would lead to equal qubit frequencies for all qubits on the grid, Eq.~\ref{eq:omegarr} shows that the bare resonator frequencies would be already slightly different due to the variation of  $C_{b,l}$. Explicitly, this means that every two adjacent resonators would have the same bare frequency (but different effective frequencies $\omega_\pm$), while every qubit would be surrounded by four resonators with different bare frequencies.

\chapter{APPLICATION: READOUT}
\label{chapt:readout}


The device presented in Chapt.~\ref{chapt:design} offers several possibilities to perform readout of the qubit. It is suitable to use the embedded resonator as a readout resonator, such that the qubit remains without direct coupling to the environment in order to protect its coherence. The resonator thus needs to be coupled to a readout cavity or a transmission line, in order to extract the information. As shown in Sec.~\ref{sec:transmissionline}, this can be done without any direct coupling between the qubit and the environment. Since the qubit can be coupled separately to several resonators (see Sec.~\ref{sec:n_resonators}), one resonator could be used for readout, while the other(s) are used to mediate the coupling to the nearby qubits (see Sec.~\ref{sec:two_blocks}). We will see that the qubit's state can be read out  using either longitudinal or transverse coupling to the readout resonator, though in a very different way. \\
Regular dispersive readout \cite{Blais2004} uses the dispersive shift that arises due to transverse coupling, which leads to a qubit state dependence of the resonator frequency. Driving the resonator at its bare frequency $\omega_r$, one can see a peak in the transmission spectrum at frequency $\omega_r \pm \chi$, where $\chi$ is the dispersive shift (see Eq.~\ref{eq:dispersive2}). While this technique is the standard measurement method in superconducting qubits, it is only approximately quantum non-demolition and has the further disadvantage that the dispersive shift makes the qubit's coherence time dependent on the finite photon lifetime. In this chapter, we will thus discuss the possibility to use longitudinal coupling for readout. As opposed to standard dispersive readout with transverse coupling, longitudinal coupling offers at least two different readout schemes, which are exactly quantum non-demolition.

\section{READOUT USING MODULATED COUPLING}


One possibility to do readout using longitudinal coupling is the flux modulation procedure introduced by Didier et al. in Ref.~\cite{Didier2015}. Their strategy is based on the modulation of the longitudinal coupling between a qubit and its resonator at the frequency of the resonator. The longitudinal coupling term $g_{zx} \, \sigma_z (a^\dagger + a)$ plays the role of a qubit-state dependent drive on the resonator. As we will see, this leads in steady-state to a qubit-state dependent displacement of the cavity field. The amplitude of this displacement is negligible in the static case, but it can be boosted by modulating the coupling in time.\\
Let us assume a system with static longitudinal coupling between qubit and resonator, that is

\begin{align}
\mathcal{H} = \hbar \,\omega_r \, a^\dagger a + \hbar \,\frac{\Delta}{2} \,\sigma_z + \hbar \, g_{zx} \, \sigma_z (a^\dagger + a).
\end{align}

As shown in Ref.~\cite{Didier2015} (and in App.~\ref{app:langevin}), the longitudinal coupling leads to a qubit state-dependent displacement of the cavity field with amplitude 

\begin{align}
\pm \frac{g_{zx}}{\omega_r + i\,\kappa/2}, 
\label{eq:sep}
\end{align}

where $\kappa$ is the coupling between the resonator and the environment (see again App.~\ref{app:langevin}). As in most cases $\omega_r \gg g_{zx}, \kappa$, this displacement is usually negligibly small and therefore does not serve for readout. However, it can be significantly boosted by modulating the coupling at the frequency of the resonator, that is

\begin{align}
g_{zx}(t) = \bar g_{zx} + \tilde g_{zx} \cos(\omega_r t),
\end{align}

where $\tilde g_{zx}$ is the modulation amplitude and $\bar g_{zx}$ a possible constant offset.
We move to a rotating frame using the unitary transformation $\mathcal{U} = \exp{(i \,\mathcal{H}_0/\hbar \,t)}$ with

\begin{align}
\mathcal{H}_0 = \hbar \, \omega_r a^\dagger a + \hbar \,\frac{\Delta}{2} \sigma_z,
\end{align}

which leads to

\begin{align}
\mathcal{H}_\text{rot} =  \mathcal{U}\, \mathcal{H}\, \mathcal{U}^\dagger + i \, \dot{\mathcal{U}} \,\mathcal{U}^\dagger = \frac{\hbar \,\tilde g_{zx}}{2} \, \sigma_z (a^\dagger + a),
\label{eq:rot}
\end{align}

where any fast rotating terms were neglected in a rotating wave approximation (see App.~\ref{app:langevin}). In this frame, the longitudinal coupling term is standing still, while all the diagonal terms are gone. We can thus adapt Eq.~\ref{eq:sep} by setting $\omega_r \to 0$ and $g_{zx} \to \tilde g_{zx}/2$, which gives a much bigger amplitude of the cavity field displacement than in Eq.~\ref{eq:sep}, namely

\begin{align}
\pm \frac{\tilde g_{zx}}{i\,\kappa}.
\label{eq:sep2}
\end{align}

As stated in Ref.~\cite{Didier2015}, this displacement should correspond to a few hundred photons and it should make the two qubit states easily distinguishable by homodyne detection. As explicitly shown there, the displacement of the cavity field goes in opposite directions for the two different qubit states (see also Eq.~\ref{eq:out_mod} in App.~\ref{app:langevin}). In the case of dispersive measurement, however, the corresponding displacement does not take the optimal path in phase space, which requires longer measurement times compared to the strategy described just now.\\
The device suggested in Ref.~\cite{Didier2015} has a lot of similarities to ours and especially leads to the same flux dependences in the transverse and longitudinal coupling (Eqs.~\ref{eq:gxx} and \ref{eq:gzx}). However, the crucial difference is that in the design of Didier et al. there are two Josephson junctions which play the roles of both qubit junction and coupling junctions at the same time, while these are separated in our design. As stated in Sec.~\ref{sec:coupling}, the additional qubit junction in our design allows us to sweep through a full cycle of the external flux at moderate changes in the qubit frequency. Just as in Didier's design, a time-dependent flux variation $\Phi_x = |\Phi_x| \cos(\omega_r t)$ will lead to time-dependent transverse and longitudinal coupling modulated with the drive frequency, though our design allows higher flux variation amplitudes. The transverse coupling term will be time-dependent both in the original and in the rotating frame of Eq.~\ref{eq:rot} and can thus be neglected in the rotating wave approximation. In Ref.~\cite{Didier2015} it is stated that spurious transverse coupling does not affect the signal-to-noise ratio of the measurement at short measurement times $\tau$, that is for $\tau \ll 1/\kappa$.


\section{READOUT USING SIDEBAND TRANSITIONS}
\label{sec:readout_billangeon}

An alternative possibility to use longitudinal coupling for readout is described by Billangeon et al. in Ref.~\cite{Billangeon2015}. Their strategy is based on the system described in Sec.~\ref{sec:scalable_architecture}, where two qubits are each coupled longitudinally to a resonator, while the two resonators are coupled via a conjugate degree of freedom. The readout scheme they conceived uses a series of sideband transitions between the qubit and its nearest  and next-nearest neighbor resonator to fill up one of the resonators with a significant number of excitations.  \\
As mentioned in Sec.~\ref{sec:scalable_architecture}, the corresponding Hamiltonian, that is Eq.~\ref{eq:long2}, can be exactly diagonalized using a series of unitary transformations, which lead to a frame where the two qubits and two resonators are uncoupled from each other. However, the coupling can be turned on again by applying a transverse drive on one of the qubits. Within the rotating wave approximation, different sideband transitions can be selected via the frequency of the drive. This renders sideband transitions between a qubit and its nearest and next-nearest neighbor resonator possible, but never induces direct coupling to the next qubit. Applying a combination of two sideband transitions fills one of the resonators with excitations, if the qubit is in the excited state. \\
Let us suppose that the two resonators are in their ground state, while the qubit we want to measure is either in its $| 0 \rangle$ or $| 1 \rangle$ state. We will apply a combination of the two sideband transitions $(\sigma_- a_1^\dagger a_2^\dagger + \sigma_+ a_1 a_2)$ and $(\sigma_- a_1^\dagger a_2 + \sigma_+ a_1 a_2^\dagger)$, where $a_1$ acts on the nearest neighbor resonator and $a_2$ acts on the next-nearest neighbor resonator of the qubit in question.
Clearly, both sideband transitions can only have an effect, if the qubit is in its $| 1 \rangle$ state. If the system is in the $| 0_q 0_{r,1} 0_{r,2} \rangle $ at the beginning, it will stay there and not be affected by neither of the sideband transitions.\\
If, however, the system starts in the state $| 1_q 0_{r,1} 0_{r,2} \rangle $, the first sideband transition will lead to 

\begin{align}
| 1_q 0_{r,1} 0_{r,2} \rangle  \to | 0_q 1_{r,1} 1_{r,2} \rangle.
\end{align}

The second sideband transition then leads

\begin{align}
| 0_q 1_{r,1} 1_{r,2} \rangle \to | 1_q 0_{r,1} 2_{r,2} \rangle.
\end{align}

After a few repetitions, the next-nearest neighbor resonator will be filled with a significant number of excitations, while the qubit and its nearest neighbor resonator simply oscillate between their ground and first excited state.\\ 
Note that this scheme relies on the fact that there are two resonators between the two qubits. The presence of the second resonator, which is not filled with excitations, is essential, as it prevents the qubit's ground state from coupling to the sideband transitions. The resonator is thus only filled with photons, if the qubit is in the excited state.\\
We should also note that this scheme using sideband transitions is possible due to the absence of the dispersive shift, as the frequency of the resonator(s) does not depend on the state of the qubit and vice versa. This means that the sideband transitions always stay resonant, even though the photon number in the resonator changes (compare again Ref.~\cite{Billangeon2015} for details).\\
The device presented in Sec.~\ref{sec:two_blocks} of this thesis is an implementation of this system with two qubits and two resonators. Placing such a device in a readout cavity or coupling it to a transmission line as described below, would allow for the detection of the photons in the resonators. In order to do the sideband transitions, we need a way to drive the qubit. This can be done simply by varying the external flux through the big qubit loop in Fig.~\ref{fig:qubit_resonator}. Setting $\varphi_{Xb} = \varphi_d \cos(\omega_d \,t)$ in Eq.~\ref{eq:pot} yields

\begin{align}
\cos(\varphi_q + \varphi_{Xb}) \approx \cos(\varphi_q) + \varphi_d \cos(\omega_d\, t) \sin(\varphi_q)
\end{align}

for the qubit term, where $\varphi_d = 2\pi \Phi_d/\Phi_0$ is the rescaled amplitude of the flux drive and $\omega_d$ is its modulation frequency. To first order in $\varphi_q$, we can identify

\begin{align}
\sin(\varphi_q) \, \hat{=} \, \sigma_x
\end{align}

(compare Eq.~\ref{eq:phiq}), which corresponds to a transverse driving term. As shown in Ref.~\cite{Billangeon2015}, sideband transitions between the qubit and its nearest and next-nearest neighbor resonator can be selected via the frequency of the drive $\omega_d$.
An alternative way to implement such a transverse drive would be an $AC$ voltage between the qubit nodes $a$ and $b$ in Fig.~\ref{fig:qubit_resonator}.

\section{COUPLING TO A TRANSMISSION LINE}
\label{sec:transmissionline}

\begin{SCfigure}[50][tb]
\centering
    \includegraphics[height=4.5cm]{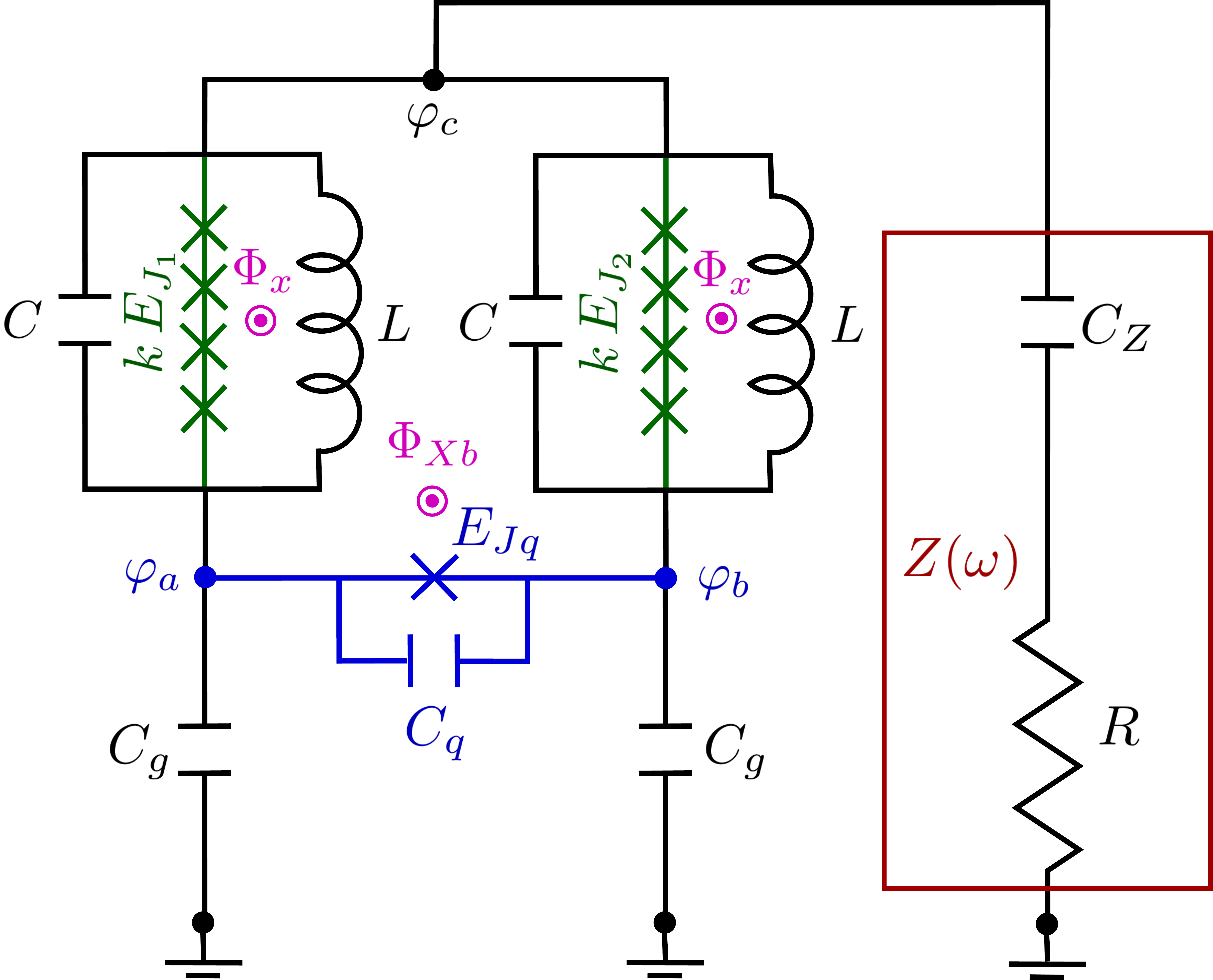}
  \caption{Qubit-resonator circuit with coupling to a transmission line, here represented as an external impedance $Z(\omega)$. The resonator acts as a readout resonator, which mediates the coupling between qubit and environment. As the circuit is symmetric with respect to the qubit junction, the qubit mode itself remains uncoupled from the transmission line. \label{fig:environment}}
\end{SCfigure}

Both readout methods described above require the resonator(s) to be coupled to a readout cavity or a transmission line, in order to extract the information. Ideally, this should be done without any direct coupling between the qubit and the readout device.
We can see our coupled qubit-resonator system as two orthogonal dipoles, which correspond to the qubit and resonator modes. The qubit consists of a dipole between the qubit nodes $\varphi_a$ and $\varphi_b$, while the dipole corresponding to the resonator mode is composed of $\varphi_a$ and $\varphi_b$ on the one hand and $\varphi_c$ on the other hand (compare Fig.~\ref{fig:environment}). This coincides with the phase variables for qubit and resonator as defined in Eq.~\ref{eq:variables}. Aligning the resonator dipole with the electric field in a cavity, will lead to coupling between the resonator and the cavity mode, while the orthogonal qubit mode will remain uncoupled. This is the case in the proposal for the physical implementation shown in Fig.~\ref{fig:phys_impl}.\\
The same is true when we couple the resonator capacitively to a transmission line as shown in Fig.~\ref{fig:environment}. The transmission line, which can be modeled as an external impedance $Z(\omega)$, is connected between node $\varphi_c$ and ground, while two identical capacitors connect the qubit nodes $\varphi_a$ and $\varphi_b$ to ground. This symmetric arrangement with respect to the qubit junction ensures that the qubit mode remains uncoupled from the transmission line, as long as the two capacitances from the qubit nodes to ground $C_g$ are equal.
Note that the transmission line is connected to our original circuit (that is Fig.~\ref{fig:qubit_resonator}) in the same way as the two blocks are coupled in Sec.~\ref{sec:two_blocks}. Just as the qubit does not couple to the transmission line in Fig.~\ref{fig:environment}, there is no direct coupling between the qubit and the next-nearest neighbor resonator in Sec.~\ref{sec:two_blocks}.

\sidecaptionvpos{figure}{c}

The coupling between the qubit-resonator system and its environment, which is indispensable for readout, can induce both relaxation and decoherence. In Ref.~\cite{Burkard2005}, Burkard et al. give expressions for the relaxation time $T_1$ and the decoherence time $T_2$. In order to describe a dissipative circuit, they construct a so-called \textit{Caldeira-Leggett} Hamiltonian, in which the circuit is split up into a dissipation-free part called \textit{system} and a \textit{bath} that interacts with the system

\begin{align}
\mathcal{H} = \mathcal{H}_{S} + \mathcal{H}_{B} + \mathcal{H}_{SB}.
\end{align}

Appendix~\ref{app:burkard} contains a detailed discussion of Burkard's method of treating dissipative circuits, using the circuit from Fig.~\ref{fig:environment} as an example. As explained there, the expressions for relaxation and decoherence times depend on the system-bath coupling, which is chosen to be of the form

\begin{align}
\mathcal{H}_{SB} = \mathbf{m} \cdot \mathbf{Q} \sum_\alpha c_\alpha x_\alpha,
\label{eq:Hsb}
\end{align}

(see Ref.~\cite{Burkard2005}), where $\mathbf{m}$ is a vector that describes the coupling between system and bath, and $\mathbf{Q}$ is the charge variable vector of the system.
For each mode in the circuit, the relaxation and decoherence times are given in terms of the matrix elements of $\mathbf{m} \cdot \mathbf{Q}$ with the states $| 0 \rangle $ and $ | 1 \rangle $ of the corresponding mode. It is

\begin{align}
\frac{1}{T_1} &= \frac{4}{\hbar} |\langle 0| \mathbf{m} \cdot \mathbf{Q} | 1\rangle|^2 J(\omega_{01}) \coth\left(\frac{\hbar \omega_{01}}{2k_B T}\right) \label{eq:T1}\\
\frac{1}{T_2} &= \frac{1}{2 T_1} + \frac{1}{T_\phi}\\
\frac{1}{T_\phi} &= \frac{1}{\hbar} |\langle 0| \mathbf{m} \cdot \mathbf{Q} | 0\rangle - \langle 1| \mathbf{m} \cdot \mathbf{Q} | 1\rangle|^2 \left. \frac{J(\omega)}{\hbar \omega} \right|_{\omega \to 0} 2k_B T,
\end{align}

where $T_\phi$ is the dephasing time, $\omega_{01}$ is the transition frequency between the ground and first excited state of the corresponding mode, and $J(\omega)$ is the spectral density of the bath (see again App.~\ref{app:burkard}). For the circuit depicted in Fig.~\ref{fig:environment}, we find that

\begin{align}
\mathbf{m} \cdot \mathbf{Q} = \frac{2 \,C_g}{C + C_g} Q_r,
\end{align}

which means that only the resonator couples to the bath, while there is no direct coupling between the qubit and the bath as desired. This means that in principle, the qubit has infinite relaxation and decoherence times, as long as the circuit is perfectly symmetric. As capacitances can be fabricated very accurately, we can assume that the influence of the system-bath coupling on the qubit is negligible. For pure longitudinal coupling between qubit and resonator, the absence of the dispersive shift further protects the qubit from decoherence.\\
The resonator, however, acquires finite $T_1$ and $T_2$ times due to its coupling to the environment. We find, however, that the dephasing time $T_\phi$ of the resonator diverges, as

\begin{align}
\langle 0| \mathbf{m} \cdot \mathbf{Q} | 0\rangle = \langle 1| \mathbf{m} \cdot \mathbf{Q} | 1\rangle = 0
\end{align}

(see App.~\ref{app:burkard} for details).
This means that the decoherence time is simply twice the relaxation time, $T_2 = 2 \,T_1$. The matrix element in Eq.~\ref{eq:T1} is however unequal to zero, as 

\begin{align}
|\langle 0| \mathbf{m} \cdot \mathbf{Q} | 1\rangle|^2 \sim |\langle 0| (a^\dagger -a) | 1\rangle|^2 = 1.
\end{align}

Naturally, $T_1$ diverges if either the capacitances from the qubit nodes to ground $C_g$ or the capacitance between the the system and the transmission line $C_Z$ go to zero (see Fig.~\ref{fig:environment}). The full expression for $T_1$ is given in App.~\ref{app:burkard}.

\chapter{CONCLUSIONS}
\label{chapt:conclusions}

In conclusion, we presented a circuit design for an inductively shunted transmon qubit that can be tuned between pure transverse and pure longitudinal coupling to an embedded resonator mode, by changing the external magnetic flux.\\
We have given an introduction to circuit quantization showing how to go from a circuit description to a second-quantized Hamiltonian in a systematic way. With the help of some concrete examples, we discussed how to include external fluxes and dissipative elements. On the Hamiltonian level, we introduced the two inherently different coupling types present in our design, that is transverse and longitudinal coupling. \\
In order to apprehend the remarkable advantages of longitudinal coupling in terms of scalability, we discussed a scalable qubit architecture conceived by Billangeon et al. in Ref.~\cite{Billangeon2015}. The crucial advantage of this architecture is that the system is exactly diagonalizable using a series of unitary transformations. In this diagonal frame, there are no dispersive shifts or residual couplings between any qubits or resonators.
Single-qubit operations and sideband transitions between a qubit and any of its resonators can be done by driving the qubit at the corresponding frequency. The coupling is strictly confined to the nearest and next-nearest neighbor resonators of each qubit; there is never any direct qubit-qubit coupling.\\
Translating this discussion from the Hamiltonian level to the language of circuit quantization, we have shown how to design circuits with specifically tailored couplings.\\
The core of this thesis consists of the circuit design of an inductively shunted transmon qubit with flux-tunable coupling to an embedded harmonic mode, which we first presented in Ref. \cite{Richer2016} and further refined and adapted in Ref. \cite{Richer2017}. Using a symmetric design, static transverse coupling terms are cancelled out, while the parity of the only remaining coupling term can be tuned via an external flux. Similarly to the fluxonium qubit, the inductive shunting protects the qubit from charge noise.
The distinctive feature of the tunable design is that the transverse coupling disappears when the longitudinal is maximal and vice versa. As opposed to other approaches, pure longitudinal coupling can be reached with moderate changes in the qubit frequency. \\
We performed quantitative analytical and numerical calculations for several qubit-resonator coupling designs, including a discussion of the effect of unwanted asymmetries. We found that by applying an additional magnetic flux through the loop of the inductively shunted qubit, both the coupling terms and the qubit anharmonicity increase significantly. Additionally, we showed that using single Josephson junctions in the qubit-resonator coupling elements is not feasible, because of the resulting large unwanted coupling terms and high resonator anharmonicity. Using junction arrays in the coupling elements is more favorable, because the ratio between the longitudinal coupling and the unwanted coupling terms can be increased by an order of magnitude, and the resonator anharmonicity is strongly suppressed. Including an additional inductance in the coupling branches helps to further increase the qubit anharmonicity and the longitudinal coupling by up to a factor of two. Finally, we proposed a prototype design based on standard circuit fabrication, integrated with high kinetic inductance elements.\\
Furthermore, we have shown that our design can be used to implement the scalable architecture conceived in Ref.~\citep{Billangeon2015}.
We presented a proposal for a circuit QED system where a qubit couples to several resonators via its longitudinal degree of freedom and every resonator is capacitively coupled to a resonator from the next unit cell. This proposal is easily scalable to any number of resonators per qubit and any number of unit cells.
Remarkably, just a single unique qubit frequency suffices for the scalability of this scheme.  The same is true for the resonators, if the resonator-resonator coupling constants are varied instead.\\
We have included a short discussion of two different readout schemes using longitudinal coupling, which could be implemented with our design. In both cases, the resonator serves as a readout resonator which mediates the coupling to the environment. We have shown that this can be implemented without any direct coupling between the environment and the qubit.

\section{OUTLOOK}
In Sec.~\ref{sec:implementation}, we have shown a prototype device for the implementation of the inductively shunted transmon qubit with tunable transverse and longitudinal coupling. This section and especially Fig.~\ref{fig:phys_impl} were provided by my co-authors from Ref.~\cite{Richer2017}, Nataliya Maleeva, Sebastian~T. Skacel and Ioan~M. Pop from \textit{Karlsruhe Institute of Technology}. The physical realization of the design presented in this thesis will surely lead to many interesting new questions and experimental challenges. I am very much looking forward to see this prototype device being realized in the near future.\\
Figure~\ref{fig:phys_impl} already contains a sketch of a scaled-up version of the prototype circuit with two qubits coupled via two resonators, as proposed in Sec.~\ref{sec:two_blocks}. This could be used to implement the two-qubit phase gate described in Sec.~\ref{sec:gate}, which employs longitudinal coupling. Hopefully, this approach could lead to high two-qubit gate fidelities, given that it works without any direct qubit-qubit coupling. 
Following the strategy discussed in Chapt.~\ref{chapt:scalability}, the circuit could be scaled up to a grid with strictly local interaction following the scheme of Ref.~\cite{Billangeon2015}. \\
The adapted circuit presented in Sec.~\ref{sec:adaptation} provides quite a lot of interesting physics, which we have not gone to explore. As mentioned there, Eq.~\ref{eq:invert} is only invertible in a certain parameter regime, out of which the potential energy becomes multi-valued. While we have chosen our parameters such that this case is avoided, it is surely an interesting question to see what happens in this multi-valued regime. Another open question is the effect of the neglected capacitances both in the adapted circuit from Sec.~\ref{sec:adaptation} and the coupling junction array from Sec.~\ref{sec:array}. While we assume that their effect is small, some future work could be done to investigate this question.

\begin{appendices}


\chapter{OPERATORS AND COMMUTATORS}
\label{app:operators}

The creation and annihilation operators for quantum harmonic oscillators are defined as

\begin{align}
a^\dagger  = \sum_{j=0}^\infty \sqrt{j + 1} \, | j + 1 \rangle \langle  j | \qquad a = \sum_{j=0}^\infty \sqrt{j + 1} \, | j \rangle \langle  j + 1 |,
\end{align}

where the Fock states $| j \rangle$ are the eigenstates of the harmonic oscillator and $a^\dagger$ and $a$ satisfy the commutation relation

\begin{align}
[a, a^\dagger ] = 1.
\end{align}

The creation and annihilation operators act on the Fock states as

\begin{align}
a^\dagger | j \rangle &= \sqrt{j + 1} | j + 1 \rangle \\
a | j \rangle &= \sqrt{j} | j - 1 \rangle,
\end{align}

while their product is known as the number operator

\begin{align}
N = a^\dagger a = \sum_{j=1}^\infty j \,| j  \rangle\langle j |.
\end{align}

The Fock states $| j \rangle$ are the eigenstates of the number operator with eigenvalue $j$

\begin{align}
N | j \rangle &= j | j \rangle.
\end{align}

If a Fock space is truncated to the two lowest levels $| 0 \rangle$ and $| 1 \rangle$, as done in the so-called two-level approximation (see Sec.~\ref{sec:qubit}) for qubits, the creation and annihilation operators transform to the qubit raising and lowering operators $\sigma^\pm$, as

\begin{align}
a^\dagger  \to  | 1 \rangle \langle  0 | = \sigma_+ = \frac{\sigma^x -  i\sigma^y}{2}
\end{align}

and 

\begin{align}
a  \to  | 0 \rangle \langle  1 | = \sigma_- = \frac{\sigma^x +  i\sigma^y}{2},
\end{align}

where $\sigma_j$ with $j = x,y,z$ are the Pauli matrices. 
These are given by 

\begin{align}
\sigma_x &= |0\rangle\langle1| + |1\rangle\langle0| \\
\sigma_y &= i(|1\rangle\langle0| - |0\rangle\langle1|) \\
\sigma_z &= |0\rangle\langle0| - |1\rangle\langle1| \\
\sigma_0 &= |0\rangle\langle0| + |1\rangle\langle1|, 
\end{align}

where we have included the identity matrix $\sigma_0$. It is

\begin{align}
\sigma_j^2 = \sigma_0  \quad \text{for} \quad j = 0,x,y,z
\end{align}

and

\begin{align}
\sigma_x \sigma_y \sigma_z = i \,\sigma_0.
\end{align}

Their commutation relation is given by

\begin{align}
[\sigma_j\,,\sigma_k] = \sigma_j \, \sigma_k - \sigma_k \, \sigma_j = 2\, i \sum_{l=x,y,z} \epsilon_{jkl}\; \sigma_l,
\end{align}

for $j,k = x,y,z$. 
Using the Pauli matrices, the number operator $N = a^\dagger a$ yields

\begin{align*}
a^\dagger a &= \sum_{j = 0}^\infty j\, | j \rangle \langle j | \to | 1 \rangle \langle 1|  \\
&=  \frac{|1 \rangle \langle 1| - |0 \rangle \langle 0|}{2} + \frac{|1 \rangle \langle 1| + |0 \rangle \langle 0|}{2}
= \frac{\sigma_z}2 + \frac{\sigma_0}{2}. \numberthis
\end{align*}

in the truncated Fock space. Note that just as $| j \rangle$ is an eigenstate of the number operator, the qubit operator $\sigma_z$ does not change the state of the qubit, as opposed to the $\sigma_x$ and $\sigma_y$ operators.

%
%
%

%
%

\chapter{VARIABLE ELIMINATION USING THE CHOLESKY DE\-COM\-PO\-SI\-TION}
\label{app:cholesky}

The so-called \textit{Cholesky decomposition}~\cite{Cholesky} of a Hermitian positive-definite matrix $\mathbf A$ consists of an upper triangular matrix $\mathbf B$ with real and positive diagonal entries and its conjugate transpose $\mathbf B^\dagger$, such that

\begin{align}
\mathbf A = \mathbf{B}^\dagger \mathbf{B}.
\end{align}

Explicitly, $\mathbf{B}$ has the form

\begin{align}
B_{jj} &= \sqrt{A_{jj} - \sum_{k=1}^{j-1} B_{kj} B_{kj}^*} & \\
B_{ji} &= \left(A_{ji} - \sum_{k=1}^{j-1} B_{ki} B_{kj}^*\right)/B_{jj} \quad &i>j \\
B_{ij} &= 0 &i>j,
\end{align}

where $B_{ij}^*$ is the complex conjugate of $B_{ij}$. For a real and symmetric $3\times 3 $ matrix $\mathbf A$ with all positive entries, the Cholesky decomposition is given by

\begin{align}
\mathbf{B} =\begin{pmatrix}
\sqrt{A_{11}} & \frac{A_{12}}{\sqrt{A_{11}}} & \frac{A_{13}}{\sqrt{A_{11}}} \\
0 & \sqrt{A_{22} - \frac{A_{12}^2}{A_{11}}} & \frac{A_{23} - \frac{A_{12}A_{13}}{A_{11}}}{\sqrt{A_{22} - \frac{A_{12}^2}{A_{11}}}} \\
0 & 0 & \sqrt{A_{33} - \frac{A_{13}^2}{A_{11}} - \left(\frac{A_{23} - \frac{A_{12}A_{13}}{A_{11}}}{\sqrt{A_{22} - \frac{A_{12}^2}{A_{11}}}}\right)^2}
\end{pmatrix}.
\end{align}

The Cholesky decomposition has oftentimes a much simpler form than a square root decomposition, making analytic calculations a lot easier. Especially the first row of $\mathbf{B}$ has a very simple form. This can be used for the elimination of unwanted variables as done in Secs.~\ref{sec:cholesky},  \ref{sec:two_blocks}, and \ref{sec:grid}.\\
As an explicit example of such a variable elimination, let us take the kinetic energy of the circuit shown in Fig.~\ref{fig:two_blocks_cc}, that is Eq.~\ref{eq:Trr0}, which can be rewritten as

\begin{align}
\mathcal T = \sum_{i=1}^2 \left(\frac{\Phi_0}{2\pi}\right)^2 \biggl(\frac{2 C_{q,i} + C_{i} + C_{g}}2 \, \dot\varphi_{q,i}^2 +\frac12 \, \boldsymbol{\dot{\varphi}}^T \mathbf{C} \, \boldsymbol{\dot{\varphi}} \biggr)
\end{align}

with $\boldsymbol{\varphi}^T = (\bar\varphi_{1}, \bar\varphi_{2},\varphi_{r,1}, \varphi_{r,2})$ (compare also Eq.~\ref{eq:T_matrix}). The capacitance matrix $\mathbf C$ in terms of these variables is given by

\begin{align}
\mathbf{C} =\begin{pmatrix}
C_{b} + 2\, C_{g,1} & - C_{b} & C_{g,1} & 0 \\
 - C_{b} & C_{b} + 2\, C_{g,2} & 0 & C_{g,2} \\
C_{g,1} & 0 & \frac{C_{1} + C_{g,1}}2 & 0 \\
0 & C_{g,2} & 0 & \frac{C_{2} + C_{g,2}}2 
\end{pmatrix}.
\end{align}

While it is not trivial to find the eigenvalues and eigenvectors of $\mathbf{C}$, its Cholesky decomposition (the first row in particular) has a quite simple form. As the two variables $\bar\varphi_i$ do not appear in the potential energy of the system (Eq.~\ref{eq:pot_two_blocks}), we can use a Cholesky decomposition to decouple them in the kinetic energy without changing the potential energy. Let us define the linear, invertible transformation 

\begin{align}
\mathbf{R}_1 =\begin{pmatrix}
1 & \frac{- C_{b}}{C_{b} + 2\, C_{g,1}} & \frac{C_{g,1}}{C_{b} + 2\, C_{g,1}} & 0  \\
0 & 1 & 0 & 0 \\
0 & 0 & 1 & 0 \\
0 & 0 & 0 & 1 
\end{pmatrix},
\end{align}

which consists of the first row of the Cholesky decomposition of $\mathbf{C}$ (re\-scaled by its first entry to be unitless) and an identity matrix for the other three rows. It is clear that the transformation $\mathbf{R}_1 \, \boldsymbol{\varphi}$ leaves the resonator variables $\varphi_{r,1}$ and $\varphi_{r,2}$ untouched and that therefore the potential energy of the system is unchanged. The transformation decouples the first variable $\bar\varphi_1$ in the capacitance matrix, which now has the form

\begin{align}
\left(\mathbf{R}_1^{-1}\right)^T \mathbf{C} \, \mathbf{R}_1^{-1}
=\begin{pmatrix}
\blacksquare & 0 & 0 & 0 \\
0 & \blacksquare & \blacksquare & \blacksquare \\
0 & \blacksquare & \blacksquare & \blacksquare \\
0 & \blacksquare & \blacksquare & \blacksquare 
\end{pmatrix}. 
\label{eq:block}
\end{align}

We can use the first line of the Cholesky decomposition of the lower block in Eq.~\ref{eq:block} (again rescaled to be dimensionless) to define

\begin{align}
\mathbf{R}_2 =\begin{pmatrix}
1 & 0 & 0 & 0  \\
0 & 1 & \frac{C_b C_{g,1}}{2 C_b (C_{g,1} + C_{g,2}) + 4 C_{g,1} C_{g,2}} & \frac{C_b C_{g,2} + 2 C_{g,1} C_{g,2} }{2 C_b (C_{g,1} + C_{g,2}) + 4 C_{g,1} C_{g,2}} \\
0 & 0 & 1 & 0 \\
0 & 0 & 0 & 1 
\end{pmatrix},
\end{align}

which decouples the next superfluous variable $\bar\varphi_2$. Note again that this transformation leaves the resonator variables $\varphi_{r,1}$ and $\varphi_{r,2}$ untouched, as it acts exclusively on $\bar\varphi_2$. The final capacitance matrix yields


\begin{align*}
\mathbf{\tilde C} &=\left(\mathbf{R}_2^{-1}\right)^T \left(\mathbf{R}_1^{-1}\right)^T \mathbf{C} \, \mathbf{R}_1^{-1} \, \mathbf{R}_2^{-1}\\
&= \begin{pmatrix}
C_{b} + 2\, C_{g,1} & 0 & 0 & 0 \\
0 & \frac{2(C_{g,1} + C_{g,2})(C_{b}+C_{g\mu})}{C_{b} + 2C_{g,1}} & 0 & 0 \\
0 & 0 & \frac{C_{1}}{2} + \frac{C_{b\mu}}4 & -\frac{C_{b\mu}}{4} \\
0 & 0 & -\frac{C_{b\mu}}{4} & \frac{C_{2}}{2} + \frac{C_{b\mu}}4 
\end{pmatrix} \numberthis
\label{eq:Ctilde}
\end{align*}

with $C_{b\mu}$ and $C_{g\mu}$ as defined in Eq.~\ref{eq:Cgmu}.
We see that both superfluous variables variables $\bar\varphi_i$ are decoupled, while there is now a direct coupling term between the resonator variables $\varphi_{r,1}$ and $\varphi_{r,2}$.
In total, the kinetic energy thus transforms to

\begin{align}
\mathcal T = \sum_{i=1,...} \left(\frac{\Phi_0}{2\pi}\right)^2 \biggl(\frac{2 C_{q,i} + C_{i} + C_{g}}2 \, \dot\varphi_{q,i}^2 +\frac12 \, \boldsymbol{\dot{\tilde\varphi}}^T \mathbf{\tilde C} \, \boldsymbol{\dot{\tilde\varphi}} \biggr)
\end{align}

with $\boldsymbol{\varphi}^T = (\varphi_{*,1}, \varphi_{*,2},\varphi_{r,1}, \varphi_{r,2})$ and $\mathbf{\tilde C}$ as given in Eq.~\ref{eq:Ctilde}. The decoupled variables can be discarded, as they do not influence the evolution of the variables we are interested in.
This method for the elimination of superfluous variables is oftentimes applicable whenever the kinetic energy includes more variables than the potential energy or vice versa.

\chapter{DISSIPATION AND READOUT}
\label{app:readout}

\section{CIRCUIT THEORY FOLLOWING BURKARD}
\label{app:burkard}

A very useful method for the treatment of dissipative circuits is presented in Refs.~\cite{Burkard2004, Burkard2005} by Burkard and others. 
In order to describe dissipation, Burkard et al. rely on the so-called \textit{Caldeira-Leggett} formalism~\cite{Caldeira1983}. In this formalism, a dissipative circuit is split up in two parts: a dissipation-free system and the environment it interacts with. This environment is depicted as a bath with infinitely many degrees of freedom. While in principle, the Hamiltonian formalism cannot capture dissipation as the energy is always conserved, the irreversible loss of energy due to dissipation can be formally described as a transition of energy from the system to the bath. A lossy element, such as an external impedance, can thus be represented as a bath of infinitely many harmonic oscillators. \\
The Caldeira-Leggett Hamiltonian of a dissipative circuit with a single external impedance can be written as

\begin{align}
\mathcal{H} = \mathcal{H}_{S} + \mathcal{H}_{B} + \mathcal{H}_{SB},
\label{eq:caldeira}
\end{align}

which is divided into the dissipation-free \textit{system}, the \textit{bath}, and the interaction between them.
Formally, the bath can be written as a sum of infinitely many harmonic oscillators, that is

\begin{align}
\mathcal{H}_B = \sum_\alpha \left(\frac{p_\alpha^2}{2 m_\alpha}  + \frac{1}{2} m_\alpha \omega_\alpha^2 x_\alpha^2 \right),
\label{eq:caldeira_bath}
\end{align}

while the system-bath coupling is chosen to be of the form

\begin{align}
\mathcal{H}_{SB} = \mathbf{m} \cdot \mathbf{Q} \sum_\alpha c_\alpha x_\alpha,
\label{eq:caldeira_coup}
\end{align}

(see Ref.~\cite{Burkard2005}), where $\mathbf{m}$ is a vector that describes the coupling between system and bath, and $\mathbf{Q}$ is the charge variable vector of the dissipation-free system. While the specific form of the bath will have no influence on the dissipation-free system, the influence of the system-bath coupling becomes evident in terms of relaxation and decoherence times (see below). \\
Burkard describes an electrical circuit as a directed graph, arbitrarily assigning directions to the branches, that is the circuit elements. Among those branches a so-called tree is chosen (as also done by Devoret in Ref.~\cite{Devoret1997}), that is a subgraph that connects all the nodes in the circuit without any closed paths. The remaining branches are called chords. Necessarily, every chord defines a unique closed path in the circuit when added to the tree.  The number of elements in the tree corresponds to the number of independent degrees of freedom. 
\\
In Ref.~\cite{Burkard2005}, the tree is chosen such that it includes all Josephson junctions, external impedances, and voltage sources and as many inductances as necessary. Capacitances are not included in the tree. These rules imply that circuits with closed paths that contain nothing else than Josephson junction, external impedances and voltage sources cannot be treated with this method. 
Apart from that, the circuit needs to have enough capacitances to ensure that the Legendre transformation (Eq.~\ref{eq:legendre}) is applicable. In the matrix representation introduced in Chapt.~\ref{chapt:quantization} (Eq.~\ref{eq:lag_matrix}), this means that the capacitance matrix needs to be invertible. In Ref.~\cite{Burkard2004}, the rules for choosing the tree are slightly different. Note that here we will adhere to the conventions from Ref.~\cite{Burkard2005}.  

\sidecaptionvpos{figure}{c}
\begin{SCfigure}[50][tb]
\centering
    \includegraphics[height=4.5cm]{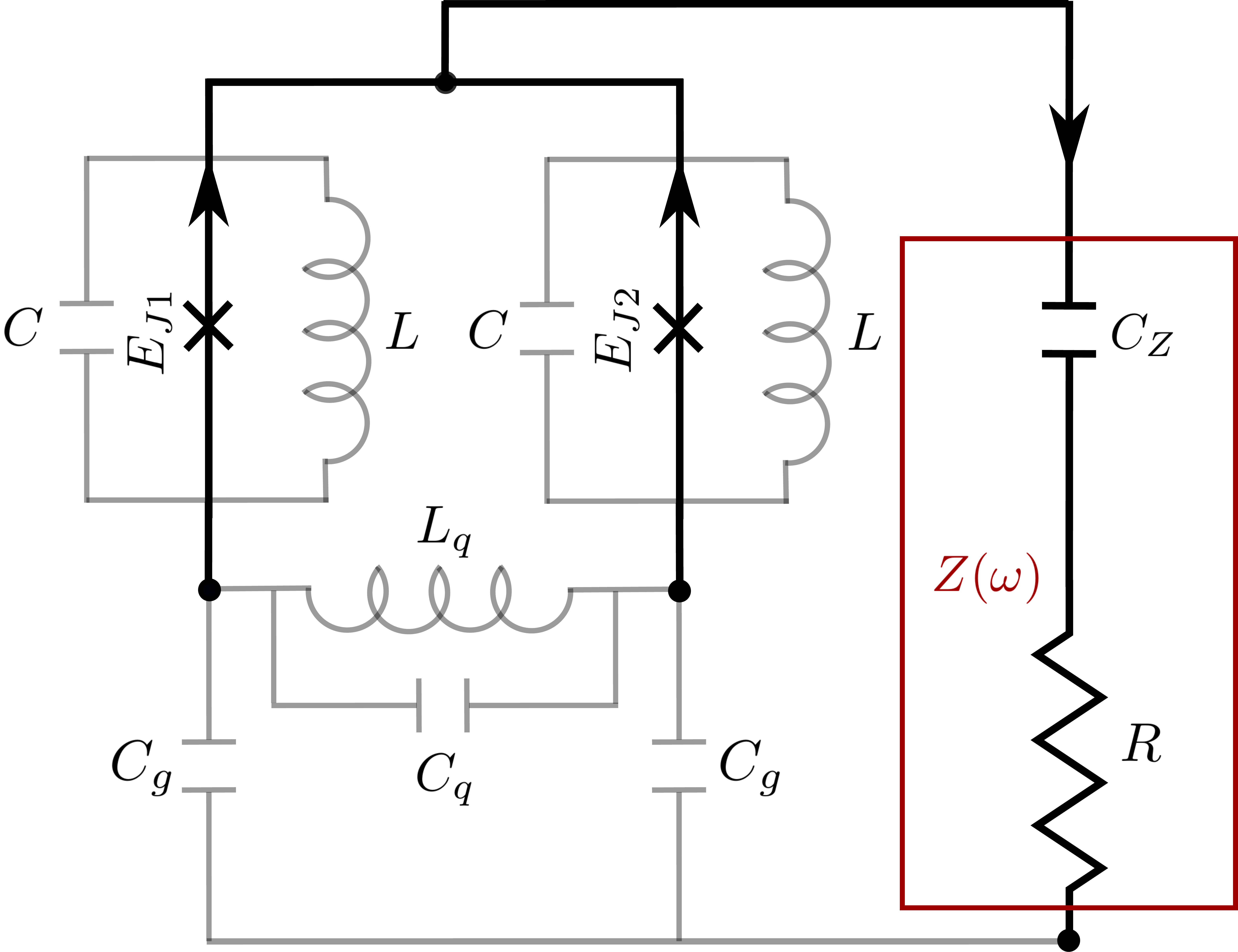}
  \caption{Qubit-resonator circuit with coupling to an external impedance $Z(\omega)$. As opposed to the circuit shown in Fig.~\ref{fig:environment}, we have substituted the qubit junction by a normal inductance in order to treat the circuit with Burkard's method. The tree is marked in thick black lines, while the chords are kept in grey. \label{fig:environment-subs}}
\end{SCfigure}

In order to treat the circuit from Sec.~\ref{sec:transmissionline} (Fig.~\ref{fig:environment}) with Burkard's method, we have to make a small adaptation, as shown in Fig.~\ref{fig:environment-subs}. We formally substitute the qubit junction by a normal inductance $L_q$, such that the tree can include all Josephson junctions without having any closed paths. In the Hamiltonian, we can later resubstitute the inductance term with a Josephson term. Apart from that we will keep to the case of single coupling junctions (compare Chapt.~\ref{chapt:design} and Sec.~\ref{sec:single}) for reasons of simplicity. The chosen tree, which includes the two coupling junctions $E_{J1}$ and $E_{J2}$ and the external impedance $Z(\omega)$, is marked in Fig.~\ref{fig:environment-subs} in thick black lines with arbitrarily chosen directions. \\
This choice of tree already defines the variables, in which the circuit is represented. Note that these are branch variables, as opposed to the node variables chosen in Chapt.~\ref{chapt:quantization} and Ref.~\cite{Devoret1997}. The two branch variables across the Josephson junctions belong to the dissipation-free \textit{system}, while the external impedance $Z(\omega)$ is represented as a \textit{bath}, as defined in Eq.~\ref{eq:caldeira}. Following Ref.~\cite{Burkard2005}, the Hamiltonian of the dissipation-free system is given by

\begin{align}
\mathcal{H}_S = \frac{1}{2}\mathbf{Q}^T \mathbf{C}^{-1} \mathbf{Q} + \mathcal U(\mathbf\Phi),
\end{align}

where $\mathbf{C}$ is the capacitance matrix and $\mathbf{\Phi}$ a vector with the conjugate flux variables to the charge variable $\mathbf{Q}$. Note that the branch variables across the Josephson junctions used here are simply the sum and the difference of the qubit and resonator variables defined in Eq.~\ref{eq:variables}. After a transformation to the variables from above, we find the same potential energy as in Eq.~\ref{eq:pot}, while in the kinetic energy is given by

\begin{align}
\mathcal{T} = \frac{Q_q^2}{2\,C_q + C+C_g}+\frac{Q_r^2}{C+C_g}.
\end{align}

This corresponds to the the Legendre transformation of Eq.~\ref{eq:kin} with the simple substitution $C \to C + C_g$ in order to account for the capacitance to ground in Fig.~\ref{fig:environment-subs}. With this substitution we can reuse all the formulas given in Chapt.~\ref{chapt:design}.\\
Following the strategy described in Ref.~\cite{Burkard2005}, the system-bath coupling for the circuit in Fig.~\ref{fig:environment-subs} can be written as

\begin{align}
\mathbf{m} \cdot \mathbf{Q} = \frac{2 \,C_g}{C + C_g} Q_r
\label{eq:mQ}
\end{align}

in terms of the variables defined in Eq.~\ref{eq:variables}. This means that only the resonator couples to the bath, while the qubit remains without direct coupling to the environment. \\
Burkard et al. give explicit formulas for the relaxation time $T_1$ and the decoherence time $T_2$ for each mode in the circuit in terms of the matrix elements of $\mathbf{m} \cdot \mathbf{Q}$ with the states $| 0 \rangle $ and $ | 1 \rangle $ of the corresponding mode. $T_1$ and $T_2$ thus depend directly on the system-bath coupling given in Eqs.~\ref{eq:caldeira_coup} and \ref{eq:mQ}. Explicitly, these are given by

\begin{align}
\frac{1}{T_1} &= \frac{4}{\hbar} |\langle 0| \mathbf{m} \cdot \mathbf{Q} | 1\rangle|^2 J(\omega_{01}) \coth\left(\frac{\hbar \omega_{01}}{2k_B T}\right)\\
\frac{1}{T_2} &= \frac{1}{2 T_1} + \frac{1}{T_\phi}\\
\frac{1}{T_\phi} &= \frac{1}{\hbar} |\langle 0| \mathbf{m} \cdot \mathbf{Q} | 0\rangle - \langle 1| \mathbf{m} \cdot \mathbf{Q} | 1\rangle|^2 \left. \frac{J(\omega)}{\hbar \omega} \right|_{\omega \to 0} 2k_B T,
\label{eq:Tphi}
\end{align}

where $T_\phi$ is the dephasing time, $\omega_{01}$ is the transition frequency between the ground and first excited state of the corresponding mode, and $J(\omega)$ is the spectral density of the bath. Following Ref.~\cite{Burkard2005} and using

\begin{align}
Z(\omega) = R + \frac{1}{i\,\omega \,C_Z},
\end{align}

we find 

\begin{align}
J(\omega) = \frac{(C+C_g)^2C_Z^2 R \, \omega}{(2C_g (C + 2 C_g) + (C+C_g)C_Z)^2 + 4C_g^2 (C + 2 C_g)^2 C_Z^2 R^2 \omega^2}
\label{eq:J}
\end{align}

%
%
for the spectral density of the bath.
As can be seen from Eqs.~\ref{eq:mQ}-\ref{eq:Tphi}, the resonator acquires finite $T_1$ and $T_2$ times due to its coupling to the environment, while for the qubits these are in principle infinite.
Looking at the quantization given for the resonator in Eq.~\ref{eq:phir}, we find $Q_r \sim (a^\dagger - a)$, which gives

\begin{align}
\langle 0| \mathbf{m} \cdot \mathbf{Q} | 0\rangle = \langle 1| \mathbf{m} \cdot \mathbf{Q} | 1\rangle = 0.
\end{align}

This means that the dephasing time $T_\phi$ of the resonator diverges, and therefore the decoherence time $T_2 = 2 \,T_1$ is simply twice the relaxation time. The matrix element in Eq.~\ref{eq:T1}, however, is given by

\begin{align*}
&|\langle 0| \mathbf{m} \cdot \mathbf{Q} | 1\rangle|^2 = \left(\frac{2 \,C_g}{C + C_g}\right)^2  |\langle 0| Q_r | 1\rangle|^2 \\
=& \frac{\hbar \, C_g^2}{(C + C_g)^2} \sqrt{\frac{(C + C_g)(1 + \eta)}{L}}  |\langle 0| (a^\dagger - a) | 1\rangle|^2\\
=& \hbar \, C_g^2 \sqrt{\frac{1 + \eta}{(C + C_g)^3 L}}, \numberthis
\label{eq:matel}
\end{align*}

where we took the expression for the quantization of the resonator as given Eq.~\ref{eq:phir}, but with $C \to C + C_g$. Using Eq.~\ref{eq:matel}, we can calculate the relaxation time of the resonator, which yields

\begin{align*}
T_1 &= \frac{1}{4\, C_g^2}  \sqrt{\frac{(C + C_g)^3 L}{1 + \eta}}  \tanh\left(\frac{\hbar }{2k_B T}\sqrt{\frac{1 + \eta}{(C + C_g) L}} \right) \times 1/J(\omega_{01}) \numberthis
\end{align*}

with the spectral density of the bath as given in Eq.~\ref{eq:J} and

\begin{align}
\omega_{01} = \omega_r = \sqrt{\frac{1 + \eta}{L \,(C + C_g)}}.
\end{align} 

Naturally, $1/T_1$ goes to zero for both $C_g \to 0$ (as the matrix element disappears) and $C_z \to 0$ (as $J(\omega)$ disappears).

\section{COUPLING TO THE BATH}
\label{app:couplingtobath}

In order to give a concrete value to the resonator-bath coupling (Eq.~\ref{eq:caldeira_coup}), we must match the bath spectral density $J(\omega)$ (Eq.~\ref{eq:J}) to the expression given in Ref.~\cite{Burkard2005}, that is

\begin{align}
J(\omega) = \frac{\pi}{2} \sum_\alpha \frac{c_\alpha^2}{m_\alpha \omega_\alpha} \delta(\omega - \omega_\alpha).
\end{align}

To do this matching, we can use that in

\begin{align*}
\frac{2}{\pi} \int_0^\infty d\omega \frac{J(\omega)}{\omega} \cos(\omega t) &= \int_0^\infty d\omega \frac{1}{\omega} \sum_\alpha \frac{c_\alpha^2}{m_\alpha \omega_\alpha} \delta(\omega - \omega_\alpha) \cos(\omega t)\\
&= \sum_\alpha \frac{c_\alpha^2}{m_\alpha \omega_\alpha^2} \cos(\omega_\alpha t) \numberthis
\label{eq:match}
\end{align*}
the integral over $\omega$ in the first line converges, giving a function of $t$ (compare Ref.~\cite{Caldeira2010}). Discretizing the integral and matching the first and the last expression in Eq.~\ref{eq:match} as a function of $t$, yields

\begin{align*}
\frac{2}{\pi} \int_0^\infty d\omega \frac{J(\omega)}{\omega} \cos(\omega t) &= \sum_\alpha \Delta \omega \frac{J(\alpha \Delta \omega)}{\alpha \Delta \omega} \cos(\alpha \Delta \omega t) \\
\to  \frac{c_\alpha^2}{m_\alpha \omega_\alpha^2} &= \frac{2}{\pi} \frac{J(\omega_\alpha)}{\omega_\alpha} \Delta \omega, \numberthis
\end{align*}

where we assumed that the frequencies of the bath modes $\omega_\alpha = \alpha \Delta \omega$ are equally spaced in steps of $\Delta \omega$. This simply means that when $\Delta \omega$ tends to zero, the sum over $\alpha$ can be rewritten as an integral

\begin{align}
\lim_{\Delta \omega \to 0} \sum_\alpha \Delta\omega f(\alpha \Delta\omega)  \to \int d\omega f(\omega).
\label{eq:limit}
\end{align}

We would now like to rewrite the full system-bath Hamiltonian in second quantization. For the bath variables, we take

\begin{align}
p_\alpha = \sqrt{\frac{\hbar m_\alpha \omega_\alpha}{2}}(b_\alpha^\dagger + b_\alpha) \qquad x_\alpha = i \sqrt{\frac{\hbar}{2\,m_\alpha \omega_\alpha}}(b_\alpha^\dagger - b_\alpha),
\label{eq:bath_quant}
\end{align}

such that Eq.~\ref{eq:caldeira_bath} yields

\begin{align}
\mathcal{H}_B = \sum_\alpha \hbar \, \omega_\alpha b^\dagger_\alpha b_\alpha.
\end{align}

For the quantization of the resonator, we can reuse Eq.~\ref{eq:phir}, where we have to make the transition $C \to C + C_g$ in order to include the capacitances to ground shown in Fig.~\ref{fig:environment}.
With this and Eq.~\ref{eq:bath_quant}, the coupling between system (that is the resonator) and bath can be written as

\begin{align}
\mathcal{H}_{SB} = \sum_\alpha \hbar \, C_g \sqrt[4]{\frac{1 + \eta}{L(C+C_g)^3}} \frac{c_\alpha}{\sqrt{2 m_\alpha \omega_\alpha}} (a - a^\dagger) (b_\alpha^\dagger - b_\alpha).
\end{align}

Neglecting the two-photon terms in a rotating wave approximation, we rewrite the coupling Hamiltonian as

\begin{align*}
\mathcal{H}_{SB} = \hbar \sum_\alpha \sqrt{\kappa_\alpha} (a^\dagger b_\alpha + a \, b_\alpha^\dagger), \numberthis
\label{eq:H_SB}
\end{align*}

where the system-bath coupling is given by

\begin{align}
\kappa_\alpha = \zeta \frac{c_\alpha^2}{m_\alpha \omega_\alpha} = \zeta \frac{2}{\pi} J(\omega_\alpha) \Delta \omega
\label{eq:kappa}
\end{align}

with

\begin{align}
\zeta = \frac{C_g^2}{2} \sqrt{\frac{1 + \eta}{L(C+C_g)^3}}.
\end{align}

In this discrete description, $\kappa_\alpha$ depends on the frequency spacing $\Delta \omega$. We will see, however, that it drops out when we do the transition to a continuous picture with $\Delta \omega \to 0$.

\section{LANGEVIN FORMALISM}
\label{app:langevin}

In order to derive Eqs.~\ref{eq:sep} and \ref{eq:sep2}, we resort to the so-called \textit{Langevin} formalism \cite{Walls1994}. We will start with a Hamiltonian with static longitudinal coupling between qubit and resonator, while the resonator is coupled to a bath

\begin{align*}
\mathcal{H} &= \hbar \,\omega_r a^\dagger a + \hbar \frac{\Delta}{2} \sigma_z + \hbar \, g_{zx} \, \sigma_z (a^\dagger + a) \\
&+ \hbar \sum_\alpha \sqrt{\kappa_\alpha} (a^\dagger b_\alpha + a \, b_\alpha^\dagger) \numberthis
\label{eq:Hstart}
\end{align*}

(compare Eq.~\ref{eq:H_SB}). The Langevin equations that describe the dynamics of $a$ and $b_\alpha$ are given by

\begin{align}
\dot a = \frac{i}{\hbar} \left[\mathcal{H}, a\right] = -i \left(\omega_r \, a + g_{zx} \, \sigma_z + \sum_\alpha \, \sqrt{\kappa_\alpha} \, b_\alpha \right)
\label{eq:a}
\end{align}

and

\begin{align*}
\dot b_\alpha &= \frac{i}{\hbar} \left[\mathcal{H}, b_\alpha \right] 
= i \, a\sum_{\alpha'} \, \sqrt{\kappa_{\alpha'}} \,  \left[b_{\alpha'}^\dagger, b_\alpha \right] + i \sum_{\alpha'} \, \omega_{\alpha'} \left[b_{\alpha'}^\dagger, b_\alpha \right] b_{\alpha'}\\
&= -i \, a\sum_{\alpha'} \, \sqrt{\kappa_{\alpha'}}  \, \delta_{\alpha,\alpha'} 
- i \sum_{\alpha'} \, \omega_{\alpha'} \delta_{\alpha,\alpha'} \, b_{\alpha'}\\
& = -i  \sqrt{\kappa_\alpha} \, a  - i \, \omega_\alpha  \, b_{\alpha}, \numberthis
\label{eq:b_alpha}
\end{align*}

where we used that $[b_\alpha, b^\dagger_{\alpha'}] = \delta_{\alpha,\alpha'} $. 
This is a system of coupled differential equations. The homogeneous equation for $b_\alpha$ is easily solved as

\begin{align}
\dot b_\alpha(t) = -i \,\omega_\alpha \,b_\alpha(t) \quad \to \quad b_\alpha (t) = b_\alpha (t_0) \, e^{-i \omega_\alpha (t-t_0)}.
\end{align}

Assuming that the inhomogeneous solution will be similar, we use $b_\alpha(t) = {b_{\alpha}}_0(t) e^{-i \omega_\alpha (t-t_0)}$ as ansatz in Eq.~\ref{eq:b_alpha} and find

\begin{align*}
\dot b_\alpha(t) &=  \left(- i \omega_\alpha \,{b_{\alpha}}_0(t) + \dot{b_{\alpha}}_0(t)\right) e^{-i \omega_\alpha (t-t_0)} \\
&= -i  \sqrt{\kappa_\alpha} \, a(t)  - i \, \omega_\alpha  \, {b_{\alpha}}_0(t) e^{-i \omega_\alpha (t-t_0)} \\
& \to  \dot{b_{\alpha}}_0(t) = -i  \sqrt{\kappa_\alpha} \, a(t) \, e^{i \omega_\alpha (t-t_0)}\numberthis
\end{align*}

and therefore

\begin{align}
b_\alpha(t) &= \left(b_\alpha(t_0)e^{i \omega_\alpha t_0}  - i  \int_0^t dt' \sqrt{\kappa_\alpha} \, a(t') \, e^{i \omega_\alpha t'}\right) \, e^{-i \omega_\alpha t}
\end{align}

as the full solution for $b_\alpha$, that is the sum of the homogeneous and the inhomogeneous solution. When we plug this result into the differential equation for $a$, we find

\begin{align*}
\dot a(t) = -i (\omega_r a(t) &+ g_{zx} \, \sigma_z)  -i \sum_\alpha \sqrt{\kappa_\alpha} \bigg(b_\alpha(t_0)e^{i \omega_\alpha t_0}  \\
& - i  \int_0^t dt' \sqrt{\kappa_\alpha} \, a(t') \, e^{i \omega_\alpha t'}\bigg) \, e^{-i \omega_\alpha t}. \numberthis
\end{align*}

At this point, we will drop last term in the first line, as it corresponds to the input field $a_\text{in} (t)$, which we take to be the vacuum here (the resonator will not be driven). Apart from that, we will use that $\kappa_\alpha = \zeta \frac{2}{\pi} J(\omega_\alpha) \Delta \omega$ (Eq.~\ref{eq:kappa}) and find 

\begin{align*}
\dot a(t) &= -i (\omega_r a(t) + g_{zx} \, \sigma_z)  - \sum_\alpha \,\kappa_\alpha  \int_0^t dt'  \, a(t') \, e^{-i \omega_\alpha (t-t')}  \\
&= -i (\omega_r a(t) + g_{zx} \, \sigma_z)  - \sum_\alpha \,\zeta \frac{2}{\pi} J(\omega_\alpha) \Delta \omega  \int_0^t dt'  \, a(t') \, e^{-i \omega_\alpha (t-t')}. \numberthis
\end{align*}

Using Eq.~\ref{eq:limit}, we assume that for $\Delta \omega \to 0$ this can be rewritten to

\begin{align*}
\dot a(t) &= -i (\omega_r a(t) + g_{zx} \, \sigma_z)  -  \int_0^\infty d\omega \,  \kappa(\omega) \int_0^t dt'  \, a(t') \, e^{-i \omega (t-t')}. \numberthis
\end{align*}

with the continuous function $\kappa(\omega) = \zeta \frac{2}{\pi} J(\omega)$. We will now assume that $J(\omega)$, and therefore also $\kappa(\omega)$, is almost constant in the frequency interval we are interested in and can thus be taken out of the integral in $\omega$. Using this and  $\int_0^\infty d\omega \, e^{-i \omega(t-t')} = \delta(t-t')$, we find the decoupled differential equation for $a$

\begin{align*}
\dot a(t) &= -i (\omega_r a(t) + g_{zx} \, \sigma_z)  - \kappa \int_0^t dt'  \, a(t') \, e^{-i \omega (t-t')} \delta(t-t')\\
&=-i \left(\omega_r a(t)  + g_{zx} \, \sigma_z\right) - \frac{\kappa}{2} a(t), \numberthis
\label{eq:a_final}
\end{align*}

where $\int_0^\infty dt'  \, a(t') \, \delta(t-t') = \frac{1}{2} a(t)$.
%
%
%
%
%
To solve the differential equation for $a$, we start again by solving the homogeneous equation, that is

\begin{align*}
\dot a(t) &= -i \left(\omega_r - i \,\frac{\kappa}{2}\right) a(t) \\
&\to  a(t) = a(t_0) \, e^{-i\left(\omega_r - i \,\frac\kappa 2\right)(t-t_0)}. \numberthis
\end{align*}

Assuming again that the inhomogeneous solution will be similar to the homogeneous one, we use $a(t) = a_0(t) e^{-i \left(\omega_r - i \,\frac\kappa 2\right) (t-t_0)}$ as ansatz in Eq.~\ref{eq:a} and find

\begin{align*}
\dot a(t) &= \left(- i \left(\omega_r - i \,\frac\kappa 2\right) \,a_0(t)  + \dot a_0(t)\right) e^{-i \left(\omega_r - i \,\frac\kappa 2\right) (t-t_0)} \\
&= - i \left(\omega_r - i \,\frac\kappa 2\right) \,a_0(t)\, e^{-i \left(\omega_r - i \,\frac\kappa 2\right) (t-t_0)} -i g_{zx} \, \sigma_z \\
&\to \dot a_0(t) =  -i g_{zx} \, \sigma_z \, e^{i \left(\omega_r - i \,\frac\kappa 2\right) (t-t_0)}. \numberthis
\end{align*}

The full solution for $a$ is thus given by

\begin{align}
a(t) = a(t_0) \, e^{-i \left(\omega_r - i \,\frac\kappa 2\right) (t-t_0)} - \frac{g_{zx}}{\omega_r - i\,\frac\kappa 2} \sigma_z.
\end{align}

We can now 
%
%
choose the constant $a(t_0)$ such that $a (t=t_0) = 0$, which leads to

\begin{align}
a (t) = -  \frac{g_{zx} }{\omega_r - i\,\frac\kappa 2}  \sigma_z \left(1 - e^{-i \left(\omega_r - i \,\frac\kappa 2\right) (t-t_0)} \right).
\label{eq:out}
\end{align}


As $g$ is in most cases much smaller than $\omega_r$, this longitudinal coupling term would lead to a negligibly small state separation in phase space. However, as proposed in Ref.~\cite{Didier2015}, we can circumvent this problem by modulating the coupling at the resonator's frequency $\omega_r$, that is

\begin{align}
g_{zx}(t) = \bar g_{zx} + \tilde g_{zx} \cos(\omega_r t).
\end{align}

We set $g_{zx} \to g_{zx} (t)$ in Eq.~\ref{eq:Hstart} move to a rotating frame using the unitary transformation $\mathcal{U} = \exp{(i \,\mathcal{H}_0/\hbar \,t)}$ with

\begin{align}
\mathcal{H}_0 = \hbar \, \omega_r a^\dagger a + \hbar \,\frac{\Delta}{2} \sigma_z + \hbar \,\omega_r b_\omega^\dagger b_\omega,
\end{align}

which leads to

\begin{align*}
\mathcal H_\text{rot} &= \hbar  \,\tilde g_{zx} \cos(\omega_r t) \, \sigma_z \left( a^\dagger e^{i\omega_r t} + a e^{-i \omega_r t}\right)\\ &+\hbar\int_0^\infty d\omega \, \sqrt{\kappa(\omega)} (a^\dagger b_\omega + a \, b_\omega^\dagger) + (\omega-\omega_r) \, b_\omega^\dagger b_\omega. \numberthis
\end{align*} 

Note that we now use the continuous version of Eq.~\ref{eq:Hstart} with $\kappa(\omega)$ instead of $\kappa_\alpha$.
When we cross out the fast-rotating terms in a rotating wave approximation, we find an Hamiltonian, that is equivalent to Eq.~\ref{eq:Hstart}, but for the diagonal terms, that is

\begin{align*}
\mathcal H_\text{RWA} &=  \frac{\hbar \,\tilde g_{zx}}{2} \, \sigma_z (a^\dagger + a) \\
&+ \hbar\int_0^\infty d\omega \, \sqrt{\kappa(\omega)} (a^\dagger b_\omega + a \, b_\omega^\dagger) + (\omega-\omega_r) \, b_\omega^\dagger b_\omega. \numberthis
\end{align*} 

The only change we have to make to the expression of the output field (Eq.~\ref{eq:out}) is thus to set $\omega_r = 0$ and $g_{zx} \to \frac{\tilde g_{zx}}{2}$. We therefore find

\begin{align}
a^\text{rot} (t) = -\frac{i \,\tilde g_{zx}}{\kappa} \sigma_z \left(1 - e^{-\frac\kappa 2 (t-t_0)} \right),
\label{eq:out_mod}
\end{align}


which gives a much larger state separation in phase space than in the static case (Eq.~\ref{eq:out}). Note that Eq.~\ref{eq:out_mod} is purely imaginary, which means that the cavity field displacement goes in opposite directions for the two different qubit states. This corresponds to the ideal path in phase space, as it yields high distinguishability even at short measurement times. Eq.~\ref{eq:out}, however, has both real and imaginary parts, such that it does not take the optimal path in phase space, which requires longer measurement times. The same is true for the dispersive case, as shown in Ref.~\cite{Didier2015}.

%
%
%
%
%

\end{appendices}

\cleardoublepage

\phantomsection
\addcontentsline{toc}{chapter}{BIBLIOGRAPHY} 
\small{\bibliographystyle{unsrtnat}
\bibliography{bib_neu}}

\begin{thebibliography}{70}
\providecommand{\natexlab}[1]{#1}
\providecommand{\url}[1]{\texttt{#1}}
\expandafter\ifx\csname urlstyle\endcsname\relax
  \providecommand{\doi}[1]{doi: #1}\else
  \providecommand{\doi}{doi: \begingroup \urlstyle{rm}\Url}\fi

\bibitem[Stein(2011)]{Stein2011}
James~D. Stein.
\newblock \emph{{Cosmic Numbers: The numbers that define our universe}}.
\newblock Basic Books, New York, 2011.
\newblock ISBN 978-0465063796.
\newblock page 103.

\bibitem[Lightman(2005)]{Lightman2005}
Alan Lightman.
\newblock \emph{{The Discoveries: Great Breakthroughs in 20th-Century
  Science}}.
\newblock Vintage, New York, 2005.
\newblock ISBN 978-0375713453.
\newblock chapter 1.

\bibitem[Gosson(2017)]{Gosson2017}
M.~A.~de Gosson.
\newblock \emph{Principles Of Newtonian And Quantum Mechanics, The: The Need
  For Planck's constant, h}.
\newblock World Scientific Publishing Company, 2017.
\newblock ISBN 978-9813200968.
\newblock page 14.

\bibitem[Planck(1900)]{Planck1900}
M.~Planck.
\newblock Zur {T}heorie des {G}esetzes der {E}nergieverteilung im
  {N}ormalspectrum.
\newblock \emph{Verhandlungen der Deutschen physikalischen Gesellschaft}, 2,
  1900.

\bibitem[Nielsen and Chuang(2000)]{Nielsen2000}
Michael~A. Nielsen and Isaac~L. Chuang.
\newblock \emph{{Quantum Computation and Quantum Information}}.
\newblock Cambridge University Press, 2000.
\newblock ISBN 978-0521635035.

\bibitem[Feynman(1982)]{Feynman1982}
Richard.~P. Feynman.
\newblock Simulating physics with computers.
\newblock \emph{International Journal of Theoretical Physics}, 21:\penalty0
  467–488, 1982.
\newblock \doi{10.1007/BF02650179}.
\newblock page 474.

\bibitem[Einstein et~al.(1935)Einstein, Podolsky, and Rosen]{Einstein1935}
A.~Einstein, B.~Podolsky, and N.~Rosen.
\newblock Can quantum-mechanical description of physical reality be considered
  complete?
\newblock \emph{Phys. Rev.}, 47:\penalty0 777--780, May 1935.
\newblock \doi{10.1103/PhysRev.47.777}.

\bibitem[Shor(1997)]{Shor1997}
Peter~W. Shor.
\newblock Polynomial-time algorithms for prime factorization and discrete
  logarithms on a quantum computer.
\newblock \emph{SIAM Journal on Computing}, 26\penalty0 (5):\penalty0
  1484--1509, 1997.
\newblock \doi{10.1137/S0097539795293172}.

\bibitem[DiVincenzo(2000)]{Divincenzo2000}
David~P. DiVincenzo.
\newblock The physical implementation of quantum computation.
\newblock \emph{Fortschritte der Physik}, 48\penalty0 (9-11):\penalty0
  771--783, 2000.
\newblock ISSN 1521-3978.
\newblock \doi{10.1002/1521-3978(200009)48:9/11<771::AID-PROP771>3.0.CO;2-E}.

\bibitem[Bouchiat et~al.(1998)Bouchiat, Vion, Joyez, Esteve, and
  Devoret]{Bouchiat1998}
V~Bouchiat, D~Vion, P~Joyez, D~Esteve, and M~H Devoret.
\newblock Quantum coherence with a single cooper pair.
\newblock \emph{Physica Scripta}, 1998\penalty0 (T76):\penalty0 165, 1998.
\newblock \doi{10.1238/Physica.Topical.076a00165}.

\bibitem[Orlando et~al.(1999)Orlando, Mooij, Tian, van~der Wal, Levitov, Lloyd,
  and Mazo]{Orlando1999}
T.~P. Orlando, J.~E. Mooij, Lin Tian, Caspar~H. van~der Wal, L.~S. Levitov,
  Seth Lloyd, and J.~J. Mazo.
\newblock Superconducting persistent-current qubit.
\newblock \emph{Phys. Rev. B}, 60:\penalty0 15398--15413, Dec 1999.
\newblock \doi{10.1103/PhysRevB.60.15398}.

\bibitem[Koch et~al.(2007)Koch, Yu, Gambetta, Houck, Schuster, Majer, Blais,
  Devoret, Girvin, and Schoelkopf]{Koch2007}
J.~Koch, Terri~M. Yu, J.~Gambetta, A.~A. Houck, D.~I. Schuster, J.~Majer,
  A.~Blais, M.~H. Devoret, S.~M. Girvin, and R.~J. Schoelkopf.
\newblock Charge-insensitive qubit design derived from the cooper pair box.
\newblock \emph{Phys. Rev. A}, 76:\penalty0 042319, Oct 2007.
\newblock \doi{10.1103/PhysRevA.76.042319}.

\bibitem[Yan et~al.(2016)Yan, Gustavsson, Kamal, Birenbaum, Sears, Hover,
  Gudmundsen, Rosenberg, Samach, Weber, Yoder, Orlando, Clarke, Kerman, and
  Oliver]{Yan2016}
Fei Yan, Simon Gustavsson, Archana Kamal, Jeffrey Birenbaum, Adam~P Sears,
  David Hover, Ted~J. Gudmundsen, Danna Rosenberg, Gabriel Samach, S~Weber,
  Jonilyn~L. Yoder, Terry~P. Orlando, John Clarke, Andrew~J. Kerman, and
  William~D. Oliver.
\newblock The flux qubit revisited to enhance coherence and reproducibility.
\newblock \emph{Nature Communications}, 7:\penalty0 12964, November 2016.
\newblock \doi{10.1038/ncomms12964}.

\bibitem[Reagor et~al.(2016)Reagor, Pfaff, Axline, Heeres, Ofek, Sliwa,
  Holland, Wang, Blumoff, Chou, Hatridge, Frunzio, Devoret, Jiang, and
  Schoelkopf]{Reagor2016}
Matthew Reagor, Wolfgang Pfaff, Christopher Axline, Reinier~W. Heeres, Nissim
  Ofek, Katrina Sliwa, Eric Holland, Chen Wang, Jacob Blumoff, Kevin Chou,
  Michael~J. Hatridge, Luigi Frunzio, Michel~H. Devoret, Liang Jiang, and
  Robert~J. Schoelkopf.
\newblock Quantum memory with millisecond coherence in circuit qed.
\newblock \emph{Phys. Rev. B}, 94:\penalty0 014506, Jul 2016.
\newblock \doi{10.1103/PhysRevB.94.014506}.

\bibitem[Minev et~al.(2016)Minev, Serniak, Pop, Leghtas, Sliwa, Hatridge,
  Frunzio, Schoelkopf, and Devoret]{Minev2016}
Z.~K. Minev, K.~Serniak, I.~M. Pop, Z.~Leghtas, K.~Sliwa, M.~Hatridge,
  L.~Frunzio, R.~J. Schoelkopf, and M.~H. Devoret.
\newblock Planar multilayer circuit quantum electrodynamics.
\newblock \emph{Phys. Rev. Applied}, 5:\penalty0 044021, Apr 2016.
\newblock \doi{10.1103/PhysRevApplied.5.044021}.

\bibitem[Ghosh et~al.(2013)Ghosh, Galiautdinov, Zhou, Korotkov, Martinis, and
  Geller]{Ghosh2013}
Joydip Ghosh, Andrei Galiautdinov, Zhongyuan Zhou, Alexander~N. Korotkov,
  John~M. Martinis, and Michael~R. Geller.
\newblock High-fidelity controlled-${\ensuremath{\sigma}}^{Z}$ gate for
  resonator-based superconducting quantum computers.
\newblock \emph{Phys. Rev. A}, 87:\penalty0 022309, Feb 2013.
\newblock \doi{10.1103/PhysRevA.87.022309}.

\bibitem[Wallraff et~al.(2004)Wallraff, Schuster, Blais, Frunzio, Huang, Majer,
  Kumar, Girvin, and Schoelkopf]{Wallraff2004}
A.~Wallraff, D.~I. Schuster, A.~Blais, L.~Frunzio, R.-S. Huang, J.~Majer,
  S.~Kumar, S.~M. Girvin, and R.~J. Schoelkopf.
\newblock Strong coupling of a single photon to a superconducting qubit using
  circuit quantum electrodynamics.
\newblock \emph{Nature}, 431\penalty0 (7005):\penalty0 162--167, September
  2004.
\newblock ISSN 0028-0836.
\newblock \doi{10.1038/nature02851}.

\bibitem[Blais et~al.(2004)Blais, Huang, Wallraff, Girvin, and
  Schoelkopf]{Blais2004}
Alexandre Blais, Ren-Shou Huang, Andreas Wallraff, S.~M. Girvin, and R.~J.
  Schoelkopf.
\newblock Cavity quantum electrodynamics for superconducting electrical
  circuits: An architecture for quantum computation.
\newblock \emph{Phys. Rev. A}, 69:\penalty0 062320, Jun 2004.
\newblock \doi{10.1103/PhysRevA.69.062320}.

\bibitem[Riste et~al.(2015)Riste, Poletto, Huang, Bruno, Vesterinen, Saira, and
  DiCarlo]{Riste2015}
D.~Riste, S.~Poletto, M.-Z. Huang, A.~Bruno, V.~Vesterinen, O.-P. Saira, and
  L.~DiCarlo.
\newblock Detecting bit-flip errors in a logical qubit using stabilizer
  measurements.
\newblock \emph{Nature Communications}, 6:\penalty0 6983--, April 2015.
\newblock \doi{10.1038/ncomms7983}.

\bibitem[Paik et~al.(2016)Paik, Mezzacapo, Sandberg, McClure, Abdo, C\'orcoles,
  Dial, Bogorin, Plourde, Steffen, Cross, Gambetta, and Chow]{Paik2016}
Hanhee Paik, A.~Mezzacapo, Martin Sandberg, D.~T. McClure, B.~Abdo, A.~D.
  C\'orcoles, O.~Dial, D.~F. Bogorin, B.~L.~T. Plourde, M.~Steffen, A.~W.
  Cross, J.~M. Gambetta, and Jerry~M. Chow.
\newblock Experimental demonstration of a resonator-induced phase gate in a
  multiqubit circuit-qed system.
\newblock \emph{Phys. Rev. Lett.}, 117:\penalty0 250502, Dec 2016.
\newblock \doi{10.1103/PhysRevLett.117.250502}.

\bibitem[Versluis et~al.(2017)Versluis, Poletto, Khammassi, Tarasinski, Haider,
  Michalak, Bruno, Bertels, and DiCarlo]{Versluis2017}
R.~Versluis, S.~Poletto, N.~Khammassi, B.~Tarasinski, N.~Haider, D.~J.
  Michalak, A.~Bruno, K.~Bertels, and L.~DiCarlo.
\newblock Scalable quantum circuit and control for a superconducting surface
  code.
\newblock \emph{Phys. Rev. Applied}, 8:\penalty0 034021, Sep 2017.
\newblock \doi{10.1103/PhysRevApplied.8.034021}.

\bibitem[Billangeon et~al.(2015)Billangeon, Tsai, and Nakamura]{Billangeon2015}
P.-M. Billangeon, J.~S. Tsai, and Y.~Nakamura.
\newblock {Circuit-QED-based scalable architectures for quantum information
  processing with superconducting qubits}.
\newblock \emph{Physical Review B}, 91:\penalty0 094517, Mar 2015.
\newblock \doi{10.1103/PhysRevB.91.094517}.

\bibitem[Didier et~al.(2015)Didier, Bourassa, and Blais]{Didier2015}
Nicolas Didier, J\'er\^ome Bourassa, and Alexandre Blais.
\newblock Fast quantum nondemolition readout by parametric modulation of
  longitudinal qubit-oscillator interaction.
\newblock \emph{Phys. Rev. Lett.}, 115:\penalty0 203601, Nov 2015.
\newblock \doi{10.1103/PhysRevLett.115.203601}.

\bibitem[Royer et~al.(2017)Royer, Grimsmo, Didier, and Blais]{Royer2017}
Baptiste Royer, Arne~L. Grimsmo, Nicolas Didier, and Alexandre Blais.
\newblock Fast and high-fidelity entangling gate through parametrically
  modulated longitudinal coupling.
\newblock \emph{{Quantum}}, 1:\penalty0 11, May 2017.
\newblock ISSN 2521-327X.
\newblock \doi{10.22331/q-2017-05-11-11}.
\newblock URL \url{https://doi.org/10.22331/q-2017-05-11-11}.

\bibitem[Geller et~al.(2015)Geller, Donate, Chen, Fang, Leung, Neill, Roushan,
  and Martinis]{Geller2015}
Michael~R. Geller, Emmanuel Donate, Yu~Chen, Michael~T. Fang, Nelson Leung,
  Charles Neill, Pedram Roushan, and John~M. Martinis.
\newblock Tunable coupler for superconducting xmon qubits: Perturbative
  nonlinear model.
\newblock \emph{Phys. Rev. A}, 92:\penalty0 012320, Jul 2015.
\newblock \doi{10.1103/PhysRevA.92.012320}.

\bibitem[Weber et~al.(2017)Weber, Samach, Hover, Gustavsson, Kim, Melville,
  Rosenberg, Sears, Yan, Yoder, Oliver, and Kerman]{Weber2017}
Steven~J. Weber, Gabriel~O. Samach, David Hover, Simon Gustavsson, David~K.
  Kim, Alexander Melville, Danna Rosenberg, Adam~P. Sears, Fei Yan, Jonilyn~L.
  Yoder, William~D. Oliver, and Andrew~J. Kerman.
\newblock Coherent coupled qubits for quantum annealing.
\newblock \emph{Phys. Rev. Applied}, 8:\penalty0 014004, Jul 2017.
\newblock \doi{10.1103/PhysRevApplied.8.014004}.

\bibitem[Richer and DiVincenzo(2016)]{Richer2016}
Susanne Richer and David DiVincenzo.
\newblock Circuit design implementing longitudinal coupling: A scalable scheme
  for superconducting qubits.
\newblock \emph{Phys. Rev. B}, 93:\penalty0 134501, Apr 2016.
\newblock \doi{10.1103/PhysRevB.93.134501}.

\bibitem[Richer et~al.(2017)Richer, Maleeva, Skacel, Pop, and
  DiVincenzo]{Richer2017}
Susanne Richer, Nataliya Maleeva, Sebastian~T. Skacel, Ioan~M. Pop, and David
  DiVincenzo.
\newblock Inductively shunted transmon qubit with tunable transverse and
  longitudinal coupling.
\newblock \emph{Phys. Rev. B}, 96:\penalty0 174520, Nov 2017.
\newblock \doi{10.1103/PhysRevB.96.174520}.

\bibitem[{Devoret}(1997)]{Devoret1997}
M.~H. {Devoret}.
\newblock {Quantum Fluctuations in Electrical Circuits}.
\newblock In S.~{Reynaud}, E.~{Giacobino}, and J.~{Zinn-Justin}, editors,
  \emph{Fluctuations Quantiques/Quantum Fluctuations}, page 351, 1997.
\newblock
  \url{http://qulab.eng.yale.edu/documents/reprints/Houches_fluctuations.pdf}.

\bibitem[Burkard et~al.(2004)Burkard, Koch, and DiVincenzo]{Burkard2004}
Guido Burkard, Roger~H. Koch, and David~P. DiVincenzo.
\newblock Multilevel quantum description of decoherence in superconducting
  qubits.
\newblock \emph{Phys. Rev. B}, 69:\penalty0 064503, Feb 2004.
\newblock \doi{10.1103/PhysRevB.69.064503}.

\bibitem[Burkard(2005)]{Burkard2005}
Guido Burkard.
\newblock Circuit theory for decoherence in superconducting charge qubits.
\newblock \emph{Phys. Rev. B}, 71:\penalty0 144511, Apr 2005.
\newblock \doi{10.1103/PhysRevB.71.144511}.

\bibitem[Bishop(2010)]{Bishop2010}
Lev~S. Bishop.
\newblock \emph{{Circuit Quantum Electrodynamics}}.
\newblock PhD thesis, Yale University, 2010.
\newblock \url{https://www.levbishop.org/thesis/}.

\bibitem[Brandt et~al.(2003)Brandt, Dahmen, and Stroh]{Brandt2003}
S.~Brandt, H.~D. Dahmen, and T.~Stroh.
\newblock \emph{{Interactive Quantum Mechanics}}.
\newblock Springer, New York, 2003.
\newblock \doi{10.1007/978-1-4419-7424-2}.

\bibitem[Josephson(1962)]{Josephson1962}
Brian~D. Josephson.
\newblock Possible new effects in superconductive tunnelling.
\newblock \emph{Physics Letters}, 1\penalty0 (7):\penalty0 251 -- 253, 1962.
\newblock ISSN 0031-9163.
\newblock \doi{10.1016/0031-9163(62)91369-0}.

\bibitem[Zagoskin(2011)]{Zagoskin2011}
A.~M. Zagoskin.
\newblock \emph{{Quantum Engineering}}.
\newblock Cambridge University Press, 2011.
\newblock \doi{10.1017/CBO9780511844157}.

\bibitem[Poletto et~al.(2012)Poletto, Gambetta, Merkel, Smolin, Chow,
  C\'orcoles, Keefe, Rothwell, Rozen, Abraham, Rigetti, and
  Steffen]{Poletto2012}
S.~Poletto, Jay~M. Gambetta, Seth~T. Merkel, John~A. Smolin, Jerry~M. Chow,
  A.~D. C\'orcoles, George~A. Keefe, Mary~B. Rothwell, J.~R. Rozen, D.~W.
  Abraham, Chad Rigetti, and M.~Steffen.
\newblock Entanglement of two superconducting qubits in a waveguide cavity via
  monochromatic two-photon excitation.
\newblock \emph{Phys. Rev. Lett.}, 109:\penalty0 240505, Dec 2012.
\newblock \doi{10.1103/PhysRevLett.109.240505}.

\bibitem[Nigg et~al.(2012)Nigg, Paik, Vlastakis, Kirchmair, Shankar, Frunzio,
  Devoret, Schoelkopf, and Girvin]{Nigg2012}
Simon~E. Nigg, Hanhee Paik, Brian Vlastakis, Gerhard Kirchmair, S.~Shankar,
  Luigi Frunzio, M.~H. Devoret, R.~J. Schoelkopf, and S.~M. Girvin.
\newblock Black-box superconducting circuit quantization.
\newblock \emph{Phys. Rev. Lett.}, 108:\penalty0 240502, Jun 2012.
\newblock \doi{10.1103/PhysRevLett.108.240502}.

\bibitem[Meissner and Ochsenfeld(1933)]{Meissner1933}
W.~Meissner and R.~Ochsenfeld.
\newblock Ein neuer {E}ffekt bei {E}intritt der {S}upraleitfähigkeit.
\newblock \emph{Naturwissenschaften}, 21:\penalty0 787–788, 1933.
\newblock \doi{10.1007/BF01504252}.

\bibitem[Aharonov and Bohm(1959)]{Aharonov1959}
Y.~Aharonov and D.~Bohm.
\newblock Significance of electromagnetic potentials in the quantum theory.
\newblock \emph{Phys. Rev.}, 115:\penalty0 485--491, Aug 1959.
\newblock \doi{10.1103/PhysRev.115.485}.

\bibitem[DiVincenzo et~al.(2006)DiVincenzo, Brito, and Koch]{Divincenzo2006}
David~P. DiVincenzo, Frederico Brito, and Roger~H. Koch.
\newblock Decoherence rates in complex josephson qubit circuits.
\newblock \emph{Phys. Rev. B}, 74:\penalty0 014514, Jul 2006.
\newblock \doi{10.1103/PhysRevB.74.014514}.

\bibitem[Caldeira and Leggett(1983)]{Caldeira1983}
A.~O. Caldeira and A.~J. Leggett.
\newblock Quantum tunnelling in a dissipative system.
\newblock \emph{Annals of Physics}, 149\penalty0 (2):\penalty0 374 -- 456,
  1983.
\newblock ISSN 0003-4916.
\newblock \doi{10.1016/0003-4916(83)90202-6}.

\bibitem[Winkler(2003)]{Winkler2003}
R.~Winkler.
\newblock \emph{{Spin-Orbit Coupling Effects in Two-Dimensional Electron and
  Hole System}}.
\newblock Springer, Berlin Heidelberg, 2003.
\newblock \doi{10.1007/b13586}.

\bibitem[Zueco et~al.(2009)Zueco, Reuther, Kohler, and H\"anggi]{Zueco2009}
D.~Zueco, G.~M. Reuther, S.~Kohler, and P.~H\"anggi.
\newblock Qubit-oscillator dynamics in the dispersive regime: Analytical theory
  beyond the rotating-wave approximation.
\newblock \emph{Phys. Rev. A}, 80:\penalty0 033846, Sep 2009.
\newblock \doi{10.1103/PhysRevA.80.033846}.

\bibitem[Richer(2013)]{Richer2013}
Susanne Richer.
\newblock {Perturbative analysis of two-qubit gates on transmon qubits}.
\newblock Master's thesis, RWTH Aachen, University, Sep 2013.
\newblock
  \url{http://www.quantuminfo.physik.rwth-aachen.de/global/show_document.asp?id=aaaaaaaaaajiobd}.

\bibitem[Sete et~al.(2014)Sete, Gambetta, and Korotkov]{Sete2014}
E.~A. Sete, J.~M. Gambetta, and A.~N. Korotkov.
\newblock Purcell effect with microwave drive: Suppression of qubit relaxation
  rate.
\newblock \emph{Phys. Rev. B}, 89:\penalty0 104516, Mar 2014.
\newblock \doi{10.1103/PhysRevB.89.104516}.

\bibitem[Bloch and Siegert(1940)]{Bloch1940}
F.~Bloch and A.~Siegert.
\newblock Magnetic resonance for nonrotating fields.
\newblock \emph{Phys. Rev.}, 57:\penalty0 522--527, Mar 1940.
\newblock \doi{10.1103/PhysRev.57.522}.

\bibitem[Lang and Firsov(1963)]{Firsov1963}
I.~G. Lang and Y.~A. Firsov.
\newblock Kinetic theory of semiconductors with low mobility.
\newblock \emph{Soviet Physics JETP}, 16:1301, 1963.

\bibitem[Manucharyan et~al.(2009)Manucharyan, Koch, Glazman, and
  Devoret]{Manucharyan2009}
Vladimir~E. Manucharyan, Jens Koch, Leonid~I. Glazman, and Michel~H. Devoret.
\newblock Fluxonium: Single cooper-pair circuit free of charge offsets.
\newblock \emph{Science}, 326\penalty0 (5949):\penalty0 113--116, 2009.
\newblock ISSN 0036-8075.
\newblock \doi{10.1126/science.1175552}.

\bibitem[Pop et~al.(2014)Pop, Geerlings, Catelani, Schoelkopf, Glazman, and
  Devoret]{Pop2014}
Ioan~M. Pop, Kurtis Geerlings, Gianluigi Catelani, Robert~J. Schoelkopf,
  Leonid~I. Glazman, and Michel~H. Devoret.
\newblock Coherent suppression of electromagnetic dissipation due to
  superconducting quasiparticles.
\newblock \emph{Nature}, 508:\penalty0 369–372, 2014.
\newblock \doi{10.1038/nature13017}.

\bibitem[Hutchings et~al.(2017)Hutchings, Hertzberg, Liu, Bronn, Keefe, Brink,
  Chow, and Plourde]{Hutchings2017}
M.~D. Hutchings, J.~B. Hertzberg, Y.~Liu, N.~T. Bronn, G.~A. Keefe, Markus
  Brink, Jerry~M. Chow, and B.~L.~T. Plourde.
\newblock Tunable superconducting qubits with flux-independent coherence.
\newblock \emph{Phys. Rev. Applied}, 8:\penalty0 044003, Oct 2017.
\newblock \doi{10.1103/PhysRevApplied.8.044003}.

\bibitem[Kou et~al.(2017)Kou, Smith, Vool, Pop, Sliwa, Hatridge, Frunzio, and
  Devoret]{Kou2017}
A.~Kou, W.~C. Smith, U.~Vool, I.~M. Pop, K.~M. Sliwa, M.~H. Hatridge,
  L.~Frunzio, and M.~H. Devoret.
\newblock Simultaneous monitoring of fluxonium qubits in a waveguide.
\newblock 2017.
\newblock URL \url{https://arxiv.org/abs/1705.05712}.

\bibitem[Chiorescu et~al.(2003)Chiorescu, Nakamura, Harmans, and
  Mooij]{Chiorescu2003}
I.~Chiorescu, Y.~Nakamura, C.~J. P.~M. Harmans, and J.~E. Mooij.
\newblock Coherent quantum dynamics of a superconducting flux qubit.
\newblock \emph{Science}, 299\penalty0 (5614):\penalty0 1869--1871, 2003.
\newblock ISSN 0036-8075.
\newblock \doi{10.1126/science.1081045}.

\bibitem[Masluk et~al.(2012)Masluk, Pop, Kamal, Minev, and Devoret]{Masluk2012}
Nicholas~A. Masluk, Ioan~M. Pop, Archana Kamal, Zlatko~K. Minev, and Michel~H.
  Devoret.
\newblock Microwave characterization of josephson junction arrays: Implementing
  a low loss superinductance.
\newblock \emph{Phys. Rev. Lett.}, 109:\penalty0 137002, Sep 2012.
\newblock \doi{10.1103/PhysRevLett.109.137002}.

\bibitem[Matveev et~al.(2002)Matveev, Larkin, and Glazman]{Matveev2002}
K.~A. Matveev, A.~I. Larkin, and L.~I. Glazman.
\newblock Persistent current in superconducting nanorings.
\newblock \emph{Phys. Rev. Lett.}, 89:\penalty0 096802, Aug 2002.
\newblock \doi{10.1103/PhysRevLett.89.096802}.

\bibitem[Pop et~al.(2010)Pop, Protopopov, Lecocq, Peng, Pannetier, Buisson, and
  Guichard]{Pop2010}
I.~M. Pop, I.~Protopopov, F.~Lecocq, Z.~Peng, B.~Pannetier, O.~Buisson, and
  W.~Guichard.
\newblock Measurement of the effect of quantum phase slips in a josephson
  junction chain.
\newblock \emph{Nat Phys}, 6\penalty0 (8):\penalty0 589--592, August 2010.
\newblock ISSN 1745-2473.
\newblock \doi{10.1038/nphys1697}.

\bibitem[Viola and Catelani(2015)]{Viola2015}
Giovanni Viola and Gianluigi Catelani.
\newblock Collective modes in the fluxonium qubit.
\newblock \emph{Phys. Rev. B}, 92:\penalty0 224511, Dec 2015.
\newblock \doi{10.1103/PhysRevB.92.224511}.

\bibitem[Hutter et~al.(2011)Hutter, Thol\'en, Stannigel, Lidmar, and
  Haviland]{Hutter2011}
Carsten Hutter, Erik~A. Thol\'en, Kai Stannigel, Jack Lidmar, and David~B.
  Haviland.
\newblock Josephson junction transmission lines as tunable artificial crystals.
\newblock \emph{Phys. Rev. B}, 83:\penalty0 014511, Jan 2011.
\newblock \doi{10.1103/PhysRevB.83.014511}.

\bibitem[Clarke and Braginski(2004)]{Clarke2004}
J.~Clarke and A.~I. Braginski.
\newblock \emph{{The SQUID Handbook}}, page~73.
\newblock WILEY-VCH, Weinheim, Germany, 2004.

\bibitem[Rotzinger et~al.(2017)Rotzinger, Skacel, Pfirrmann, Voss, M\"unzberg,
  Probst, Bushev, Weides, Ustinov, and Mooij]{Rotzinger2017}
H~Rotzinger, S~T Skacel, M~Pfirrmann, J~N Voss, J~M\"unzberg, S~Probst,
  P~Bushev, M~P Weides, A~V Ustinov, and J~E Mooij.
\newblock Aluminium-oxide wires for superconducting high kinetic inductance
  circuits.
\newblock \emph{Superconductor Science and Technology}, 30\penalty0
  (2):\penalty0 025002, 2017.
\newblock \doi{10.1088/0953-2048/30/2/025002}.

\bibitem[Annunziata et~al.(2010)Annunziata, Santavicca, Frunzio, Catelani,
  Rooks, Frydman, and Prober]{Annunziata2010}
Anthony~J Annunziata, Daniel~F Santavicca, Luigi Frunzio, Gianluigi Catelani,
  Michael~J Rooks, Aviad Frydman, and Daniel~E Prober.
\newblock Tunable superconducting nanoinductors.
\newblock \emph{Nanotechnology}, 21\penalty0 (44):\penalty0 445202, 2010.
\newblock \doi{10.1088/0957-4484/21/44/445202}.

\bibitem[Samkharadze et~al.(2016)Samkharadze, Bruno, Scarlino, Zheng,
  DiVincenzo, DiCarlo, and Vandersypen]{Samkharadze2016}
N.~Samkharadze, A.~Bruno, P.~Scarlino, G.~Zheng, D.~P. DiVincenzo, L.~DiCarlo,
  and L.~M.~K. Vandersypen.
\newblock High-kinetic-inductance superconducting nanowire resonators for
  circuit qed in a magnetic field.
\newblock \emph{Phys. Rev. Applied}, 5:\penalty0 044004, Apr 2016.
\newblock \doi{10.1103/PhysRevApplied.5.044004}.

\bibitem[Vissers et~al.(2010)Vissers, Gao, Wisbey, Hite, Tsuei, Corcoles,
  Steffen, and Pappas]{Vissers2010}
M.~R. Vissers, J.~Gao, D.~S. Wisbey, D.~A. Hite, C.~C. Tsuei, A.~D. Corcoles,
  M.~Steffen, and D.~P. Pappas.
\newblock Low loss superconducting titanium nitride coplanar waveguide
  resonators.
\newblock \emph{Applied Physics Letters}, 97\penalty0 (23):\penalty0 232509,
  2010.
\newblock \doi{10.1063/1.3517252}.

\bibitem[Dunsworth et~al.(2017)Dunsworth, Megrant, Quintana, Chen, Barends,
  Burkett, Foxen, Chen, Chiaro, Fowler, Graff, Jeffrey, Kelly, Lucero, Mutus,
  Neeley, Neill, Roushan, Sank, Vainsencher, Wenner, White, and
  Martinis]{Dunsworth2017}
A.~Dunsworth, A.~Megrant, C.~Quintana, Zijun Chen, R.~Barends, B.~Burkett,
  B.~Foxen, Yu~Chen, B.~Chiaro, A.~Fowler, R.~Graff, E.~Jeffrey, J.~Kelly,
  E.~Lucero, J.~Y. Mutus, M.~Neeley, C.~Neill, P.~Roushan, D.~Sank,
  A.~Vainsencher, J.~Wenner, T.~C. White, and John~M. Martinis.
\newblock Characterization and reduction of capacitive loss induced by
  sub-micron josephson junction fabrication in superconducting qubits.
\newblock \emph{Applied Physics Letters}, 111\penalty0 (2):\penalty0 022601,
  2017.
\newblock \doi{10.1063/1.4993577}.

\bibitem[Wu et~al.(2017)Wu, Long, Ku, Lake, Bal, and Pappas]{Wu2017}
X.~Wu, J.~L. Long, H.~S. Ku, R.~E. Lake, M.~Bal, and D.~P. Pappas.
\newblock Overlap junctions for high coherence superconducting qubits.
\newblock \emph{Applied Physics Letters}, 111\penalty0 (3):\penalty0 032602,
  2017.
\newblock \doi{10.1063/1.4993937}.

\bibitem[Gr\"unhaupt et~al.(2017)Gr\"unhaupt, von L\"upke, Gusenkova, Skacel,
  Maleeva, Schl\"or, Bilmes, Rotzinger, Ustinov, Weides, and
  Pop]{Grunhaupt2017}
Lukas Gr\"unhaupt, Uwe von L\"upke, Daria Gusenkova, Sebastian~T. Skacel,
  Nataliya Maleeva, Steffen Schl\"or, Alexander Bilmes, Hannes Rotzinger,
  Alexey~V. Ustinov, Martin Weides, and Ioan~M. Pop.
\newblock An argon ion beam milling process for native alox layers enabling
  coherent superconducting contacts.
\newblock \emph{Applied Physics Letters}, 111\penalty0 (7):\penalty0 072601,
  2017.
\newblock \doi{10.1063/1.4990491}.

\bibitem[Brecht et~al.(2017)Brecht, Chu, Axline, Pfaff, Blumoff, Chou,
  Krayzman, Frunzio, and Schoelkopf]{Brecht2017}
T.~Brecht, Y.~Chu, C.~Axline, W.~Pfaff, J.~Z. Blumoff, K.~Chou, L.~Krayzman,
  L.~Frunzio, and R.~J. Schoelkopf.
\newblock Micromachined integrated quantum circuit containing a superconducting
  qubit.
\newblock \emph{Phys. Rev. Applied}, 7:\penalty0 044018, Apr 2017.
\newblock \doi{10.1103/PhysRevApplied.7.044018}.

\bibitem[Rosenberg et~al.(2017)Rosenberg, Kim, Das, Yost, Gustavsson, Hover,
  Krantz, Melville, Racz, Samach, Weber, Yan, Yoder, Kerman, and
  Oliver]{Rosenberg2017}
D.~Rosenberg, D.~Kim, R.~Das, D.~Yost, S.~Gustavsson, D.~Hover, P.~Krantz,
  A.~Melville, L.~Racz, G.~O. Samach, S.~J. Weber, F.~Yan, J.~Yoder, A.~J.
  Kerman, and W.~D. Oliver.
\newblock 3d integrated superconducting qubits.
\newblock \emph{npj Quantum Information}, 3:\penalty0 42, 2017.
\newblock \doi{10.1038/s41534-017-0044-0}.

\bibitem[Cho()]{Cholesky}
In Mathematica, the Cholesky decomposition of a Matrix \textbf{A} is predefined
  as \textbf{CholeskyDecomposition[A].}

\bibitem[Caldeira(2010)]{Caldeira2010}
A.~Ordacgi Caldeira.
\newblock {C}aldeira-{L}eggett model.
\newblock \emph{Scholarpedia}, 5\penalty0 (2):\penalty0 9187, 2010.
\newblock \doi{10.4249/scholarpedia.9187}.

\bibitem[Walls and Milburn(1994)]{Walls1994}
D.~F. Walls and G.~J. Milburn.
\newblock \emph{{Quantum Optics}}.
\newblock Springer, Berlin Heidelberg, 1994.
\newblock \doi{10.1007/978-3-540-28574-8}.

\end{thebibliography}

\chapter*{List of Publications}
\addcontentsline{toc}{chapter}{LIST OF PUBLICATIONS}

\begin{itemize}

\item Susanne Richer and David DiVincenzo. Circuit design implementing longitudinal coupling: A scalable scheme for superconducting qubits. \textit{Phys. Rev. B}, 93:134501, Apr 2016. \doi{10.1103/PhysRevB.93.134501}

\vspace*{2mm}
\hspace*{2mm} \textit{We present a circuit construction for a new fixed-frequency superconducting qubit and show how it can be scaled up to a grid with strictly local interactions. The circuit QED realization we propose implements $\sigma_z$-type coupling between a superconducting qubit and any number of $LC$ resonators. The resulting \textit{longitudinal coupling} is inherently different from the usual $\sigma_x$-type \textit{transverse coupling}, which is the one that has been most commonly used for superconducting qubits. In a grid of fixed-frequency qubits and resonators with a particular pattern of always-on interactions, coupling is strictly confined to nearest and next-nearest neighbor resonators; there is never any direct qubit-qubit coupling. We note that just four distinct resonator frequencies, and only a single unique qubit frequency, suffice for the scalability of this scheme. A controlled phase gate between two neighboring qubits can be realized with microwave drives on the qubits, without affecting the other qubits. This fact is a supreme advantage for the scalability of this scheme.}

\item Susanne Richer, Nataliya Maleeva, Sebastian T. Skacel, Ioan M. Pop and David DiVincenzo. Inductively shunted transmon qubit with tunable transverse and longitudinal coupling. \textit{Phys. Rev. B}, 96:174520, Nov 2017. \doi{10.1103/PhysRevB.96.174520}

\vspace*{2mm}
\hspace*{2mm} \textit{We present the design of an inductively shunted transmon qubit with flux-tunable coupling to an embedded harmonic mode. This circuit construction offers the possibility to flux-choose between pure transverse and pure longitudinal coupling, that is coupling to the $\sigma_x$ or $\sigma_z$ degree of freedom of the qubit. While transverse coupling is the coupling type that is most commonly used for superconducting qubits, the inherently different longitudinal coupling has some remarkable advantages both for readout and for the scalability of a circuit. Being able to choose between both kinds of coupling in the same circuit provides the flexibility to use one for coupling to the next qubit and one for readout, or vice versa. We provide a detailed analysis of the system's behavior using realistic parameters, along with a proposal for the physical implementation of a prototype device. }

\end{itemize}

\end{document}